\documentclass[a4paper,11pt]{article}
\usepackage[german,american]{babel} 
\usepackage[T1]{fontenc} 
\usepackage[utf8]{inputenc} 
\usepackage[acronym,toc]{glossaries}
\usepackage{longtable}
\usepackage[a4paper,top=3.5cm,bottom=3.5cm,left=3.0cm,right=3.0cm,marginparwidth=1.75cm]{geometry}
\usepackage{longtable}
\usepackage{array}
\usepackage{ragged2e} %

\usepackage{amsmath}
\usepackage{amsfonts}
\usepackage[acronym,toc]{glossaries}
\usepackage{amssymb} % Loaded twice, kept once
\usepackage{amsxtra}
\usepackage{bm}
\usepackage{mathtools}
\usepackage[OMLmathsfit]{isomath} % For \mathsfit, \mathsfbfit

\usepackage{graphicx}
\usepackage[font=small,labelfont=bf]{caption}
\usepackage{subcaption}
\usepackage{float}
\DeclareCaptionLabelFormat{mylabel}{#1 #2\hspace{1.5ex}}
\usepackage{lscape}
\usepackage{verbatim}
\usepackage{enumitem}
\usepackage{longtable}
\usepackage{booktabs} % Loaded twice, kept once
\usepackage{array}
\usepackage{soul} % For \ul, \so, \hl etc.
\usepackage{ulem} % For \sout, \xout etc. (often used for revisions)

\usepackage{xcolor} % Loaded multiple times implicitly/explicitly, keep once early
\definecolor{darkgreen}{rgb}{0.0,0.5,0.0}
\definecolor{BurntOrange}{rgb}{0.8,0.3,0.0}
\definecolor{mygray}{gray}{0.5}
\definecolor{darkred}{rgb}{0.8,0,0}
\definecolor{darkblue}{rgb}{0,0,0.8}
\definecolor{mauve}{rgb}{0.58,0,0.82}
\usepackage[colorlinks=true, allcolors=blue]{hyperref} % Load hyperref late, but before packages that patch it (like cleveref, if used)

\usepackage[numbers,sort&compress]{natbib}

\usepackage{authblk} % <<<< ADDED FOR AUTHOR MANAGEMENT
\newcommand{\tochange}[1]{{\color{red} #1}}

\providecommand{\keywords}[1]
{
  { % Use braces for local scope
  \small
  \textbf{\textit{Keywords---}} #1
}}

\makeglossaries

\newacronym{mar}{MAR}{Missing At Random}
\newacronym{loess}{LOESS}{Locally Estimated Scatterplot Smoothing}
\newacronym{aode}{AODE}{Averaged One-Dependence Estimators}
\newacronym{gbn}{GBN}{Generative Bridging Network}
\newacronym{cart}{CART}{Classification and Regression Trees}
\newacronym{id3}{ID3}{Iterative Dichotomiser 3}
\newacronym{gbm}{GBM}{Gradient Boosting Machine}
\newacronym{knn}{k-NN}{k-Nearest Neighbors}
\newacronym{lvq}{LVQ}{Learning Vector Quantization}
\newacronym{som}{SOM}{Self-Organizing Map}
\newacronym{pca}{PCA}{Principal Component Analysis}
\newacronym{pls}{PLS}{Partial Least Squares}
\newacronym{mds}{MDS}{Multidimensional Scaling}
\newacronym{rbm}{RBM}{Restricted Boltzmann Machine}
\newacronym{dbn}{DBN}{Deep Belief Network}
\newacronym{svm}{SVM}{Support Vector Machine}
\newacronym{rbf}{RBF}{Radial Basis Function}
\newacronym{AI}{AI}{Artificial Intelligence}
\newacronym{ML}{ML}{Machine Learning}

\newglossaryentry{apriori}{
    name=Apriori,
    description={An algorithm for frequent item set mining and association rule learning over transactional databases}
}
\newglossaryentry{eclat}{
    name=Eclat,
    description={Equivalence Class Clustering and bottom-up Lattice Traversal; an algorithm for mining frequent item sets}
}
\newglossaryentry{perceptron}{
    name=Perceptron,
    description={The simplest type of artificial neural network, used for binary classifiers}
}
\newglossaryentry{hopfield}{
    name=Hopfield Network,
    description={A form of recurrent artificial neural network that serves as content-addressable memory systems}
}
\newglossaryentry{stackedautoencoders}{
    name=Stacked Autoencoders,
    description={A deep learning model composed of multiple layers of autoencoders}
}
\newglossaryentry{convnet}{
    name=Convolutional Neural Network,
    description={A class of deep neural networks, most commonly applied to analyzing visual imagery}
}
\newglossaryentry{boosting}{
    name=Boosting,
    description={An ensemble technique that combines the outputs of several weak learners to produce a strong learner}
}
\newglossaryentry{bagging}{
    name=Bagging,
    description={Bootstrap Aggregating; an ensemble method that improves the stability and accuracy of machine learning algorithms}
}
\newglossaryentry{adaboost}{
    name=AdaBoost,
    description={Adaptive Boosting; an ensemble learning method that combines multiple weak classifiers to create a strong classifier}
}
\newglossaryentry{lasso}{
    name=LASSO,
    description={Least Absolute Shrinkage and Selection Operator; a regression analysis method that performs both variable selection and regularization}
}
\newglossaryentry{ridgeregression}{
    name=Ridge Regression,
    description={A technique for analyzing multiple regression data that suffer from multicollinearity}
}

\title{Artificial Intelligence in Materials Science and Engineering: Current Landscape, Key Challenges, and Future Trajectories.}

\author[1,2]{Iman Peivaste}
\author[1]{Salim Belouettar\thanks{Corresponding Author: \texttt{salim.belouettar@list.lu}}}
\author[3]{Francesco Mercuri}
\author[4]{Nicholas Fantuzzi}
\author[1]{Hamidreza Dehghani}
\author[1]{Razieh Izadi}
\author[1]{Halliru Ibrahim}
\author[1]{Jakub Lengiewicz}
\author[5]{Maël Belouettar-Mathis}
\author[1]{Kouider Bendine}
\author[1]{Ahmed Makradi}
\author[6]{Martin Hörsch}
\author[7]{Peter Klein}

\author[1]{Mohamed El Hachemi}
\author[8]{Heinz A. Preisig}
\author[9]{Yacine Rezgui}
\author[10]{Natalia Konchakova} 
\author[11]{Ali Daouadji}
\affil[1]{Luxembourg Institute of Science and Technology (LIST), 5, Avenue des Hauts-Fourneaux, L-4362 Esch-sur-Alzette, Luxembourg}
\affil[2]{Department of Physics and Materials Science, University of Luxembourg, L-4365 Esch-sur-Alzette, Luxembourg}
\affil[3]{Istituto per lo Studio dei Materiali Nanostrutturati (ISMN), Consiglio Nazionale delle Ricerche (CNR), Bologna 40129, Italy}
\affil[4]{Department of Civil, Chemical, Environmental and Materials Engineering, University of Bologna, Italy}
\affil[5]{CPGE, Lycée Fabert, 57000 Metz, France}
\affil[6]{Norwegian University of Life Sciences (NMBU), Ås, Norway}
\affil[7]{Fraunhofer Institute for Industrial Mathematics (ITWM), Kaiserslautern, Germany}
\affil[11]{Ecole Centrale de Lyon, Ecully, France}
\affil[8]{Norwegian University of Science and Technology (NTNU), Trondheim, Norway}
\affil[9]{School of Engineering, Cardiff University, Cardiff CF24 3AA, United Kingdom}
\affil[10]{INSA Lyon, GEOMAS, 20 Av. Albert Einstein, 69100 Villeurbanne}

\date{\today}

\begin{document}
\parskip4pt
\parindent0pt

\maketitle
\begin{abstract} 

Artificial Intelligence is rapidly transforming materials science and engineering, offering powerful tools to navigate complexity, accelerate discovery, and optimize material design in ways previously unattainable. Driven by the accelerating pace of algorithmic advancements and increasing data availability, AI is becoming an essential competency for materials researchers. This review provides a comprehensive and structured overview of the current landscape, synthesizing recent advancements and methodologies for materials scientists seeking to effectively leverage these data-driven techniques. We survey the spectrum of machine learning approaches, from traditional algorithms to advanced deep learning architectures, including CNNs, GNNs, and Transformers, alongside emerging generative AI and probabilistic models such as Gaussian Processes for uncertainty quantification.
The review also examines the pivotal role of data in this field, emphasizing how effective representation and featurization strategies, spanning compositional, structural, image-based, and language-inspired approaches, combined with appropriate preprocessing, fundamentally underpin the performance of machine learning models in materials research. Persistent challenges related to data quality, quantity, and standardization, which critically impact model development and application in materials science and engineering, are also addressed.
Key applications are discussed across the materials lifecycle, including property prediction at multiple scales, high-throughput virtual screening, inverse design, process optimization, data extraction by large language models, and sustainability assessment. Critical challenges such as model interpretability, generalizability, and scalability are addressed, alongside promising future directions involving hybrid physics-ML models, autonomous experimentation, collaborative platforms, and human-AI synergy.

\end{abstract}
\noindent

%%%%%%%%%%%%%%%%%%%%%%%%%%%%%%%%
%%%%%%%%%%%%%%%%%%%%%%%%%%%%%%%%
%\section*{List of Acronyms}\label{acronyms}

%%%%%%%%%%%%%%%%%%%%%%%%%%%%%%%%

\section*{List of Acronyms and Brief Definitions} 

\begin{itemize}[itemsep=2pt, parsep=0pt, topsep=4pt, partopsep=0pt]
    \small
    \item \textbf{AdaBoost} – Adaptive Boosting: An ensemble method that combines multiple weak learners to create a strong classifier.
    \item \textbf{AFLOW} – Automatic flow for materials discovery: A computational framework used to discover and design new materials.
    \item \textbf{AODE} – Averaged One-Dependence Estimators: A machine learning algorithm that uses a probabilistic framework to perform classification.
    \item \textbf{ANN} – Artificial Neural Network: A computational model inspired by the way biological neural networks process information.
    \item \textbf{Apriori} – Apriori Algorithm: A classic algorithm used for mining frequent item sets and generating association rules in databases.
    \item \textbf{Backpropagation} – Backward Propagation of Errors: A method used in training artificial neural networks by adjusting weights based on error gradients.
    \item \textbf{BERT} – Bidirectional Encoder Representations from Transformers: A pre-trained transformer model used for NLP tasks like text classification and question answering.
    \item \textbf{BNNs} – Bayesian Neural Networks: A type of neural network that incorporates uncertainty by modeling parameters with probability distributions.
    \item \textbf{CBFVs} – Composition-Based Feature Vectors: A representation of chemical compositions used for material property predictions.
    \item \textbf{CART} – Classification and Regression Trees: A decision tree algorithm used for both classification and regression tasks.
    \item \textbf{CIF} – Crystallographic Information File: A standard text file format for storing information about crystal structures.
    \item \textbf{CNNs} – Convolutional Neural Networks: A deep learning architecture primarily used for analyzing visual data such as images.
    \item \textbf{CPS} – Cyber-Physical Systems: Systems that integrate computational algorithms with physical processes.
    \item \textbf{DL} – Deep Learning: A subset of machine learning that uses neural networks with many layers to model complex patterns in data.
    \item \textbf{DBN} – Deep Belief Network: A type of generative model made of stacked layers of restricted Boltzmann machines.
    \item \textbf{DPP} – Digital Product Passport: A digital representation of a product's lifecycle information for sustainability and traceability.
    \item \textbf{DNN} – Deep Neural Network: A neural network with multiple layers used to model complex relationships in data.
    \item \textbf{Eclat} – Equivalence Class Transformation: An algorithm for mining frequent item sets using a vertical data format.
    \item \textbf{FAIR} – Findable, Accessible, Interoperable, Reusable: A set of principles for data management and stewardship in research.
    \item \textbf{MDS} – Multidimensional Scaling: A technique used to visualize the similarity or dissimilarity of data in a lower-dimensional space.
    \item \textbf{GANs} – Generative Adversarial Networks: A machine learning framework consisting of two networks that compete to generate realistic data.
    \item \textbf{GBN} – Gaussian Bayesian Network: A type of probabilistic graphical model that represents dependencies using Gaussian distributions.
    \item \textbf{GBM} – Gradient Boosting Machine: An ensemble technique that builds models sequentially to improve prediction accuracy.
    \item \textbf{GATGNN} – Graph Attention GNN: A variant of graph neural networks that uses attention mechanisms to focus on important nodes.
    \item \textbf{GNoME} – Graph Networks for Materials Exploration: A framework that uses graph neural networks to predict material properties.
    \item \textbf{GNN} – Graph Neural Network: A type of neural network that operates on graph structures, often used for relational data.
    \item \textbf{GP} – Gaussian Processes: A method for probabilistic modeling, often used in regression and classification tasks.
    \item \textbf{GPT} – Generative Pretrained Transformer: A large language model trained on a massive corpus of text data to generate human-like text.
    \item \textbf{KRR} – Kernel Ridge Regression: A non-linear regression technique that uses kernel methods to map data to higher-dimensional spaces.
    \item \textbf{k-NN} – k-Nearest Neighbors: A simple algorithm used for classification and regression based on the closest training examples in the feature space.
    \item \textbf{LASSO} – Least Absolute Shrinkage and Selection Operator: A regression method that performs variable selection and regularization.
    \item \textbf{LIME} – Local Interpretable Model-Agnostic Explanations: A technique for explaining predictions of machine learning models in a human-understandable way.
    \item \textbf{LLMs} – Large Language Models: Models like GPT and BERT that are trained on vast text corpora to understand and generate human language.
    \item \textbf{LSTM} – Long Short-Term Memory: A type of recurrent neural network designed to learn and retain information over long sequences.
    \item \textbf{LVQ} – Learning Vector Quantization: A type of artificial neural network used for classification tasks, similar to k-NN but with a neural structure.
    \item \textbf{LOESS} – Locally Estimated Scatterplot Smoothing: A non-parametric method for regression that fits local polynomials to the data.
    \item \textbf{MARS} – Multivariate Adaptive Regression Splines: A regression technique that models non-linear relationships between variables using piecewise linear splines.
    \item \textbf{ML} – Machine Learning: A field of artificial intelligence that focuses on the development of algorithms that allow computers to learn from data.
    \item \textbf{MLE} – Maximum Likelihood Estimation: A statistical method used to estimate the parameters of a probability distribution that maximize the likelihood of observed data.
    \item \textbf{MLIP} – Machine-Learned Interatomic Potential: A machine learning method used to predict interatomic potentials in material science.
    \item \textbf{MOEA} – Multi-Objective Evolutionary Algorithm: An algorithm used to optimize multiple objectives simultaneously, often in optimization and design tasks.
    \item \textbf{MOFormer} – MOF-focused Transformer-based Model: A transformer-based model designed to predict properties of Metal-Organic Frameworks (MOFs).
    \item \textbf{MPNN} – Message Passing Neural Network: A neural network that propagates information between nodes in a graph to learn representations.
    \item \textbf{NaiveBayes} – Naive Bayes: A statistical method for classification based on Bayes' theorem, assuming independence between features.
    \item \textbf{NLP} – Natural Language Processing: A field of AI that focuses on the interaction between computers and human languages.
    \item \textbf{OLS} – Ordinary Least Squares: A method used in linear regression to minimize the sum of squared residuals between the observed and predicted values.
    \item \textbf{OWL} – Web Ontology Language: A language used for defining and instantiating ontologies in the semantic web.
    \item \textbf{PCA} – Principal Component Analysis: A dimensionality reduction technique that transforms data into a new coordinate system.
    \item \textbf{PLS} – Partial Least Squares: A statistical method that finds the fundamental relations between two matrices.
    \item \textbf{PGO} – Precipitate Geometry Ontology: An ontology used to represent the geometry of precipitates in materials science.
    \item \textbf{PMDco} – PMD Core Ontology: A foundational ontology for the representation of materials data.
    \item \textbf{QM/QC} – Quantum Mechanics/Quantum Computing: Fields of physics and computer science that use quantum theory to model and simulate systems.
    \item \textbf{RAG} – Retrieval-Augmented Generation: A technique for combining retrieval-based and generation-based models to enhance text generation.
    \item \textbf{RBF} – Radial Basis Function: A kernel function used in machine learning algorithms, such as support vector machines.
    \item \textbf{RBM} – Restricted Boltzmann Machine: A type of neural network used for unsupervised learning and dimensionality reduction.
    \item \textbf{RDF} – Resource Description Framework: A framework for representing and exchanging data on the web.
    \item \textbf{RNNs} – Recurrent Neural Networks: A class of neural networks that are effective for sequence prediction and time-series data.
    \item \textbf{SAT} – Self-Attention-based Transformer: A mechanism used in transformers to weigh the importance of different parts of input data.
    \item \textbf{SOM} – Self-Organizing Map: A type of unsupervised neural network used to map high-dimensional data onto a lower-dimensional grid.
    \item \textbf{SHAP} – Shapley Additive Explanations: A method for explaining machine learning models based on game theory.
    \item \textbf{SISSO} – Sure Independence Screening and Sparsifying Operator: A method for finding simple linear models from high-dimensional data.
    \item \textbf{SMILES} – Simplified Molecular Input Line Entry System: A notation system for representing chemical structures as text strings.
    \item \textbf{SVM} – Support Vector Machine: A supervised learning algorithm used for classification and regression tasks.
    \item \textbf{TTO} – Tensile Test Ontology: An ontology used to represent data from tensile testing in materials science.
    \item \textbf{VAE} – Variational Autoencoders: A generative model that uses variational inference to approximate complex data distributions.
    \item \textbf{XAI} – Explainable AI: AI methods that aim to make machine learning models more transparent and interpretable.
    \item \textbf{XML} – Extensible Markup Language: A markup language designed to store and transport data in a structured format.
    \item \textbf{XGBoost} – Extreme Gradient Boosting: A machine learning algorithm based on gradient boosting that is optimized for speed and performance.
   
\end{itemize}

%%%%%%%%%%%%%%%%%%%%%%%%%%%%%%%%
\section{Introduction}\label{sec1}

%%%%%%%%%%%%%%%%%%%%%%%%%%%%%%%%
Materials science and engineering stand at the core of technological innovation, underpinning advancements across virtually every sector of modern society, from energy conversion and storage to healthcare, transportation, information technology, and sustainable manufacturing \cite{bojan_podgornik_2023, colin_tong_2019, raabe2023accelerating, advanced_materials_for_societal_implementation_2023}. The discovery, design, and optimization of materials with tailored properties has historically been a time-consuming, resource-intensive process, often relying on empirical approaches, trial-and-error experimentation, and incremental improvements to existing systems. This traditional paradigm is increasingly challenged by the accelerating pace of technological development, growing sustainability imperatives, and the expanding complexity of performance requirements that modern materials must satisfy \cite{himanen2019data, rodrigues2021big, bauer2024roadmap, carla_p__gomes__2019, michael_mckerns_2018}.

The past decade has witnessed a remarkable evolution in our approach to materials research, catalyzed by the convergence of three transformative trends: the exponential growth in computational power, the increasing availability of materials data from both experiments and simulations, and the rapid development of artificial intelligence (AI) and machine learning (ML) methodologies \cite{materials_zhu_2022, materials_liu_2017, machine_gubernatis_2018, recent_choudhary_2022, big_zhou_2019, opportunities_morgan_2020, machine_butler_2018, machine_gu_2019}. This synergy has catalyzed the growth of materials informatics, an interdisciplinary field situated at the intersection of materials science, data science, and artificial intelligence \cite{machine_ramprasad_2017, klenam2023data}. This confluence has created unprecedented opportunities to revolutionize how we discover, design, characterize, and deploy advanced materials. What once took decades of experimental iteration can now potentially be achieved in months or even weeks through sophisticated computational approaches augmented by AI \cite{dft_schleder_2019, exploiting_anand_2022, evolution_gupta_2023}.

AI and ML offer powerful frameworks for navigating the vast, high-dimensional spaces that characterize materials science problems. Unlike traditional modeling approaches that rely on predetermined physical equations, ML algorithms can identify complex, non-linear patterns in data, enabling them to construct predictive models that capture subtle structure-property relationships which might otherwise remain hidden \cite{saal2020machine, zhu2022materials}. From property prediction and high-throughput virtual screening to inverse design and autonomous experimentation, these data-driven approaches are dramatically accelerating the materials development cycle while simultaneously reducing costs and resource consumption. The urgency of this transformation cannot be overstated. Global challenges in energy, healthcare, transportation, and sustainability demand novel materials with increasingly specialized and often competing performance metrics \cite{cai2020machine, decost2020scientific}.

For materials scientists and engineers, proficiency in AI and ML is rapidly transitioning from an optional specialization to a core competency. The ability to effectively leverage these data-driven tools is becoming essential for maintaining research competitiveness and addressing complex materials challenges \cite{national2019decadal, wang2020machine}. However, this transition presents significant barriers, including the need to develop new skill sets, understand the capabilities and limitations of different ML approaches, and integrate these methods with domain-specific knowledge and established physics-based modeling techniques.

The rise of AI in materials science has coincided and been accelerated by several concurrent developments: the establishment of large-scale materials databases \cite{agrawal2016perspective, butler2018machine}, advances in high-throughput computational and experimental methods \cite{curtarolo2013high, stein2019progress}, and the growing ecosystem of open-source software tools for materials informatics \cite{ong2013python, himanen2020dscribe}. These elements collectively provide the infrastructure and foundation for data-driven discovery at unprecedented scales. Materials databases now house millions of computed structures and properties, while frameworks enable streamlined data processing, feature generation, and workflow automation.
At the same time, the field faces substantial challenges. Materials data remains heterogeneous, often sparse, and unevenly distributed across chemical and structural spaces. ML models frequently struggle with extrapolation beyond known domains, and many critical materials properties lack adequate training data. Moreover, the "black box" nature of advanced algorithms can obscure physical insights, complicating their acceptance and integration with traditional scientific approaches.
Despite these challenges, the evidence for AI's transformative impact on materials science is compelling. There have been many achievements that underscore the message that resonates throughout this review: the judicious application of AI is rapidly becoming not merely advantageous but essential for materials innovation in the 21st century.

As we navigate this evolving landscape, we emphasize that ML complements rather than replaces domain expertise. The most powerful approaches combine data-driven insights with physical understanding, leveraging the strengths of both paradigms. For materials scientists, the opportunity lies not just in adopting new computational tools, but in reimagining how these tools can augment scientific intuition, accelerate discovery cycles, and enable the exploration of previously inaccessible regions of materials design space. Furthermore, AI is becoming increasingly pivotal in advancing sustainability and lifecycle assessment (LCA) in materials science \cite{ghoroghi2022advances, akhshik2022prediction}. Digital product passports can leverage AI to track and document materials' lifecycles from production to recycling, enabling more transparent and efficient resource utilization. By integrating real-time data with predictive ML models, these tools facilitate decision-making that minimizes environmental impacts, optimizes recycling processes, and aligns with circular economy principles \cite{lin2023advancing}.

This paper provides a comprehensive orerview of how AI and ML are transforming the field of materials science and engineering. It explores a wide range of ML methodologies and techniques used for materials discovery, modelling, and design, from established algorithms to cutting-edge deep learning and generative AI approaches. A significant portion of the paper is dedicated to the critical role of data in this field, addressing how material data is represented, preprocessed, and the persistent challenges related to its quality and standardisation. The review highlights key applications of AI across the entire materials lifecycle, including predicting material properties, virtual screening for new materials, inverse design to create materials with desired characteristics, optimising manufacturing processes, extracting knowledge from scientific literature using large language models, and assessing the sustainability of materials. Our approach bridges theory and practice, examining both the mathematical foundations of these methods and their real-world implementation across diverse materials systems and applications. Furthermore, the paper addresses crucial challenges facing the integration of AI in materials science, such as the interpretability and generalisability of models, their scalability, and issues related to data availability and incorporating existing scientific knowledge. Finally, the authors outline promising future directions for the field, including the development of hybrid models combining physics and machine learning, autonomous experimentation, collaborative research platforms, and enhanced collaboration between humans and AI. UBy providing this integrated perspective, we aim to equip researchers, engineers, and decision-makers with the knowledge necessary to navigate this rapidly evolving field and harness its transformative potential.

The review is organized as follows:

Section \ref{sec-2} examines ML methodologies in materials modeling, from traditional approaches to deep learning and generative AI, with emphasis on transformers and generative models for materials discovery.
Section \ref{sec-3} explores data's critical role in materials informatics, covering representation strategies, preprocessing techniques, and challenges in data quality and standardization.
Sections \ref{sec-4}, \ref{sec-5} and \ref{sec-6} discuss key ML applications across the materials lifecycle, including property prediction, virtual screening, inverse design, process optimization, and sustainability assessment, with case studies on polymers, metals, composites, and high-entropy alloys.
Section \ref{sec-7} addresses challenges in applying ML to materials science, focusing on interpretability, generalizability, scalability, and data limitations.
Section \ref{sec-8} outlines future directions, including hybrid physics-ML models, autonomous experimentation, collaborative platforms, and enhanced human-AI synergy.
Section \ref{sec-9} provides concluding perspectives on AI's evolving role in materials research, emphasizing the integration of data-driven methods with physical understanding to accelerate materials discovery and innovation.

This review serves as both an introduction for newcomers to the field and a reference for experienced practitioners, providing a roadmap for leveraging ML in materials modeling and design while acknowledging the current limitations and future possibilities of this rapidly advancing domain. As we stand at the threshold of a new era in materials innovation, driven by the synergistic combination of human expertise and artificial intelligence, this review offers a timely roadmap for those seeking to accelerate materials discovery and design in service of addressing our most pressing technological challenges. The integration of AI into materials research represents more than a methodological evolution—it heralds a fundamental transformation in how we discover, understand, and deploy the materials that underpin technological progress.

\section{Methodologies and Techniques in ML for Materials Modeling and Design} \label{sec-2}

ML approaches have brought about a significant transformation in materials modelling and design. They enable researchers to identify meaningful patterns within extensive datasets and to make precise predictions regarding material properties and behaviours. These approaches can be broadly classified into three learning strategies: supervised learning, where models are trained on labeled datasets with known material properties; unsupervised learning, which identifies patterns in unlabeled data; and reinforcement learning, where an agent learns through interaction with an environment by optimizing a reward function \cite{jordan2015machine}. The choice of a specific algorithm is heavily influenced by the particular materials science problem being addressed.   Within each of these learning strategies, researchers can apply various algorithms, ranging from traditional statistical methods to advanced neural networks and generative models. The selection of an appropriate algorithm depends on several factors, including dataset size and quality \cite{butler2018machine}, interpretability requirements \cite{machine_ramprasad_2017}, computational resources \cite{opportunities_morgan_2020}, and the specific materials science problem being addressed. To facilitate algorithm selection, Table~\ref{tab:ml_comparison} provides a comparative summary of several machine learning techniques commonly employed in materials modeling and design. The table outlines the learning type, primary advantages, potential drawbacks, and scenarios where alternative approaches might be more appropriate for each method.

\begin{figure}[h]
        \centering
        \includegraphics[width=1\linewidth]{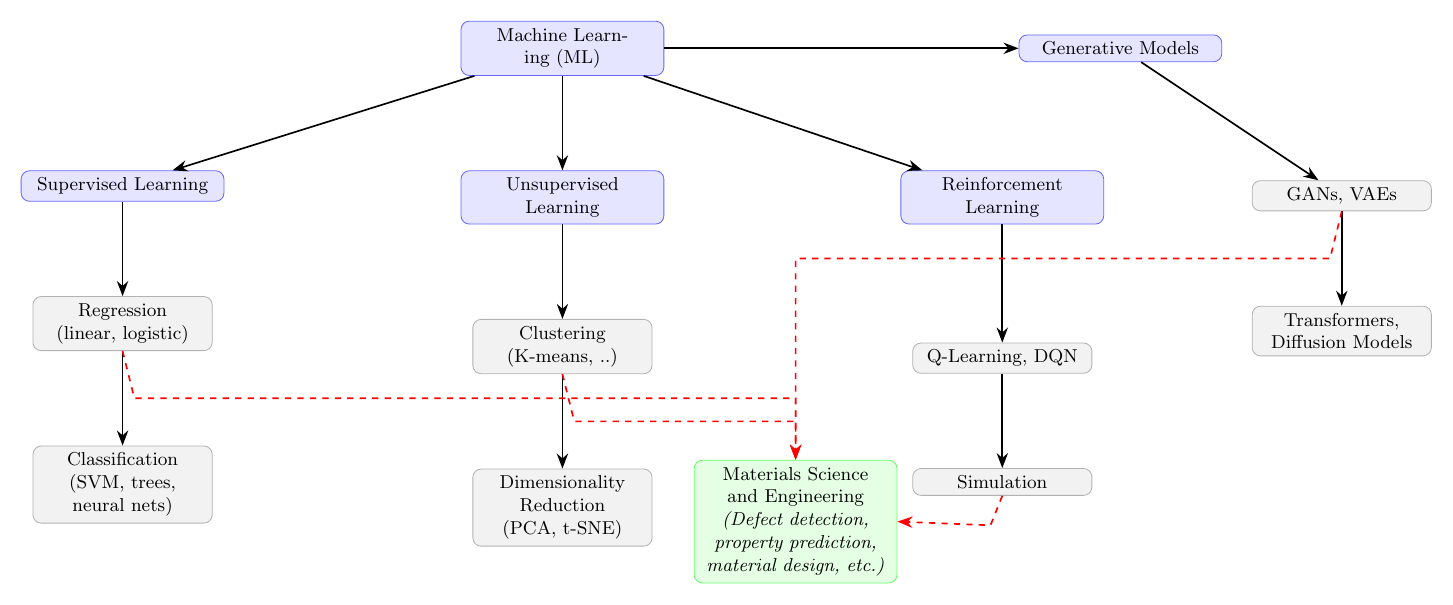}
         \caption{Classification of Machine Learning Techniques and their Applications in Materials Science and Engineering}
  \label{fig:ml_class-1}
\end{figure}

\begin{figure}[!h]
        \centering
        \includegraphics[width=0.85\linewidth]{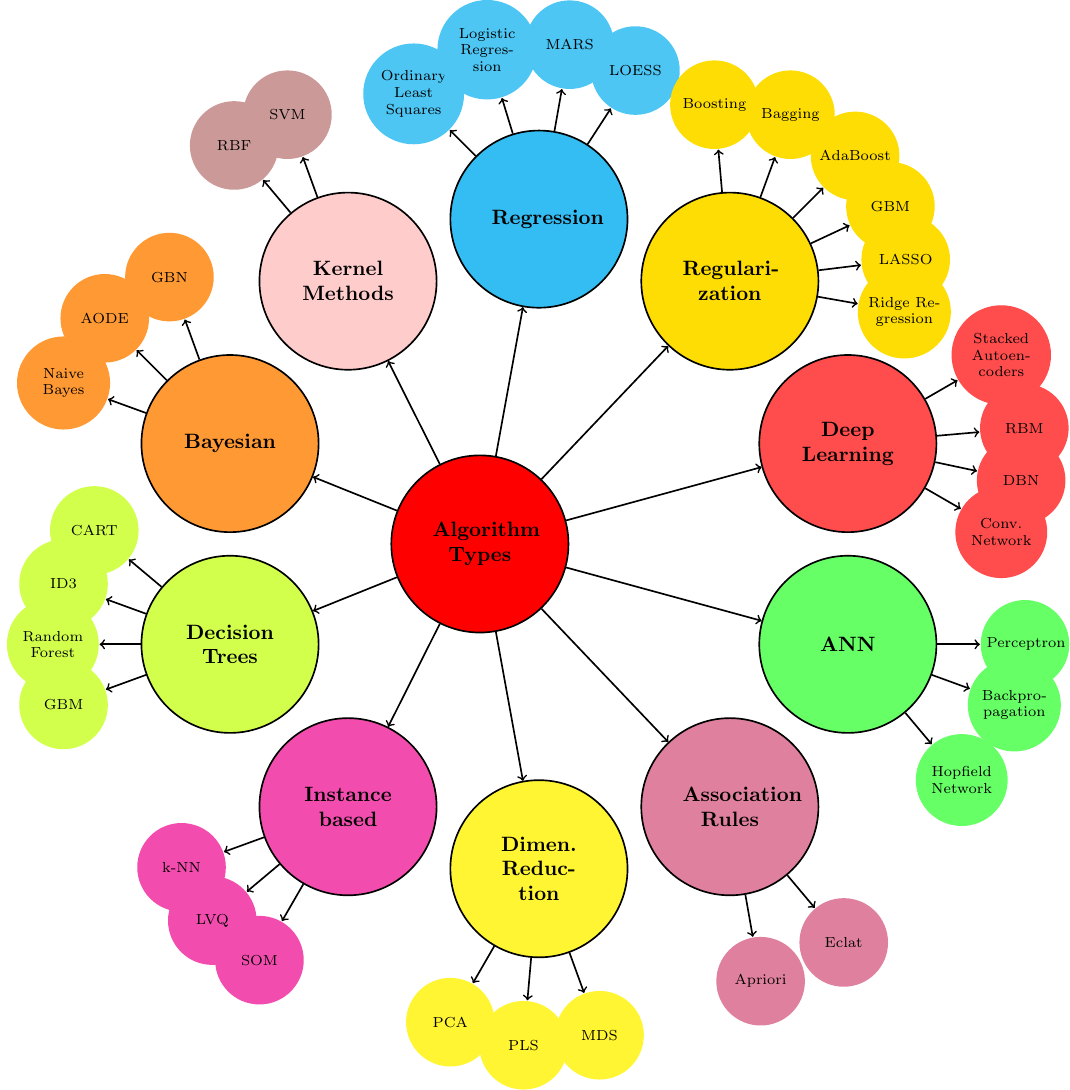}
         \caption{A diagram illustrating various types of machine learning algorithms, categorized into different groups such as Regression, Regularization, Deep Learning, and others. Each group is further subdivided into specific methods or techniques, demonstrating the hierarchical structure of machine learning approaches. Each group node is color-coded and spatially arranged to reflect its conceptual relation to the central theme of algorithm types.}
  \label{fig:ml_class-2}
\end{figure}

This section presents a structured overview of ML methodologies applied to materials modeling and design, beginning with traditional ML techniques in Section \ref{sec: methods: traditional ML}, followed by deep learning approaches in Section \ref{sec: methods: deep learning}, and emerging techniques in Section \ref{sec:emerging_tech}. Figure \ref{fig:ml_class-1} and Figure \ref{fig:ml_class-2} illustrate a general classification of different machine learning techniques used in material science. 

%%%%%%%%%%%%%%%%%%%%%%%%%%%%%%%%
\subsection{Traditional ML Techniques}
\label{sec: methods: traditional ML}

Traditional ML approaches play a major part in data-driven materials science, offering effective tools for predictive modelling and exploratory analyses \cite{opportunities_morgan_2020}.  Unlike deep neural architectures, these methods typically rely on carefully engineered domain-specific features (domain-specific features are explained in section \ref{domain-specific}) to predict properties of materials or categorize materials into classes based on their observed behaviors \cite{fang2022machine, agrawal2016perspective}. Traditional approaches remain indispensable when data availability is moderate, interpretability is critical, and computational resources are limited \cite{sparks2016data}.  Traditional ML techniques in materials informatics can be broadly classified into three categories: tree-based methods, kernel-based methods, and classical approaches. Each has distinct strengths and practical considerations for materials science applications. 

\subsubsection{Tree-Based Methods}\label{sec: methods: traditional ML:Tree-Based Methods}

Tree-based methods construct predictions using a hierarchy of decision rules, with internal nodes dividing data based on specific features and leaf nodes assigning outputs. These methods excel at handling non-linear property relationships while providing interpretable feature rankings that connect directly to materials characteristics \cite{butler2018machine}.

Random Forest and Gradient Boosting are two powerful ensemble learning techniques widely adopted in materials modeling and design due to their robust performance and interpretability. Random Forest builds an ensemble of decision trees, each trained on a bootstrapped subset of the data with randomly selected features at each split, which helps reduce overfitting and improves generalisation while also providing estimates of feature importance that can offer insights into the underlying physical processes \cite{kim2017materials}. Boosted tree models, also known as XGBoost, iteratively refine residual errors, resulting in high accuracy in high-throughput materials screening tasks \cite{chen2016xgboost}. Tree-based methods are particularly valuable for their interpretability, ability to handle nonlinear effects, and high performance in data-driven materials research.

\subsubsection*{Decision Trees}
Decision trees are one of the basic supervised algorithms, with each internal node dividing the training set based on a feature threshold \cite{breiman2017classification}. This flowchart-like structure ends in leaf nodes that contain predictions, whether they are continuous values (for regression) or class labels (for classification) \cite{kingsford2008decision, sparks2016data}. In materials research, early studies used single decision trees to determine which compositional variables had the most effect on polymer dielectric properties \cite{machine_ramprasad_2017}, ceramic microstructure characteristics \cite{Lookman_ActiveLearning_2019}, phase stability in high-entropy alloys (HEAs) \cite{machine_butler_2018}, and chemical and mechanical planarization \cite{li2019prediction}.  Despite their transparency and ease of interpretation, decision trees are frequently subject to overfitting, particularly when data is limited or high-dimensional \cite{rokach2005decision}. While they are almost always used in supervised settings, rare adaptations (e.g., hierarchical clustering trees) are found in unsupervised explorations of material families \cite{huo2019semi}.
 
\subsubsection*{Random Forest}
Random forests build on the decision tree concept by training multiple trees on independent subsets of data and features and then combining their results to reach a final prediction \cite{breiman2001random}. This ensemble strategy effectively reduces variance while avoiding overfitting. In materials science, random forests have produced reliable predictions for various properties \cite{pilania2013accelerating, ghiringhelli2015big, guo2021artificial}. Their ability to rank feature importance is particularly valuable for identifying the key drivers of material behaviors \cite{tabor2018accelerating}. While Random Forests primarily serve supervised learning tasks, variations such as unsupervised Random Forests can assist in clustering unlabeled data \cite{shi2006unsupervised}, though these applications are less common in materials informatics.
\paragraph{Gradient Boosting:}
Gradient boosting algorithms, symbolized as XGBoost, enhance the ensemble concept by sequentially focusing on residual errors left by previous learners \cite{chen2016xgboost}. Each new weak learner, typically a shallow decision tree, fine-tunes predictions where the ensemble failed, gradually improving accuracy. This method has proven particularly useful in materials discovery studies, where even minor accuracy improvements result in significant savings in experimental or computational screening. Recent research has used XGBoost to predict thermoelectric performance over large chemical spaces and to optimize catalytic activity for energy applications \cite{wang2022machine}. Gradient boosting, like random forests, is almost exclusively supervised and used for regression or classification tasks; few attempts have been made to incorporate it into reinforcement learning or fully unsupervised frameworks. Moreover, XGBoost’s capability to provide detailed information insights enhances model interpretability.

\subsubsection{Kernel-Based Methods}
\label{sec: methods: traditional ML:Kernel-Based Methods}

Kernel-based approaches transform input data into higher-dimensional feature spaces, capturing complex relationships that may appear non-linear in the original domain \cite{cortes1995support}.  In materials science, these methods excel at handling high-dimensional or data-limited scenarios while providing uncertainty quantification for experimental applications \cite{Lookman_ActiveLearning_2019}.

In materials science, Support Vector Machines (SVMs), as a subset of kernel-based methods, use this principle for classification and regression, suitable for high-dimensional or data-limited scenarios \cite{xu2023small}. Gaussian Processes (GPs) extend kernel concepts into a probabilistic framework, producing predictive distributions with uncertainty estimates, an important feature for experimental exploration \cite{liu2020gaussian,bostanabad2019globally}. While GPs can be computationally expensive, they succeed in active learning workflows, especially when datasets are small but predictions need to be very accurate \cite{kanagawa2018gaussian}. Because of their ability to model non-linearities and evaluate uncertainty, kernel-based methods are good candidates for detecting subtle structure-property correlations.

\paragraph{Support Vector Machines: }
Support Vector Machines (SVMs) try to maximize the margin between data classes or, in the case of regression, to minimize error within a "tube" around the function \cite{hearst1998support}. Kernel functions, such as the radial basis or polynomial kernel, enable nonlinear separation without explicitly transforming feature spaces, allowing SVMs to capture complex relationships in material data \cite{lu2013using}. Applications include classifying crystal structures based on X-ray diffraction patterns \cite{himanen2019data} and predicting mechanical moduli using compositional descriptors \cite{khakurel2021machine}. While SVMs perform effectively in high-dimensional, small-sample contexts, their performance depends on careful hyperparameter selection and kernel tuning \cite{tabor2018accelerating}. Besides traditional supervised classification and regression, the One-Class SVM variant \cite{scholkopf2001estimating} provides an unsupervised approach to outlier detection, which has been used to identify anomalous experiments or unexpected phases in materials databases \cite{ccevik2015support}.

\paragraph{Gaussian Processes}
Gaussian Processes (GPs) provide a probabilistic kernel-based framework that is especially useful when data is scarce or noisy \citep{Rasmussen2005GP}. Instead of learning fixed parameters, GPs construct a prior over functions and use observed data to generate a posterior distribution that estimates both mean and uncertainty in predictions \cite{seeger2004gaussian}. This built-in uncertainty quantification is important in active learning pipelines, where newly chosen experiments seek to reduce model uncertainty or maximize expected improvements \cite{rasmussen2003gaussian}. Despite these advantages, GPs work poorly with large datasets due to the computational complexity of matrix operations and are typically limited to supervised regression or extended to Gaussian process classification. 

\paragraph{Kernel Ridge Regression:}
Kernel Ridge Regression (KRR) combines ridge regression with the kernel trick, offering a non-probabilistic alternative to Gaussian Processes \cite{murphy2012machine}. By mapping input features to a higher-dimensional space, KRR captures non-linear relationships while maintaining computational efficiency for moderately sized datasets. In materials science, KRR has been successfully applied to predict electronic, mechanical, and thermodynamic properties from compositional and structural descriptors.

\subsubsection{Regression and Instance-Based}
\label{sec: methods: traditional ML:Regression and Instance-Based}
Classical ML methods, which include both parametric (linear/logistic regression) and instance-based (k-nearest neighbors) approaches, are frequently used as baseline models or for rapid prototyping in materials design \cite{cheng2009combining}.

Linear regression is still one of the most interpretable models, allowing researchers to identify simple trends, such as how changing composition affects a material's property \cite{wang2019symbolic}. Although these methods have limitations when dealing with complex, non-linear relationships, their simplicity and interpretability make them valuable components in materials data analysis, particularly for identifying preliminary trends or establishing performance benchmarks \cite{sathish2020analysis, li2018discrimination}.

\paragraph{Linear and Logistic Regression}
Their simplicity and interpretability make them valuable components in materials data analysis, particularly for identifying preliminary trends or establishing performance benchmarks. In materials science, linear regression is still quite popular to relate compositional relationships with properties like mechanical \cite{chaabene2020machine} and fracture properties \cite{liu2020machine}, as well as electrical \cite{pentos2022evaluation} and thermal properties \cite{li2022estimating}. Logistic regression maps the features to class membership probabilities using the logistic (sigmoid) function. It has been used to identify stable and unstable compounds and specific crystalline phases \cite{spellings2018machine}. While these methods are computationally inexpensive and simple to interpret, they are limited by their strictly linear (or log-linear) decision boundaries unless polynomial or interaction terms are included. 

\paragraph{k-Nearest Neighbors}
k-Nearest Neighbors (k-NN) is an instance-based algorithm that derives a new sample's label (or continuous value) from the labels (or values) of its nearest neighbors in the training set \cite{cover1967nearest}. This method requires no formal training phase; instead, it stores the entire dataset and defines similarity through distance metrics, typically Euclidean or cosine distance. Because of its simplicity, k-NN has been used in materials science to make quick baseline predictions, such as comparing the similarity of newly proposed compounds to known reference points in multidimensional descriptor space \cite{sharma2020prediction, arunadevi2023comparison}. While k-NN can model non-linear boundaries when local neighborhoods are well-defined, it becomes computationally demanding for large datasets. It is highly sensitive to feature scaling and appropriate neighbour count selection (k) selection. Like linear and logistic regression, k-NN is primarily used in supervised learning, with little application in unsupervised tasks \cite{kramer2013k}.

\subsubsection{Probabilistic Modeling for Materials Design}

Probabilistic modeling offers a powerful framework for addressing uncertainty and guiding decision-making in materials design. Unlike deterministic models, which provide single-value predictions, probabilistic approaches characterize the full distribution of possible outcomes \cite{samui2019handbook}. This capability is essential for risk assessment, efficient experimentation, and robust optimization in complex materials systems.

In materials science, uncertainty arises from inherent randomness and limitations in our knowledge. Probabilistic models typically distinguish between:
\begin{itemize}
    \item \textbf{Aleatoric uncertainty}: the intrinsic variability present in measurements and natural processes. This type of uncertainty is irreducible but must be accurately characterized to understand the variability in material properties.
    \item \textbf{Epistemic uncertainty}: uncertainty stemming from incomplete information or limited data. This uncertainty can be reduced through additional experimentation or by incorporating more detailed models.
\end{itemize}
Quantifying both forms of uncertainty is crucial for reliable predictions and for designing experiments that effectively reduce unknowns.

\paragraph{Parameter estimation and model calibration.}
Central to probabilistic modeling is parameter estimation, with Maximum Likelihood Estimation (MLE) being a widely used approach. MLE identifies parameter values that maximize the probability of observing the experimental data, providing a rigorous basis for model calibration. This well-defined likelihood function becomes a key component in more advanced probabilistic frameworks.

\paragraph{Bayesian inference.}
Bayesian inference builds on the principles of MLE by incorporating prior knowledge about model parameters. Through Bayes’ theorem, the posterior distribution of the parameters is derived as
\begin{equation}
    p(\theta \mid D) \propto p(D \mid \theta)\, p(\theta),
\end{equation}
where $\theta$ denotes the model parameters and $D$ represents the observed data. This posterior distribution not only provides updated estimates but also offers a natural quantification of uncertainty given data. By combining prior beliefs with observed data, Bayesian methods yield data-informed and physically consistent predictions.

One of the cornerstones of Bayesian methods in materials science is the Gaussian Process (GP) model~\citep {Rasmussen2005GP}. As non-parametric models, GPs define a distribution over functions and deliver both mean predictions and uncertainty estimates. They may serve as efficient surrogate models for expensive simulations or experiments, allowing researchers to explore the design space with significantly reduced computational cost while retaining a measure of confidence in the predictions.

To handle the high-dimensional complexity of modern materials data, uncertainty is integrated directly into deep learning architectures, leading to Bayesian Neural Networks (BNNs)~\citep{JospinEtAl_BayesianNN_2022}. Since exact Bayesian inference in deep networks is computationally intractable, approximate methods--such as variational inference--are employed. These techniques optimize a tractable lower bound on the model evidence, enabling scalable Bayesian learning that retains robust uncertainty quantification, even in large-scale settings.

\paragraph{Active learning and Bayesian optimization.}
Probabilistic models naturally lend themselves to active learning strategies~\citep{Lookman_ActiveLearning_2019}, which aim to maximize the information gained from each experiment or simulation. By identifying regions in the design space with high uncertainty, active learning methods can prioritize acquiring new data most likely to improve model performance. Bayesian optimization~\citep{Frazier_BayesianOptimiz_2016} is one such strategy: it uses probabilistic surrogate models (often based on GPs) to balance exploration of unknown regions with exploitation of known high-performing areas. This approach is particularly valuable in materials design, where experiments can be costly and time-consuming.

\paragraph{Emerging Challenges and Future Directions.}
Probabilistic modeling is a versatile and robust approach that equips researchers with the tools to navigate uncertainty, integrate prior knowledge, and optimize experimental strategies. However, although it has already shown great promise in materials design, several challenges remain:
\begin{itemize}
    \item \textbf{Scalability.} Developing algorithms that efficiently perform Bayesian inference in high-dimensional spaces and large datasets is an ongoing research challenge.
    \item \textbf{Multi-fidelity integration.} Combining data from diverse sources (e.g., low-fidelity simulations versus high-fidelity experiments) within a unified probabilistic framework is essential for a more comprehensive understanding of material behavior.
    \item \textbf{Embedding physical constraints.} Incorporating known physical laws and domain-specific constraints into probabilistic models can further enhance their predictive power and reliability.
\end{itemize}
Addressing these issues will allow the full potential of probabilistic modeling to be attained in the next generation of materials discovery and design.

%%%%%%%%%%%%%%%%%%%%%%%%%%%%%%%%
\subsection{Deep Learning Approaches}
\label{sec: methods: deep learning}

Deep Learning (DL) approaches involve artificial neural networks (ANNs) with multiple layers that progressively extract higher-level features from raw input data \cite{LeCun2015, Goodfellow-et-al-2016}.
While the theoretical foundations of ANNs were established in the 1980s, with early works demonstrating their universal approximation capabilities \cite{Hornik1989, Cybenko1989}, the practical breakthrough for deep learning occurred in the late 2000s.

Arguably, the first remarkable practical demonstration of the superiority of deep learning approaches was in image recognition in 2012, when the convolutional neural network (CNN) architecture AlexNet~\citep{krizhevsky2012imagenet} trained on large-scale datasets with GPUs outperformed other models at the ImageNet Large Scale Visual Recognition Challenge. This was a turning point from which deep learning approaches dominated many application domains~\citep{LeCun2015, Goodfellow-et-al-2016}. The following drastic improvement in state-of-the-art computer vision~\citep{redmon_YOLO_2016}, automated speech recognition~\citep{HintonEtAlSpeechRecognition_2012, DahlEtAl_DNNsSpeechRecognition_2012} and natural language processing~\citep{vaswani2017attention}  (NLP) had a huge impact on all other domains. 
A stark example for this was the evolution in NLP, in which the idea of attention mechanism~\citep{bahdanau2016} and the Transformer architecture~\citep{vaswani2017attention} gave rise not only to well-known generative pretrained transformer (GPT) models~\citep{Brown_GPT3_NEURIPS_2020}, but also made a huge impact beyond the NLP domain, for instance leading to breakthroughs in computational biology, as exemplified by the 2024 Noble Prize winning AlphaFold model~\citep{Jumper_AlphaFold_2021}. Quite naturally, DL methods also found a broad area of application to materials modeling and design, as discussed later in this work. 
In materials science, early applications of neural networks included developing ANN-based constitutive models \cite{Ghaboussi1991, Lefik2003} and homogenization schemes \cite{Unger2009, Lefik2009}. However, recent advances in hardware capabilities (particularly GPUs), the availability of large-scale datasets, and sophisticated network architectures have significantly expanded their application scope.

\subsubsection{Architecture Selection and Network Design}

\begin{figure}[h!]
        \centering
        \includegraphics[width=1\linewidth]{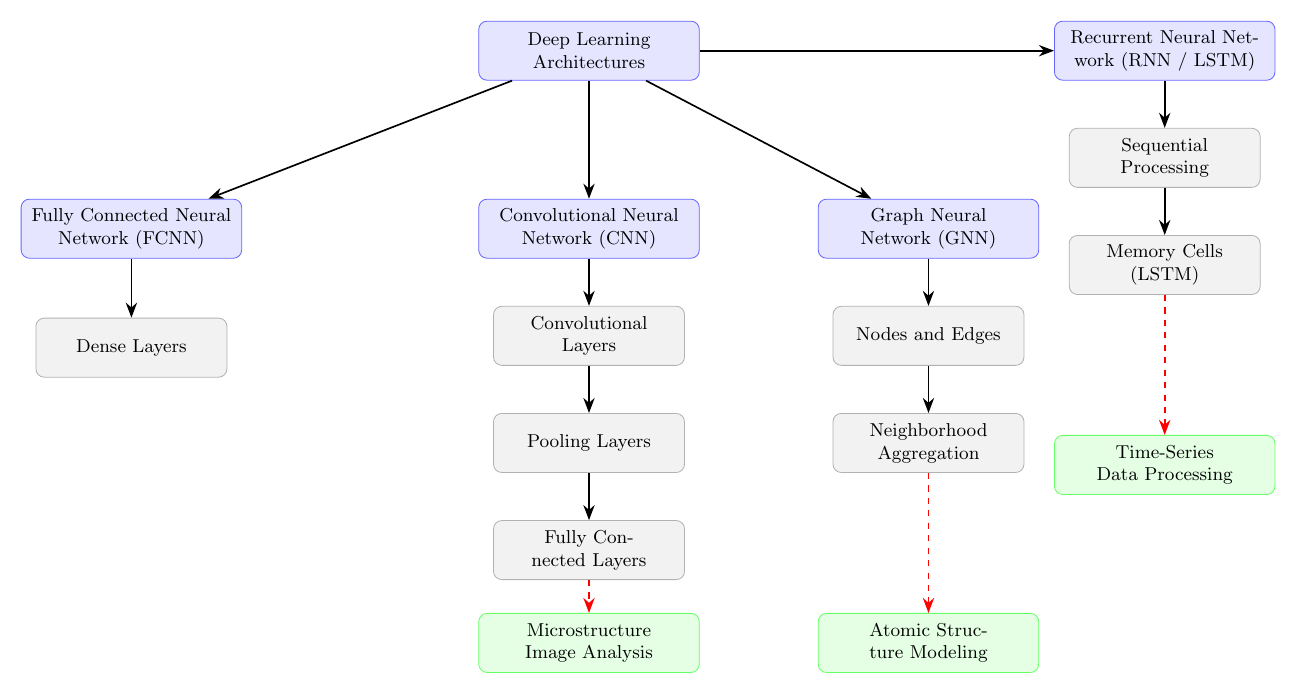}
         \caption{Comparative architectures of deep neural networks and their applications in materials science. FCNNs consist of fully connected layers suitable for tabular data. CNNs utilize convolutional and pooling layers for spatial data, such as microstructure images. GNNs operate on graph structures, making them ideal for modeling atomic structures. RNNs, including LSTMs, are designed for sequential data, enabling analysis of time-series measurements.}
  \label{fig:DL-archi}
\end{figure}

Deep learning encompasses a variety of architectures, each designed to handle specific types of data and tasks, Figure \ref{fig:DL-archi}. Fully Connected Neural Networks (FCNNs) rely on dense layers where each neuron is connected to every neuron in the preceding and succeeding layers, making them suitable for general-purpose learning but often less efficient for structured data. Convolutional Neural Networks (CNNs) incorporate convolutional and pooling layers that enable spatial feature extraction, making them particularly effective for image-based applications such as microstructure analysis. Graph Neural Networks (GNNs) are designed to operate on graph-structured data, using node and edge representations along with neighborhood aggregation mechanisms to model relationships within atomic structures. Recurrent Neural Networks (RNNs), including their more advanced variant Long Short-Term Memory (LSTM) networks, are tailored for sequential data processing, utilizing memory cells to retain contextual information over time, which is especially useful for time-series analysis. Each architecture plays a crucial role in advancing deep learning applications in materials science by aligning model structure with data characteristics.

Deep neural networks offer superior performance compared to shallow networks for several fundamental reasons:

\begin{enumerate}
    \item Hierarchical Feature Representation: Multiple layers enable progressively more abstract and high-level representations of material properties and structures
    \item Representational Efficiency: Deep architectures can represent certain functions more efficiently, requiring exponentially fewer neurons than shallow networks for equivalent performance \cite{bengio2009learning, telgarsky2015representationbenefitsdeepfeedforward}
    \item Transfer Learning Potential: Features learned at intermediate layers can transfer between related tasks, enabling effective knowledge sharing across materials science problems \cite{YosinskiEtAl_TransferLearning_2014}
\end{enumerate}

The selection of appropriate network architecture depends critically on data structure:

\begin{itemize}
    \item Fully-Connected Networks (Multilayer Perceptrons): Suitable for tabular data without clear spatial or temporal structure
    \item Convolutional Neural Networks (CNNs) ~\citep{krizhevsky2012imagenet, redmon_YOLO_2016, Alzubaidi_CNN_Review_2021}: Optimal for spatial data such as material microstructures, materials field variables or diffraction patterns
    \item Recurrent Neural Networks (RNNs) and Long Short-Term Memory (LSTM) networks~\citep{Hochreiter_LSTM_1997, Sak_LSTM_2014, Sherstinsky_RNN_LSTM_2020}: Designed for sequential or time-series data in materials processing
    \item Graph Neural Networks (GNNs)~\citep{Kipf_classGCN_2017, Velickovic_GraphAttention_2018, Wu_GNNs_Review_2021, schutt2018schnet, xie2018crystal, shen2025MLMBD}: Specialized for material structures with important connectivity information, such as molecular configurations or crystal lattices
    \item Transformer-Based Architectures~\citep{vaswani2017attention}: Increasingly applied to capture complex dependencies across multiple material properties and structures
\end{itemize}

Advanced training techniques have made deep learning practical for materials science applications, including improved weight initialization \cite{GlorotBenigo_Initialization_2010, He_Initialization_2015}, batch normalization \cite{Ioffe_BatchNormal_2015}, residual connections \cite{He_ResNets_2015}, and specialized optimizers \cite{Kingma_ADAM_2017}.

\subsubsection{Convolutional Neural Networks in Materials Informatics}

CNNs are a specialized class of neural networks designed to handle data with spatial (or spatiotemporal) structure, originally popularized in computer vision. CNNs apply trainable filters (kernels) over local regions of the input, often images or image-like tensors, capturing hierarchical patterns such as edges, shapes, and higher-order features. This localized, weight-sharing strategy reduces the number of trainable parameters compared to fully connected layers, making CNNs more efficient and translation-invariant.
A typical CNN architecture consists of:

\begin{enumerate}
    \item Convolutional Layers: Small filters (e.g., 3×3 or 5×5) slide across the input feature map, producing feature maps highlighting local patterns.
    \item Nonlinear Activations: Nonlinearities (e.g., ReLU) are applied after each convolution, increasing the representational capacity of the network.
    \item Pooling Layers: Spatial pooling (e.g., max or average pooling) down-samples the feature maps, reducing dimensionality and encouraging feature invariance to small translations.
    \item Fully Connected or Global Pooling: Toward the end, features are aggregated, either via fully connected layers or global average pooling, to produce a final prediction or embedding.
\end{enumerate}
In materials science, CNNs excel at:

\begin{itemize}
    \item Microstructure characterization: Automatically classifying material phases, detecting defects, quantifying morphological features, and tracking the evolution of morphological features. 
    \item Structure-property mapping: Establishing direct relationships between microstructural images and material properties and representing crystal structures as 3D voxelized images or 2D projections for property prediction. CNNs are also well-suited to structured grids that arise in many computational materials problems. By viewing discretized field variables (e.g., stress, temperature, concentration) as multichannel images, CNNs can approximate solutions to partial differential equations (PDEs) or emulate continuum-scale simulations \cite{guo2016convolutional, peivaste2022machine, choi2024accelerating, peivaste2025teaching}.
    \item Diffraction pattern analysis: Identifying crystal structures and phases from X-ray or electron diffraction data
    \item Data Augmentation and Inverse Design: generative CNNs, such as autoencoders or Generative Adversarial Networks (GANs) with convolutional layers (autoencoders and GANS are described in section \ref{generative_ai}) can synthesize realistic microstructures for data augmentation or perform inverse design (e.g., searching for microstructures that yield specific properties) 
    \cite{ohno2022training, brzin2024using}.

    \item Data representation for atomic structures: Although graph neural networks (GNNs) are often used for atomic-scale connectivity (Section~\ref{graph_structural}), CNNs can also encode 3D voxel grids of atomic positions or 2D projections of crystal structures. For instance, by dividing the crystal lattice into small voxel “bins,” CNNs can discover emergent spatial patterns relevant to energy or stability \cite{peivaste2024rapid, wallach2015atomnet}. Such volumetric approaches sidestep the need for manual features, letting the network learn crucial structural features.

    \item Topology Optimization: CNNs have also been used to predict the “optimal” distribution of material under load constraints, accelerating iterative design processes \cite{seo2023topology}.
\end{itemize}
Recent work has also demonstrated the versatility of CNNs in representing atomic structures through spatial occupancy grids. 
The effectiveness of CNNs stems from their architectural advantages, including parameter sharing (reducing model complexity), translation invariance (capturing the same features regardless of position), and hierarchical feature learning (progressively abstracting from simple edges to complex structural patterns).
Beyond microstructure analysis, CNNs have been adapted to represent crystal structures by converting atomic arrangements into 3D volumetric grids or 2D projections. CNNs thus play a multifaceted role in materials informatics, from image-based microstructure analysis to PDE surrogate modeling on structured grids. Their ability to automatically learn meaningful hierarchical features makes them indispensable for tasks that involve spatially resolved data, whether at the micro- or macro-scale. As the field continues integrating domain knowledge (e.g., physics-informed constraints, data fusion from multiple imaging modalities), CNN-based architectures are poised to remain essential in data-driven and simulation-driven materials research.
While CNNs excel at processing data with grid-like structures, many materials science problems involve irregular, graph-like structures where GNNs offer significant advantages.

Models like 3D-CNN \cite{peivaste2024rapid} and AtomNet \cite{wallach2015atomnet} transform crystal structures into spatial occupancy grids where each voxel contains information about atom types and local environments. This approach enables direct learning from structural data without requiring hand-crafted descriptors, proving effective for predicting formation energies, bandgaps, and mechanical properties from crystal structures alone. Unlike GNNs that explicitly model atomic connections, these CNN-based methods implicitly capture structural patterns through spatial convolutions, offering complementary strengths in crystallographic representation.

\subsubsection{Graph Neural Networks for Materials Representation} \label{gnn}

Graph neural networks (GNNs) offer a powerful framework for modeling materials at multiple scales by representing atomic or spatial relationships within a graph structure. In a typical GNN, nodes correspond to relevant entities (atoms, discretized mesh points, etc.), and edges encode interactions or connectivity, such as chemical bonds or neighboring cells in a mesh. Unlike other neural architectures that assume grid-like or sequential data, GNNs naturally handle variable-sized, irregular graphs—common in fields ranging from molecular systems to continuum modeling of complex geometries. A key concept in GNNs is the message-passing paradigm:

\begin{enumerate}
    \item Initialization: Each node begins with an embedding (vector) reflecting its intrinsic features (e.g., atomic number, local environment).

    \item Message Passing: Nodes exchange “messages” with neighbors at each layer. These messages encapsulate edge features (bond types, distances, etc.) and the neighbor’s embedding.

    \item Aggregation: Each node aggregates incoming messages (via sum, mean, or attention-based pooling) and updates its embedding.

    \item Readout: The final node embeddings can be aggregated to predict bulk material properties (e.g., bandgap, formation energy) or used individually for node-level tasks (e.g., local atomic charges).
\end{enumerate}

This structure makes GNNs particularly well-suited for graph-based feature characterisation of materials rather than relying solely on hand-engineered features.

In materials science, GNNs offer several advantages:

\begin{itemize}
    \item Invariance to permutation: The properties of materials typically do not depend on the order in which atoms are listed
    \item Locality exploitation: Many material properties emerge from local atomic environments and interactions
    \item Scalability: GNNs can process materials with varying numbers of atoms using the same model architecture
\end{itemize}

GNNs learn both chemical and structural constraints end to end. Examples include SchNet \cite{schutt2018schnet} and Crystal Graph Convolutional Neural Network (CGCNN) \cite{xie2018crystal}, which have demonstrated state-of-the-art performance with minimal manual feature engineering.

Beyond atomic-scale modeling, GNNs have proven effective in surrogate modeling for PDEs. In continuum mechanics, fluid dynamics, or other physics-based simulations, the domain is often discretized into grids or unstructured meshes, which can be naturally viewed as graphs. In mesh-based GNNs \cite{gladstone2024mesh} or in some neural operators \cite{li2020multipole}, each mesh cell or vertex becomes a node, and edges represent spatial adjacency (i.e., shared faces or edges). By unifying atomic-scale and continuum-scale modeling under the GNN umbrella, researchers can tackle various material problems, from predicting molecular interactions to simulating thermomechanical fields at engineering scales.

\subsubsection {Transformer-Based Models in Materials Informatics}
Transformer architectures, which revolutionized natural language processing through their attention mechanisms \cite{vaswani2017attention}, have recently been adapted for materials science applications. The attention mechanism allows these models to focus on relevant portions of the input data when making predictions, capturing long-range dependencies and complex interactions between material constituents.
Materials Transformers represent materials as sequential data, with attention mechanisms identifying relationships between elements, functional groups, or structural components. This approach has proven effective for property prediction across diverse material classes, from small molecules to complex alloys and polymers.
Examples include Crystal Transformer \cite{wei2022crystaltransformerselflearningneural} for crystal structure representation and Crystal Edge Graph Attention Neural Network (CEGANN) \cite{banik2023cegann} for global, structure-level representation, both demonstrating how architectures based on attention mechanisms can integrate heterogeneous materials data for improved prediction accuracy.

% CNNs, GNNs, RNNs (LSTM), Transformer-based, tailored to materials modeling... TBC
The choice of DNN architecture is closely tied to data structure. If the data does not possess a clear spatial, temporal, or relational structure (as in many tabular datasets), fully-connected networks (i.e, multilayer perceptrons, MLPs) or traditional ML techniques presented in Section~\ref{sec: methods: traditional ML} can be equally or even more effective than sophisticated DNN architectures. Additionally, MLPs are often employed as the final classification or regression layer in more specialized architectures. 
For sequential data (e.g., text or time series), recurrent neural networks (RNNs) such as long short-term memory (LSTM) networks~\citep{Hochreiter_LSTM_1997, Sak_LSTM_2014, Sherstinsky_RNN_LSTM_2020}, or Transformer-based approaches~\citep{vaswani2017attention} are commonly used because they efficiently handle sequences and temporal dependencies. For 2D and 3D images and other spatial data, convolutional neural networks (CNNs)~\citep{krizhevsky2012imagenet, redmon_YOLO_2016, Alzubaidi_CNN_Review_2021} are preferred, as they leverage spatial patterns and local connectivity. The generalization of CNNs to manifolds led to the concept of geometric deep learning~\citep{Bronstein_GometricDL_2017}, while extending them to more arbitrary graph structures resulted in graph neural networks (GNNs)~\citep{Kipf_classGCN_2017, Velickovic_GraphAttention_2018, Wu_GNNs_Review_2021, schutt2018schnet, xie2018crystal, shen2025MLMBD} and neural operators~\citep{KovachkiJMLR2023NeuralOperator}. Importantly, Transformer-based architectures have demonstrated the ability to efficiently learn complex dependencies within a wide variety of data types, including those encountered in computer vision~\citep{Dosovitskiy_VisionTransformer_2021} and multimodal domains~\citep{jaegle_PerceiveriIO_2022, Xu_MultimodalTransformer_2023}.

% RNN and LSTM~\citep{Hochreiter_LSTM_1997, Sak_LSTM_2014, Sherstinsky_RNN_LSTM_2020},

% AlexNet CNN (2012)~\citep{krizhevsky2012imagenet},
% CNN review (2021)~\citep{Alzubaidi_CNN_Review_2021},
% YOLO real-time object detection (2016)~\citep{redmon_YOLO_2016}, Classification with Graph Convolutional Networks (2017)~\citep{Kipf_classGCN_2017}, 
% Geometric Deep Learning (2017)~\citep{Bronstein_GometricDL_2017}, Graph Attention Networks (2018)~\citep{Velickovic_GraphAttention_2018}, 
% Perceiver IO: Transformer-based multi-modal model for structured data (2022)~\citep{jaegle_PerceiveriIO_2022},     
%All these advances have also inspired the adoption of deep learning in materials research, where DNNs offer powerful tools for capturing complex relationships in data, enabling more accurate predictions and accelerating materials discovery and design.

%%%%%%%%%%%%%%%%%%%%%%%%%%%%%%%%
\subsection{Emerging Techniques} \label{sec:emerging_tech}

The rapid evolution of ML methodologies continues to drive transformative advances in materials modeling and design. Beyond traditional approaches, emerging techniques such as generative AI, transfer learning, federated learning,
and large language models (LLMs), and self-supervised learning
are redefining the capabilities of materials informatics. These techniques address persistent challenges in data scarcity \cite{chang2022, tian2022}, model interpretability \cite{dean2023interpretable, kailkhura2019reliableexplainablemachinelearning}, and scalability \cite{rodrigues2021big, Liu2023}, facilitating accelerated materials discovery and optimization across diverse domains, including composites, alloys, and electronic materials.

\subsubsection{Generative AI} \label{generative_ai}
\begin{figure}[h!]
        \centering
\includegraphics[width=0.7\linewidth]{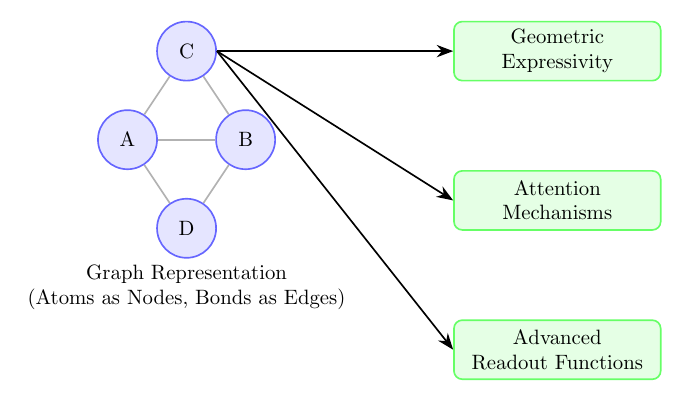}
\includegraphics[width=0.2\linewidth]{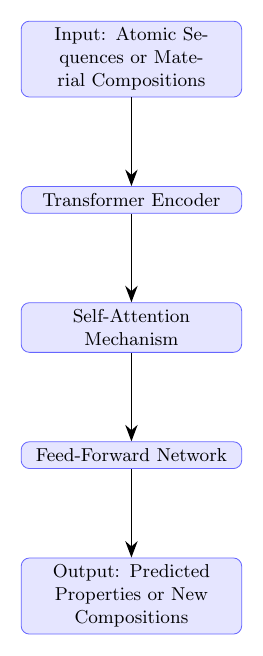}
  
         \caption{This diagram illustrates the Transformer model's architecture adapted for materials informatics applications. It showcases the flow from input representations—such as atomic sequences or material compositions—through embedding layers and positional encodings, into the encoder comprising multi-head self-attention and feed-forward networks. The decoder, equipped with masked self-attention and encoder-decoder attention mechanisms, processes this information to generate outputs like predicted material properties or novel compositions. This architecture enables the modeling of complex relationships within material data, facilitating tasks like property prediction and generative material design}
  \label{fig:Gen}
\end{figure}

Generative AI, encompassing techniques such as Generative Adversarial Networks (GANs) (Figure \ref{fig:Gen}) \cite{GANs2014}, Variational Autoencoders (VAEs) \cite{kingma2022autoencodingvariationalbayes, MAL_056}, diffusion models \cite{Diffusion2015, NEURIPS2020_4c5bcfec}, and generative Transformers \cite{vaswani2017attention}, has become an increasingly valuable tool for reconstructing complex material microstructures from microscopic images \cite{Lee16092024} and materials discovery and design, from continuum micromechanical scale \cite{HENKES2022115497} to the molecular and atomic level \cite{D3DD00137G, Kamal2024}. By generating synthetic data, these models alleviate experimental burdens and facilitate enhanced virtual prototyping. These approaches enable researchers to explore material configurational spaces efficiently and generate novel candidates with targeted properties.

\paragraph{GANs:}
GANs consist of two competing neural networks, a generator and a discriminator, that operate in opposition. The generator produces synthetic (fake) data, such as images or microstructures, while the discriminator evaluates whether the generated data is real or fake. Through iterative training, the generator progressively enhances its ability to create realistic outputs that can deceive the discriminator. In contrast, the discriminator simultaneously improves its ability to differentiate between real and fake.

In materials science, GANs have proven valuable for:

\begin{itemize}
    \item Data augmentation: Expanding limited experimental datasets to improve the training of other ML models \cite{ohno2022training}.

    \item Inverse design: Enabling property-driven material discovery by generating structures with targeted properties \cite{jabbar2022recent}

    \item Discovery of new materials: Exploring novel material configurations beyond those found in training data, potentially identifying materials with unprecedented property combinations (discussed in detail in Section \ref{sec-4})
    \item Microstructure generation: Creating realistic, synthetic microstructures for three-phase solid oxide fuel cell electrodes \cite{Hsu2021_} and heterogeneous energetic materials \cite{Chun_2020}

    \item Grain structure simulation: Developing 3D grain shapes in polycrystalline structures, such as WStyleGANs for crystalline titanium grains \cite{Jangid2022}

\end{itemize}

GANs have emerged as a powerful tool for generating realistic microscopic images of materials with heterogeneous microstructures, such as composite materials and alloys. By leveraging their ability to learn from existing material data, GANs hold significant potential for advancing material design and simulation, as evidenced by several successful examples in the literature. For instance, the generation of 3D microstructures for three-phase solid oxide fuel cell (SOFC) electrodes \cite{Hsu2021_} highlights GANs' ability to effectively learn and reproduce realistic, complex multiphase microstructures in both two and three dimensions. Additionally, physics-based simulations of electrochemical performance reveal that the generated microstructures closely replicate the characteristics of the original cases.
Other significant contributions utilizing GANs include the generation of microstructures for heterogeneous energetic materials \cite{Chun_2020} and the development of WStyleGANs for creating 3D grain shapes in polycrystalline structures, such as crystalline titanium grains \cite{Jangid2022}. These applications underscore the versatility of GANs in generating high-fidelity material microstructures.

\begin{figure}[h!]
    \centering
    \includegraphics[width=0.8\linewidth]{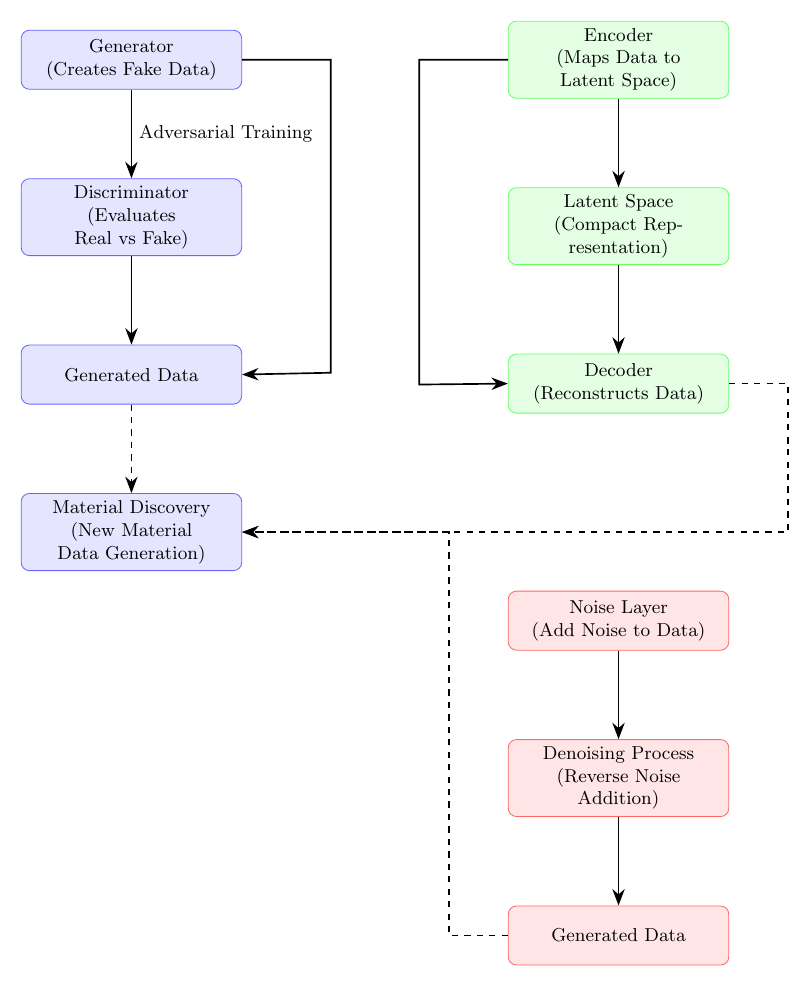}
    \caption{Schematic overview of generative AI models applied to materials discovery. This diagram depicts the operational flow of three key generative models: (1) Generative Adversarial Networks (GANs), where a generator creates synthetic data and a discriminator evaluates its authenticity through adversarial training. (2) Variational Autoencoders (VAEs), which learn compact latent representations of data via an encoder-decoder structure to generate new material samples. (3) Diffusion Models, which progressively add noise to data and then reverse this process to generate realistic new samples. These models facilitate the generation of novel materials data, contributing to the discovery of new materials with enhanced properties.}
    \label{fig:generative}
\end{figure}

Generative AI models, such as Generative Adversarial Networks (GANs), Variational Autoencoders (VAEs), and Diffusion Models,Figure \ref{fig:generative}, are powerful tools that can be leveraged for materials discovery. These models work by learning complex data distributions from existing material datasets and generating new, synthetic material data. GANs operate by training two neural networks in opposition: the generator creates fake data, while the discriminator assesses the authenticity of this data, refining the generator’s ability to produce realistic samples over time. VAEs, on the other hand, map data to a lower-dimensional latent space, allowing for the generation of new material samples by decoding this compressed representation. Diffusion Models add noise to data and then reverse the process to recover the original data, enabling the generation of novel samples through denoising.

While GANs produce high-fidelity (quality) results with fast generation, they also come from notable downsides. One significant challenge is that the adversarial nature of GANs can lead to unstable training dynamics, often resulting in non-convergence or mode collapse, where the generator produces a limited variety of outputs \cite{Ahmad2024}. Additionally, the training procedure of GANs is often resource-intensive. It requires large, high-quality datasets, which can be a limitation in materials science, where experimental data is frequently scarce or noisy. GANs also lack inherent density estimation, making it difficult to evaluate how well the generated distribution matches the real data distribution \cite{Taylor2022}. Finally, GANs are not invertible, meaning they lack an inverse mapping from generated data back to the latent space, complicating tasks such as feature manipulation and inverse material design.

\paragraph{VAEs and VAE-GANs}

VAEs are a class of generative models that learn efficient, probabilistic representations of complex data by encoding it into a continuous latent space (encoder) and then reconstructing the input from this representation using a generator (decoder) network. During training, VAEs optimize a combined loss function comprising a reconstruction loss, which ensures fidelity between the original and generated data, and a Kullback-Leibler (KL) divergence term, which regularizes the latent space to follow a standard Gaussian distribution.

This latent space is a lower-dimensional, organized representation of the input data, offering a compact description of materials' representations. It enables:

\begin{itemize}
    \item Material design space exploration: Efficiently navigating potential material configurations \cite{sardeshmukh2024}
    \item Structure reconstruction: Generating realistic material structures from learned representations \cite{zhang2023, hashemi2024}
    \item Inverse problem-solving: Identifying material designs that satisfy specified property targets \cite{ATTARI2023119204, zeng2024, Wang_2020}
\end{itemize}

This latent space is a lower-dimensional, organized representation of the input data, offering a compact description of materials representations and allowing for interpolation, exploration, and manipulation of data \cite{kingma2022autoencodingvariationalbayes, MAL_056}.
The latent space is particularly valuable for tasks like material design \cite{sardeshmukh2024}, microstructure reconstruction \cite{zhang2023, hashemi2024}, and inverse problem-solving \cite{ATTARI2023119204, zeng2024, Wang_2020}. By leveraging this latent space, researchers can efficiently explore extensive design spaces, optimize material properties, and generate novel material configurations, significantly reducing the reliance on trial-and-error and computationally expensive simulations.

Despite their versatility, VAEs often generate outputs that are blurrier or less detailed compared to other generative models, such as GANs, due to their reliance on Gaussian assumptions in the latent space \cite{cai2017}. Moreover, the inherent trade-off between reconstruction accuracy and latent space regularization can result in suboptimal representations, particularly for complex or high-dimensional data \cite{asperti2020}.

\paragraph{Diffusion models}

A diffusion probabilistic model (commonly referred to as a “diffusion model”) \cite{Diffusion2015} generates structured outputs by iteratively transforming random noise into meaningful data through a process of adding and removing noise. These models employ a forward Markov chain that adds noise incrementally to destroy the signal and a reverse Markov chain that removes the noise to reconstruct the data. 
Further improved by Denoising Diffusion Probabilistic Models (DDPMs)\cite{NEURIPS2020_4c5bcfec}, diffusion models have become powerful tools for generating high-dimensional data, including molecular structures and material microstructures. 

Recent studies have highlighted the significant potential of diffusion probabilistic models in reconstructing and generating the microstructures of polycrystalline alloys, carbonates, ceramics, copolymers, and fiber composite materials \cite{lee2023}. Furthermore, a data-driven framework \cite{lee2023_2} leveraging diffusion models was developed for the inverse design of multifunctional composites, enabling the creation of materials with tailored properties by optimizing their microstructural features. Additionally, \cite{Li_2025} introduced a novel approach to crystal structure generation by combining diffusion models with point cloud representations, facilitating the generation of energetically stable crystal structures while addressing challenges related to structural periodicity \cite{xie2022crystaldiffusionvariationalautoencoder}.

Diffusion models excel in generating high-quality, diverse data (outperforming VAEs) and exhibit superior stability during training (outperforming GANs) due to their likelihood-based formulation \cite{NEURIPS2020_4c5bcfec}. However, they are computationally more expensive and have slower sampling speeds compared to GANs, which can generate results in a single forward pass. Additionally, while diffusion models are highly flexible and robust, they require careful tuning of noise schedules and longer training times than VAEs.
Although the intermediate steps of progressive noise refinement in diffusion models can be considered as distributed latent-like representations, they sacrifice the compactness and interpretability of the latent space in VAEs. Finally, we highlight that hybrid approaches such as Latent Diffusion Models \cite{rombach2022highresolutionimagesynthesislatent} show promise in overcoming the limitations of these approaches.

\textbf{Generative Transformers:}
Transformer models, initially developed for NLP tasks,
are increasingly being adapted in domains like materials discovery and design. Central to their architecture is the self-attention mechanism \cite{vaswani2017attention}, a process that enables the model to selectively focus on relevant parts of input sequences when generating output. Unlike traditional models that rely on fixed-length contexts or simple recurrent architectures, transformers can dynamically assess and capture long-range dependencies within data, allowing for greater flexibility and scalability in handling complex sequences. These models work by encoding input sequences into high-dimensional representations using multi-head attention layers and decoding these representations to generate coherent and contextually accurate outputs. 
These models leverage the powerful architecture of transformers to generate novel material compositions and predict properties. By representing materials as sequences of atomic or molecular structures, generative transformers facilitate both forward and inverse design, significantly accelerating the discovery and development of new materials with targeted functionalities \cite{KOROLEV2023100803, huang2023materialsinformaticstransformerlanguage, wei2022crystaltransformerselflearningneural, chen2024mattergptgenerativetransformermultiproperty, Kamal2024, fu2022materialstransformerslanguagemodels}.
Furthermore, unlike some black-box generative models, transformers inherently provide attention maps that reveal which atomic or molecular features contribute most significantly to their predictions \cite{KOROLEV2023100803, huang2023materialsinformaticstransformerlanguage}. This interpretability enhances the trustworthiness of generative outputs, enabling researchers to gain insights into the underlying principles of material design.
One of the most promising applications of generative transformers lies in predicting properties and designing crystalline materials. Models like \textit{Crystal Transformer} \cite{wei2022crystaltransformerselflearningneural} leverage the transformer’s attention mechanism to generate stable crystal structures by filling "blank spaces" in material compositions, analogous to the approach used by masked language models in NLP. Similarly, transformer-based models like MatInFormer \cite{huang2023materialsinformaticstransformerlanguage, KOROLEV2023100803} learn the "grammar" of crystallography, enabling them to predict material properties with high accuracy.
Generative transformers have also demonstrated success in the design of 2D materials. In \cite{Dong2023}, the authors introduce a material transformer generator (MTG) which utilizes a transformer-based neural language model to generate hypothetical 2D compositions. These compositions are then evaluated for thermodynamic stability using Density Functional Theory (DFT) computations, showcasing the potential of generative transformers in discovering novel 2D and functional materials.
Inverse design, where desired material properties guide the generation process, represents another critical application area for generative transformers. Models like AtomGPT \cite{chen2024mattergptgenerativetransformermultiproperty} and MatterGPT \cite{Kamal2024} exemplify this approach by leveraging transformer-based architectures to predict structures of interest. AtomGPT generates atomic structures that meet specific performance criteria, such as superconducting transition temperatures, formation energies, and electronic bandgaps. Additionally, it efficiently predicts material properties based on a given atomic structure description. Similarly, MatterGPT employs transformer-based architectures to design crystal structures with targeted single properties (both formation energy or bandgap) and multiple properties, enabling multi-objective inverse design of crystals.

\begin{figure}
    \centering
    \includegraphics[width=0.9\linewidth]{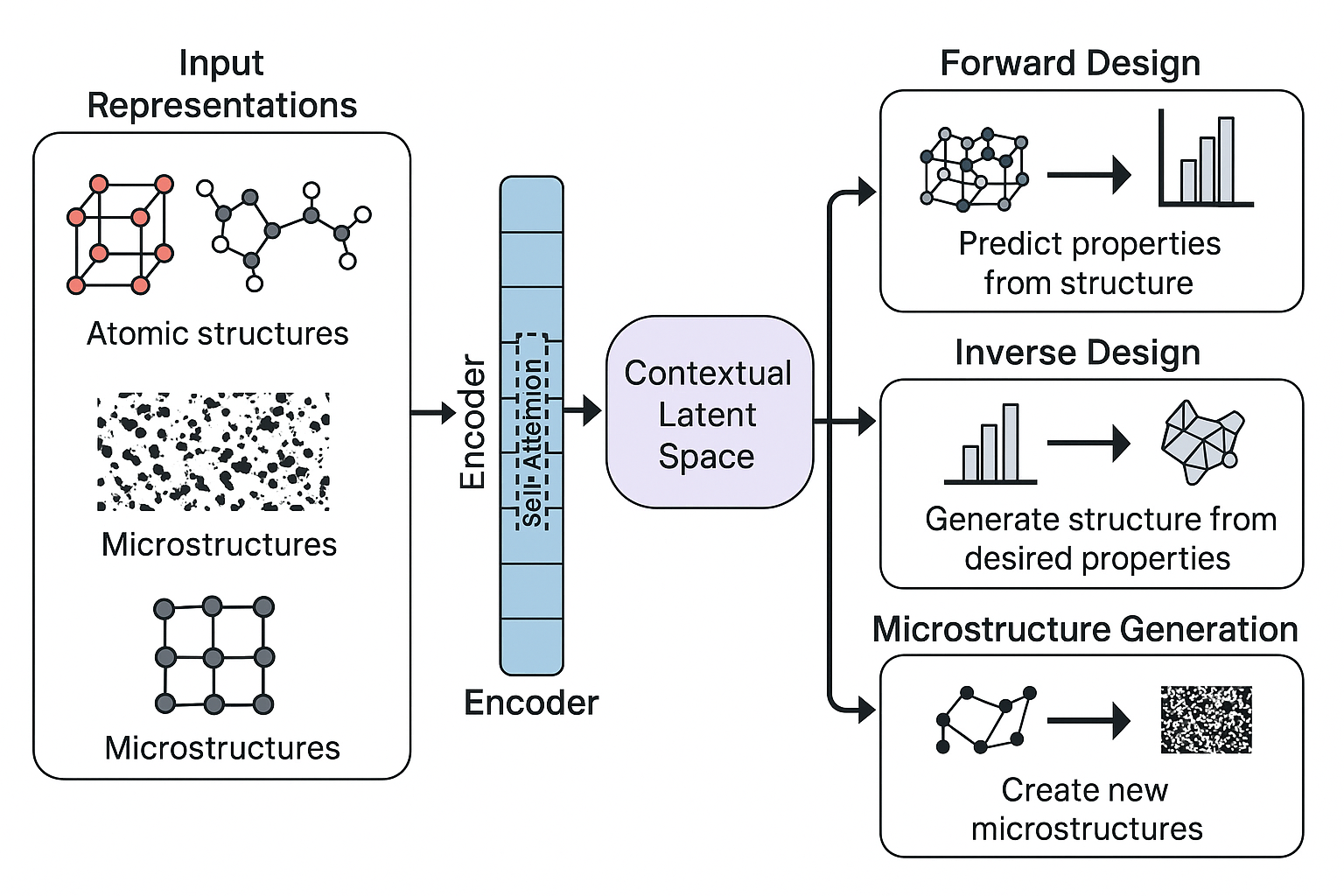}

    \caption{Illustration of a generative transformer workflow for materials discovery and design. The input material microstructure and representations (e.g., atomic or molecular sequences) are processed by an encoder with self-attention mechanisms to create a contextual latent embedding. This embedding is then decoded to generate new material structures or compositions. Applications include forward design (predicting properties from structures), inverse design (generating structures from target properties), and microstructure generation}
    \label{fig:enter-label}
\end{figure}

\subsubsection{Transformer-Based Approaches in Materials Informatics} \label{transformer}
Building on the deep learning approaches discussed previously, transformer architectures have emerged as a powerful paradigm in materials informatics, bringing innovations from natural language processing to material property prediction and design. Transformers \cite{vaswani2017attention} were originally introduced for NLP tasks but have rapidly gained traction in materials science. Their core innovation is the attention mechanism, which allows the model to weigh the importance of each element (or “token”) in the input sequence relative to every other element—thus capturing both local and long-range dependencies. Multi-head attention extends this idea by computing multiple parallel attention distributions (i.e., “heads”), each focusing on different aspects of the sequence or input representation. These heads operate simultaneously and then aggregate their information, enabling the Transformer to learn a richer and more nuanced representation of complex data than earlier sequence-to-sequence architectures.

While Transformers were initially designed to process textual data, researchers have adapted them to various materials-related challenges by customizing how material structures and properties are represented.  A significant early application is language-inspired featurization, where molecules or crystal structures are encoded as text-like sequences (e.g., SMILES or SELFIES). Models such as ChemBERTa \cite{chithrananda2020chemberta}, MatInFormer \cite{huang2023materialsinformaticstransformerlanguage}, and Crystal Transformer \cite{wei2022crystaltransformerselflearningneural} leverage the Transformer’s attention blocks to map sequential “tokens” (atoms, functional groups, elemental symbols, etc.) into embeddings. This capability facilitates tasks like property prediction and classification while also providing interpretability: attention maps can reveal which parts of the sequence drive the model’s prediction. As this initiative showed promising results, the transformer-based featurization will be later fully discussed in the section Language-Inspired Featurization \label{lang_inspired_feature}.

The same underlying architecture underpins large language models (LLMs) now used for knowledge extraction from scientific literature: by treating materials-relevant documents as input text, these models can identify chemical names, parse reaction conditions, and synthesize structured information \cite{ansari2023agentbasedlearningmaterialsdatasets, Vangala2024}. These LLMs treat materials-specific documents as input sequences and use the multi-head attention layers to locate domain-specific keywords or phrases. This approach underpins automated literature mining systems that rapidly identify material-property relationships, enabling accelerated data curation and hypothesis generation in materials science. The use of LLMs for knowledge extraction will be discussed in detail in section \label{knowledge_extraction}.

Transformers also feature prominently among emerging generative techniques in materials informatics, building on concepts from Generative Pretrained Transformer (GPT) models \cite{Brown_GPT3_NEURIPS_2020}. Recent work has showcased how attention-driven decoder blocks can propose novel material formulas or crystal structures, conditioned on targeted properties or "masked" tokens \cite{Dong2023, Kamal2024, chen2024mattergptgenerativetransformermultiproperty}. In these settings, Transformers serve as powerful agents for inverse design, autonomously suggesting candidates that meet user-specified constraints such as bandgaps, formation energy, or mechanical performance. Unlike some black-box generative models, Transformers often provide attention maps with interpretability, indicating which atomic or compositional features drive the prediction.
Overall, Transformer-based models excel in three major facets of materials modeling: (1) featurization, where sequential data representations harness attention mechanisms for improved predictive accuracy; (2) generative modeling, which leverages masked language modeling or decoder-based architectures for inverse design; and (3) knowledge extraction, where large-scale LLMs parse scientific text or patents to unearth hidden relationships in vast materials datasets. As Transformer research advances, we can expect ever deeper integration with graph representations, physics constraints, and multimodal learning, further broadening their impact on materials informatics. This continued adaptation will likely broaden their impact even more, from automated discovery pipelines to closed-loop experimentation and beyond \cite{jiang2025applications, jia2024llmatdesign}.

\subsection{Summary Comparison of Core Algorithms}

Having explored the landscape of traditional machine learning techniques (Section \ref{sec: methods: traditional ML}) and deep learning architectures (Section \ref{sec: methods: deep learning}) applied to materials modeling and design, we now provide a consolidated comparison. Table~\ref{tab:ml_comparison} summarizes the core algorithms discussed, contrasting their learning types, key advantages, potential disadvantages, and specific scenarios where caution might be warranted, serving as a quick reference guide.

\begin{longtable}{@{} >{\raggedright}p{2.5cm} >{\raggedright}p{2cm} >{\raggedright}p{2.5cm} >{\raggedright}p{2.5cm} >{\raggedright\arraybackslash}p{3.5cm} @{}}
\caption{Comparative Overview of Key Machine Learning Algorithms in Materials Informatics, Detailing Learning Type, Pros, Cons, and Applicability Considerations.} \label{tab:ml_comparison}\\ 
\toprule
\textbf{Algorithm} & \textbf{Type} & \textbf{Pros} & \textbf{Cons} & \textbf{When NOT to Use} \\
\midrule
\endfirsthead

\multicolumn{5}{c}%
{{\bfseries \tablename\ \thetable{} -- continued from previous page}} \\
\toprule
\textbf{Algorithm} & \textbf{Type} & \textbf{Pros} & \textbf{Cons} & \textbf{When NOT to Use} \\
\midrule
\endhead

\midrule
\multicolumn{5}{r}{{Continued on next page}} \\
\midrule
\endfoot

\bottomrule
\endlastfoot

Linear Regression & Supervised & Simple, interpretable, fast & Sensitive to outliers, non-linear limits & Data with strong non-linearity \\
\midrule
Logistic Regression & Supervised & Probabilistic, interpretable & Weak with non-linear boundaries & Data is highly non-linear \\
\midrule
Decision Tree & Supervised & Easy to interpret & Overfitting, unstable & Noisy or complex datasets \\
\midrule
Random Forest & Supervised & High accuracy, robust & Slower, less interpretable & Need real-time results \\
\midrule
Gradient Boosting & Supervised & State-of-the-art accuracy & Overfitting, needs tuning & When interpretability matters \\
\midrule
SVM & Supervised & Works in high dimensions & Slow on large data & Large noisy datasets \\
\midrule
KNN & Supervised & Simple, no training phase & Slow, noisy, sensitive & High-dimensional noisy data \\
\midrule
K-Means & Unsupervised & Fast, easy to implement & Needs K, sensitive to scale & Non-spherical data \\
\midrule
Neural Networks (MLP) & Supervised & Non-linear learning power & Needs large data \& tuning & Small data, low compute \\
\midrule
CNN & Supervised & Excellent for images and grid-like spatial data & High resource demand & Sequence/text data \\
\midrule
GNNs & Supervised & Handles non-Euclidean data, captures relations, permutation invariant & Scalability issues, over-smoothing, complex tuning & Data without clear graph structure, very large graphs \\
\midrule
Transformer  & Supervised \\/Self supervised & Long context, fast & Heavy compute, large model & Small projects \\
\midrule
Autoencoders & Unsupervised & Effective denoising & Can overfit, black-box & When no compression needed \\
\end{longtable}

\subsection{\textcolor{blue}{Gap Analysis}}
The analysis of Machine Learning (ML) algorithms  reveals significant trade-offs between performance, interpretability, scalability, and domain applicability. These challenges have direct implications for their deployment in materials modelling and design, especially in resource-constrained, physics-constrained, or uncertainty-sensitive settings. 
\begin{enumerate}
    \item \textbf{Interpretability vs. Predictive Power:} Algorithms such as \textit{linear regression}, \textit{logistic regression}, and \textit{decision trees} offer high interpretability but suffer from limited capacity to capture complex non-linear behaviour—common in microstructure–property relationships. On the other hand, high-capacity models like \textit{neural networks}, \textit{CNNs}, and \textit{transformers} offer state-of-the-art predictive performance but operate as black boxes, limiting trust and explainability in scientific discovery. This creates a fundamental gap between model performance and the actionable insights required by domain experts.

\item \textbf{Scalability and Computational Constraints:} \textit{Transformers}, \textit{CNNs}, and \textit{GNNs} show promise for handling complex, high-dimensional, and non-Euclidean materials data (e.g., microstructure images, atomistic graphs), yet their computational costs remain prohibitive for large-scale or real-time applications. \textit{Support Vector Machines (SVMs)} and \textit{KNNs} also become inefficient as data volumes grow. Efficient model scaling and acceleration, particularly for deployment in inverse design and uncertainty quantification loops, remain an open challenge.

\item \textbf{Data Volume and Quality Sensitivity:} Several algorithms—including \textit{MLPs}, \textit{CNNs}, and \textit{gradient boosting}—require large, clean datasets to achieve generalisable performance. However, materials datasets are often limited, sparse, or noisy due to experimental and simulation constraints. Algorithms sensitive to outliers or high noise levels (e.g., \textit{decision trees}, \textit{KNNs}) perform poorly under such conditions, revealing the need for robust learning under data scarcity and uncertainty.

\item \textbf{Unsupervised and Generative AI:} While unsupervised learning is crucial for tasks such as anomaly detection, clustering of material families, or latent space discovery, its use remains underexploited. \textit{K-means} clustering assumes spherical distributions and scales poorly with dimensionality, while \textit{autoencoders} can suffer from overfitting and lack interpretability. There is a pressing need for principled unsupervised learning techniques that can operate effectively under the physical constraints and data limitations of materials informatics.

\item \textbf{Mismatch Between Algorithm Design and Materials Representation:} Not all algorithms are equally suited for structured materials data. For example, \textit{graph neural networks (GNNs)} are well adapted for atomistic or crystalline data due to their relational inductive biases, yet suffer from issues such as over-smoothing and complex hyperparameter tuning. Conversely, applying conventional \textit{CNNs} or \textit{MLPs} to non-Euclidean domains often results in poor performance or information loss. This highlights a representational mismatch and the need for adaptive architectures that align with the physics and topology of materials data.

\item \textbf{Data Scarcity and Quality:} One of the primary bottlenecks in data-driven approaches is the lack of high-quality, domain-specific data. Experimental data in materials science are often expensive, time-consuming to generate, and not readily shareable due to intellectual property constraints or lack of standardisation. Moreover, data are frequently sparse, noisy, or heterogeneous, which impedes the training of robust ML models. This issue is particularly critical in high-fidelity simulations or rare-event scenarios such as failure mechanics or fatigue prediction.

\end{enumerate}

The gap analysis and algorithm-specific challenges underscore the importance of tailoring ML workflows to the unique characteristics of materials data, selecting models not only based on predictive power but also on their physical interpretability, scalability, and compatibility with domain-specific constraints.

%%%%%%%%%%%%%%%%%%%%%%%%%%%%%%%%
\section{Data and Data Representation }\label{sec-3}
The effectiveness of the ML methodologies discussed in the previous sections critically depends on appropriate data representation. While selecting suitable algorithms is important, transforming raw materials data into meaningful features often represents the more challenging aspect of materials informatics.
Data scientists commonly estimate that around 80\% of the workload in an ML project is dedicated to data preparation, while only 20\% focuses on selecting and implementing the ML model \cite{felice2025boosting}. This principle holds true for any data-driven field, where the quality and availability of data play a critical role. Materials modeling and design rely on diverse data sources, broadly categorized into experimental data, computational data, and hybrid approaches that integrate both \cite{zhou2019big}. 
Effective data preprocessing and quality control are essential for ensuring the reliability and usability of materials' datasets.

\subsection{Data Preprocessing and Quality Control}
Raw data, whether obtained from experiments or computational simulations, often contains errors, inconsistencies, and missing values that can compromise the accuracy of predictive models. Data cleaning is a crucial preprocessing step that addresses these issues \cite{corrales2018address}. The specific cleaning techniques employed may vary depending on the type of data and research objectives, but commonly include:

\begin{itemize}
\item Handling missing values through imputation techniques, \cite{mohit2023handling}
\item Standardizing units for consistency across datasets, \cite{lawson2021clarity}
\item Eliminating duplicate, redundant, or highly correlated entries
\item Identifying and removing erroneous data points through outlier detection, \cite{hodge2004survey}
\end{itemize}

Normalization and scaling techniques, such as min-max scaling and standardization, ensure that numerical values across datasets remain comparable and do not disproportionately affect ML models. Normalization is particularly essential for gradient-based and distance-based algorithms, including ANNs, logistic regression, SVMs, and KNN. However, normalisation is generally unnecessary for tree-based models like Random Forests and XGBoost, as these models are less sensitive to feature scaling \cite{sujon2024use}.

Beyond numerical transformations, proper labeling and annotation are equally important. Inconsistent property definitions or unreliable human annotations can introduce bias, reducing model reliability. Addressing these issues through standardization and comprehensive metadata documentation improves the interpretability and interoperability of materials data, ensuring that datasets can be effectively utilized across different studies and applications.

Another important issue is that materials science datasets often suffer from long-tail distributions, where well-studied materials dominate while rare or novel compounds remain underrepresented. This issue can pose significant challenges for ML models, as overrepresented materials tend to dominate learned representations, while rare or novel materials receive inadequate attention \cite{zhang2025systematic}. To mitigate this, data-centric strategies such as automated data curation \cite{hart2024trust}, clustering-based rebalancing \cite{liu2024clustering}, and data augmentation \cite{lin2024data, ma2020data, yang2023two} can help create more representative datasets. Hierarchical k-means clustering can ensure that diverse material compositions, structures, and properties are equally represented, preventing dominant classes from overshadowing rare ones. Additionally, oversampling underrepresented materials, undersampling dominant ones, and leveraging synthetic data generation techniques, such as GANs, can improve class balance \cite{dan2020generative}. Multi-modal data integration, combining experimental measurements, computational simulations, and literature-extracted data, further enhances the dataset's diversity, reducing biases introduced by a single data source.

Beyond data preprocessing, algorithmic and model-centric techniques are crucial in addressing bias. Cost-sensitive learning, where the loss function is adjusted to assign higher importance to underrepresented materials, helps balance model predictions \cite{brownlee2020imbalanced}. Self-supervised pretraining on automatically curated datasets, followed by fine-tuning on specific material classes, can enhance generalization while preserving balanced feature representations. Additionally, ensemble learning and meta-learning approaches improve robustness by aggregating predictions from multiple models trained on different subsets of data \cite{sun2021fingerprinting}. Fairness-aware modeling, such as adversarial debiasing and domain adaptation techniques, further ensures that ML models do not disproportionately favor certain data \cite{bhattacharya2024representation}. By integrating these strategies, researchers can develop more equitable and generalizable models, accelerating materials discovery while minimizing dataset-induced biases.

\subsection{Data Representation}

\begin{figure}[h!]
    \centering
    \includegraphics[width=1\linewidth]{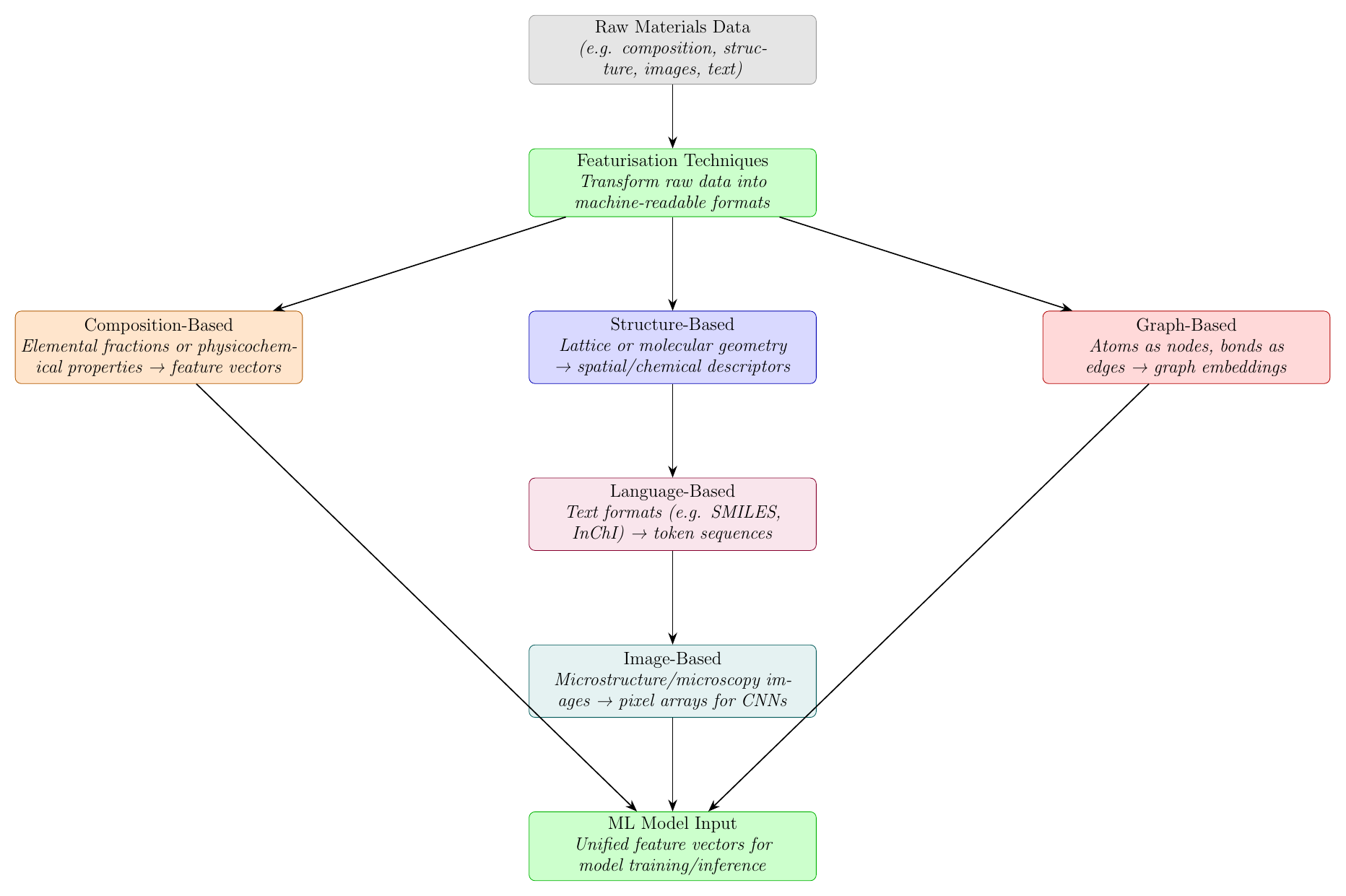}
    \caption{Illustration of various data representation (featurisation) techniques used in materials science for machine learning applications. Raw materials data, including composition, structure, images, and textual information, is transformed through multiple featurisation strategies. These include composition-based vectors, structural descriptors, graph representations, language-inspired formats such as SMILES, and image-based encodings. The resulting feature sets are used as inputs for training or inference in machine learning models, enabling data-driven materials design and discovery.}
    \label{fig:feature}
\end{figure}

Data representation and featurization in materials science transforms raw material data into numerical or categorical descriptors suitable for ML models \cite{johnson2020invited}, Figure \ref{fig:feature}. This critical step bridges the diversity and complexity of material characteristics (such as structure, composition, and properties) and computational ML models designed for prediction and analysis. Effective featurization captures the essence of material behavior in a way that enhances the accuracy, performance, and interpretability of ML algorithms \cite{dean2023interpretable, butler2018machine}.
Whether derived from experimental observations or computational simulations, data on materials is inherently complex and multidimensional. For example, electronic structure calculations using DFT generate outputs such as band structures, total energies, and atomic positions. Transforming these raw outputs into standardized, machine-readable datasets involves generating descriptors or features encapsulating critical material attributes. These features can be hand-crafted, such as those representing chemical composition or structural arrangement, or they can be learned automatically through representation learning techniques, such as autoencoders \cite{Goodfellow-et-al-2016}.
The effectiveness of featurization lies in its ability to transform complex material characteristics into actionable insights. Features may be numerical (e.g., atomic radii or bandgaps), categorical (e.g., crystal structures), or ordinal, each requiring specific preprocessing methods. Advanced techniques, such as one-hot encoding or feature transformations, are often employed to ensure compatibility with ML algorithms.

The categorization of featurization methods depends on the context and objectives of the analysis. In this review, we classify featurization methods as follows:

\subsubsection{General and Domain-Agnostic Features}  \label{domain-agnostic features}

General and domain-agnostic features are derived from featurization techniques requiring minimal domain knowledge, making them broadly applicable across various material systems. These features offer simple yet effective representations, enabling ML models to process material data without relying on detailed scientific insights. Their inherent flexibility and scalability make them particularly well-suited for large-scale datasets and exploratory studies.

These features can be created through approaches such as the random assignment of numerical values (vectors) to elements or encoding the fractional composition of materials. In random featurization, numerical values are assigned arbitrarily to elements in the dataset. For instance, Dhaliwal et al. \cite{dhaliwal2022machine} employed this method to develop computationally efficient ML-based interatomic potentials. Alternatively, fractional representation involves calculating the fraction of each element in a material, with each element assigned a unique identifier. A prominent example of this approach is ElemNet \cite{jha2018elemnet}, which demonstrated the effectiveness of deep learning in predicting material properties directly from elemental compositions, bypassing the need for manual feature engineering.

These techniques, which do not rely on additional physical or chemical property information, are particularly effective in scenarios involving large datasets. Their ability to provide general-purpose descriptors makes them valuable tools for diverse material modeling tasks, especially when detailed domain-specific features are unavailable or impractical to generate.

\subsubsection{Domain-Specific Features} \label{domain-specific}

Domain-specific features are tailored representations that leverage expert knowledge from chemistry, physics, and materials science. These features explicitly encode materials' physical, chemical, structural, or process-dependent properties, ensuring that the resulting descriptors are meaningful and interpretable in material behavior. Unlike general or domain-agnostic features, domain-specific features are customized to capture the intricate details of material systems. This customization makes them particularly effective in scenarios where data is limited or the properties of interest require detailed representation.

One common example of domain-specific features is composition-based feature vectors (CBFVs), which aggregate elemental properties such as mean or variance and apply empirical rules to generate rich and informative representations. These features have been instrumental in modeling various systems of materials.
An illustrative case is the design of compositionally complex alloys (CCAs), where elemental properties, empirical rules, and statistical formulas are applied to generate features for each observation, yielding a richer and more informative representation \cite{peivaste2023data}. These features are highly interpretable, with various studies identifying specific features as particularly impactful for design and modeling. For instance, valence electron conductivity (VEC), a feature often used for phase prediction in high-entropy alloys (HEAs), is calculated using the rule of mixtures. 
VEC is expressed as 
\begin{equation}
X = \sum_{i=1}^{n} c_i(X_i),    
\end{equation}
where \( X_i \) represents the VEC of each component, \( c_i \) denotes the composition of each component, and \( n \) is the total number of components in the alloy. 
Similarly, atomic size difference, another crucial feature, quantifies size mismatch effects and is calculated as 
\begin{equation}
\delta r = 100 \sqrt{\sum_{i=1}^{n} c_i\left(1-\frac{r_i}{r_{\text{mean}}}\right)^2},    
\end{equation}
where \( r_i \) is the atomic radius of each component, and \( r_{\text{mean}} \) is the average atomic radius. In addition to these, mixing enthalpy, given by (\( \Delta H_{\text{mix}} = 4\sum_{i=1, i\neq j}^{n} c_i c_j H_{ij} \)), reflects the energetic interactions between alloy components and plays a pivotal role in understanding alloy stability and behavior. These descriptors provide interpretable insights into phase stability and material properties. For example, VEC has been widely recognized as a key predictive feature in several studies \cite{yifan_wang_phase_2023, yong_liu_focused_2021, wei_zhong_insights_2022, ujjawal_kumar_jaiswal_machine_2021}, while atomic size difference \cite{wanlin_guo_machine-learning_2019, zhu2022phase} and mixing enthalpy \cite{seok_yeong_byeon_deep_2021, mitra2022machine} have also been highlighted for their importance.
The diversity of these features underscores the multifaceted nature of featurization in predicting the behavior of CCAs. Each descriptor offers unique insights into phase stability and material properties, collectively advancing our understanding of these complex systems. Beyond CCAs, this method of featurization has also been applied in developing predictive tools for other materials, such as perovskite solar cells \cite{hu2024interpretable, li2019computational}, superconductors \cite{meredig2018can}, and catalysts \cite{sun2020machine, sun2020covalency}.
Several tools have been developed to work with compositional properties, including Magpie \cite{ward2016general}, Jarvis \cite{choudhary2018machine}, and Oliynyk \cite{oliynyk2016high}. These tools demonstrate the versatility and effectiveness of compositional property-based featurization in advancing materials science.

Domain-specific features are not limited to compositional data but include crystallographic and electronic properties. Structural descriptors such as lattice parameters, space groups, bond lengths, and electronic properties like bandgap energy, density of states (DOS), and HOMO-LUMO gaps for molecules are widely used in materials modeling. These features are often derived from computational methods like DFT or online repositories like the Materials Project Database \cite{jain2013commentary}. For example, rotational angles, bond lengths, and interlayer distances have been used to model transition metal dichalcogenides \cite{ge2020predicted}, while electron affinity, ionization potential, and optical gaps of organic molecules have been employed to predict electronic properties in other studies \cite{li2021combining, bai2019accelerated}. The availability of such features from computational and experimental sources underscores their versatility and significance in representing material.

Experimental parameters are another important type of domain-specific feature, capturing the effects of synthesis or operational conditions on material behavior. Features such as temperature, pressure, pH, and other synthesis conditions provide essential information about how external factors influence material properties. For example, studies such as \cite{gao2021machine, fathinia2016development} utilized experimental parameters to incorporate information about photocatalytic oxide TiO\(_2\) into machine learning models. Similarly, Karim et al. \cite{karim2020coupling} employed control parameters to optimize the synthesis of platinum group metal-free materials.

While developing some domain-specific features can be computationally or labor intensive, these features offer valuable insights into the fundamental properties of materials, making them highly effective for targeted applications.

\subsubsection{Graph and Structural Features} \label{graph_structural}
Graphs and structural representations encode materials as networks, where atoms are represented as nodes and their interactions (such as bonds, distances, and angles) are edges. These representations capture the spatial and topological arrangement of atoms within a material, making them particularly suited for property prediction, structural stability analysis, and material discovery. As discussed in section \ref{gnn}, GNNn architectures are specifically designed for such data, with nodes representing atoms or ions and edges representing bonds, interatomic distances, or angular relationships. Foundational GNN architectures tailored for materials include SchNet \cite{schutt2018schnet} and Crystal Graph Convolutional Neural Networks (CGCNN) \cite{xie2018crystal}. These models refine the representation of each atom by considering its neighboring atoms and the bond lengths connecting them, aggregating the updated atom representations into a comprehensive structure-level representation.

Since their introduction in 2018, SchNet and CGCNN have spurred a surge of advancements and innovations in GNN architectures for materials science applications. The development of GNNs for atomic structures has followed three key trends. 

The first trend focuses on enhancing geometric expressiveness by incorporating more detailed geometric features beyond bond lengths to achieve greater local descriptive power. Examples include models that explicitly encode angle information, such as DimeNet \cite{gasteiger2021gemnet}, GemNet \cite{gasteiger2020directional}, Atomistic Line Graph Neural Network (ALIGNN) \cite{choudhary2021atomistic}, and M3GNet \cite{chen2022universal}. Higher-order geometric information is incorporated in ALIGNN-d \cite{hsu2022efficient} and BOTNet \cite{batatia2022design}. Additionally, directional information is captured using relative vectors in equivariant representations like E3NN \cite{geiger2022e3nn}, NequIP \cite{batzner20223, musaelian2023learning}, and Equiformer \cite{liao2022equiformer}. Other models, such as Learnable Structural and Positional Representations (LSPE) \cite{dwivedi2021graph} and Structure-aware Transformer (SAT) \cite{chen2022structure}, integrate canonical positional and structural information.

The second trend involves incorporating attention mechanisms into GNNs to enhance the graph convolution operation, enabling effective modeling of chemical bonding strengths between atoms. Examples of models utilizing attention mechanisms include SAT \cite{chen2022structure}, GATGNN \cite{louis2020graph}, Matformer \cite{yan2022periodic}, and CEGANN \cite{banik2023cegann}.

The third trend focuses on advancing readout functions beyond traditional average pooling. More sophisticated approaches, such as global attention mechanisms, facilitate global information exchange when aggregating atomic data. Notable examples include \cite{chen2019graph, chen2022universal, louis2020graph, wu2021representing}. These advancements in GNN architectures aim to enhance local and global modeling capabilities to represent atomic structures accurately. However, as Gong et al. \cite{gong2023examining} observed, while GNNs effectively capture basic local structural descriptors, they struggle to represent the periodicity inherent in crystal structures.

Beyond GNNs, other methods, such as Coulomb Matrices \cite{rupp2012fast}, have been developed to encode materials. Coulomb Matrices represent materials as matrices of interatomic Coulombic interactions, where rows and columns correspond to atoms, and elements encode Coulombic interactions based on atomic charges and distances. The sorted Coulomb Matrix \cite{montavon2013machine}, which sorts eigenvalues for permutation invariance, improves prediction stability.

Additional structural encoding methods include Smooth Overlap of Atomic Positions (SOAP) \cite{Bartok2012OnRC, de2016comparing} and Behler–Parrinello Atom-Centered Symmetry Functions (ACSF) \cite{behler2007generalized, behler2011atom, behler2014representing}. These are widely used in machine learning potentials to predict atomic forces and energies. SOAP describes atomic environments by computing the overlap of Gaussian-smeared atomic densities, encoding similarities between local environments and commonly being used in kernel-based methods for structural comparisons. ACSF encodes local environments using radial and angular symmetry functions. While both SOAP and ACSF are local descriptors that do not represent entire structures, information from multiple local sites can be averaged or combined using kernel functions \cite{bartok2017machine}.

\subsubsection{Language-Inspired Featurization} \label{lang_inspired_feature}

Language-inspired featurization adapts techniques from NLP to represent molecules and materials, primarily using string-based representations such as chemical formulas, Simplified Molecular Input Line Entry System (SMILES) strings, and Self-Referencing Embedded Strings (SELFIES). These methods conceptualize molecules as "sentences" and their substructures, such as functional groups or atoms, as "words." By leveraging this analogy, advanced NLP models can generate embeddings that capture materials' chemical and structural nuances.

RNNs and their variants, such as LSTM networks, have been widely employed to encode sequences of polymers and copolymers \cite{simine2020predicting, webb2020targeted, patel2022featurization, goswami2021deep}. For instance, Bhattacharya et al. utilized RNNs with sequence embedding techniques to predict the aggregate morphology of macromolecules, demonstrating the effectiveness of this approach for capturing sequential dependencies in complex molecular systems \cite{bhattacharya2022predicting}.
NLP-inspired methods like Word2Vec \cite{mikolov2013efficient} and transformers \cite{vaswani2017attention} have further advanced the field by encoding molecular substructures into high-dimensional vector spaces. Mol2vec \cite{Jaeger2018Mol2vecUM}, for example, treats substructures derived using the Morgan algorithm as "words" and entire compounds as "sentences." Applying the Word2Vec algorithm to a corpus of compounds generates embeddings that cluster chemically related substructures in the same regions of the vector space, facilitating insights into molecular similarity and behavior.

In recent years, transformers as discussed in section \ref{transformer}  have emerged as a transformative technology, demonstrating exceptional performance across a range of applications. By representing materials as sequences of atomic or molecular structures, generative transformers facilitate both forward and inverse design, significantly accelerating the discovery and development of new materials with targeted functionalities \cite{KOROLEV2023100803, huang2023materialsinformaticstransformerlanguage, wei2022crystaltransformerselflearningneural, chen2024mattergptgenerativetransformermultiproperty, Kamal2024, fu2022materialstransformerslanguagemodels}. Furthermore, unlike some black-box generative models, transformers inherently provide attention maps that reveal which atomic or molecular features contribute most significantly to their predictions \cite{KOROLEV2023100803, huang2023materialsinformaticstransformerlanguage}.

ChemBERTa \cite{chithrananda2020chemberta, ahmad2022chemberta2chemicalfoundationmodels}, a transformer model inspired by BERT (Bidirectional Encoder Representations from Transformers), effectively learns robust molecular embeddings from sufficient data. Reinforcing the utility of such models in materials science, various transformer architectures have been successfully applied to encode small organic molecules \cite{Honda2019SMILESTP, ying2021transformers, irwin2022chemformer, magar2022crystal}. This methodology has also been adapted for larger structures, with PolyBERT \cite{kuenneth2023polybert} and TransPolymer \cite{xu2023transpolymer} serving as examples for encoding polymer sequences.

More recently, transformers have been extended to structure-agnostic applications in materials modeling. For instance, they have been used to predict the properties of metal-organic frameworks (MOFs) from text string representations. Cao et al. introduced MOFormer, a transformer-based model that achieved state-of-the-art results in property prediction tasks for MOFs \cite{cao2023moformer}. These advancements highlight the growing potential of language-inspired featurization in addressing complex challenges in materials science. 

\subsubsection{Image-Based Featurization} \label{image_base_fea}

Image-based featurization derives meaningful representations of materials from image data, playing a critical role in integrating machine learning (ML) with materials science. Images in materials science come from diverse sources, with microscopy marking a transformative advancement in visualizing material structures. Recent developments in super-resolution microscopy, such as stimulated emission depletion microscopy (STED) and photoactivated localization microscopy (PALM)/stochastic optical reconstruction microscopy (STORM), have revolutionized imaging capabilities, allowing the visualization of sub-nanoscale structures \cite{blom2014sted}. These techniques provide unprecedented insights into critical material properties, including crystal structures, phase structures, defects, and nanoscale features, using tools like transmission electron microscopy (TEM) and scanning electron microscopy (SEM) \cite{dadsetan2022real, lu2021situ}. Atomic force microscopy (AFM) has also been instrumental in advancing nanofabrication processes and analyzing chemical compositions \cite{che2023deep}.

In image-based featurization, materials are typically represented as arrays of pixels in 2D, 3D, or even 4D formats, where the additional dimension might represent attributes such as RGB layers or time steps. Beyond physical imaging, computational data—such as arrays of field variables derived from simulations—can also be treated as image-like data, enabling similar analysis techniques \cite{peivaste2024rapid, peivaste2022machine}. This flexibility expands the applicability of image-based featurization across diverse domains.

With increasing complexity, three primary approaches are used to construct features from images. The simplest involves treating raw pixel intensities as features, where each pixel corresponds to a grayscale or color intensity value \cite{kim2014material}. A more advanced approach applies classical image processing techniques to extract patterns such as edges, textures, and morphological features. Edge detection methods \cite{vincent2009descriptive, rong2014improved, rani2020edge}, texture analysis \cite{venkatesh2022analysis, webel2018new, velichko2008unambiguous, decost2017exploring}, and morphological operations \cite{chen2022application, jivani2022skeletal, polak2023structural} enable the characterization of shapes, sizes, and spatial distributions of microstructures. 

The most sophisticated approach leverages deep neural networks (DNNs), such as convolutional neural networks (CNNs), to automatically learn hierarchical features from image data. CNNs are particularly effective at processing and analyzing spatial information, with convolutional layers extracting localized patterns (e.g., edges or corners) and deeper layers identifying abstract features like shapes and textures. Fully connected layers aggregate these features for downstream tasks such as property prediction \cite{chollet2016convolutional}. Additionally, CNNs can be integrated with architectures like autoencoders or variational autoencoders (VAEs), which map input data through a reduced-dimensional bottleneck before reconstructing it. This process maximizes the information encoded within latent variables, creating compact and meaningful representations \cite{ji2024towards}.

Recent studies have demonstrated the application of 2D and 3D CNNs to atomistic configurations and crystal structures. For instance, 2D images have been used to represent atomic configurations \cite{mianroodi2022lossless}, while 4D arrays encode 3D crystal structures with the fourth dimension representing atom types \cite{peivaste2024rapid}. These approaches effectively leverage the strengths of CNNs in capturing spatial and structural relationships within materials data.

In continuum modeling, physical phenomena such as stress, temperature, or pressure are typically represented as continuous field variables over spatial or temporal domains. These variables, governed by partial differential equations (PDEs), are discretized onto grids or meshes for numerical simulations. These discretized fields can be reformulated as structured arrays of data points, analogous to images, where spatial coordinates map to grid points. Additional variables or time steps can be incorporated as channels or dimensions, similar to RGB channels in image processing \cite{smith2024convolutional}. CNNs are particularly adept at handling such multi-dimensional tensors, capturing spatiotemporal correlations and enabling accurate representation of physical phenomena. This technique has been applied in solid mechanics \cite{mianroodi2021teaching, kapoor2022surrogate, khorrami2023artificial}, phase-field modeling \cite{choi2024accelerating, peivaste2022machine, montes2021accelerating, farizhandi2023spatiotemporal, alhada2024transfer}, and fluid dynamics \cite{guo2016convolutional, kochkov2021machine, thuerey2020deep, kim2019deep, bhatnagar2019prediction}.

\subsubsection{Hybrid Features}

Hybrid features may combine various types of features, including manually crafted descriptors derived from domain-specific or domain-agnostic knowledge, as well as learned representations (embeddings) generated by advanced models such as GNNs, CNNs, transformers, autoencoders and VAEs \cite{mai2022machine, Goodfellow-et-al-2016, maqsood2024future}. This methodology enables researchers to capture diverse aspects of material systems, spanning them. Researchers often employ feature selection and dimensionality reduction techniques to address these challenges in micro-scale interactions, while image-based features capture microstructural characteristics critical for predicting material properties. This hybrid approach has succeeded in various applications \cite{lin2020directly, generalpurpose_ward_2016, peivaste2023data}.
Despite its advantages, hybrid featurization presents challenges. Generating and integrating features from multiple methods can be resource-intensive and lead to complex, high-dimensional datasets that are difficult to manage and interpret. Including too many features can increase model complexity, potentially causing overfitting and reducing predictive performance \cite{geron2022hands}. To address these challenges, researchers often employ feature selection and dimensionality reduction techniques to optimize the feature set.

Feature selection aims to identify a sparse subset of features that improve model predictivity while simplifying interpretation. Two commonly used strategies are down-selection and dimensionality reduction \cite{mai2022machine}. Down-selection reduces a large pool of descriptors to a manageable subset using statistical methods. For instance, adding an \( L1 \) regularization term to a regression model penalizes less relevant features, shrinking their coefficients to zero, as seen in the least absolute shrinkage and selection operator (LASSO) \cite{hastie2017elements}. More recently, the sure independence screening and sparsifying operator (SISSO) has gained popularity for handling large and correlated feature spaces, identifying optimal feature combinations relevant to specific material properties \cite{ouyang2018sisso}. Additionally, tree-based algorithms such as RF and XGBoost are widely used to assess the importance of descriptors during model training, providing robust tools for feature selection \cite{louppe2013understanding}.
Dimensionality reduction offers another effective strategy by projecting high-dimensional features into a lower-dimensional space, generating new descriptors as linear combinations of the original ones. Principal component analysis (PCA) \cite{song2010feature} is one of the most commonly used techniques for this purpose. PCA identifies principal components—orthogonal vectors that maximize the variance in the data—thereby reducing dimensionality while retaining essential information. This approach mitigates the curse of dimensionality and accelerates ML model computation, making it particularly valuable for complex datasets. PCA has been extensively applied in materials science, including heterogeneous catalysts, photovoltaics, supramolecular materials, and high-entropy alloys, where simplifying analysis without losing critical insights is essential \cite{liu2020scalable, li2020adaptive, garcia2019statistical, yu2022accelerating, bundela2022application, dai2020using}.

Hybrid features provide a powerful framework for representing complex material systems. This approach enables more accurate and interpretable ML models while addressing the challenges posed by high-dimensional and heterogeneous datasets, accelerating advancements in materials discovery and design.

\begin{longtable}{@{} >
{\raggedright}p{2.5cm} >{\raggedright}p{3cm} >{\raggedright}p{3cm} >{\raggedright}p{3cm} >{\raggedright\arraybackslash}p{2.5cm} @{}}
\caption{Overview of Key Data Aspects in Materials Informatics} \label{tab:data_challenges}\\ 
\toprule
\textbf{Data Aspect} & \textbf{Description} & \textbf{Advantages / Strengths} & \textbf{Limitations / Challenges} & \textbf{When NOT to Use (Considerations)} \\
\midrule
\endfirsthead

\multicolumn{5}{c}%
{{\bfseries \tablename\ \thetable{} -- continued from previous page}} \\
\toprule
\textbf{Data Aspect} & \textbf{Description} & \textbf{Advantages / Strengths} & \textbf{Limitations / Challenges} & \textbf{When NOT to Use (Considerations)} \\
\midrule
\endhead

\midrule

\endfoot

\textbf{Data Representation (Featurisation)} & Transforms raw materials data into structured numerical/categorical features suitable for ML models. & Enables learning and predictive modelling of material properties; improves model interpretability and accuracy. & Poor feature engineering can lead to suboptimal or misleading results. & When selected features fail to capture the relevant mechanisms or physics. \\

\midrule
\textbf{Domain-Agnostic Features} & General-purpose features requiring minimal domain knowledge; broadly applicable. & Easy to compute; transferable across systems. & May overlook critical domain-specific phenomena and constraints. & When specialised structural or mechanistic information is essential. \\

\midrule
\textbf{Composition-Based Feature Vectors (CBFVs)} & Derived from elemental composition (e.g., VEC, $\delta r$, $\Delta H_{\text{mix}}$). & Fast to compute; useful for phase stability and general trends; human-interpretable. & Ignores atomistic arrangement and bonding complexity. & When atomic-scale structure or topology significantly influences behaviour. \\

\midrule
\textbf{Structure-Based Descriptors} & Capture atomic or crystalline structure, e.g., GNNs, Coulomb matrices, SOAP, ACSF. & Effective for modelling atomic environments and non-Euclidean geometry; GNNs encode topological information. & May struggle with periodicity (GNNs) or long-range order (SOAP/ACSF). & When global structural features or periodic interactions dominate material behaviour. \\

\midrule
\textbf{Language-Inspired Featurisation} & Treats materials as text using NLP-inspired encodings like SMILES, SELFIES, Word2Vec, Transformers. & Leverages sequence models to learn chemical patterns; scalable with deep learning. & Loses spatial/geometric information; harder to interpret physically. & When 3D structure and spatial context are central to material properties. \\

\end{longtable}

\subsection{Gaps Analysis in Data and Data Representation }
The successful application ML in materials science fundamentally depends on the quality, representation, and processing of data. As highlighted in Table~\ref{tab:data_challenges}, the journey from raw materials data to reliable ML predictions encompasses several non-trivial challenges that, if overlooked, can undermine model accuracy, interpretability, and generalisability. One of the foremost issues is the \textbf{central importance of data preparation}. Contrary to common belief, model selection or algorithmic sophistication alone cannot compensate for poor data quality or insufficient pre-processing. In practice, a significant proportion of the effort in ML projects is devoted to curating, cleaning, and structuring datasets. The process of \textbf{data representation}, also known as \textit{featurisation}, is especially critical. Transforming materials data—such as atomic structures, compositions, or crystallographic features—into machine-readable numerical descriptors determines the learning potential of ML algorithms. Approaches range from simple domain-agnostic descriptors to highly domain-specific representations such as composition-based feature vectors (CBFVs) or atomistic descriptors like SOAP, ACSF, or graph-based encodings via Graph Neural Networks (GNNs). While general features offer simplicity and transferability, they may miss vital mechanistic subtleties. Conversely, advanced features offer rich information but may struggle with scalability or periodicity, particularly in crystalline solids. Another promising but complex strategy involves \textbf{language-inspired featurisation}. By drawing from natural language processing (NLP), representations such as SMILES or SELFIES encode molecular or materials structures as sequences. These approaches are particularly effective for organic compounds and small molecules, yet they risk discarding important spatial or three-dimensional information that is critical in extended systems.

The role of \textbf{data pre-processing}—including outlier detection, scaling, imputation, dimensionality reduction, and dataset balancing is equally important. Improper or overly aggressive pre-processing can introduce bias or inadvertently remove important signals. Techniques like Principal Component Analysis (PCA), Synthetic Minority Over-sampling Technique (SMOTE), and generative approaches (e.g., GANs) can help, but their applicability depends heavily on the nature and structure of the data. Persistent challenges include \textbf{data sparsity, heterogeneity, and lack of standardisation}. Many datasets in materials science remain incomplete, inconsistent, or lack structural annotations. Furthermore, the absence of community-wide standards for data reporting impedes interoperability, validation, and the development of generalisable models. To overcome these obstacles, future efforts should prioritise the development of open, high-quality, and standardised datasets, along with domain-adapted feature engineering techniques and preprocessing protocols tailored to materials data. Only through rigorous data-centric practices can ML methods achieve robust and reliable results in materials design and discovery.

%%%%%%%[[[[[]]]]]

%%%%%%%
\subsection{Explainable Machine Learning}
Explainable Machine Learning (XAI) is a rapidly emerging field that has the potential to transform how machine learning models are applied in materials science. With the increasing use of ML models, particularly Deep Neural Networks (DNNs), in materials research, there is a growing need to understand how these models arrive at their predictions. XAI aims to provide insights into the decision-making processes of complex models, which is essential for scientific exploration and hypothesis generation.

While ML models such as DNNs can achieve remarkable predictive accuracy, they are often viewed as "black boxes", making it difficult for researchers to understand the underlying causal relationships. This lack of transparency can limit the usefulness of ML models in fields like materials science, where understanding the reasoning behind model predictions is crucial. "Explainable Machine Learning in Materials Science" provides a comprehensive overview of the current state of XAI and its applications in materials science. Several challenges in applying these techniques, particularly in domains like materials science, where data is often complex, sparse, and lacks ground truth \cite{zhong2022explainable}. Indeed, while simple models such as linear regression and decision trees are inherently transparent, more complex models like Support Vector Machines (SVMs) and DNNs often operate as "black boxes"  often operate in a way that is difficult to interpret.
To address these issues, various levels of explainability have been identified, which can guide the understanding of machine learning models:

\begin{itemize}
    \item \textbf{Simulatability:} The model is simple enough for a human to understand entirely (e.g., linear regression).
    \item \textbf{Decomposability:} Only certain parts of the model are explainable, such as individual parameters or specific functions.
    \item \textbf{Algorithmic Explainability:} The learning algorithm itself is simple and converges to a single, interpretable solution.
\end{itemize}

Good explanations are characterized by four key features:
\begin{itemize}
    \item \textbf{Contrastive:} They explain why one decision was made over another.
    \item \textbf{Selective:} They simplify the model and focus on revealing the main causes.
    \item \textbf{Causal:} They emphasize causal relationships rather than probabilistic ones.
    \item \textbf{Social:} The explanation serves as an interaction, facilitating understanding between parties.
\end{itemize}

To enhance the interpretability of machine learning models in materials science, several techniques have been developed, ranging from post-hoc to ante-hoc methods. Post-hoc approaches, such as feature importance, example-based explanations, and surrogate models, aim to explain the behavior of complex models after they have been trained. In contrast, ante-hoc techniques focus on designing models that are inherently interpretable. These methods offer valuable insights into model predictions and help bridge the gap between predictive accuracy and the need for transparency in fields like materials science.

\begin{itemize}

\item  \textbf{Feature importance} is one of the most widely used post-hoc techniques, where the contributions of different features are visualized, often using heatmaps. This is particularly useful in models like DNNs that process complex data, such as X-ray Diffraction (XRD) spectra.
    
\item \textbf{Example-Based Explanations}
ML models can also be explained by providing examples of similar data points that contributed to a specific prediction. This method is useful for understanding how instrumental variations, such as those found in Scanning Electron Microscopy (SEM) images, affect material property predictions.

\item \textbf{Inherently Interpretable Models}
Ante-hoc techniques focus on building models that are designed for interpretability from the outset. For example, SchNet is a DNN model designed to predict quantum properties of atomistic systems while incorporating domain-specific knowledge to improve explainability.

\end{itemize}

Despite the potential benefits of XAI in materials science, several challenges remain. One significant issue is the lack of ground truth data, as materials data is often sparse and difficult to validate, complicating the process of evaluating model explanations. Additionally, the current evaluation methods for XAI are often subjective, with varying criteria depending on the technique used. This highlights the need for standardized evaluation frameworks to assess the quality of explanations across different models and applications.

%%%%%%%%%%%%%%%%%%%%%%%%%%%%%%%%
%%%%%%%%%%%%%%%%%%%%%%%%%%%%%%%%
\section{Applications of ML-based Materials Charaterisation}\label{sec-4}

 There is a critical need to develop data-driven approaches, such as ML models and workflows, to enable rapid and automated interpretation of characterisation results. This demand is driven by the growing interest in AI-guided materials design, as well as high-throughput synthesis and measurement. Traditional analysis in material science and engineering relies on experts with domain knowledge and experience to interpret qualitative and quantitative information. When the volume of data being generated becomes too fast and too large for manual analysis, there is a clear need to develop ML workflows that allow for rapid and objective analysis of structural characterisation data. indeed, the growth of materials data from experiments and simulations surpasses what can be manually handled. This makes it essential to develop new data-driven methods for discovering patterns across multiple length and time scales, as well as for identifying structure-property relationships. Advanced imaging technologies play a crucial role in material science research by providing rich insights into material processes and properties. However, image data interpretation heavily relies on the "intuition" of experienced researchers. As a result, many deep graphical features derived from these tools remain underutilised due to difficulties in data processing and correlation identification. Deep learning can effectively address these challenges. ML algorithms, which have made considerable progress in image processing, are capable of extracting comprehensive information directly from large sets of images. Combining image-based characterisation techniques with ML algorithms can yield insight into the structures and properties of advanced materials with the advantages of automation, precision, and high throughput. ML can also assist in automating high-throughput scattering measurements.

 \begin{figure}[h]
    \centering
    \includegraphics[width=0.90\linewidth]{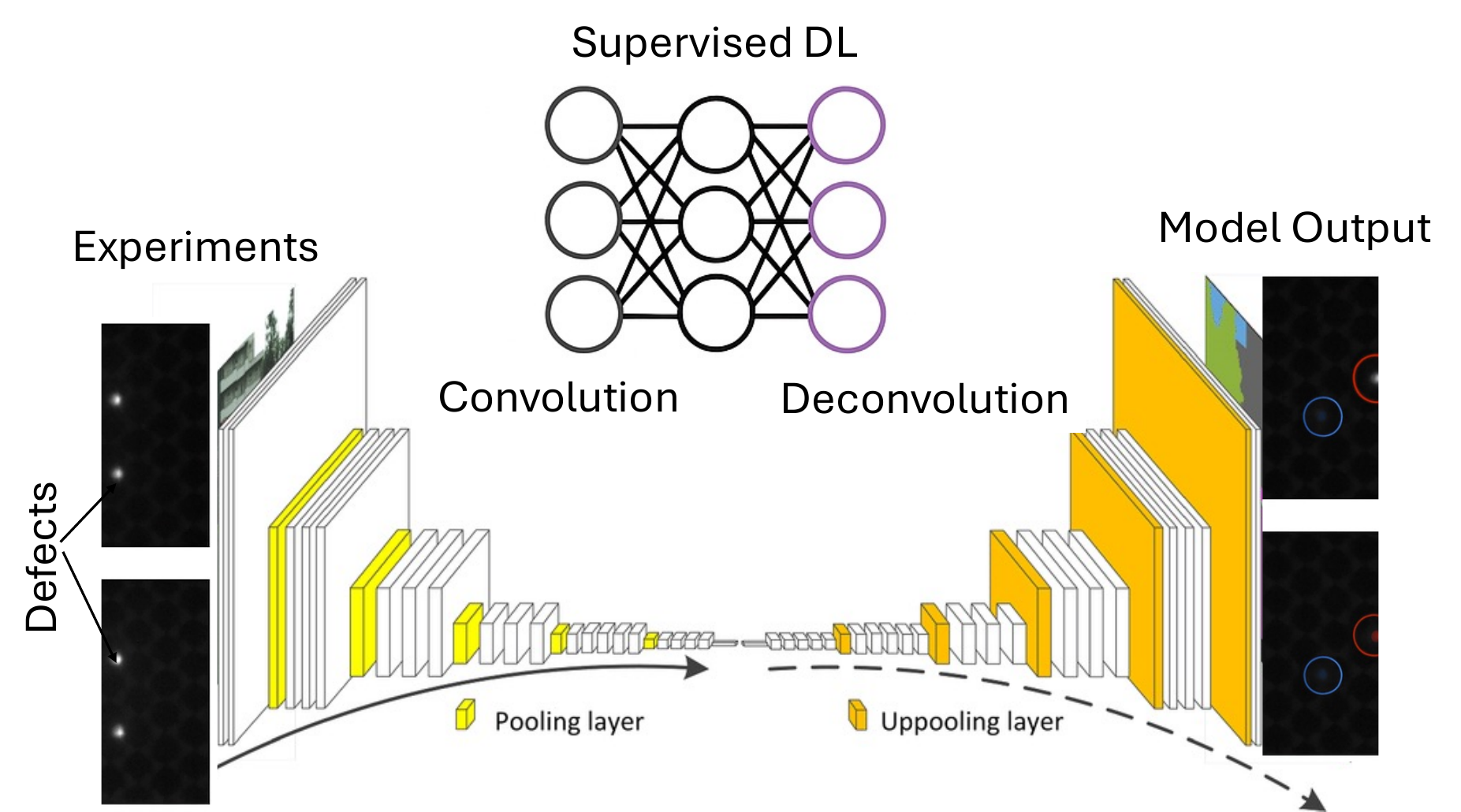}
    \caption{ML and domain knowledge based material characterisation \cite{pilania2021machine}.}
    \label{fig:ML-knoweldge}
\end{figure}

\subsection{Machine Learning in Polymer Characterisation}
Machine learning (ML) plays an increasingly important role in polymer characterisation by providing data-driven approaches to accelerate and automate the interpretation of results and establish structure-property relationships. The literature highlights several key roles. ML is used to quickly and automatically analyse results from common structural characterisation techniques for polymer materials, such as microscopy, scattering, and spectroscopy. Traditionally, interpreting such results relies on expert researchers. ML offers attractive alternatives to overcome challenges related to reference spectra and speeds up data analysis. In microscopy, ML models, especially convolutional neural networks (CNNs), are used to classify microscopy images based on morphology (lamellar vs. spherical), domain shapes (circular, elliptical, fractal), or particle orientation. Transfer learning—where a pre-trained model is fine-tuned with specific experimental data—is a viable approach for small microscopy datasets. Examples include classifying nanowire morphologies from TEM images and polymer blend miscibility. ML has been successfully used for particle detection, segmentation, and shape analysis in microscopy images. Deep neural networks are applied for atomic detection in STEM images and for tracking atomic reorientation. ML is also employed to detect and segment nanoparticles in a polymer matrix. Computational methods are used to automate image acquisition prior to analysis. High-throughput TEM experiments combined with automated image analysis have been used to map the phase diagram of block copolymer amphiphiles. Statistical shape analysis quantifies the size and shape of block copolymer assemblies. ML-based regression methods and deep learning approaches are compared for predicting polymer properties (e.g., glass transition temperature) from AFM microscopy data. Support vector regression models are used to identify structural features from AFM images. In the context of scattering, ML can automate the classification of measured scattering data, linking the measured profile to a specific morphology class (spheres, rods, etc.). CNNs and convolutional autoencoders have been used to classify X-ray scattering patterns of self-assembled polymer films, nanoparticles, and other structures. For qualitative (identifying dominant domain shapes) and quantitative (calculating shape and size distributions) interpretation, ML models are developed to input SAXS and SANS profiles and directly output the most likely shapes of domain structures. Approaches like CREASE (Computational Reverse Engineering Analysis of Scattering Experiments) use optimisation algorithms, such as genetic algorithms, potentially enhanced by ML surrogate models, for interpreting scattering profiles. ML assists in automating high-throughput scattering measurements, as demonstrated by the Autonomous Formulation Lab (AFL) which integrates robotic and software systems for automated synthesis and characterisation using X-ray and neutron scattering techniques. In spectroscopy, unsupervised methods such as PCA, VAEs and t-SNE are used to classify X-ray absorption and emission spectra. ML can also classify functional groups in small molecules using infrared spectra. Bayesian optimisation is applied to guide the design and testing of next-generation biomacromolecules using spectroscopic data. ML models, such as variational autoencoders (VAEs) and generative adversarial networks (GANs), are used to reconstruct and generate characterisation data from complementary techniques. For instance, VAEs can link and reconstruct microscopy and scattering data, while GANs are used for 3D reconstruction from 2D slices.

\subsection{Machine Learning for the Characterisation of Metallic Materials}

For metallic systems, traditional experimentation and physical metallurgy approaches are resource-intensive and time-consuming. This necessitates the development of efficient data-driven workflows that can link structure and properties effectively. ML techniques have been applied to enhance the interpretation and prediction of results from traditional destructive testing methods for metals.  Support Vector Machines (SVM) have been used for image-based corrosion detection in pipelines, where surface texture features are extracted for classification \cite{svmcorrosion}. Comparative studies have explored other ML algorithms such as Artificial Neural Networks (ANN), decision trees, and k-Nearest Neighbours (k-NN) for detecting pitting corrosion in stainless steel 316L, incorporating environmental variables as model inputs \cite{corrosioncomparison}. Multivariate linear regression and ANN models have been applied using input features such as chemical composition, processing parameters, and fatigue strength to predict material fatigue resistance. Ensemble methods have proven especially effective in this domain \cite{fatigueprediction}. ML algorithms, including Random Forests (RF), ANN, and k-NN, have been employed to predict tensile behaviour of cast alloys based on microstructure images. These models are capable of performing pattern recognition tasks on photomicrographs \cite{tensileprediction}. ML applications to SPT data, including ANN coupled with Finite Element Modelling (FEM), have gained popularity for estimating material properties such as ultimate tensile strength (UTS). Simple ML models, including linear regression and RF, have also been used effectively \cite{sptprediction}.

ML is becoming a powerful ally in the characterisation of metallic materials. Its contributions span high-throughput data analysis, mechanical property prediction, defect detection such as corrosion, and the optimisation of traditional testing methods. Nevertheless, building robust and generalisable ML models requires the availability of larger and diverse datasets. Community-wide efforts are needed to share both successful and negative results to strengthen the reliability of ML in metallurgy.  Another  major limitation in applying ML in materials science is the scarcity of large, labelled datasets. High-quality experimental and simulation data are often costly and time-consuming to obtain, which restricts the success of complex ML models. Therefore, there is an ongoing need to develop algorithms tailored for small data regimes and to adopt strategies such as transfer learning and synthetic data generation. The long-term vision  includes enabling and accelerating new materials discovery by integrating ML with high-performance computing and automation. This includes high-throughput identification and quantification of key material properties and the establishment of structure-property-performance relationships. ML helps map often non-linear correlations between properties and underlying microstructural features, allowing researchers to gain insights across different length and time scales.

\subsection{Machine Learning for the Characterisation of Composite Materials}

The application of machine learning (ML) in the field of composite materials is experiencing rapid growth, opening new avenues for characterisation and composite material design. With the exponential increase in data generated from experiments and simulations, ML-based methods have become essential tools to uncover complex relationships between structure, manufacturing processes, and material properties. ML methods are now employed for the analysis of microstructural characterisation data in composite materials. 

Techniques such as microscopy, diffraction, spectroscopy and tomography generate vast quantities of data, whose interpretation can be automated and accelerated through ML. Convolutional neural networks (CNNs) and other computer vision architectures are used for feature extraction from microstructural images, phase segmentation, and morphology classification. These approaches enable quantitative microstructure analysis and the establishment of links with macroscopic properties. In composites, ML facilitates damage characterisation, particularly under compression. ML models can be trained using large virtual datasets generated by finite element (FE) simulations to predict and characterise the mechanical response of composites under compressive loads. For instance, recurrent neural networks with Long Short-Term Memory (LSTM) architecture have been used to solve direct FE problems and compare their predictive capabilities with inverse approaches. ML also supports the generation of cross-modal characterisation data. By using complementary data obtained from different characterisation techniques on the same system, ML models can learn to infer one type of data from another. This is useful for filling in missing information or establishing correlations across scales and modalities.

\begin{longtable}[c]{|p{4cm}|p{5cm}|p{5cm}|}
\caption{Overview of ML-based methods for Materials Characterisation} \label{tab:ML-characterisation}\\
\hline
\textbf{ML Technique} & \textbf{Application in Materials Characterisation} & \textbf{Use Case Example} \\
\hline
\endfirsthead
\hline
\textbf{ML Technique} & \textbf{Application in Materials Characterisation} & \textbf{Use Case Example} \\
\hline
\endhead
\hline
\endfoot
\hline
\endlastfoot
CNNs & Image-based feature extraction, phase segmentation, morphology classification & Classifying TEM images for nanowire morphology and polymer blend miscibility \\
\hline
Support Vector Machines (SVM) & Image-based classification and detection & Corrosion detection in pipelines based on surface texture features \\
\hline
Artificial Neural Networks (ANN) &  Defect detection, classification & Tensile behaviour of cast alloys from microstructure images \\
\hline
k-Nearest Neighbours (k-NN) & Classifying material states and defect detection & Pitting corrosion detection in stainless steel and tensile strength prediction \\
\hline
Recurrent Neural Networks (RNN) / Long Short-Term Memory (LSTM) & Time-series prediction, sequence modelling for materials response & Mechanical responses of composites \\
\hline
Generative Adversarial Networks (GANs) & 3D reconstruction from 2D slices, data augmentation & Generating missing data in microscopy and scattering results \\
\hline
Variational Autoencoders (VAE) & Data compression, unsupervised feature extraction, generation of synthetic data & Reconstructing scattering data from microscopy images \\
\hline
Principal Component Analysis (PCA) & Dimensionality reduction, feature extraction for spectral data & Classifying X-ray absorption spectra and functional group identification \\
\hline
Bayesian Optimisation & Efficient material design, parameter optimisation & Guiding the design and testing \\

\hline

\end{longtable}

\subsection{Gaps Analysis in ML-based Materials Characterisation}

Despite recent advances in machine learning (ML) for materials characterisation, several key limitations persist in current research and practice:

\begin{itemize}
    \item \textbf{Lack of Standardised and Open Datasets:} 
    Publicly available and well-curated datasets for characterisation tasks are scarce. This limits model training, benchmarking, and reproducibility, especially in fields like microscopy or spectroscopy where annotated data is expensive to generate.

    \item \textbf{Limited Integration of Multi-modal and Multi-scale Data:} 
    Most ML applications focus on single-modality datasets. There is a need for ML frameworks capable of learning from heterogeneous, multi-scale data (e.g., linking microscopy and spectroscopy) to capture the complexity of structure-property relationships.

    \item \textbf{Insufficient Incorporation of Domain Knowledge:} 
    Many ML models lack physical interpretability. Embedding domain knowledge—such as symmetry constraints, conservation laws, or physics-based priors—remains underutilised and is essential for model reliability.

    \item \textbf{Challenges in Uncertainty Quantification (UQ):} 
    UQ is critical for assessing model reliability, especially in high-stakes applications. Most current pipelines do not quantify predictive uncertainty or epistemic variability, which hinders deployment in experimental settings.

    \item \textbf{Data Scarcity and Small-sample Learning:} 
    Experimental data in materials science is often limited. Approaches such as transfer learning, active learning, few-shot learning, and synthetic data generation should be further developed to address data scarcity.

    \item \textbf{Automation and Real-time Feedback:} 
    ML is mostly used for offline data analysis. Real-time integration of ML into experimental workflows (e.g., adaptive microscopy or in-situ feedback) remains underexplored and could significantly improve throughput and discovery.

    \item \textbf{Benchmarking and Evaluation Metrics:} 
    There is a lack of standard evaluation metrics and public benchmarks tailored to characterisation tasks (e.g., phase segmentation, morphology classification). This limits objective comparison between ML methods.

    \item \textbf{Model Generalisation Across Materials Systems:} 
    Many ML models show poor generalisation to unseen materials systems or experimental conditions. Improving transferability and robustness is key to broader adoption in materials research.
\end{itemize}

Addressing these gaps will require interdisciplinary collaboration and the development of interpretable, uncertainty-aware, and physics-informed ML models tailored for materials characterisation tasks.

%%%%%%%%%%%%%%%%%%%%%%%%%%%%%%%%
%%%%%%%%%%%%%%%%%%%%%%%%%%%%%%%%
\section{Applications of ML in Materials Modeling and Design}\label{sec-5}

The integration of AI, particularly ML, into materials science and engineering has significantly advanced traditional design, analysis, and prediction methodologies. This integration provides novel avenues for addressing complex, multidisciplinary challenges. Advanced AI techniques, offer powerful tools for modeling and predicting mechanical, electronic, thermal, chemical, and process-related properties of complex material systems with enhanced accuracy, efficiency, and reliability. This section details key application areas where ML is making substantial contributions: property prediction across various length scales, multiscale modeling, materials discovery, process optimization, and sustainability assessment.

\subsection{ML-based Surrogate Modelling for Materials Design}\label{application-3}

\begin{figure}[h!]
    \centering
    \includegraphics[width=0.8\linewidth]{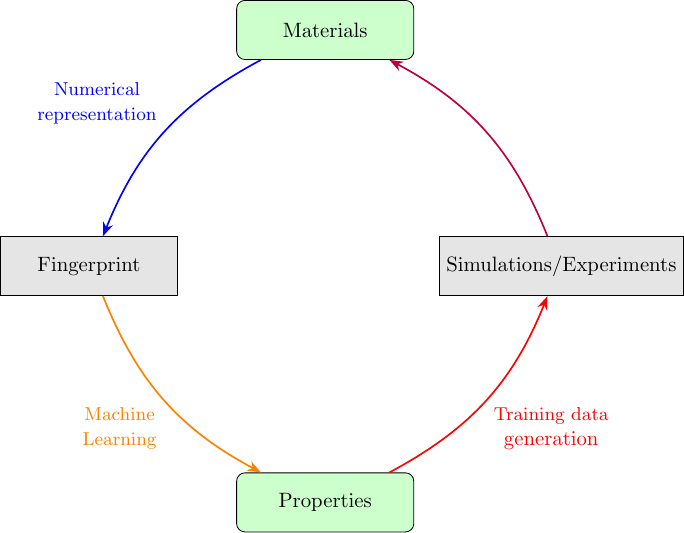}
   \caption{Surrogate model development workflow in materials science. The circular diagram illustrates the iterative process involving: (i) \textbf{Materials} design, (ii) generation of a numerical \textbf{Fingerprint} (descriptor), (iii) prediction of \textbf{Properties} via machine learning models, and (iv) generation of training data through high-fidelity \textbf{Simulations/Experiments}. Colored curved arrows highlight the information flow and interdependence among steps.}
    \label{fig:Surrogate-ML}
\end{figure}

Surrogate modelling has emerged as a powerful alternative to traditional high-fidelity simulations in computational mechanics, particularly when dealing with complex, multiscale systems such as composite materials. 
Surrogate models serve as efficient approximations for high-fidelity computational models, enabling the rapid prediction of structural behaviours at a fraction of the computational cost. The development of surrogate models for materials modeling and design involves an iterative, data-driven workflow aimed at enabling efficient property prediction and performance optimization, Figure \ref{fig:Surrogate-ML}. At the core of this approach lies the transformation of material microstructure or molecular material systems into structured, machine-readable representations known as fingerprints or descriptors. These descriptors are designed to encapsulate microstructure feature or  physical, chemical, and structural characteristics of the materials, thereby enabling algorithms to perform on these "systems". Once the fingerprint is established, machine learning techniques are employed to model the relationships between material structure and targeted properties ranging from mechanical strength to electronic behavior or thermal conductivity. These predictive models are trained on datasets derived from either high-throughput simulations or experimental measurements. As predictions are generated, they inform the selection or design of new candidate materials, which are then subjected to additional simulations or experiments to validate and expand the training dataset. This results in a self-improving, closed-loop process, where new data continually refine the surrogate model, enhancing its accuracy and generalisability. The integration of this iterative methodology not only reduces computational and experimental costs but also enables rapid screening of vast compositional and structural spaces.

Integrating artificial intelligence (AI) and machine learning (ML) into materials science involves a structured, iterative workflow that mirrors the scientific discovery process, Figure \ref{fig:AI-worflow}. It typically begins with the collection of data from various sources, including experiments, simulations, and existing databases. This raw data must be cleaned and preprocessed to ensure quality and consistency, forming the foundation for training reliable ML models. Once trained, these models can predict key properties or performance metrics of materials, enabling informed decision-making in the design of new materials. The proposed designs are then validated through further experiments or simulations. Importantly, the results of these validations are fed back into the system, allowing the models to be refined and improved over time. 

\begin{figure}[h]
    \centering
    \includegraphics[width=0.9\linewidth]{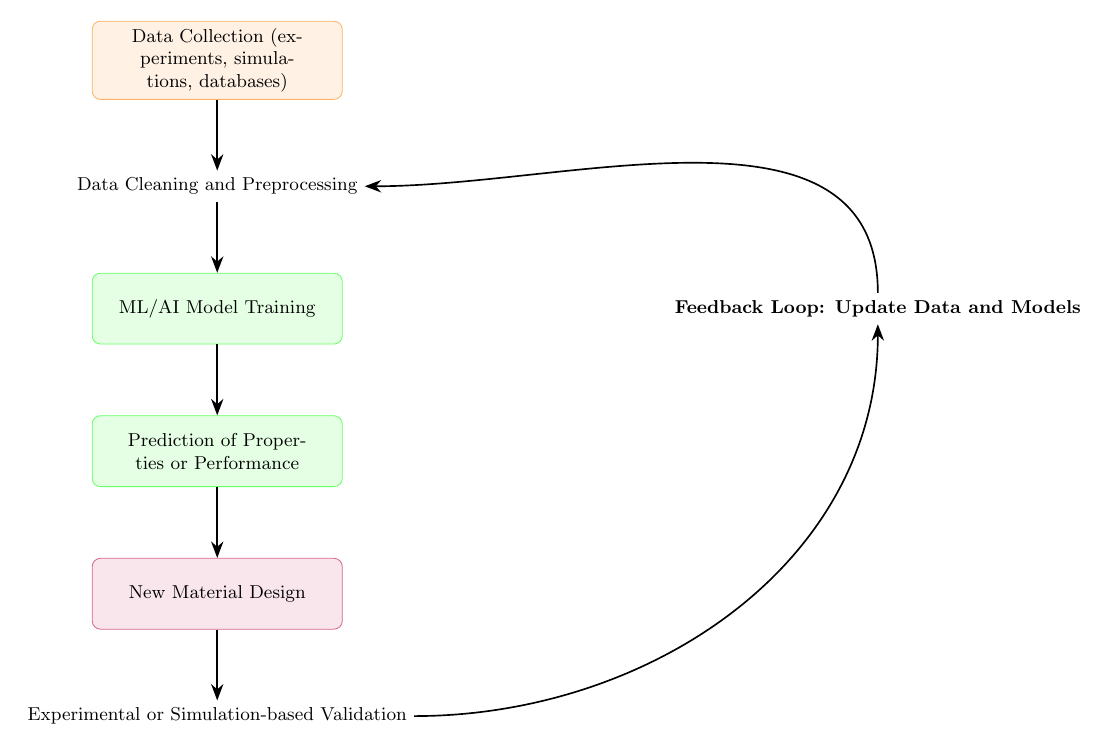}
    \caption{Typical workflow for integrating artificial intelligence (AI) and machine learning (ML) into materials science. The process begins with data collection from experiments, simulations, or databases, followed by data cleaning and preprocessing. ML/AI models are then trained and used to predict material properties or performance, guiding the design of new materials. The workflow is completed with experimental or simulation-based validation, which feeds back into the process to refine models and improve future predictions.}
    \label{fig:AI-worflow}
\end{figure}

  Techniques such as Convolutional Neural Networks (CNNs) and architectures like U-Net have shown great promise in this domain \cite{Deshpande_CNN_UNet_2022}. These models are particularly advantageous when dealing with resource-intensive simulations, making them indispensable for iterative design processes and optimization tasks. Leveraging machine learning (ML) and deep learning (DL) techniques, surrogate models aim to approximate the outputs of computationally expensive simulations at a fraction of the cost. These models are trained on existing data and provide rapid predictions, making them suitable for tasks such as optimisation, design exploration, and real-time simulations \cite{willard2020integrating, wang2021deep}.
  
One of the key applications of surrogate modelling is in replacing classical constitutive models with artificial neural networks (ANNs). These models are trained in a purely data-driven manner by minimising the discrepancy between the network’s output and reference simulation or experimental data. Such approaches are particularly attractive for complex material behaviours where analytical models fall short \cite{bessa2017framework, ghavamian2020multiscale}. Neural networks are widely employed to approximate the solutions of high-fidelity simulations, addressing tasks such as uncertainty quantification (UQ), Bayesian inference, digital twin development, and optimal experimental design \cite{raissi2019physics}. This approximation reduces computational overhead while maintaining acceptable accuracy. PINNs incorporate known physical laws—typically expressed as partial differential equations (PDEs)—directly into the training process. This ensures that the surrogate model's predictions remain consistent with the underlying physics, even in data-sparse regimes. PINNs have been effectively applied to forward and inverse problems, viscoelasticity, and solid mechanics \cite{raissi2019physics, haghighat2021physics}. Graph Transformer Networks have been proposed as a surrogate modelling strategy for long-term prediction of physics-based systems defined on meshes. GTNs aggregate information from neighbouring nodes using attention mechanisms and are trained with multi-step loss formulations. They have shown improved robustness for stress prediction in solid mechanics when coupled with novel activation functions such as symlog and symexp \cite{li2022graph}. Hybrid models combine the rigour of physics-based formulations with the flexibility of data-driven approaches. These models embed learnable parameters within established constitutive laws and allow them to evolve through a data-driven operator. The result is a system where internal state variables act as memory, enabling path-dependent behaviour to emerge naturally \cite{huang2020learning}.

\paragraph{Surrogate Modelling with Uncertainty:} Deploying surrogate models in high-stakes decision-making scenarios requires rigorous treatment of uncertainty. Traditional ML models are often overconfident and prone to overfitting, especially when trained on limited, noisy scientific data. Thus, uncertainty quantification (UQ), verification and validation (V\&V), and model credibility assessments are essential \cite{ohlson2022assessing, huang2021learning}. A major challenge lies in the scarcity and incompleteness of high-fidelity simulation data. This scarcity hinders generalisation and calls for UQ-aware training approaches. Classical maximum likelihood training often leads to overconfident models with poor uncertainty calibration. Robust VVUQ frameworks are crucial when integrating ML into physics-based models (scientific ML or SciML). As surrogate models become more prevalent in composite design and evaluation, systematic VVUQ strategies help ensure predictive reliability \cite{willcox2021role}. Bayesian frameworks are a cornerstone for handling uncertainty in surrogate modelling. Bayesian neural networks (BayesNNs) infer distributions over model parameters and allow for principled uncertainty estimates in predictions. These models mitigate overfitting and quantify epistemic uncertainty, especially in small data regimes \cite{blundell2015weight, gal2016dropout}. OPAL-surrogate (Occam Plausibility Algorithm for Surrogate models) provides a systematic approach to selecting credible surrogate models under uncertainty. It combines hierarchical Bayesian inference with model validation tests to balance the trade-off between model complexity, predictive accuracy, and uncertainty \cite{ohlson2022assessing}. Although PINNs have proven valuable for embedding physics into ML models, their integration with UQ remains challenging. Recent approaches explore encoder-decoder convolutional architectures for UQ, Bayesian formulations for input uncertainty, and reduced-order models for high-dimensional random fields. These methods aim to rigorously quantify both training and generalisation errors \cite{zhang2022quantifying}. In the context of composite materials, surrogate models powered by PINNs have been employed to solve stochastic PDEs that arise from random material properties and geometries. These models yield statistical measures (e.g., mean, standard deviation, probability density functions) that closely match those obtained from Monte Carlo finite element simulations, demonstrating excellent agreement \cite{yang2021physics}.

\begin{longtable}[c]{|p{4cm}|p{5cm}|p{5cm}|}
\caption{ML Techniques for Surrogate Modelling in Material Design} \label{tab:ml_surrogate_materials} \\
\hline
\textbf{ML Technique} & \textbf{Application in Surrogate Modelling for Material Design} & \textbf{Use Case Example} \\
\hline
\endfirsthead

\hline
\multicolumn{3}{|c|}{\textbf{Table \thetable\ (continued): ML Techniques for Surrogate Modelling}} \\
\hline
\textbf{ML Technique} & \textbf{Application in Surrogate Modelling for Material Design} & \textbf{Use Case Example} \\
\hline
\endhead

\hline
\endfoot

\hline
\endlastfoot

Physics-Informed Neural Networks (PINNs) & Embed physical laws into ML models, solve stochastic PDEs, uncertainty quantification & Solving stochastic PDEs in composite materials, generating statistical measures for material properties \\
\hline
Occam Plausibility Algorithm for Surrogate Models (OPAL) & Systematic model selection under uncertainty, combining hierarchical Bayesian inference with model validation & Selecting credible surrogate models balancing complexity, accuracy, and uncertainty for material design \\
\hline
Gaussian Process Regression (GPR) & Provide probabilistic predictions, estimate uncertainty, perform regression with uncertainty quantification & Used to predict material properties in composite design with uncertainty estimates for optimal material selection \\
\hline
Random Forests (RF) with UQ-Aware Training & Incorporate uncertainty quantification into ensemble methods, robust predictions for design & Predicting composite material properties with uncertainty estimates in high-dimensional parameter spaces \\
\hline
Reduced-Order Models (ROMs) & Dimensionality reduction for high-dimensional stochastic problems, accelerate simulations & Applied to composite material simulations with high-dimensional random fields, reducing computational costs \\
\hline
Bayesian Optimization & Efficient search for optimal material properties, balancing exploration and exploitation & Guiding the design of novel materials by incorporating uncertainty quantification into the optimization process \\
\hline
Monte Carlo Methods & Statistical sampling to estimate uncertainty, used for assessing variability in material properties & Used in conjunction with surrogate models to perform uncertainty quantification in material property predictions \\
\hline
Variational Inference & Approximate posterior distributions for uncertain model parameters, reduce computational burden & Applied to surrogate modelling in high-stakes material design, handling large-scale parameter spaces with uncertainty \\
\hline
Neural Tangent Models (NTMs) with UQ & Provide closed-form analysis of model uncertainty, incorporate random processes for robust predictions & Applied in material design, particularly in high-dimensional, uncertain systems for predicting mechanical properties \\
\hline
Bayesian Neural Networks (BayesNNs) & Handle uncertainty in model parameters, infer distributions for uncertainty estimation & Mitigating overfitting and quantifying epistemic uncertainty in small data regimes \\
\hline
\end{longtable}

\subsubsection{Gap Analysis for Materials ML-based surrogate Modelling} 
Despite the considerable progress made in surrogate modelling and uncertainty quantification (UQ), several fundamental challenges persist, particularly in the context of multiscale simulations and complex materials such as composites. These limitations must be addressed to enable the wider adoption and credibility of machine learning (ML)-enhanced computational tools in industrial applications. 
\begin{enumerate}

   \item  \textbf{Data Scarcity and Quality:} A primary bottleneck in the development of accurate surrogate models is the scarcity of high-fidelity training data. Generating this data through detailed simulations or physical experiments is expensive and time-consuming. Moreover, in many cases, the data are noisy, incomplete, or not representative of the full design space. This is especially problematic in composite materials, where uncertainties in fibre orientation, voids, and manufacturing defects significantly affect performance but are difficult to characterise comprehensively. \textbf{Gap: Data-efficient Learning: } Current ML methods often rely on large datasets to generalise well. There remains a critical need for data-efficient learning algorithms, such as transfer learning, active learning, and few-shot learning, that can operate effectively in low-data regimes relevant to engineering. 

   \item  \textbf{Physical Inconsistency and Interpretability:} While surrogate models such as deep neural networks can approximate complex behaviours, they frequently lack adherence to physical principles. This leads to physically inconsistent outputs, particularly when extrapolating beyond the training data. In addition, the black-box nature of these models hinders interpretability, making them less trustworthy for safety-critical applications. \textbf{Gap: Physics-guided and Interpretable ML.}
Despite advances in physics-informed neural networks (PINNs), there is a lack of standardised frameworks to guarantee physical consistency and enforce conservation laws across all surrogate modelling approaches. Moreover, few models provide interpretable latent representations that align with physical concepts such as stress, strain, or energy dissipation.

   \item  \textbf{Uncertainty Quantification under High-dimensionality:} Quantifying uncertainty in high-dimensional spaces, as encountered in multiscale modelling, remains an open challenge. Classical UQ techniques become computationally intractable, and surrogate models often underestimate uncertainty, especially when extrapolating or encountering out-of-distribution inputs. \textbf{Gap: Scalable and Reliable UQ Methods:} Current Bayesian approaches, including BayesNNs and Gaussian Processes, scale poorly with dimensionality and dataset size. Efficient and reliable UQ methods that balance scalability, fidelity, and robustness are still lacking. Furthermore, few existing methods offer comprehensive treatment of both aleatoric and epistemic uncertainty in real-world composite simulations.

   \item  \textbf{Generalisation and Extrapolation Capabilities:} Surrogate models trained in narrow design spaces often fail when applied to new geometries, loading conditions, or boundary conditions. This lack of generalisation undermines their applicability to digital twins, robust design, and optimisation tasks. \textbf{Gap: Domain Generalisation and Adaptivity.} There is a pressing need for adaptive surrogate models that can generalise across domains or automatically adapt to new conditions without retraining from scratch. Meta-learning and domain adaptation techniques are promising but underexplored in the context of physics-based simulations.

   \item  \textbf{Verification, Validation, and Trustworthiness:} The integration of ML into engineering workflows necessitates rigorous verification and validation (V\&V). However, unlike traditional solvers, ML models lack formal guarantees, and their credibility is difficult to assess systematically. The absence of standardised V\&V methodologies for ML-based surrogates impedes their adoption in certification-sensitive sectors such as aerospace and civil engineering. \textbf{Gap: Standardised VVUQ Frameworks.} Although recent frameworks such as OPAL-surrogate provide tools for surrogate model credibility, comprehensive, domain-specific guidelines are still missing. A unified VVUQ methodology tailored for hybrid ML-physics systems is crucial for ensuring reliability in industrial use cases.

   \item  \textbf{Integration into Multiscale and Multi-physics Framework:s} Embedding surrogate models within existing multiscale and multi-physics simulation frameworks presents architectural and methodological challenges. Discrepancies in spatial and temporal scales, interface mismatches, and instability in coupled solvers can undermine the effectiveness of surrogates. \textbf{Gap: Hybrid Coupling Strategies.} Robust and flexible coupling schemes that integrate ML surrogates into multiscale solvers—while ensuring stability, consistency, and efficiency—are still under active development. The lack of modular and interoperable software tools exacerbates this issue.

\end{enumerate}

%%%%%%%%%%%%%%%%%%%%%%%%%%%%%%%%
\subsection{ML-based Property Prediction} \label{property_pred}

In materials research, property prediction aims to estimate material behaviour without performing new physical experiments. Traditional approaches based on first-principles, semi-empirical, or phenomenological models combined with multiscale bridging techniques. While these methods remain essential, they are time-consuming and require substantial computational resources due to the complexity and high dimensionality involved \cite{chong2024advances}. Machine learning (ML) offers a powerful alternative by leveraging large datasets to uncover complex structure-property relationships, enabling faster and more cost-effective predictions.  ML-based methods represent a paradigm shift in this context, enabling the exploration of these intricate relationships through statistical learning and data-driven modelling techniques that can reveal patterns beyond the reach of traditional theoretical frameworks~\cite{damewood2023representations}. This subsection reviews the application of ML methods for predicting material properties at three primary length scales: (1) Atomistic scale, concerning atoms, molecules, and crystal unit cells; (2) Micro/Meso scale, involving microstructures such as grains, phases, and pores, typically ranging from nanometers to microns; and (3) Macro scale, pertaining to the bulk material behavior of components or structures.

\subsubsection{Atomistic-Scale Property Prediction with ML}

At the atomistic scale, fundamental material properties such as formation energies, band gaps, elastic moduli, and optical spectra are often investigated using computationally intensive methods like \textbf{ab initio} calculations (e.g. DFT) or high-throughput experiments. While traditional computational techniques provide high accuracy, their significant computational cost restricts applicability to large systems or long-duration simulations. ML has emerged as a viable alternative, particularly through the development of ML-driven surrogate models, trained on data from high-fidelity calculations, which offer reliable interpolation of properties at a fraction of the computational expense \cite{peivaste2024rapid}.

\begin{figure}[h]
    \centering
    \includegraphics[width=1\linewidth]{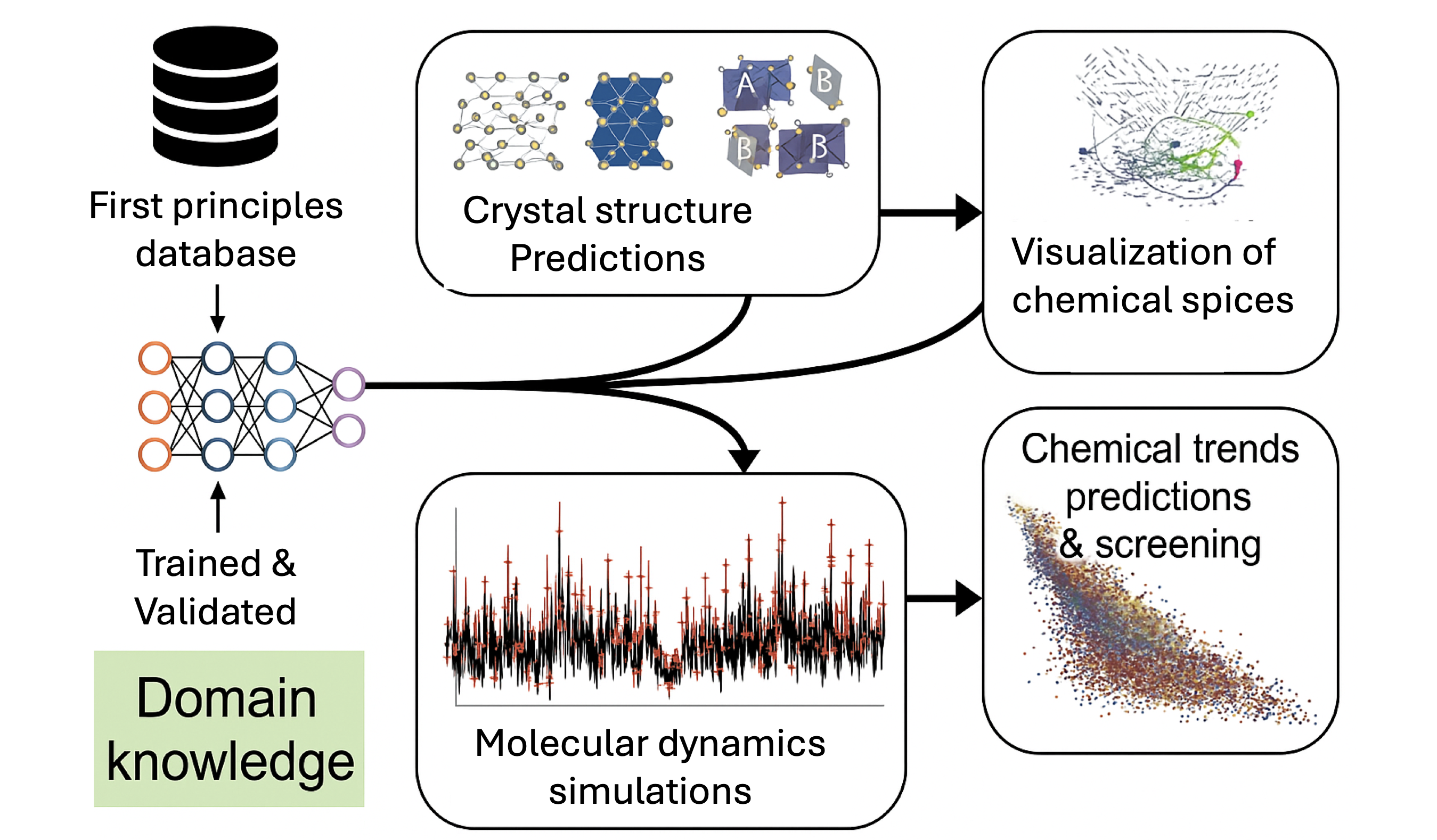}
    \caption{Schematic representation of a machine learning (ML)-driven materials discovery workflow. First-principles databases are used to train and validate ML-based force fields, integrating domain knowledge. \cite{pilania2021machine}.}
    \label{fig:AI-worflow}
\end{figure}

Furthermore, large databases (discussed in section \ref{platforms}) provide tens of thousands to millions of entries of materials with computed properties. This wealth of data is fertile ground for ML models. Early approaches in applying ML at the atomic scale used manually crafted features like general and domain-specific features (refer to sections \ref{domain-specific} and \ref{domain-agnostic features}).  For example, J. Ward et al. \cite{goodall2020predicting} showed that using a set of composition-based features in a Random Forest model could predict band gaps for crystals without requiring the crystal structure as input, achieving decent accuracy for many inorganic compounds. Nyshadham et al. \cite{nyshadham2019machine} developed a surrogate model utilizing several domain-specific featurization methods and DNNs to accurately predict material properties such as elastic constants, enthalpy, and band-gap, achieving precision similar to that of ab initio methods.

In recent years, advances in deep learning have enabled automatic learning of representations such as image-based, graph-based, and sequence-based featurization (sections \ref{graph_structural} and \ref{image_base_fea}). Such featurization methods have achieved significant accuracy in predicting atomistic-scale properties. For instance, Mianroodi et al. \cite{mianroodi2022lossless} utilized a 2D image-based featurization technique to capture atomistic structural images and correlate them with the elasticity tensor, which was determined through molecular statics simulations. In \cite{peivaste2024rapid} a 3D image-based representation was used to make a surrogate model that speeds up atomistic simulation for mechanical property calculation up to 2000 times. In \cite{xie2018crystal} a surrogate model based on graph-based featurization was developed for predicting DFT-calculated properties across eight different crystal properties with various structural types.

Beyond predicting static properties (like a single value for a given material), ML is making enormous strides in predicting the potential energy surfaces that govern atomic interactions. Machine-learning interatomic potentials (MLIPs) have emerged as powerful tools that essentially predict a material’s energy (and forces) given the positions of atoms \cite{Unke2021}. Once an MLIP is trained for a material or class of materials, one can predict various properties: e.g. compute the thermal conductivity via molecular dynamics, predict defect formation energies by evaluating different atomic configurations, or determine mechanical strength by simulating deformation, all using the MLIP to provide fast energy/force evaluations \cite{wang2024machine}.

ML at the atomistic scale not only reproduces known properties efficiently but also guides the discovery of novel materials. For instance, in the field of 2D materials, ML has been employed to screen vast numbers of potential compounds for promising electronic or optical properties. Despite these successes, challenges persist. Despite the successes, challenges remain at the atomistic scale. One key issue is generalization: ML models can sometimes achieve low error on test data similar to training data, yet fail to extrapolate to new types of chemistries or structures. Another challenge is data scarcity for certain properties. While some properties have thousands of known data points (e.g. band gaps, thanks to DFT databases), others like dielectric breakdown strength or crack toughness at the atomic scale are very sparse, limiting direct ML application. Lastly, the interpretability and trust of atomistic ML models are crucial for adoption. Researchers often hesitate to trust a black-box model’s predictions for a new material unless there are ways to explain why the model predicts a certain value.

\subsubsection{Micro- and Meso-Scale Property Prediction}

Many material properties of engineering interest, such as yield strength, fracture toughness, thermal conductivity of composites, magnetic coercivity, or optical scattering, are profoundly influenced by the material’s microstructure \cite{biner2017programming}.  Traditionally, linking microstructure to properties required homogenization methods, empirical models, or finite element simulations on representative volume elements, methods that are often computationally expensive and lack broad applicability \cite{peivaste2025teaching}. ML now offers a data-driven alternative to capture complex, non-linear correlations between microstructure and properties \cite{montes2021accelerating, peng2024can}.

Image-based featurization has been particularly impactful for analyzing microstructures represented as 2D / 3D images or tensors of field variables. By learning hierarchical features from these images, CNNs can statically or dynamically predict properties like effective stiffness, stress-strain response, or ionic conductivity from micrographs. These models have outperformed traditional descriptors in several studies, especially when trained on synthetic datasets generated from simulation or processing models \cite{peng2024can}. For example, Eidel \cite{eidel2023deep} utilized image-based featurization and CNNs to dynamically predict the effective stiffness of 3D random heterogeneous multiphase materials, using microstructure images as inputs. Graph-based featurization also shows promise; Peng and Xu \cite{peng2024unraveling} employed a graph-based featurization and GNNs to predict the effective ionic conductivity of polycrystalline ceramics by capturing features of grains, grain boundaries, and their interactions. Other studies  \cite{chen2022multiscale, dai2021graph, pahlavani2022deep} have applied ML to determine effective properties of microstructured materials, such as the elastic properties of metamaterials, the Young’s modulus of metal foams, and the magnetostriction of polycrystalline materials

While CNNs and GNNS  are powerful, they require large labeled datasets (which can be a limiting factor if simulation or experiments to get labels are expensive). Thus, in parallel, researchers have explored using domain-specific features as inputs to traditional ML models \cite{xu2015machine}.  
Dimensionality reduction techniques and autoencoders help distill high-dimensional microstructure data into latent representations that can be mapped to properties via regression models. These surrogate models enable rapid property estimation.

However, microstructure-based ML modeling is not without limitations. The high dimensionality of image data, scarcity of labeled examples, and sensitivity to experimental noise can impede model accuracy and transferability. Furthermore, ensuring ML models produce physically plausible outputs, especially in spatially resolved predictions like stress fields, is an ongoing concern. Embedding physical constraints into training or combining ML with partial differential equation solvers represents a promising direction. Moreover, generalizability remains a major issue—models trained on specific material systems or morphologies may fail when applied to unfamiliar ones. Addressing these challenges requires more comprehensive datasets, improved feature representations, and potentially, the integration of multi-fidelity or multi-task learning strategies that leverage data at different resolutions or from different sources.

\subsubsection{Macro-Scale Property Prediction and Structural Materials}

At the macro scale, we are concerned with the properties and performance of bulk materials (or even entire components) as they would be used in real-world applications. Relevant properties include:
\begin{itemize}
    \item Mechanical properties of structural materials (yield strength, ultimate tensile strength, fracture toughness, fatigue life, creep life, Young’s modulus on bulk scale, etc.).
    \item Thermal properties at component level (thermal conductivity of bulk composites, heat capacity of a fully dense material, behavior of thermal barrier coatings, etc.).
    \item Electrical and Optical performance of devices (e.g., efficiency of a solar cell, breakdown voltage of an insulator, reflectance of a coated surface).
    \item Multi-physics performance (e.g., corrosion rate of an alloy in a certain environment, which couples chemistry and mechanics, or the cycle life of a battery which involves mechanical swelling, electrical degradation, etc.).
\end{itemize}
Compiling data at the macro scale often means aggregating information from published literature, databases, simulation, or standardized testing results
ML models can then be trained to predict these outputs for new combinations of inputs.

Features at this scale commonly include composition features, processing features, microstructure features, and service condition parameters. An example is HEAs, as mentioned earlier, these are macroscopically tested for hardness, strength, etc. Researchers have created models that input the elemental fractions and maybe some known phase info, and output properties like hardness or strength \cite{chang2019prediction}. Another example is additive manufacturing: a model might take input as laser power, scan speed, hatch spacing (process parameters), and output as-built density or ultimate strength of a printed part \cite{qin2022research}. 

Additionally, Chen et al. \cite{chen2024deep} have highlighted that ML models have achieved promising results in predicting the multiaxial fatigue life of metals. Fatigue life, which is the number of cycles a material can endure under cyclic loading before failure, is notoriously variable and depends on factors such as material purity, surface finish, and microstructure. Traditional approaches use empirical models like Basquin’s law and Coffin-Manson \cite{suresh1998fatigue}, fitted for each material. On the infrastructure side, materials like concrete have benefitted from ML in predicting strength based on mixture proportions and curing conditions. The macro property during a long time can be predicted by models taking in cement, water, aggregate ratios, and admixture details \cite{kazemi2024machine}. For instance, in ref \cite{deifalla2022machine} authors concentrated on employing ML models to precisely calculate the ultimate torsion strength of concrete beams reinforced with externally bonded fiber-reinforced polymer. This capability of long-term prediction can be used to predict the performance of devices such as batteries and solar cells.  For example, the capacity fade rate of a lithium-ion battery cell is determined by materials (cathode, anode, electrolyte) and usage conditions. ML models have been trained on cycling data to predict battery lifetime from the first few cycles of data, effectively treating the battery as a material whose “property” is cycle life \cite{fei2021early, yang2021machine}. 

Another significant advance is the development of machine-learned constitutive models. In computational mechanics, a constitutive model describes how a material responds to loads (stress–strain behavior). Traditionally, constitutive models are formula-based (e.g., the Ramberg-Osgood equation for stress–strain curve, or more complex plasticity models with parameters). With the rise of data-driven modeling, there’s a push to learn constitutive relations directly from data (from experiments or detailed simulations) using ML, effectively treating the stress–strain curve (or surface) as a target for prediction.  ML models can learn these directly from experimental or simulated data, capturing complex, history-dependent behavior such as plasticity or viscoelasticity \cite{tasdemir2022strategy}. Such models have been successfully integrated into finite element solvers, providing a data-driven alternative to traditional constitutive equations \cite{hussain2024machine}. While they offer flexibility and improved accuracy in many regimes, ensuring that these models obey thermodynamic principles and remain stable under extrapolation remains a key research focus.

Nevertheless, ML models at the macro scale face significant challenges. The dependencies between variables are often complex and highly nonlinear, with interactions between composition, processing, and usage conditions that are difficult to disentangle. Moreover, the scarcity of standardized, high-quality datasets limits the scope and reliability of predictions. Extrapolation outside the domain of the training data is particularly hazardous, as errors at this scale can have costly consequences in real-world applications. To address these issues, researchers are increasingly turning to hybrid models that integrate physics-based constraints into the learning process, and to ensemble methods or Bayesian frameworks that provide uncertainty estimates alongside predictions.  One example of such efforts is physics-informed neural networks (PINNs) that embed the governing equations into the loss function \cite{karniadakis2021physics, cuomo2022scientific, raissi2019physics}. Another direction is connecting digital twin frameworks with materials ML \cite{rathore2021role}. A digital twin of a component could use ML models for the material behavior that update in real-time with sensor data \cite{ali2020deep}.

\begin{longtable}{|p{2.5cm}|p{2.5cm}|p{4cm}|p{4.5cm}|}
\caption{Machine Learning Techniques and Challenges in Property Prediction Across Scales} \label{tab:ml_property_prediction} \\
\hline
\textbf{Property}  & \textbf{Concepts} & \textbf{ML Techniques} & \textbf{Limitations} \\
\hline
\endfirsthead

\hline
\multicolumn{4}{|c|}{\textbf{Table \thetable\ (continued): ML Techniques  in Property Prediction Across Scales}} \\
\hline
\textbf{Property Prediction Area}  & \textbf{Key Concepts} & \textbf{ML Techniques} & \textbf{Challenges and Limitations} \\
\hline
\endhead

\hline
\endfoot

\hline
\endlastfoot

\hline
\textbf{Atomistic-Scale Property Prediction} & Focus on fundamental material properties. & ML-driven surrogate models; Deep Learning (DNNs, CNNs); Graph-based featurization; MLIPs (Machine Learning Interatomic Potentials) & Generalisation to new microstructures; Data scarcity for certain properties; Interpretability of black-box models; High computational cost of high-fidelity calculations \\
\hline
\textbf{Micro/Meso-Scale Property Prediction} & PSP linkages (e.g., yield strength). & Image-based featurization (e.g., CNNs); Graph-based featurization (e.g., GNNs); Dimensionality reduction (e.g., autoencoders) & High dimensionality of microstructure data; Data scarcity and noise; Physical plausibility of outputs; Transferability across different material systems \\
\hline
\textbf{Macro-Scale Property Prediction} & Bulk material behavior (e.g., mechanical properties, thermal conductivity) & Composition and processing-based features; Additive manufacturing models; Hybrid models with physics-based constraints; Digital twin frameworks & Complex, nonlinear relationships between state variables; Scarcity of high-quality datasets; Extrapolation risks outside training data domain; Adherence to thermodynamic principles; Long-term performance predictions (e.g., battery life, fatigue life) \\
\hline
\textbf{Machine-Learned Constitutive Models} & Learning constitutive relationships from data (e.g., stress-strain behavior) & Data-driven constitutive models using ML (e.g., plasticity, viscoelasticity); Integration with finite element solvers & Ensuring stability under extrapolation; Integration with existing computational frameworks; Ensuring thermodynamic consistency; Predicting complex, history-dependent behaviors \\
\hline

\end{longtable}

\subsubsection{Gap Analysis}

Several gaps remain that hinder the widespread deployment and generalisation of these models in materials science and engineering workflows.

\begin{itemize}
    \item \textbf{Data Scarcity and Quality:} A persistent limitation across all scales—from atomistic to macro—is the scarcity of high-quality, consistent datasets. While large databases exist for certain properties (e.g., formation energies from DFT calculations), other key properties such as fracture toughness, dielectric strength, or long-term performance under cyclic loading remain underrepresented. Furthermore, experimental datasets are often heterogeneous, with inconsistent formats and missing metadata, limiting their direct utility for supervised learning.

    \item \textbf{Multi-Scale Integration:} There is a lack of robust frameworks that seamlessly integrate ML predictions across scales. Most existing models are trained and validated within a specific scale (e.g., atomic or mesoscopic), without mechanisms for propagating uncertainties or dependencies across scales. The absence of hierarchical or hybrid modelling pipelines that link microstructural features to macroscopic performance limits the real-world applicability of ML-based approaches.

    \item \textbf{Physical Consistency and Interpretability:} Although ML models can capture complex, nonlinear patterns, they often behave as black boxes. Ensuring that the predictions adhere to known physical laws (e.g., conservation of mass/energy, thermodynamic consistency) remains a critical challenge, particularly in constitutive modelling. In parallel, the interpretability of ML outputs—vital for domain trust and scientific insight—is often insufficient.

    \item \textbf{Generalisation and Extrapolation:} Many ML models exhibit poor generalisation when applied to materials or structures outside their training domain. This is particularly problematic for materials discovery tasks, where models are expected to predict the behaviour of novel chemistries, phases, or morphologies. The lack of robust extrapolation mechanisms or uncertainty quantification exacerbates the risk of misleading predictions.

    \item \textbf{Integration with Legacy Models and Digital Twins:} There is a growing need to integrate data-driven models with legacy physics-based solvers (e.g., finite element methods) and digital twin platforms. However, compatibility in terms of data structures, temporal resolution, and feedback loops is still underdeveloped. Moreover, many ML models lack the ability to evolve over time with continuous data streams—essential for digital twins and adaptive control.

    \item \textbf{Benchmarking and Reproducibility:} Standardised benchmarks and protocols for evaluating ML models in materials science are lacking. Model performance is often reported on disparate datasets with varying preprocessing and evaluation metrics, impeding reproducibility and fair comparison. Open-access repositories, curated benchmarks, and community standards are essential to address this issue.
\end{itemize}

\bigskip

Addressing these gaps requires a coordinated effort that combines innovations in data curation, uncertainty quantification, interpretable AI, and hybrid modelling strategies. It also necessitates interdisciplinary collaboration between materials scientists, data scientists, and software engineers to develop robust, trustworthy, and generalisable tools for the materials of the future.

\subsection{ML-based Materials Multiscale Modelling}
Multiscale modeling is an efficient technique in computational mechanics and material design where information is transferred between different time and length scales. The use of multiscale methods is motivated as many problems in science and engineering has multiscale features that stem from different spatial and temporal scales with distinct governing physics. Fine-scale physical models such as ab-initio, DFT, or molecular dynamics simulations can provide deep insight into the material behavior, however, as the size of the system increases severe computational expenses are imposed on the problem. By use of multiscale modeling, we can preserve the accuracy of fine-scale simulations while we still benefit from the efficiency of larger-scale modeling, such as continuum descriptions \cite{elliott2011novel, izadi2021bending, ghavanloo2018computational, rafii1998multi, li2005multiscale}.
A growing number of studies have implemented multiscale modelling successfully in predicting intricate phenomena in various fields, including fluid mechanics \cite{martys2002multiscale, chen2014multiscale}, biomechanics \cite{bhattacharya2017multiscale, rezaei2024equivalent, rezaei2024developing, tawhai2009multiscale}, nanomechanics \cite{ghoniem2003multiscale, izadi2021torsional, izadi2021combined, weinberger2016multiscale}, materials design \cite{fish2021mesoscopic, sivapuram2016simultaneous}, fracture of materials \cite{holian1995fracture,izadi2024fracture, swadener2002molecular}, manufacturing \cite{horstemeyer2010multiscale, rezaei2024hierarchical, markl2016multiscale} and process modeling \cite{gawad2013hierarchical, markl2016multiscale}.
Although ML has been widely used for computational modelling of materials at each of the macro, micro, and atomistic scales, its potential for multiscale modeling remains to be fully explored \cite{gunasegaram2021towards, nguyen2023challenges, bhatia2023confluence, sanderse2024scientific}. 

Until now, few works have incorporated ML for multiscale modeling of materials while considering different spatial and temporal scales \cite{chapman2020multiscale, han2021artificial, karapiperis2021data, mortazavi2020machine}. Bishara et al \cite{bishara2023state} provides an overview of the integration of ML in multiscale material modelling covering different simulation techniques at the fine scale such as molecular dynamics and DFT. In \cite{alber2019integrating, peng2021multiscale} the potential of incorporating ML in multiscale modeling in biomedical sciences is addressed. Jackson et al. \cite{jackson2019recent} provide the state of the art on integrating machine learning in the multiscale modelling of soft materials.

The integration of multiscale modelling and ML can be used as an up-scaling technique as well as for the homogenization of composite or porous materials \cite{arbabi2020linking}.
Figure \ref{fig:multiscale}  shows a schematic for integrating ML into multiscale modelling as an up-scaling method where ML can play the bridging role in the multiscale framework.

\begin{figure}[h!]
        \centering
\includegraphics[width=1\linewidth]{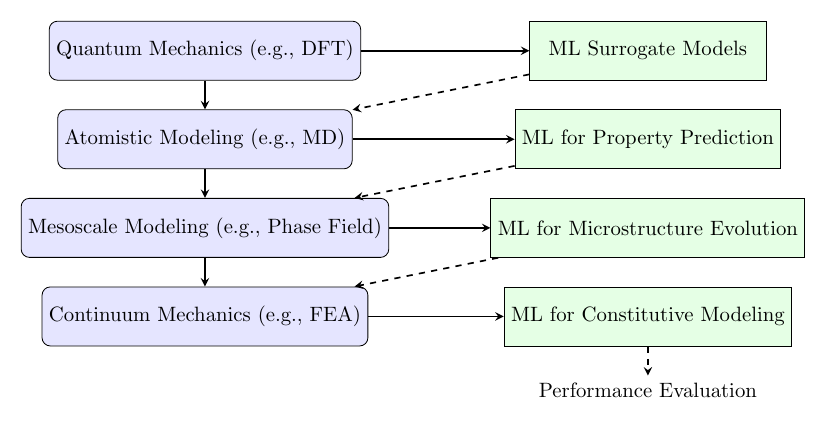}

\caption{Integration of AI/ML into multiscale materials modeling workflows. Each modeling scale—from quantum mechanics to continuum mechanics—is augmented with machine learning modules that serve as surrogate models, property predictors, or tools for microstructure evolution and constitutive modeling. Solid arrows indicate the traditional flow of information across scales, while dashed arrows represent the feedback loops enabled by ML, facilitating accelerated and enhanced multiscale simulations.}
  \label{fig:ml_classification-3}
\end{figure}

An early ML-based multiscale work has been performed in \cite{hambli2011multiscale} for a biomechanical study which has tried to bridge between the mesoscale and FEM at macroscale. In another study, Xiao et al \cite{xiao2020machine} used ML to enhance a hierarchical multiscale method for bridging molecular dynamics simulation and continuum mechanics. MD simulations were used to generate a dataset representing physical phenomena at the nanoscale, including stress–strain relations, dislocation phenomenon, and failure occurrence. The collected data set was further used to train two models based on SVM and a single-layer NN. The trained models are used in FEM at the macroscale to evaluate stresses and determining material failure modes in a molecule chain and an aluminum crystalline solid.
In \cite{hsu2020using}, a multi-scale approach is proposed for predicting fracture patterns in crystalline solids. For training the ML model, a large dataset embedding the spatial and temporal behavior of cracking has been constructed by performing MD simulations. The trained ML model based on the deep learning method is used for predicting the crack’s propagation where different crystalline orientations are set. 

Multiscale homogenization techniques have been developed through decades for studying the behavior of composites and porous materials \cite{rezaei2025non}. However, it has been shown that the integration of ML has efficiently accelerated the time required for solving homogenization problems \cite{arbabi2020linking, liu2020intelligent}. 

For the homogenization of heterogeneous materials, the deep learning network model is an ideal tool to simulate the nonlinearity of porous and composite materials \cite{unger2009neural, lu2019data, wang2019meta}, however, self-consistent clustering analysis (SCA) method \cite{LIU2016319, bessa2017framework} and the deep material network (DMN) have been also used in literature for multiscale homogenization \cite{nguyen2020deep, wu2020bayesian}.
 As an example, in \cite{le2015computational}, Le et al. used neural networks to approximate a surface response that represents the corresponding effective potential of heterogeneous materials from which the homogenized stress and effective tangent modulus were obtained. 
 In another work, Wang et el \cite{wang2018multiscale} proposed to combine supervised machine learning-based models with classic constitutive models to simulate porous materials with various size of pores.

\begin{longtable}{|p{3.2cm}|p{4cm}|p{6cm}|}
\caption{Summary of ML-Integrated Multiscale Modelling Approaches in Materials Science} \label{tab:ml_multiscale_summary} \\

\hline
\textbf{Scale Bridged} & \textbf{ML Technique} & \textbf{Contribution / Description} \\
\hline
\endfirsthead

%\hline
%\textbf{Application Area} & \textbf{Scale Bridged} & \textbf{ML Technique} & \textbf{Contribution / Description} \\
%\hline
%\endhead

\hline
\endfoot

\hline
\endlastfoot

 Meso to Macro (Biomechanics) & Neural Networks & Early integration of ML with multiscale modelling: bridging mesoscale to finite element macro models. \\
\hline

Nano to Macro & SVM, Single-layer NN & Hierarchical multiscale modelling combining MD data with ML models for macro-level FEM simulation and failure analysis. \\
\hline

 Nano to Macro (temporal). Fracture Mechanics & Deep Learning & Prediction of crack propagation in crystalline solids using DL models trained on MD simulation data capturing spatio-temporal fracture behaviour. \\
\hline

 Micro to Macro (Homogenisation)& Deep Learning Networks & Efficient approximation of nonlinear effective properties in porous and composite materials, speeding up traditional homogenisation tasks. \\
\hline

Micro to Macro \newline (Composite Modelling) & Neural Networks  & Approximation of effective potential surfaces for stress and tangent modulus prediction in heterogeneous materials. \\
\hline

 Micro to Macro (Porous Materials) & Supervised ML + Constitutive Laws & ML-enhanced simulation of porous materials with varying pore sizes, merging data-driven and physics-based models. \\
\hline

 Micro to Macro (Homogenisation) & Deep Material Network (DMN) & Application of hierarchical ML architectures for efficient multiscale material response prediction and clustering-based homogenisation. \\
\hline

\end{longtable}

\subsection{Gap Analysis: ML for Multiscale modelling}

Despite the increasing interest in integrating ML into multiscale modelling and simulation workflows, several significant gaps remain that limit its widespread adoption and full potential in materials science.

\begin{itemize}

\item \textbf{Limited Integration Across Scales:} 
Current ML approaches often remain confined to single-scale modelling, typically focusing on either atomistic (e.g., DFT, MD) or continuum (e.g., FEM) simulations. Few frameworks truly bridge multiple spatial and temporal scales in a unified ML-assisted architecture. This gap leads to inefficient upscaling or downscaling strategies and reduces the physical interpretability of the surrogate models employed.

\item \textbf{Insufficient Physical Consistency:}
A critical limitation in many ML-based multiscale approaches is the lack of embedded physical constraints. While surrogate models offer fast approximations, they frequently ignore thermodynamic consistency, conservation laws, or material symmetries, leading to physically implausible results when extrapolated beyond the training data domain.

\item \textbf{Sparse Datasets Across Scales:}
Multiscale modelling inherently involves data generation across distinct regimes (e.g., atomic, microstructural, macroscopic), where high-fidelity simulations such as MD or DFT are computationally expensive. The resulting data sparsity hampers the training of robust ML models, especially those requiring large, diverse, and labelled datasets.

\item \textbf{Scalability and Generalisation:}
While deep learning techniques can approximate complex mappings, their scalability to multiscale configurations and generalisation across different material systems remains underexplored. Many current models are trained on specific case studies, limiting their applicability to other materials, loading conditions, or structural configurations.

\item \textbf{Temporal Bridging Remains Underdeveloped:}
Most ML-based multiscale frameworks focus on spatial scale bridging (e.g., linking microstructure to macroscale properties), with limited work on bridging temporal scales. The challenge lies in the long time scales associated with phenomena such as creep, fatigue, or phase transitions, which are not easily captured through short-time atomistic simulations or standard ML time-series models.

\item \textbf{Uncertainty Quantification Deficiency:}
There is a clear lack of rigorous uncertainty quantification (UQ) in existing ML-assisted multiscale workflows. Given the inherent approximation and extrapolation in ML models, especially in high-stakes engineering applications, quantifying epistemic and aleatoric uncertainties is essential for model reliability and decision-making.

\item \textbf{Lack of Hybrid Modelling Frameworks:}
Hybrid approaches that combine physics-based models with data-driven ML components (e.g., physics-informed neural networks, operator learning) have shown promise but remain nascent in the context of multiscale materials modelling. The development of modular, interpretable, and adaptive hybrid pipelines that span multiple scales is still in early stages.

\item \textbf{Computational Infrastructure and Workflow Integration:}
Multiscale modelling workflows often involve multiple tools, codes, and platforms. Integrating ML components into these existing workflows is non-trivial, especially for legacy simulation tools. The lack of standardisation in data formats, interoperability, and ML-ready pipelines remains a practical barrier to adoption.

\item \textbf{Data-driven Bridging Operators:}
Bridging operators (e.g., homogenisation, localisation, and coarse-to-fine mapping operators) are key elements in multiscale simulations that enable the transfer of information across scales. Traditional operators are typically derived analytically or numerically based on simplified assumptions. ML-based surrogate operators have shown potential to accelerate or enhance this process, but robust data-driven formulations of bridging operators that ensure accuracy, generalisability, and physical interpretability are still lacking. In particular, the development of invertible and bi-directional operators, capable of both upscaling and downscaling, is an open challenge in the community.

\end{itemize}

%%%%%%%%%%%%%%%%%%%%%%%%%%%%%%%%
\subsection{ML-based Materials Discovery}
%% High-throughput screening, inverse design, and quantum-based approaches. !!Sorry for interfering hhh -Kouider!!

%The materials discovery process typically consists of four key phases: (1) defining the research objective, (2) gathering and analyzing existing data, (3) formulating a hypothesis, and (4) conducting experiments and tests to validate the proposed ideas. Integrating artificial intelligence (AI) into this process significantly enhances its efficiency, accuracy, and scalability. For example, Natural Language Processing (NLP) can analyze scientific papers, patents, and databases to uncover trends, identify knowledge gaps, and highlight emerging challenges in materials science \cite{olivetti2020data,lee2023natural} . Data mining and ML algorithms efficiently extract and process relevant information from experimental results, simulations, and open-access repositories, enabling deeper insights into material properties and behaviors \cite{liu2017materials,lu2017data}. Additionally, Reinforcement Learning (RL) can optimize testing protocols, reducing time and resources required to identify materials with desired properties \cite{raccuglia2016machine,sui2021deep}.\cite{pyzer2022accelerating}

AI has emerged as a transformative force in materials discovery, reshaping traditional paradigms in materials modeling and accelerating innovation. This section explores three major aspects of AI-driven materials discovery: (i) Inverse Design and Materials Generation, where AI methods inversely map desired properties to novel material structures; (ii) High-Throughput Virtual Screening and Experimental Design, involving efficient navigation of large chemical and physical spaces through computational predictions and autonomous experimentation; and (iii) Knowledge Extraction and Scientific Discovery, which leverages AI to distill scientific insights and patterns from expansive literature and experimental datasets. Despite remarkable advances, significant challenges remain, including data scarcity and bias, interpretability of AI models, accurate uncertainty quantification, limitations in extrapolating beyond known datasets, and effective integration of AI methodologies with established domain knowledge. 

\begin{figure}[h]
    \centering
    \includegraphics[width=0.65\linewidth, , angle=-90]{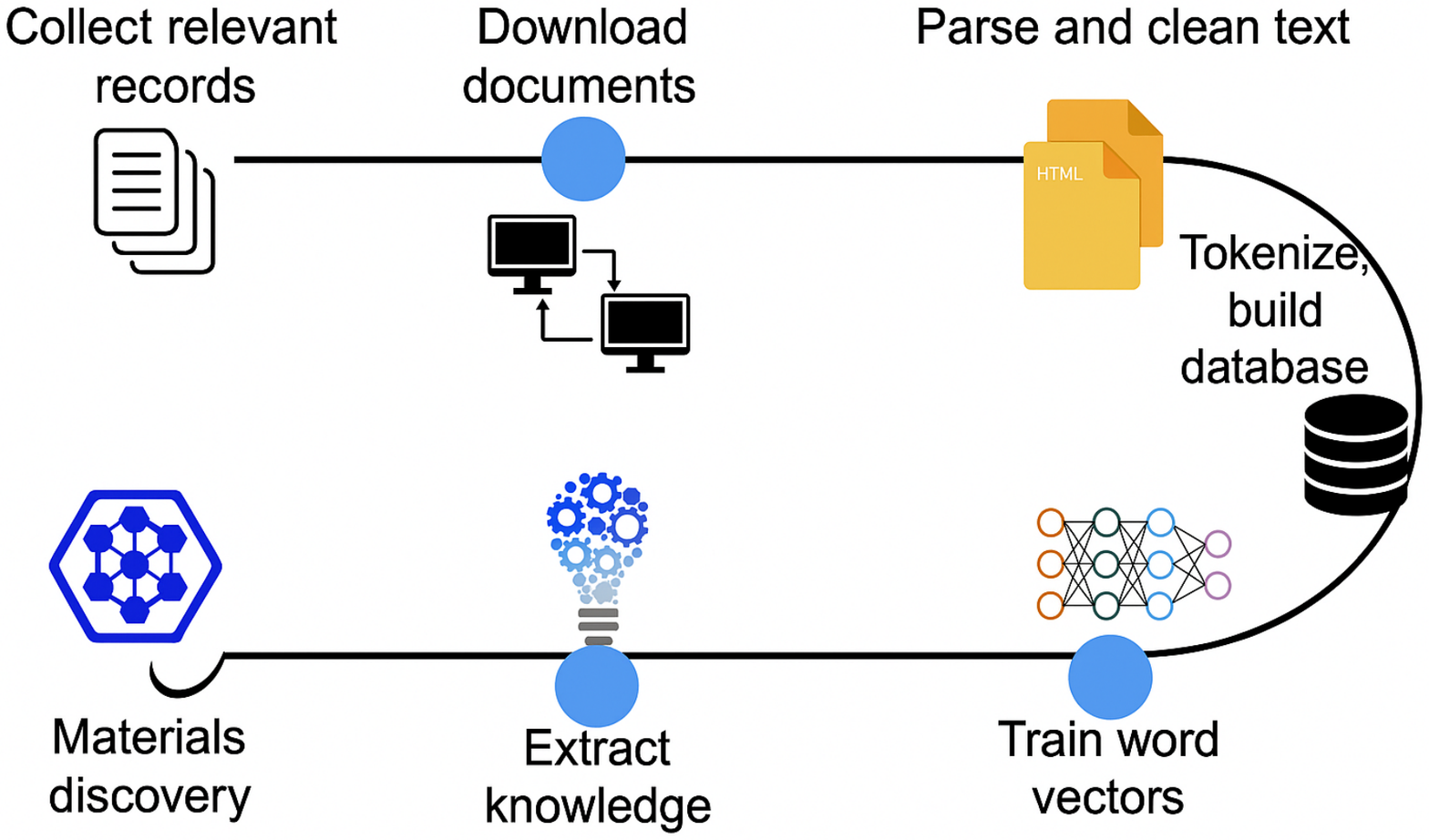}
    \caption{Overview of a text-mining and machine learning pipeline for materials discovery. The workflow includes collecting and downloading relevant documents, parsing and cleaning text, tokenizing and building a database, training word vectors, extracting knowledge, and leveraging insights for materials discovery. \cite{pilania2021machine}.}
    \label{fig:AI-worflow}
\end{figure}

\subsubsection{Inverse Design and Materials Generation}
Modern AI methods contribute at every stage of materials discovery. In silico, ML models can predict a material’s properties (explained in section \ref{property_pred}) without needing lengthy lab synthesis \cite{belle2021machine, moosavi2020role}. Inverse design approaches invert the traditional materials discovery process by starting from target properties and working backward to find candidate structures \cite{cheng2025ai}. Instead of screening existing compounds, these AI-driven methods generate novel material formulas or structures likely to exhibit the desired performance. Modern algorithms leverage deep generative models, such as VAEs, GANs, and diffusion models, to map relationships between structures and properties and sample new materials conditioned on target criteria \cite{gangwal2024generative, miret2024perspective}. One of the first studies in this regard is the development of iMatGen \cite{noh2019inverse} that implied VAE to discover the inverse design of solid materials. It transforms unit cells into 3D grid-based representations, utilizing spherical linear interpolation and Gaussian random sampling to explore the latent space and generate materials. One recent example in generative AI for materials is the development of MatterGen, a generative model for inorganic materials design \cite{zeni2025generative}. MatterGen explores a vastly larger space of inorganic materials than previous screening methods and can be guided by human-readable prompts. Generative models have also been used to discover 2D materials \cite{lyngby2022data} or produce realistic and high-quality structures that remain physically accurate across a range of pressure conditions \cite{luo2024deep}. In \cite{xiao2023invertible} a system named "simplified line-input crystal-encoding" was developed that successfully recovered 94.95\% of more than 40,000 crystal structures with diverse structural and chemical compositions. An example of Generativ AI in designing provskite is in \cite{chenebuah2024deep, chenebuah2023evolutionary}. Diffusion probabilistic models are a newer class of generative models that have shown state-of-the-art performance in materials generation. Diffusion models can sample high-quality candidates from a learned distribution by learning to reverse a noising process. Recent breakthroughs include a Crystal Diffusion Variational Autoencoder (CDVAE) \cite{xie2021crystal} for periodic material structures and diffusion models for alloy design. The CDVAE workflow begins by generating the lattice unit cell, initial chemical composition, and atom count using a VAE. Then, the diffusion model simultaneously refines atomic types and coordinates to produce the final crystal structure.

An emerging trend involves repurposing LLMs, originally trained on text, for materials design. LLMs show potential as agents for interactive material optimization. For example, LLMatDesign \cite{jia2024llmatdesign}, an LLM-driven framework for interactive materials discovery, facilitates iterative design and evaluation to achieve desired properties, such as band gaps and stability. In ref \cite{gruver2024fine}, researchers have fine-tuned LLMs on text representations of inorganic crystal structures, effectively treating material formulas or serialized structures as a “language.” Remarkably, a fine-tuned 70-billion-parameter LLM could generate new inorganic compounds in text format, with about 90\%  of sampled structures satisfying charge balance and atomic packing constraint. While unconventional, LLM-based generation can become a promising complement to physics-specific generative models. These successes mark a paradigm shift: instead of searching within known materials, AI generative models enable inverse design, proposing novel molecules or structures that are designed to have specific properties or performance.

\subsubsection{High-Throughput Virtual Screening and Experimental Design}

AI is revolutionizing the exploration of the immense space of potential materials through high-throughput virtual screening. ML models, trained on computational or experimental datasets, are efficient surrogates for resource-intensive, physics-based calculations. These surrogate models rapidly evaluate vast numbers of candidate materials, enabling millions of hypothetical structures to be efficiently assessed in silico for stability and performance.
For instance, researchers at DeepMind introduced GNoME (Graph Networks for Materials Exploration) \cite{merchant2023scaling} in 2023, an AI framework designed to predict the stability of inorganic crystals. GNoME screened approximately 2.2 million new inorganic structures, identifying around 380,000 highly stable materials. This single AI-driven study expanded humanity’s catalog of predicted materials by a magnitude equivalent to roughly 800 years of manual discovery, highlighting AI's profound impact on accelerating materials science research. Many of these materials were flagged as promising candidates for future technological applications \cite{szymanski2023autonomous}.

The efficiency and effectiveness of AI-driven materials discovery are further enhanced by transfer learning, which applies insights gained from one domain or large dataset to another smaller dataset. Pre-trained models on extensive datasets comprising millions of structures can be fine-tuned effectively with minimal additional data, maintaining high predictive accuracy. A notable example is the deep learning-based inverse design framework for composite plates introduced in \cite{yang2025transfer}, which efficiently generates structural layouts from specified thermo-mechanical fields. Utilizing transfer learning, this approach drastically reduces the required training data while achieving high accuracy.
AI-driven materials discovery methods also benefit from active learning, Bayesian optimization, and RL, which efficiently explore the materials design space, often in closed-loop integration with experimental validation. Unlike generative models that propose candidate materials in a single step, these iterative approaches progressively refine candidates based on experimental or simulated feedback, balancing exploration and exploitation. The CAMEO platform \cite{kusne2020fly}, for example, integrated an AI planner with a synchrotron X-ray instrument, leading to the discovery of a new phase-change memory material in a closed-loop manner. Such autonomous laboratory systems combine robotics for experimental execution with ML models for data analysis, guided by AI-driven planners that determine subsequent experiments. This significantly accelerates each cycle; proposal, experimentation, and analysis, compared to human-driven methods \cite{angello2024closed}.

Integrating computational predictions with automated experimentation serves as a powerful catalyst, markedly accelerating the discovery, characterization, and optimization of novel materials \cite{szymanski2023autonomous, li2020ai}. One example of feedback from simulation is developing amorphous polymers with high thermal conductivity using RL. The researchers in \cite{ma2022exploring} trained an ML model on thermal conductivity data from molecular dynamics simulations and used a recurrent neural network to generate new polymer structures.

Materials discovered using AI methods have the potential to lead significant breakthroughs in critical areas such as superconductors and advanced battery electrodes, demonstrating deep learning’s unmatched capability in exploring expansive chemical spaces rapidly \cite{stanev2018machine, gashmard2024predicting}. Additionally, AI drives automation in "self-driving" laboratories, autonomously planning, executing, and analyzing experiments. AI systems equipped with computer vision, spectroscopy, and other analytical methods enable a fully integrated, closed-loop discovery process \cite{angello2024closed, kusne2020fly}.
Furthermore, recent advancements include integrating LLMs to enhance experimental workflows. For example, the AI system "Coscientist," powered by GPT-4, autonomously designs, plans, and executes complex chemical experiments, successfully completing diverse tasks such as synthesizing compounds and optimizing reactions. This approach integrates multiple capabilities—web searches, coding, and experimental automation—demonstrating sophisticated reasoning and robust experimental design \cite{boiko2023autonomous}.

Overall, AI methods profoundly enhance numerous aspects of materials discovery and management, spanning initial screening and design through manufacturing, monitoring, operation, and maintenance. They facilitate real-time modeling, rapid fault detection, cost optimization, efficiency improvements, rigorous quality control, and the identification of superior performance materials \cite{yazdani2023artificial}. One approach is to construct materials knowledge graphs that integrate disparate data sources and encode relationships between materials, processes, and properties.

\subsubsection{Knowledge Extraction and Scientific Discovery} \label{knowledge_extraction-1}

Beyond generating and screening materials, AI assists scientists in navigating and synthesizing the vast, fragmented body of existing materials knowledge residing in literature and databases. Knowledge extraction techniques leverage NLP and data mining to gather insights from literature, databases, and experiments that would be impossible to unify manually. For example, millions of materials science papers and patents contain valuable data on compositions, properties, and synthesis conditions, much of it buried in text. AI-driven text mining can automatically extract this information (e.g. identifying material names, measured properties, and experimental parameters) and consolidate it into structured forms \cite{schilling2024text, dagdelen2024structured}. Recent reports increasingly indicate that LLMs are being used for data extraction in both the fields of chemistry \cite{zhang2024fine, zheng2023chatgpt} and materials science \cite{shetty2021automated, polak2024extracting, gupta2024data, gupta2022matscibert}. In \cite{tshitoyan2019unsupervised}, a model read millions of scientific abstracts and learned relationships like the periodic table of elements and structure–property correlations without any human labeling. Impressively, the algorithm could recommend new thermoelectric materials years before they were experimentally discovered, implying that clues to these discoveries were hidden in past publications.
Furthermore, materials knowledge graphs have been created to capture the symbolic and data-driven relationships among materials properties, to illustrate the interconnections between various sources of materials data, to integrate multiple data streams, and to encode the links among factual knowledge, analytical models, and domain experts \cite{mrdjenovich2020propnet, zhang2017mmkg, ye2024construction, bai2025construction}. These knowledge graphs serve as semantic networks, nodes representing entities like compounds or properties, and edges denoting relationships (synthesis steps, property trends, etc.),  which enable advanced querying and pattern discovery. In a recent study in ref \cite{bai2025construction}, a knowledge graph was built by analyzing over 100,000 articles, resulting in 2.53 million nodes and 4.01 million relationships. Following its creation, the graph was applied to enhance data retrieval, data mining, and the development of advanced question-answering systems. The resulting graph encodes a material's ontology (e.g., linking the material name and formula to its application and related properties). It supports algorithms for link prediction to suggest new connections or hypotheses \cite{ye2024construction}. In essence, such AI-driven knowledge frameworks can comb through multidisciplinary studies to uncover hidden relationships

\begin{longtable}{|>{\RaggedRight\arraybackslash}p{3cm}|>{\RaggedRight\arraybackslash}p{3cm}|>{\RaggedRight\arraybackslash}p{3.8cm}|>{\RaggedRight\arraybackslash}p{3.8cm}|}
\caption{Machine Learning Techniques and Challenges in Materials Discovery} \label{tab:ml_materials_discovery} \\
\hline
\textbf{Aspect} & \textbf{Key Concepts} & \textbf{ML Techniques} & \textbf{Challenges and Limitations} \\
\hline
\endfirsthead
\hline
\multicolumn{4}{|c|}{\textbf{Table \thetable\ (continued)}} \\
\hline
\textbf{Aspect of Materials Discovery} & \textbf{Key Concepts} & \textbf{ML Techniques} & \textbf{Challenges and Limitations} \\
\hline
\endhead
\hline
\endfoot
\hline
\endlastfoot
\hline
\textbf{Inverse Design and Materials Generation} & AI-driven reverse process of material discovery based on desired properties. & VAEs (Variational Autoencoders); GANs (Generative Adversarial Networks); Diffusion models; Generative models for material design. & Data scarcity for diverse material types; Difficulty in ensuring physical feasibility; Lack of generalization across material systems. \\
\hline
\textbf{High-Throughput Virtual Screening} & Rapid screening of large material spaces for stability and performance. & Graph-based models (e.g., GNNs); Surrogate models; Deep learning for property prediction; Transfer learning. & High computational cost for large databases; Risk of false positives/negatives; Difficulty in extrapolating from existing data. \\
\hline
\textbf{Experimental Design and Optimization} & AI integration with experimental systems to optimize materials testing. & Bayesian optimization; Reinforcement learning; Active learning for iterative testing. & Experimental validation challenges; High cost of experimental setups; Difficulty in achieving real-time feedback. \\
\hline
\textbf{Knowledge Extraction and Scientific Discovery} & Extraction of insights from vast literature and experimental data. & Natural Language Processing (NLP); Data mining; Knowledge graphs. & Difficulty in extracting meaningful patterns from unstructured data; Ambiguity in text mining; Limited generalization across domains. \\
\hline
\textbf{AI-driven Material Knowledge Graphs} & Representing relationships between material properties and synthesis conditions. & Knowledge graphs; Semantic networks; Data mining for relationship discovery. & Data sparsity in material databases; Integration of diverse data sources; Ensuring scalability for large datasets. \\
\hline
\end{longtable}

\subsubsection{\color{blue}Gap Analysis on ML-based Materials Discovery}

The application of AI in materials discovery has shown great promise, but significant gaps remain in its integration, scalability, and generalization across material systems. By addressing these gaps through continued research and the development of advanced ML techniques, the field can unlock more efficient and effective pathways for discovering new materials and enhancing material properties.

\begin{longtable}{|>{\RaggedRight\arraybackslash}p{3cm}|>{\RaggedRight\arraybackslash}p{5cm}|>{\RaggedRight\arraybackslash}p{6cm}|}
\caption{Research gaps and future directions for the integration of AI in materials science workflows} \label{tab:ai_gaps_materials} \\
\hline
\textbf{Area} & \textbf{Gap} & \textbf{Needed Developments} \\
\hline
\endfirsthead

\hline
\multicolumn{3}{|c|}{\textbf{Table \thetable\ (continued): Research gaps and future directions}} \\
\hline
\textbf{Area} & \textbf{Gap} & \textbf{Needed Developments} \\
\hline
\endhead

\hline \multicolumn{3}{r}{{Continued on next page}} \\
\endfoot

\hline
\endlastfoot

\textbf{Inverse Design} & 
AI models can predict new material structures based on desired properties using GANs and VAEs. Some methods have discovered novel materials, but generalization to new systems is still limited. & 
Autonomously generate novel, physically feasible materials across broad systems. Models should be guided by performance targets. \\
\hline

\textbf{High-Throughput Screening} &
AI accelerates screening of material spaces (e.g., inorganic crystals, alloys), but computational costs remain high for large, diverse datasets. & 
Fully automated systems for screening millions of candidates with minimal human input. Accurate predictions of stability, performance, and synthesis feasibility with reduced cost. \\
\hline

\textbf{Knowledge Extraction} & 
Text mining and NLP extract information from literature and patents, but results are often incomplete or inconsistent. & 
Automated, accurate extraction of structured knowledge from scientific literature and datasets. Advanced NLP models for synthesizing insights from complex technical documents. \\
\hline

\textbf{Material Knowledge Graphs} &
Graphs exist to connect material properties, processes, and applications but often lack completeness and predictive integration. & 
Create comprehensive, interconnected knowledge graphs integrated with predictive models and hypothesis generators. \\
\hline

\textbf{Experimental Design} &
AI models are being integrated with experiments for adaptive testing, but real-time feedback loops are not yet widespread. & 
Combine reinforcement learning and active learning for autonomous experimental workflows. Link experimental and predictive models in real-time. \\
\hline

\textbf{Refinement of Inverse Design Models} &
Generative models need better understanding of physical constraints. Existing models may not always generate feasible structures. & 
Develop robust models incorporating physical laws and feasibility constraints to generate viable material candidates. \\
\hline

\textbf{Optimization of Screening Systems} &
High-fidelity models are computationally expensive. Training datasets are not always diverse enough. & 
Use surrogate models and expand dataset diversity to enhance accuracy and reduce computational demands. \\
\hline

\textbf{Advancement of Knowledge Extraction Techniques} &
Current models struggle with the complexity and heterogeneity of scientific texts. Integration of multiple sources remains difficult. & 
Develop AI frameworks capable of multi-source synthesis for knowledge-driven material discovery. \\
\hline

\textbf{Expansion and Enhancement of Knowledge Graphs} &
Graphs need to scale with more data and link across domains. Hypothesis generation is not yet fully developed. & 
Build global, scalable material knowledge graphs with integrated AI for property prediction and discovery acceleration. \\
\hline

\end{longtable}

\subsection{Example of Applications of ML-based Multiscale Modellind and Design Across Material Systems}

\subsubsection{Application for Composite Materials}

The application of machine learning (ML) in composite materials has garnered significant interest due to its ability to handle the complex, nonlinear behaviour of advanced composites.  Machine learning, with its ability to process large volumes of data and learn from past observations, provides the flexibility needed to model and optimise the behaviour of these materials \cite{liang2025review, sharma2022advances}. The application of ML in composites can be explored under three main aspects: constitutive modelling, prediction of mechanical properties, and design and optimisation.

\begin{itemize}
    \item \textbf{Constitutive Modelling for Composites:} Traditional models often rely on explicit mathematical functions that require calibration based on experimental data. However, the complexity of composite materials, such as isotropic and anisotropic properties, demands more sophisticated modelling approaches. 

 \item \textbf{Non linear and time depend behaviour}  ML, particularly Artificial Neural Networks (ANN), has proven effective at approximating complex nonlinear relationships in constitutive curves. Several studies have employed Backpropagation ANN (BPANN) to construct constitutive models for metal matrix composites, using inputs like temperature, strain rate, and strain to predict stress \cite{yang2021deep}. While multi-layered ANNs typically offer better generalisation, determining their optimal structure (e.g., the number of neurons, transfer functions, and optimisation algorithms) can be time-consuming. Bayesian Neural Networks (BNN) have been used in some studies to handle smaller datasets and model the uncertainty in weights, showing better performance in such scenarios \cite{zhou2025machine}.

\item \textbf{Management of the anisotropy}
 Predicting the stress-strain curves for anisotropic composites requires more sophisticated ML models, such as Convolutional Neural Networks (CNN) and Recurrent Neural Networks (RNN). The complexity of the mechanical behaviour of these materials necessitates the use of multiscale models combined with ML techniques \cite{yang2021deep}. Data acquisition for training ML models remains a significant challenge, leading to the use of techniques such as transfer learning and integration with other modelling approaches like finite element analysis (FEA) \cite{ogaili2022integration}.

\item \textbf{Structure-Based Prediction:} The relationship between the microstructure of composites and their mechanical or physical properties is another critical area where ML has been applied. ANN models, often combined with representative volume element (RVE) analysis, have been used to identify relationships between structure and performance \cite{Zhou2018}. Additionally, genetic algorithms and CNNs have been employed to predict mechanical properties based on structural images \cite{Yang2020}.

\item \textbf{Prediction Based on Environmental Factors}
Environmental factors such as temperature, humidity, and exposure to aggressive chemicals can significantly affect the mechanical properties of composites. ML models are  applied to predict the degradation of composite materials under various environmental conditions \cite{Wang2021}.

\item \textbf{Design and Optimisation of Composites:} Machine learning is also being applied in the inverse design of composites, where desired properties are used to optimise material structure and manufacturing processes.

\item \textbf{Property-Targeted Design: } In property-targeted design, ML models are constructed with the target property as input and the design parameters as output to predict the optimal configurations \cite{Lee2019}. 

\item \textbf{Design Based on Known Properties}

In contrast, design based on known properties follows a three-step process: generation and reduction of descriptor dimensionality, construction and validation of the ML model, and prediction of new materials with experimental validation \cite{Xie2018}. This process can be coupled with optimisation techniques such as genetic algorithms to further refine the design.

\item \textbf{Direct Use of Inverse ML Methods}
Inverse ML techniques, such as Generative Adversarial Networks (GANs) and Graph Neural Networks (GNNs), have been employed to directly explore and optimise material structures \cite{Nguyen2021}. These methods, when combined with feedback loops from physics-based models, allow for iterative improvement of the designs.

\end{itemize}

\subsection{ Case Studies on High-Entropy Alloys}
High-entropy alloys (HEAs) are a novel class of materials recognized for their exceptional structural and functional properties. However, the vast combinatorial design space and intricate phase formation dynamics pose significant challenges for rational design. In this case study, ML is employed to construct predictive models for phase formation. Model interpretability is leveraged to gain insight into the physicochemical factors governing phase stability, thereby aiding targeted alloy design. The code is available at: \href{https://github.com/Iman-Peivaste/ML_Phase}{github.com/Iman-Peivaste/ML\_Phase}.

The dataset, compiled from experimental records, contains 11,252 entries, later refined to 5,692 after data cleaning. It spans 50 elements, with a focus on Fe, Ni, Cr, Co, Al, Cu, Ti, and Mn. The dataset includes 11 phase categories, such as BCC, FCC, IM, AM, and combinations thereof. It is accessible at: \href{https://github.com/Iman-Peivaste/ML_HEAs_Phase_Dataset}{github.com/Iman-Peivaste/ML\_HEAs\_Phase\_Dataset}.

Feature engineering relied on Composition-Based Feature Vectors (CBFVs), which integrate domain knowledge (see Section~\ref{domain-specific}). Multiple algorithms were applied:
\begin{itemize}
    \item \textbf{Support Vector Classifier (SVC)} with a 6-degree polynomial kernel,
    \item \textbf{Random Forest (RF)} with 100 decision trees,
    \item \textbf{Extreme Gradient Boosting (XGBoost)},
    \item \textbf{Artificial Neural Networks (ANN)} with four hidden layers trained using the ADAM optimizer.
\end{itemize}

The dataset was split 80/20 for training/validation. Data normalization and Adaptive Synthetic Sampling (ADASYN) were used to address class imbalance. Hyperparameters were optimized via grid search cross-validation. Evaluation metrics included accuracy, standard deviation, and k-fold validation.

XGBoost and RF achieved the highest overall accuracy (97\%) for solid solution phases. Accuracy dropped to 86\% when including AM and IM phases, with XGBoost still outperforming other models. ANN achieved 95\% for AM phases. Feature importance analyses identified mixing entropy, electronegativity deviation, and valence electron concentration (VEC) as most relevant. Different models prioritized different features: SVC on melting point, RF on electronegativity, and XGBoost on mixing entropy. High Al favored IM formation; high Ni promoted FCC and AM phases. Differentiating BCC and AM required atomic radius and other auxiliary features.

\subsection{Material Elastic Large-Deformation}

Surrogate models offer efficient approximations of complex simulations, crucial for real-time applications such as design optimization and multi-scale simulations~\citep{Lefik2009, Krokos2021, KROKOS2024112545, STORM2024117001, Unke2021}. This case study focuses on deep learning surrogates for predicting full-field responses of elastic solids under large deformation.

\paragraph{Data Generation and }Data is synthetically generated from Finite Element (FE) simulations. Inputs are surface load or body-force vectors; outputs are 2D or 3D full-field displacements. Inputs are represented as 2D/3D vectors, potentially extended to include location data. These are formatted into mesh structures for CNNs or Graph Neural Networks (GNNs). Outputs are displacement fields, also formatted as meshes or latent representations via autoencoders.

\paragraph{ML Models:} The topological structure of the data motivated the use of: \textbf{CNN U-Nets}~\citep{MENDIZABAL2020101569, Deshpande_CNN_UNet_2022}, \textbf{Graph U-Nets}~\citep{Deshpande_MagNET_2024},  \textbf{Attention-based models} such as Perceiver IO~\citep{Deshpande_ConvAggrAtt_2023},  \textbf{Probabilistic models} via Variational Bayes~\citep{Deshpande_CNN_UNet_2022} and Gaussian Processes~\citep{Deshpande_GP_2025}.

\paragraph{Data and Model Repositories:}  \href{https://github.com/saurabhdeshpande93/gp-auto-regression}{gp-auto-regression}, \href{https://github.com/saurabhdeshpande93/MAgNET}{MAgNET}, \href{https://github.com/saurabhdeshpande93/convolution-aggregation-attention}{conv-agg-attn} and \href{https://doi.org/10.5281/zenodo.7585319}{Zenodo dataset}.

\paragraph{Training and Validation:}\textbf{Standard training using ADAM optimizer:} Learning rate schedules were adapted per case. Ablation studies explored architectural sensitivity. For probabilistic models: CNN Variational Bayes models used the Flipout estimator~\citep{wen2018flipout} and GP models were trained by maximizing the log marginal likelihood~\citep{Rasmussen2005GP}.

\paragraph{Analysing Results}
All models showed strong predictive performance. Relative average error $<$ 0.1\%. Speedup over FE solvers: CNN U-Net: $>$30x (CPU), $>$300x (GPU), Graph U-Net: $\sim$15x (CPU) and Perceiver IO: $\sim$400x.

Probabilistic models effectively captured epistemic and aleatoric uncertainty. These tools allow detection of out-of-distribution predictions and estimation of input noise effects.

These last two use-case studies demonstrate how machine learning accelerates discovery and simulation in materials science. From interpreting complex phase diagrams in HEAs to replacing expensive mechanical simulations with efficient surrogates, ML not only enhances predictive accuracy but also provides deeper physical understanding and design capabilities.

%%%%%%%%%%%%%%

\section{ML-based for Life Cycle Assessment} \label{sec-6}
Sustainability has become a fundamental pillar of modern manufacturing, driven by pressing environmental challenges, regulatory frameworks, and growing consumer awareness \cite{smith2012steps, kazakova2022sustainable, bjornbet2021life}. While traditionally associated with environmental preservation, sustainability now plays a crucial role in long-term value creation within industries \cite{rashid2013resource}. The Eco-design for Sustainable Products Regulation (ESPR) represents a major step forward in embedding circular economy principles and lifecycle thinking into legislation. These efforts, supported by mandatory Life Cycle Assessment (LCA) practices, provide stakeholders with actionable insights that facilitate environmentally conscious decision-making \cite{ahmed2024advancing}. Consumers also benefit by making more informed choices based on transparent sustainability metrics.

Despite their benefits, traditional sustainability assessment methods, particularly LCAs, face challenges in materials modeling. Scientists and engineers must navigate data gaps, uncertain process parameters, and rapidly evolving manufacturing methods, making comprehensive assessments difficult \cite{bassey2024ai}. Recent advancements in machine learning (ML) offer solutions by automating data collection, uncovering patterns in large datasets, and improving predictive accuracy. Meanwhile, emerging frameworks such as Digital Product Passports (DPPs) promise greater transparency across a product’s entire lifecycle, from raw material sourcing to end-of-life disposal. DPPs create a foundation for robust sustainability strategies when combined with ML-driven analytics by continuously tracking product-specific information and linking it through secure digital platforms. This section explores how ML methodologies enhance LCA, leading to more precise, efficient, and actionable approaches in sustainable manufacturing.

\subsection{ML-based Life Cycle Assessment (LCA)}

\begin{figure}[h!]
    \centering
    \includegraphics[width=0.8\linewidth]{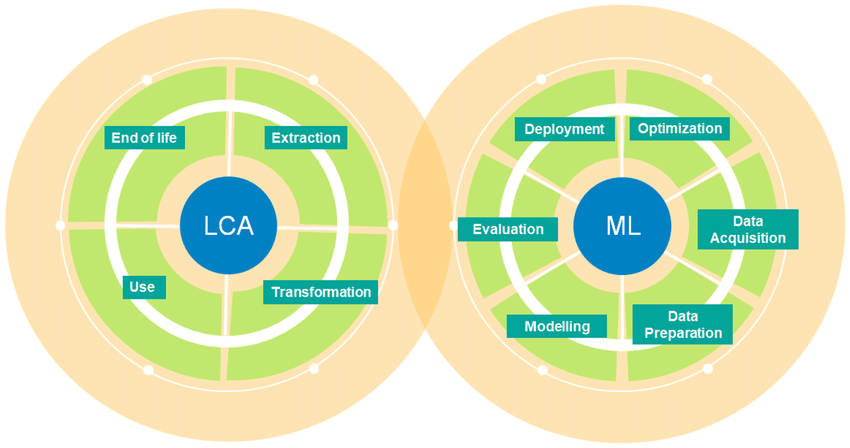}
    \caption{ML and LCA deployment processes \cite{ghoroghi2022advances}}
    \label{fig:ML-LCA}
\end{figure}

LCA is a systematic process used to assess the input and output of materials and energy, as well as their environmental impacts throughout the life cycle of a product or system. ML techniques have the potential to address the complexities and limitations inherent in traditional LCA methods, offering solutions for improving data accuracy, prediction, and decision-making in sustainability assessments, Figure \ref{fig:ML-LCA}. 

ML provides several key advantages for LCA, including its ability to fill gaps in data, predict impact parameters, optimise processes, and integrate and contextualise inventory databases.

\begin{itemize}
    \item \textbf{Prediction of Missing Data:} ML methods can be used to estimate missing data in Life Cycle Inventories (LCI), improving the quality and completeness of the assessments. For instance, algorithms such as k-Nearest Neighbours (KNN) have been successfully applied to estimate missing data for chemicals \cite{aydilek2012novel}, and Artificial Neural Networks (ANNs) have been used to estimate missing eco-indicators for materials in electronic products, as well as integrate recycling scenarios \cite{ghoroghi2022advances}.
    
    \item \textbf{Forecasting Impact Parameters:} ML enables the prediction of environmental impacts, either directly by forecasting emissions or indirectly by modelling contributing factors. Regression models based on design factors, along with ML techniques such as Random Forests (RF) and Multi-Layer Perceptrons (MLP), have been applied to predict carbon emissions from buildings. ANNs have also been used to predict energy demand and LCA for buildings.
    
    \item \textbf{Process Optimisation:} ML can be integrated into LCA process optimisation to identify the most environmentally friendly alternatives. Real-time ML applications can evaluate production processes and suggest adjustments to reduce environmental impact. Hybrid algorithms combining ANNs with Genetic Algorithms (GA) have been used to optimise materials and product envelope design, considering energy consumption and environmental impacts.
    
    \item \textbf{Integration of Inventory Databases:} ML techniques can integrate existing inventory databases, even when they are incomplete, thereby providing a solid foundation for streamlining the LCA process. Conceptual frameworks for integrating AI into LCI have been developed to enhance data completeness and utility.
    
    \item \textbf{Decision Support Tools:} ML allows for the development of tools that model and predict the environmental impacts of products from the design phase, facilitating decision-making for more sustainable choices. ANNs have been used to create decision support tools that assess the performance of buildings rapidly.
    
    \item \textbf{Uncertainty Analysis:} ANNs can also be used to propagate variability in input data and perform quantitative uncertainty analysis in LCA models, which is critical in assessing the robustness of LCA results and making informed decisions under uncertainty.
\end{itemize}

Several ML techniques are applied in LCA, each with its strengths and weaknesses, offering flexibility in addressing various environmental impact prediction challenges.

\begin{itemize}
    \item \textbf{Artificial Neural Networks (ANNs):} Inspired by the human brain, ANNs are efficient at modelling complex and non-linear relationships between input data and environmental impacts. They are widely used for prediction, optimisation, and uncertainty analysis in the LCA of buildings, chemicals, and agricultural processes. Deep Neural Networks (DNNs) have demonstrated the ability to outperform traditional methods in certain LCA applications.
    
    \item \textbf{Support Vector Machines (SVM):} SVMs are particularly useful for classification and regression in high-dimensional data spaces and with smaller sample sizes. While less commonly used than ANNs in LCA, SVMs have shown good performance in predicting the aquatic toxicity of pesticides and modelling fuel consumption in trucks.
    
    \item \textbf{Random Forests (RF):} RF, an ensemble method, combines multiple decision trees to improve prediction accuracy and robustness. In LCA, RFs have been used to predict the yield and characteristics of biomass under different pyrolysis conditions and evaluate the environmental and economic performance of these processes. RFs have proven effective for process classification and material selection in eco-design.
    
    \item \textbf{Hybrid and Ensemble Methods:} The combination of different ML techniques and soft computing methods can lead to superior results by leveraging the strengths of each method. Hybrid approaches, such as integrating ANNs with genetic algorithms for optimisation, or combining fuzzy clustering with Extreme Learning Machines (FCM-ELM) for evaluating the environmental performance of buildings, have shown promise. Ensemble methods, which utilise multiple decision trees, increase model accuracy.
\end{itemize}

\subsection{ML-based Life Cycle Inventory (LCI)}
The life cycle inventory (LCI) stage forms the backbone of LCA, compiling all relevant input and output data, including raw materials, energy consumption, emissions, and by-products \cite{ISO14040_2006, ISO14044_2006}. Constructing an accurate LCI can be particularly challenging for novel materials, where data availability often lags behind technological innovations. As a result, practitioners frequently rely on generic or geographically mismatched databases \cite{kalverkamp2020impacts}. ML-based methods offer a promising solution by efficiently filling these data gaps and improving the quality and quantity of inventory data.

A growing body of research has demonstrated the effectiveness of supervised learning algorithms, such as neural networks and random forests, in estimating missing product characteristics or process yields. 
For instance, Cheng et al. \cite{cheng2020slow} utilized a random forest model to predict biochar yields and properties, feeding these estimates into energy use and greenhouse gas emissions calculations within an LCA framework. 
Similarly, Liao et al. developed an ML model to forecast activated carbon yield, enabling rapid quantification of associated energy and carbon footprints. In the construction sector, Thilakarathna et al. and Naseri et al. applied ML techniques to predict the compressive strength of concrete mixtures, linking model outputs directly to embodied carbon calculations \cite{thilakarathna2020embodied, naseri2020designing}.

Several studies \cite{meng2019data, nguyen2019surrogate, thilakarathna2020embodied, liao2019generating} have employed ML approaches, including linear regression models, ANNs, and random forest algorithms, to estimate environmental emissions. These techniques effectively quantify impacts such as greenhouse gas emissions or the release of organic chemicals during the operational phase of products. Furthermore, data-driven models are increasingly used to predict missing product characteristics, such as biobased chemical properties derived from hydrothermal treatment. By leveraging ML, researchers can improve the accuracy and completeness of LCA datasets, leading to more reliable sustainability assessments \cite{romeiko2024review}. These machine-generated estimates populate the foreground system of the inventory, significantly reducing manual data collection efforts.

In cradle-to-grave LCAs, obtaining timely and comprehensive data remains a challenge. Traditional workflows rely on spreadsheets, manual surveys, and site audits, leading to inefficiencies and extended turnaround times. Additionally, complex supply chains and proprietary manufacturing processes hinder access to granular data, such as emissions from raw material extraction or water usage in specific production facilities. The emergence of large language models (LLMs) offers an opportunity to automate and standardize data acquisition \cite{gachkar2025text, cornago2023can}.

Recent research \cite{preuss2024large, cornago2023can, zhu2023chatgpt, wang2023genai4sustainability} has explored the capabilities of LLMs in assisting LCA practitioners. Findings suggest that LLMs can enhance LCI development by automating data retrieval, summarizing complex information, and improving communication of LCA results. For example, Gachkar et al. \cite{gachkar2025text} developed an NLP-based algorithm to automate LCI analysis in the building sector, significantly reducing data collection time while improving accuracy by extracting material data from bills of quantities (BoQs). Similarly, Chen et al. \cite{chen2024advancing} trained LLMs on research papers focused on LCA for proton exchange membrane water electrolysis and ML applications in LCA, demonstrating that customized LLMs can help address data inaccuracies and information gaps in LCA studies.

Researchers can significantly streamline data collection and enhance assessment accuracy by integrating ML and LLMs into LCA workflows, particularly in advanced material systems. Once trained and validated, ML models can rapidly generate extensive inventory datasets, enabling sensitivity analyses and scenario comparisons with minimal manual effort.

\subsection{ML for Life Cycle Impact Assessment and Interpretation}
Once inventory data are compiled, life cycle impact assessment (LCIA) translates each input or emission into impact categories, such as global warming potential, toxicity, or eutrophication \cite{bassey2024ai}. Traditional tools like SimaPro \cite{SimaPro_2024} and openLCA \cite{GreenDelta_openLCA} facilitate these conversions, but ML presents an opportunity to enhance or accelerate the process. Depending on the goal and scope of the LCA, ML-based approaches can take various forms, including classification models for impact category predictions \cite{kaab2019combined, sun2022improved} or regression models for estimating characterization factors such as ecotoxicity \cite{hou2020estimate}. ML significantly reduces computational time by bypassing traditional numerical simulations, making it particularly useful for multi-scenario evaluations in material selection and process optimization \cite{asif2019integrated}. Beyond impact quantification, LCIA ultimately informs decision-making. ML-based analytics help analyze complex datasets, identify hotspots, optimize trade-offs, and assess uncertainties in environmental impacts. One key advantage of ML is interpretability, allowing researchers to quantify the contribution of different factors to specific sustainability metrics. For instance, Sun et al. \cite{sun2022improved} applied Shapley additive explanations (SHAP) to assess the relative contributions of various impact categories, including climate change, particulate matter formation, terrestrial acidification, freshwater ecotoxicity, human toxicity, and metal depletion \cite{martinez2024frameworks}. Additionally, ML techniques such as random forests and adaptive neuro-fuzzy inference systems have been used to identify key drivers of carbon emissions, providing actionable insights for emissions reduction \cite{mousavi2017combined}. ML also enables the search for low-impact material configurations, balancing sustainability with economic and performance constraints \cite{sharif2019developing}.

\begin{longtable}{|>{\RaggedRight\arraybackslash}p{4cm}|>{\RaggedRight\arraybackslash}p{4.2cm}|>{\RaggedRight\arraybackslash}p{6cm}|}
\caption{Machine learning applications in the context of Life Cycle Assessment (LCA) and environmental sustainability} \label{tab:ml_lca} \\
\hline
\textbf{ML Application Area} & \textbf{Techniques Used} & \textbf{Purpose and Example Use Cases} \\
\hline
\endfirsthead

\multicolumn{3}{c}%
{{\bfseries Table \thetable\ (continued): Machine learning applications in LCA}} \\
\hline
\textbf{ML Application Area} & \textbf{Techniques Used} & \textbf{Purpose and Example Use Cases} \\
\hline
\endhead

\hline \multicolumn{3}{r}{{Continued on next page}} \\
\endfoot

\hline
\endlastfoot

\textbf{Prediction of Missing Data} & 
K-Nearest Neighbours (KNN), Artificial Neural Networks (ANNs) & 
Estimate missing values in Life Cycle Inventories (LCIs), e.g., chemical properties, eco-indicators, and recycling scenarios. \\
\hline

\textbf{Forecasting Environmental Impacts} & 
Random Forests (RF), Multi-Layer Perceptrons (MLP), ANNs & 
Predict emissions or energy demand based on design and process parameters; applied in building energy models and carbon footprint estimation. \\
\hline

\textbf{Process and Product Optimisation} & 
Hybrid models (ANN + Genetic Algorithms), Fuzzy Clustering + Extreme Learning Machines (FCM-ELM) & 
Identify optimal configurations to reduce environmental impact; applied in material selection and envelope design. \\
\hline

\textbf{Integration of Inventory Databases} & 
Framework-level ML approaches & 
Merge and harmonise incomplete or heterogeneous LCI databases to streamline LCA workflows. \\
\hline

\textbf{Decision Support and Uncertainty Analysis} & 
ANNs, Deep Neural Networks (DNNs) & 
Model uncertainty propagation in LCA results; support sustainability decisions under uncertain inputs. \\
\hline

\textbf{LCI Data Estimation} & 
RF, ANN, Linear Regression & 
Predict missing process yields or material properties (e.g., biochar, activated carbon) to enrich inventory datasets. \\
\hline

\textbf{Automated LCI Acquisition} & 
Natural Language Processing (NLP), Large Language Models (LLMs) & 
Extract LCI data from unstructured sources (e.g., bills of quantities, research papers), reducing manual workload and improving data quality. \\
\hline

\textbf{Life Cycle Impact Assessment (LCIA)} & 
RF, Support Vector Machines (SVM), Adaptive Neuro-Fuzzy Inference Systems & 
Map emissions and resource use to impact categories (e.g., global warming potential, eutrophication) and support scenario comparisons. \\
\hline

\textbf{Impact Category Attribution and Interpretation} & 
SHAP (Shapley Additive Explanations), RF & 
Quantify the contribution of various input factors to impact categories such as climate change or ecotoxicity; identify emission drivers and hotspots. \\
\hline

\textbf{Safe and Sustainable by Design (SSbD)} & 
Support Vector Machines (SVM) & 
Classify materials or chemicals for toxicity and sustainability using limited and high-dimensional data. \\
\hline

\end{longtable}

\subsection{\color{blue}Gap Analysis for ML-based Sustainability}

Despite the significant advantages, there are several challenges and limitations in applying ML in LCA.

\begin{itemize}
    \item \textbf{Cost and Availability of Data:} LCA and the training of powerful ML models require substantial amounts of structured, high-quality training data, which can be costly and time-consuming to obtain. The lack of real-world, high-quality data remains a significant barrier to the effective application of ML algorithms in LCA.
    
    \item \textbf{Complexity and Interpretability of Models:} Some ML models, such as DNNs, RFs, and SVMs, are "black-box" models, meaning it is difficult to understand how they arrive at their predictions and to validate their performance in specific contexts. The lack of transparency can make it challenging to trust and interpret the recommendations provided by these models.
    
    \item \textbf{Limited Information in Early Design Phases:} Early design stages often lack the detailed information necessary for comprehensive LCA, making it difficult to apply detailed LCA during initial product development. This limits the application of ML for rapid decision-making in conceptual phases.
    
    \item \textbf{Need for Standardisation:} To ensure compatibility between predictive modelling and experimental studies, standardisation of conditions, experiments, and reporting is necessary to ensure consistency and reproducibility of results.
\end{itemize}

\subsubsection*{Future Research Directions in ML-Driven Sustainable Materials and Manufacturing Systems}

Future research should increasingly focus on the deployment of machine learning (ML) technologies for real-time monitoring, optimisation, and control of materials and manufacturing systems. These applications can enable adaptive process adjustments, predictive maintenance, and resource-efficient operation across diverse production environments. Particular attention should be given to the development of transparent and interpretable ML models such as physics-informed ML or explainable AI methods to foster trust and facilitate adoption in life cycle assessment (LCA) and other sustainability-related evaluations. Indeed, Hybrid modelling approaches, combining data-driven and physics-based techniques, represent a promising pathway to address the limitations of individual models when applied to complex manufacturing scenarios. Additionally, integrating ML with dynamic, time-dependent datasets (e.g., from IoT sensors or real-time process logs) can support continuous environmental impact assessment throughout the product life cycle. A particularly promising direction for sustainable materials engineering lies in the convergence of ML, LCA, and Digital Product Passports (DPPs). While LCA provides a robust framework for quantifying environmental impacts, ML can automate data extraction, reduce uncertainties, and enhance predictive accuracy---especially when inventory data is incomplete or inconsistent. These refined assessments can be embedded into DPPs, which act as comprehensive digital records tracing materials and products across their entire life cycle---from raw material sourcing to end-of-life treatment. By harnessing ML to continuously update and validate DPP data, stakeholders across the value chain including manufacturers, regulators, and end-users can access near real-time insights into the environmental performance of materials and processes. This integration enhances transparency, supports regulatory compliance, and facilitates circular economy strategies. Ultimately, such an approach promotes more informed decisions regarding material selection, process design, and waste valorisation, accelerating the transition toward a more sustainable and resilient manufacturing paradigm.

%%%%%%%%%%%%%%%%%%%%%%%%%%%%%%%%
%%%%%%%%%%%%%%%%%%%%%%%%%%%%%%%%
\section{ML in Materials and Engineeting: Challenges and Opportunities} \label{sec-7}

Despite the transformative potential of machine learning (ML) in advancing materials design, modelling, and discovery, its practical implementation is constrained by several critical challenges. These limitations span multiple dimensions, including:

\begin{itemize}
    \item \textbf{Data availability and quality:} ML models require large volumes of high-quality, diverse, and representative data. In materials science, such datasets are often scarce, incomplete, or inconsistently reported, especially for novel or complex materials systems.
    
    \item \textbf{Model interpretability and reliability:} Many ML algorithms function as “black boxes,” offering limited insight into the underlying physical phenomena. This lack of interpretability hinders scientific understanding and can reduce confidence in model predictions.
    
    \item \textbf{Computational scalability:} The training and deployment of advanced ML models, particularly deep learning architectures, can demand significant computational resources, which may limit their accessibility and applicability at scale.
    
    \item \textbf{Generalisation and transferability:} ML models trained on specific datasets often exhibit limited ability to generalise to unseen material systems or processing conditions, posing challenges for robust deployment in real-world applications.
    
    \item \textbf{Security and robustness:} ML models can be vulnerable to adversarial attacks or input perturbations, raising concerns about their reliability and security in critical materials design and manufacturing workflows.
\end{itemize}

The following paragraphs provide a discussion of these limitations, along with emerging strategies to overcome them.

\subsection{Data availability and quality} 

Data serves as the foundation of successful machine learning (ML) applications in materials science~\cite{liang2022advances}. Both the \textit{quantity} and \textit{quality} of available data play pivotal roles in determining model accuracy, generalisability, and robustness. While large datasets enable models to capture complex relationships and handle variability, high-quality data ensures the learning of meaningful patterns, reduces bias, and enhances prediction reliability. Striking the right balance is essential: insufficient or low-quality data can lead to underperformance, overfitting, or misleading conclusions~\cite{sambasivan2021everyone}. Experimental datasets are frequently incomplete, non-uniform, or missing critical features such as atomic structure, synthesis history, or measurement conditions~\cite{hashimoto2025materials}. This incompleteness undermines ML model performance and generalisation, especially in tasks like inverse design or phase prediction~\cite{lu2024unleashing}. A recent survey indicated that 96\% of enterprises face issues related to training data quality and labelling in ML projects~\cite{businesswire2019}. IBM also emphasises the importance of addressing data-related challenges, highlighting that the success of ML initiatives is often limited by poor data practices~\cite{ibm2020}. As a result, model training often occurs in narrowly defined domains, which impedes extrapolation to novel compositions or processing regimes. Several technical and structural barriers continue to limit the effectiveness of data-driven modelling:

\begin{itemize}
    \item \textbf{Experimental inconsistencies and lack of standardisation:} Data acquired under varying conditions across laboratories often lack reproducibility, hindering dataset interoperability~\cite{xu2023small, butler2024setting}.

    \item \textbf{Methodological biases in simulations:} Computed data, while scalable, come with their own inaccuracies (e.g., DFT underestimation of band gaps), which can propagate bias into ML models~\cite{rodrigues2021big}.

    \item \textbf{Dataset imbalance and missing negatives:} Published datasets tend to overrepresent positive or successful outcomes, while failed experiments or unstable configurations are rarely included~\cite{merchant2023scaling}. This skews the model’s perception of the underlying distribution and can impair decision boundaries.

    \item \textbf{Heterogeneity and noise:} Differences in formats, units, naming conventions, and uncertainty reporting create significant friction in combining datasets across sources~\cite{morgan2020opportunities}.
\end{itemize}

To overcome these limitations, the field is progressively moving towards more collaborative and systematic data strategies:

\begin{itemize}
    \item \textbf{Collaborative data curation:} Joint efforts between experimentalists, theorists, and ML experts can enhance descriptor relevance, detect data artefacts, and support uncertainty quantification.

    \item \textbf{Active and adaptive data collection:} ML-guided experimental design, through active learning or Bayesian optimisation, can help identify the most informative data points and reduce experimental costs~\cite{xu2023small}.

    \item \textbf{Data standardisation and FAIR principles:} Frameworks like the FAIR (Findable, Accessible, Interoperable, Reusable) guidelines and repositories such as \textit{NOMAD}~\cite{nomad}, \textit{Citrination}~\cite{citrine}, and the \textit{Materials Data Facility}~\cite{blaiszik2016materials} promote transparency and accessibility across the community.

    \item \textbf{Uncertainty-aware and robust learning:} Incorporating noise models, uncertainty quantification, and probabilistic frameworks improves model trustworthiness and decision-making in real-world applications.

    \item \textbf{Synthetic data generation:} Simulations and generative models (e.g., GANs, VAEs) can augment underrepresented data regimes, particularly when guided by physical constraints.
\end{itemize}

%%%%%%%%%
%%%%%%%%%
\subsection{Model Interpretability and Reliability}

\textit{Model Interpretability} refers to the ability to understand, explain, and trust the decisions or predictions made by a machine learning (ML) model. In materials science and engineering, this means being able to link the model's output to the physical phenomena that govern material behaviour.   \textit{Model Reliability} in the context of materials modeling with ML refers to the consistency, robustness, and trustworthiness of a model's predictions across a wide range of materials systems. A reliable model should not only provide accurate predictions but also generalize well to new, unseen systems and maintain its predictive performance under varying conditions.  While interpretability and reliability address different facets of model quality, they are deeply interconnected achieving trustworthy predictions in materials science requires not only understanding the rationale behind model outputs but also ensuring their robustness across the vast and diverse materials design space.

A major challenge in the deployment of ML models lies in \textbf{robust generalisation}, i.e., the ability to make reliable predictions for unseen or out-of-distribution (OOD) data~\cite{butler2018machine, von2020exploring}. The materials design space is vast and sparse: even for common alloy systems, the potential compositional permutations can reach astronomical numbers (e.g., $\sim10^{50}$ combinations for high-entropy alloys)~\cite{rao2022machine}. For organic materials, the situation is even more extreme. The number of synthetically accessible, drug-like molecules is estimated to exceed $10^{60}$~\cite{bohacek1996art}, and for polymers, factors such as sequence, topology, and molecular weight add further complexity~\cite{audus2017polymer}. Given that typical datasets only sample a minuscule portion of this space, extrapolation to novel chemistries or structures remains a formidable challenge. In such extrapolative regimes, model confidence often degrades rapidly, potentially yielding misleading or erroneous predictions.

Another common concern is \textbf{overfitting}, where models memorise training data rather than learning generalisable patterns. This leads to inflated in-sample accuracy but poor performance on unseen data. Conversely, underfitting may occur when models are overly simplistic or fail to capture essential physical trends~\cite{schmidt2019recent}. Both phenomena limit a model’s capacity to reliably predict material properties under varying conditions or compositions.

\textbf{Interpretability} also plays a central role in the trustworthiness of ML. Many high-performing models, especially deep learning architectures, act as “black boxes,” making it difficult to rationalise their decisions in physically meaningful terms. This opaqueness hinders scientific insight and limits model adoption in domains requiring accountability or regulatory compliance. Approaches such as symbolic regression, attention mechanisms, feature attribution techniques (e.g., SHAP, LIME), or hybrid physics-ML frameworks have been proposed to enhance model interpretability, though trade-offs with complexity and accuracy often arise~\cite{evolution_gupta_2023}.

Critically, most ML models lack an explicitly defined \textbf{Domain of Applicability (DoA)} the subset of input space for which predictions can be considered reliable~\cite{meredig2018can}. This absence complicates the assessment of model trustworthiness, especially in high-stakes scenarios such as energy storage materials or biomedical applications. Without well-characterised applicability domains or robust uncertainty quantification (UQ), users may inadvertently place unwarranted confidence in predictions made far from the training distribution.

\textbf{Uncertainty quantification} methods offer a potential remedy, providing confidence estimates that flag high-risk predictions. Techniques such as Bayesian neural networks, Gaussian processes, ensemble averaging, and dropout-based UQ have seen increasing application in materials ML. Additionally, integrating physical constraints—e.g., known conservation laws, thermodynamic principles, or symmetry considerations—into ML architectures can improve generalisation and reliability by embedding inductive biases aligned with materials theory~\cite{evolution_gupta_2023}.

%%%%%%%%%%%%%%%%%%%%%%%%%%%%%%%%
\subsection{Model interpretability and reliability} 

\textbf{Model Interpretability} refers to the ability to understand, explain, and trust the decisions or predictions made by a machine learning (ML) model. In materials science, this means being able to link the model's output to the physical phenomena that govern material behavior. Interpretability is crucial because, for materials modeling and design, researchers need to know not only what a model predicts, but also why it makes those predictions, to ensure that the model aligns with known physical principles and provides insights that can guide further research or practical applications.

\textbf{Model Reliability} in the context of materials modeling with ML refers to the consistency, robustness, and trustworthiness of a model's predictions across a wide range of materials systems. A reliable model should not only provide accurate predictions but also generalize well to new, unseen systems and maintain its predictive performance under varying conditions. It should be resilient to noise or uncertainties in the data and maintain its robustness across diverse experimental or computational scenarios.

While ML models can achieve impressive predictive power in specific cases, their reliability can be limited when applied to new materials or systems outside the training data's scope. This is often due to a lack of generalizability, meaning that the model may perform well on the types of data it has seen but poorly on new, unseen data. Additionally, the models might struggle to provide physical insights, which are necessary for understanding the underlying material mechanisms and for making informed decisions about material design.

To ensure \textbf{high confidence levels} and \textbf{robust predictive performance}, it is important to refine ML methodologies. This involves addressing knowledge gaps, improving model generalization, ensuring scalability, and creating frameworks for model validation that integrate physical principles and experimental data. This effort is critical in materials science, where the reliability and interpretability of models directly affect the effectiveness and success of computational materials design and discovery.

While ML has substantially augmented the capabilities within materials modeling and design, the models themselves possess inherent limitations. Despite demonstrating impressive predictive power and efficiency in specific contexts, ML models often encounter difficulties related to generalization across diverse materials systems, providing interpretable physical insights, and ensuring computational scalability for complex problems. These limitations can impede their reliability and applicability, particularly in scenarios demanding high confidence levels and robust predictive performance. A critical understanding of these shortcomings is necessary to refine existing methodologies, address knowledge gaps, and guide the effective integration of ML into the materials science workflow.

A primary challenge for ML models in materials science is achieving robust generalization, the ability to make accurate predictions on new, unseen data that may differ significantly from the training data \cite{butler2018machine, von2020exploring}. This limitation stems from material systems' immense diversity and complexity, encompassing vast and highly variable design spaces. For instance, even when limited to commonly used elements in the periodic table, the compositional space for designing HEAs includes approximately \(10^{50}\) possible alloy combinations \cite{rao2022machine}. Similarly, the estimated number of potential drug-like molecules (with molecular weight \( \leq 500 \)) reaches around \(10^{60}\) \cite{bohacek1996art}. The chemical space for polymers is even more expansive, incorporating additional variables such as sequence, molecular weight distribution, branching, and activity \cite{audus2017polymer}. This sheer diversity presents a significant challenge for ML models, which are typically trained on datasets that cover only a small portion of the materials space. In such extrapolative regimes, where the input differs significantly from the training distribution, predictions often come with high uncertainty and reduced accuracy. This is particularly problematic in materials discovery tasks, where the goal is to explore uncharted regions of the design space.

One common issue in ML is overfitting, where models perform exceptionally well on training data but fail to generalize to test data or real-world applications. This happens when the model essentially "memorizes" the training data rather than genuinely learning underlying patterns, leading it to pick up on noise or spurious correlations. On the other hand, overly simplistic models can result in underfitting, where important trends in the data are missed entirely \cite{schmidt2019recent}.

Furthermore, many ML models lack explicit mechanisms for defining their Domain of Applicability (DoA); the specific range of inputs for which the model's predictions can be considered reliable \cite{meredig2018can}.  Without clear boundaries indicating where the model’s predictions are reliable, users risk over-relying on predictions in inappropriate contexts \cite{meredig2018can}. This lack of DoA is a critical gap, especially for high-stakes applications such as developing energy materials or biomedical devices, where predictive errors can have significant consequences.
Efforts to address these challenges include integrating domain-specific knowledge into ML models, developing robust uncertainty quantification (UQ) methods, and incorporating physical constraints into the learning process. These strategies are essential to improve the generalization capabilities of ML models and ensure their reliability in diverse and complex material systems 
\cite{evolution_gupta_2023}.

\subsubsection{Interpretability, Transparency, and Causality}

The lack of interpretability and transparency in ML remains a significant challenge in materials science, particularly when predictions are derived from complex latent representations rather than domain-specific features \cite{dean2023interpretable}. Often extracted through deep learning techniques, latent representations encapsulate high-dimensional abstractions that are challenging to connect to physical phenomena or material properties \cite{schmidt2019recent}. Unlike domain-specific features, which are grounded in established scientific principles (explained in section \ref{domain-specific}), latent features offer limited insight into the underlying mechanisms governing the observed predictions. This "black-box" nature makes it difficult for researchers to validate or trust the outcomes, especially in high-stakes applications \cite{wang2024xelemnet}.
A lack of interpretability hinders the ability to derive structure-property relationships, a cornerstone of materials science \cite{dai2020method, wang2024xelemnet}. For instance, while ML models may accurately predict a material's bandgap, they may fail to provide an explanation rooted in atomic or electronic interactions. This limitation extends to multi-scale modeling scenarios, where latent representations often conflate features from different length and time scales, making distinguishing between atomic-level interactions and mesoscale phenomena challenging. For example, ML models might correctly identify stable structures in crystal structure prediction but fail to elucidate the underlying energy landscape that governs phase stability.

Transparency is equally critical for addressing biases or errors within ML models. When the decision-making process is opaque, it becomes challenging to identify why a model fails for certain materials or classes of compounds \cite{barocas2023fairness, kumar2022machine}. This issue is exacerbated in cases where latent features dominate the model's predictions, as these features are often abstract and lack direct physical interpretability \cite{ras2022explainable}. For example, latent variables derived from various learning algorithms may reveal patterns in material compositions. Still, without a clear understanding of their connection to measurable properties, they cannot reliably inform materials design \cite{xu2021machine}. Moreover, the challenge becomes particularly acute in industrial applications, where regulatory requirements often mandate explainable decision-making processes, especially for safety-critical materials in aerospace or biomedical applications \cite{hoenig2024explainable, he2021challenges}.

Another major limitation is the inability of many ML models to address causality and quantify uncertainties in their predictions \cite{draper1995assessment}. ML predictions are typically correlation-based and do not establish cause-and-effect relationships. This is particularly problematic in materials science, where understanding the causal mechanisms behind material behavior is essential for tasks such as tailoring mechanical properties or predicting reaction pathways or synthesis routes \cite{moosavi2020role, han2024multifunctional}. Furthermore, the lack of robust uncertainty quantification methods makes it difficult to assess prediction reliability, especially when guiding expensive experimental campaigns or when working with limited, noisy datasets typical in materials research \cite{thebelt2022maximizing}.

Several strategies are under development to address these challenges. Model-agnostic interpretability methods, such as SHAP (Shapley Additive Explanations) and LIME (Local Interpretable Model-Agnostic Explanations), provide post-hoc explanations by quantifying the contributions of individual features to a model's predictions \cite{zhong2022explainable}. However, these methods are less effective when features are derived from latent representations, as they do not inherently capture the physical or chemical context of the data \cite{oviedo2022interpretable}. Neural networks, in particular, require extrinsic explanations due to their nonlinear and opaque nature. The interpretation of such models often depends on analyzing the relationship between inputs and outputs. Neural networks are best explained using local explanations, as each input interacts uniquely with the model to produce specific outputs. This localized approach aligns with the complexity of materials science, where predictions often depend on intricate and system-specific relationships \cite{gilpin2018explaining}. Additionally, physics-informed machine learning models integrate domain knowledge into the learning process, improving interpretability and capturing causal relationships \cite{karniadakis2021physics}. Finally, leveraging graphical models, causal inference techniques, or incorporating explainability constraints during model training provides promising pathways to bridge the gap between predictions and actionable physical understanding \cite{scholkopf2021toward, kim2019learning}.

%%%%%%%%%%%%%%%%%%%%%%%%%%%%%%%%
\subsection{Computational scalability:}
Applying ML in materials science is often constrained by scalability challenges and the substantial computational demands associated with large-scale problems. As the complexity and dimensionality of materials systems increase, spanning vast compositional spaces, intricate structures, and multiple physical phenomena, ML models encounter difficulties in maintaining computational efficiency, predictive accuracy, and practical feasibility. These challenges are particularly acute in tasks like high-throughput virtual screening, comprehensive multiscale simulations, iterative active learning workflows, and the deployment of computationally intensive deep learning algorithms.

\subsubsection{Computational Demands}
Algorithmic complexity and optimization challenges in large-scale materials modeling stem largely from the feature spaces' high dimensionality. Materials design problems often consider a vast range of input parameters, including chemical compositions, atomic structures, and various external conditions, all of which combine into an exponentially growing parameter space \cite{peivaste2022machine, peivaste2023data, zhang2020machine}. This phenomenon, sometimes called the “curse of dimensionality,” makes brute-force exploration or naive optimization prohibitively expensive \cite{lin2024empowering, poggio2017and}. Compounding matters, navigating such non-convex, high-dimensional landscapes typically requires advanced, and often computationally heavy, search and optimization strategies \cite{Lookman2016InformationSF, schmidt2019recent}. Parallelization across multi-node or GPU-enabled clusters can partially alleviate these constraints, but doing so introduces additional synchronization and communication overhead \cite{peivaste2024rapid}. Moreover, iterative and adaptive workflows, where ML models and simulation steps alternate in refining each other, further increase the overall computational burden by necessitating repeated cycles of high-dimensional search and sophisticated parameter tuning \cite{himanen2019data}.  The complexity of resource scheduling and queue management in such distributed systems adds another layer of overhead, requiring careful job scheduling optimisation across multiple users and projects.

A second layer of complexity arises from multi-scale modeling requirements, which demand simultaneous consideration of physical processes across radically different lengths and timescales \cite{wang2024multi}. Integrating these diverse scales into a single ML framework often entails creating deeply nested computational graphs, each incorporating specialized solver routines that handle distinct physical processes. Where multi-physics coupling is involved, for instance, combining thermal, mechanical, and chemical effects, models become even more resource-intensive \cite{peng2021multiscale, wang2024multi, mianroodi2022lossless}. Each additional coupling or scale adds layers of algorithmic overhead, requiring complex scheduling to ensure that data and partial results flow coherently. The system architecture's topology and interconnect bandwidth can significantly impact performance at this stage, particularly in distributed computing environments. Adaptive mesh refinement, dynamic time stepping, or similar techniques for zooming in and out of different lengthscales can drastically alter computational loads on the fly. Efficient load balancing then becomes a non-trivial task, calling for robust distribution strategies that reassign compute resources when new scales are triggered or refined. These interconnected factors drive the need for holistic, infrastructure-aware solutions to manage algorithmic complexity and keep computational costs within feasible limits.

\subsubsection{Hardware Constraints and Resource Allocation}
Hardware constraints and resource allocation challenges become especially evident when scaling ML solutions for large-scale materials modeling and design. Many of these workflows benefit significantly from parallelized hardware, most notably GPUs, Field-Program-mable Gate Arrays (FPGAs), application-specific integrated circuits (ASICs), and tensor processing units (TPUs), which excel at matrix operations critical to training deep neural networks or running computationally demanding simulations \cite{ben2019demystifying, jouppi2017datacenter}. However, the cost and logistical complexity of acquiring and managing these specialized resources can be substantial, particularly in research environments where budgets and allocations compete with other large-scale projects \cite{harris2022physics}. Even when the hardware is available, users must navigate queueing systems, scheduling policies, and varying availability across compute clusters, all of which introduce overhead in both time and effort. Insufficient availability of GPU or TPU clusters can bottle-neck work by forcing researchers to either downsize their experiments or wait in long queues, slowing iterative design cycles.

Beyond raw compute capacity, memory and storage infrastructures can become significant limiting factors \cite{wadhwani2017big}. Large-scale materials models often require running multiple simulations in parallel or examining massive parameter spaces. These tasks produce copious intermediate outputs, which must be efficiently stored, retrieved, and processed \cite{vescovi2022linking}. As a result, the performance of an ML-driven approach may be constrained not only by the GPUs or TPUs themselves but also by the memory bandwidth and storage I/O in place. If the system’s memory or storage layers are not optimized for these workflows, the benefits of high-powered computing hardware can be drastically reduced. Moreover, deploying ML-based materials solutions to high-performance computing (HPC) environments requires careful consideration of distributed file systems, load balancing across nodes, and data transfer protocols. Each of these elements can become a bottleneck if not managed effectively, and their complexity grows in tandem with the size and ambition of the materials modeling task at hand.

%%%%%%%%%%%%%%%%%%%%%%%%%%%%%%%%
\subsection{Security and Privacy}

AI-driven materials research often relies on large, high-quality datasets that are frequently proprietary. Industrial laboratories and companies treat materials data (e.g. novel compounds, processing parameters) as valuable intellectual property, which means such data often remain siloed in private databases \cite{himanen2019data}. This lack of open data sharing creates a significant privacy barrier, not in terms of personal data, but in guarding trade secrets and competitive advantages. As a result, collaborative efforts can be stifled, and AI models may be trained on limited public data, impacting their performance and generalizability. Ensuring that sensitive experimental data and simulation results are protected (through agreements or anonymization) is, therefore, a key challenge in materials AI research.
Furthermore, like all ML systems, AI models for materials design are vulnerable to adversarial threats. Malicious actors could attempt to “poison” training data or introduce adversarial inputs to skew predictions, potentially leading a model to recommend flawed or unsafe material formulations \cite{esezoobo2023integrating}. There is also the risk of unauthorized access or theft of AI models themselves. Model stealing attacks have been demonstrated, where an attacker can extract a deployed model’s parameters or behavior without direct access, undermining intellectual property and exposing any sensitive data learned by the model. Such attacks compromise the competitive edge of the model’s owners and make the stolen models more susceptible to further manipulation \cite{santos2024beyond}. These security concerns highlight the need for robust access controls, encryption of model parameters, and continuous monitoring for anomalies in model outputs or performance.

The application of AI in materials discovery raises new ethical questions and the need for domain-specific regulations. One concern is dual-use discovery. An AI model might propose a material or chemical composition that, while novel, could be hazardous or easily weaponized \cite{trotsyuk2024toward}. For example, adversaries could misuse generative models to design toxic substances or bypass chemical safety regulations, prompting calls to oversee how AI is deployed in chemistry and materials science. Researchers have noted that current AI tools are often released without sufficient evaluation of such misuse risks, and there is a “lack of awareness and consensus on how to prevent or control these risks” in scientific AI \cite{he2023control}. This gap points to the need for ethical guidelines and possibly new regulations tailored to AI-driven research, such as ensuring compliance with export controls or environmental safety standards when using AI to design advanced materials. Additionally, questions of accountability arise (e.g., who is responsible if an AI-guided experiment causes harm, or who owns an AI-discovered material innovation), further underlining that ethics and policy must evolve alongside technical advances in materials AI.

Addressing materials AI's privacy and security challenges requires technical and organizational solutions. On the data side, adopting privacy-preserving ML techniques can protect sensitive information; for instance, federated learning allows multiple parties to train models on distributed data without exposing proprietary datasets \cite{carter2023advanced}. Such approaches help safeguard intellectual property while still benefiting from collective data, and they have been identified as crucial for scenarios where data sharing is precluded due to confidentiality. Researchers are also exploring using encryption and secure hardware (e.g. trusted execution environments) to prevent unauthorized model access during computation. To counter adversarial threats, best practices include rigorous model validation, stress-testing with adversarial examples, and continuous monitoring of model outputs for signs of anomaly or attack. Equally important are policy-level measures: clear data governance frameworks and legal agreements can facilitate safe collaboration between industry and academia, enabling data sharing without compromising proprietary rights \cite{trotsyuk2024toward}. In summary, a combination of technical safeguards (for data and models) and sound policies (for ethical use and IP protection) is essential to ensure privacy and security in AI-enhanced materials research, building trust in these advanced tools within the scientific community.

\subsection{Generalisation and Transferability}
 Generalisation refers to the model's capacity to make accurate predictions on unseen data sampled from the same distribution as the training dataset. In the context of materials science, this could mean predicting the behaviour of a new alloy composition, defect type, or processing parameter not explicitly represented in the training set. Due to the high-dimensional and often sparse nature of materials datasets, overfitting is a common issue, where models learn spurious correlations that do not translate to real physical behaviour. Several strategies have emerged to enhance generalisation in ML models for materials applications:

\begin{itemize}
  \item \textbf{Data augmentation:} Introducing variations in the input space, such as applying symmetry operations to crystal structures or transformations to microstructural images, helps prevent overfitting and improves robustness.
  \item \textbf{Regularisation:} Techniques such as dropout, weight decay, and early stopping limit model complexity, forcing the model to learn more meaningful representations.
  \item \textbf{Physics-informed learning:} Embedding physical constraints or conservation laws directly into the loss function or model architecture helps guide the model toward physically plausible solutions \cite{karniadakis2021physics}.
  \item \textbf{Cross-validation and ensemble methods:} Employing robust cross-validation and combining multiple models can improve predictive stability and reduce variance.
\end{itemize}

Transferability pertains to the ability of a model trained on one dataset or domain to perform well on a different but related domain. In materials science and engineering, this may involve transferring a model trained on a high-throughput simulation dataset (e.g., density functional theory) to experimental data, or using a model trained on one class of materials (e.g., steels) to predict properties in another (e.g., titanium alloys). Challenges in transferability often stem from domain mismatch, where feature distributions, noise characteristics, and physical mechanisms differ significantly across datasets. To mitigate these issues, several techniques are employed:

\begin{itemize}
  \item \textbf{Transfer learning:} A model pretrained on a source task (e.g., structure recognition from synthetic data) is fine-tuned using a smaller, labelled target dataset, often yielding better performance with less data \cite{ward2017including}.
  \item \textbf{Domain adaptation:} Techniques such as adversarial training or feature alignment aim to reduce distributional discrepancies between source and target domains \cite{tshitoyan2019unsupervised}.
  \item \textbf{Multi-fidelity modelling:} Combining data of varying resolutions or accuracies (e.g., combining DFT with molecular dynamics or experimental data) allows the model to learn across scales while retaining physical meaning \cite{ghaboussi2006knowledge}.
  \item \textbf{Few-shot and active learning:} When data is limited, these strategies help prioritise the acquisition of the most informative data points for improving model performance on new domains \cite{lookman2019active}.
\end{itemize}

A promising direction in enhancing both generalisation and transferability lies in the development of universal material representations—feature encodings that capture the essential physics and chemistry across multiple material classes. Graph-based neural networks, descriptor-free learning, and hybrid physics–ML approaches are gaining traction in this regard \cite{xie2018crystal}. These methods aim to develop scalable and interpretable models that can inform design decisions across diverse material systems with minimal retraining.

\begin{longtable}{|>{\RaggedRight\arraybackslash}p{0.33\textwidth}|>{\RaggedRight\arraybackslash}p{0.33\textwidth}|>{\RaggedRight\arraybackslash}p{0.28\textwidth}|}
\caption{Examples of strategies to improve generalisation and transferability in ML for materials science} \label{tab:ml_cases} \\
\hline
\textbf{Problem} & \textbf{Strategy} & \textbf{Example} \\
\hline
\endfirsthead

\multicolumn{3}{c}%
{{\bfseries \tablename\ \thetable{} -- continued from previous page}} \\
\hline
\textbf{Problem} & \textbf{Strategy} & \textbf{Example} \\
\hline
\endhead

\hline \multicolumn{3}{r}{{Continued on next page}} \\
\endfoot

\hline
\endlastfoot

Predicting mechanical properties from microstructure images & 
Convolutional Neural Networks (CNNs) with data augmentation and regularisation & 
Generalisation from steel to aluminium microstructures \\
\hline

Using simulation data to train defect detection models & 
Transfer learning from synthetic to experimental domains & 
DFT-trained model adapted to
TEM images \\
\hline

Lack of experimental data for new alloy systems & 
Few-shot learning or active learning to reduce data requirements & 
Model suggests the most informative next experiment \\
\hline

\end{longtable}

\subsection{Robustness in Materials-aware Machine Learning}
Robustness refers to the model’s ability to maintain stable performance under noisy, incomplete, or unexpected input conditions. As ML models are increasingly deployed in design loops and experimental workflows, ensuring robustness is key to building trustworthy and reproducible tools. In real-world settings, data acquired from experiments or imaging devices (e.g., SEM, TEM) may be subject to noise, artefacts, or systematic bias, making it crucial for ML models to remain reliable under such uncertainty. Robust ML models are particularly essential for high-stakes materials applications such as structural health monitoring, safety-critical alloy design, or defect detection in composite manufacturing. Several strategies have been developed to enhance robustness in the context of materials informatics:

\begin{itemize}
  \item \textbf{Noise-aware training:} Explicitly injecting noise into training data or modelling label uncertainty can make models more resilient to real-world variability.

  \item \textbf{Adversarial training:} While more common in computer vision, adversarial techniques can be adapted to materials domains to improve model behaviour under worst-case perturbations.

  \item \textbf{Bayesian and probabilistic ML:} Methods such as Gaussian Processes or Bayesian Neural Networks provide uncertainty quantification, allowing users to assess confidence levels in predictions and reduce over-reliance on noisy inputs \cite{kendall2017uncertainties}.

  \item \textbf{Hybrid physics--ML models:} Integrating known physical constraints (e.g., symmetry, conservation laws) can reduce the model’s sensitivity to noise and enforce physically valid outputs.

  \item \textbf{Ensemble methods:} Aggregating multiple models trained under varying initial conditions or noise levels tends to improve robustness and reduce prediction variance.
\end{itemize}

%%%%%%%%%%%%%%%%%%%%%%%%%%%%%%%%
%%%%%%%%%%%%%%%%%%%%%%%%%%%%%%%%
\section{Future Directions}\label{sec-8}
A promising frontier is emerging at the intersection of Artificial Intelligence (AI) and materials science. Building on the advancements and challenges outlined in the previous parts of this review, the following discussion aims to highlight key trends and transformative paradigms expected to shape the field in the coming years. The rapid evolution of AI algorithms, combined with the growing availability of high-quality materials data, is set to unlock unprecedented opportunities for researchers.

Among the most exciting developments are hybrid modelling approaches that combine physics-based knowledge with the predictive power of machine learning (ML), leading to more robust and interpretable outcomes. Equally transformative is the rise of autonomous experimentation, where AI systems increasingly steer and accelerate the materials discovery process. The integration of human intuition with AI’s analytical strengths is also gaining recognition as a powerful driver of scientific innovation. Key themes likely to be addressed include:

\begin{itemize}

    \item The creation and deployment of universal material representations capable of capturing essential physical and chemical properties across diverse material families, thereby enhancing generalisation and transferability in ML models.

    \item The growing emphasis on interpretable and explainable AI (XAI), which provides transparency and physical insight into model predictions, promoting trust and advancing scientific understanding.
    
    \item The application of knowledge graphs to unify disparate materials data sources and enable data-driven reasoning and hypothesis generation for accelerated discovery.
    
    \item Efforts towards rigorous data standardisation and the implementation of FAIR (Findable, Accessible, Interoperable, Reusable) principles to ensure the quality, accessibility, and long-term reusability of materials datasets—critical for effective AI deployment in the field.
    
    \item The development of Open Materials Acceleration Platforms (O-MAPs) that seamlessly integrate high-throughput experimentation, computational methods, and AI to compress the innovation cycle.
\end{itemize}

\subsection{Creation and Deployment of Universal Material Representations}

The development of universal material representations capable of capturing essential physical and chemical properties across diverse material families is a critical challenge in the advancement of machine learning (ML) applications in materials science. These representations, often referred to as "universal descriptors," aim to standardise the way materials are represented in ML models, ensuring that they can generalise across different material families and improve the transferability of models to new, unseen materials. To fully realise the potential of ML in materials science, several research and development needs must be addressed in this area:

\begin{itemize}
    \item \textbf{Integration of Multiscale and Multi-modal Data:} Materials exhibit properties that depend on interactions between scales and physics. Future research should focus on developing representations that can effectively integrate data across these scales and physics, enabling ML models to capture the full complexity of material behaviour.
    
    \item \textbf{Data-Driven Universal Descriptors:} A key research direction is the development of data-driven universal descriptors that are learned directly from experimental data or simulations. These descriptors should allow generalisation across different material classes, such as polymers, metals, ceramics, and composites. Techniques such as graph neural networks (GNNs) and deep learning-based feature extraction could play a role in automating the discovery of these descriptors.
    
    \item \textbf{Improvement of Transfer Learning Techniques:} Transfer learning can improve the generalisation of ML models across material families. Better transfer learning techniques are needed, including the creation of representations that are both general and detailed enough to capture the unique characteristics of each material class.
    
    \item \textbf{Incorporating Physical and Chemical Insights into Representations:} Future work should blend domain-specific knowledge from physics and chemistry with data-driven approaches. This could include incorporating bonding interactions, atomic environments, or crystallographic information into material representations to ensure ML models adhere to known physical laws.
    
    \item \textbf{Scalability and High-Throughput Generation of Representations:} Given the vast diversity of materials, automating the generation of material representations for large datasets is crucial. Future research should focus on scalable algorithms for high-throughput generation, enabling efficient integration of materials science with other fields such as robotics and sustainability studies.
    
    \item \textbf{Benchmarking and Validation of Universal Representations:} Rigorous validation methods are needed to assess the accuracy and generality of universal material representations. Standard datasets and benchmarking protocols will ensure that ML models based on these representations are both reliable and widely applicable.
    
\end{itemize}

\subsection{Knowledge Graphs for Materials Data Integration}
Knowledge graphs, which represent relationships between entities in a structured format, can integrate data from various sources such as experimental data, simulations, and theoretical models. The use of KGs in materials science can provide a powerful framework for understanding complex material behaviours and discovering novel materials with desired properties. Several key research and development needs must be addressed to fully realise the potential of KGs in materials science:

\begin{itemize}
    \item \textbf{Integration of Heterogeneous Data Sources:} One of the major challenges in materials science is the integration of data from diverse sources, including experimental data, computational simulations, and literature. Research is needed to develop methods for effectively integrating these heterogeneous data sources into unified knowledge graphs that maintain data quality and consistency across multiple domains.
    
    \item \textbf{Linking Physical and Chemical Properties:} Future research should focus on linking physical and chemical properties of materials within knowledge graphs. This would allow the establishment of relationships between different properties (e.g., conductivity, elasticity, strength), facilitating the prediction of material behaviours and enabling the generation of new hypotheses based on these connections.
    
    \item \textbf{Semantic Enrichment of Data:} Knowledge graphs can provide semantic enrichment to raw materials data by adding context and metadata that enhance the interpretability and usability of data. Future research should explore techniques for automatically enriching materials data with relevant semantic information, such as contextualising data with manufacturing conditions or environmental factors.
    
    \item \textbf{Automated Hypothesis Generation:} One of the key advantages of KGs is their ability to facilitate automated hypothesis generation. By exploring the relationships and dependencies within the graph, KGs can be used to generate new hypotheses about material behaviours, novel material combinations, or previously unobserved phenomena, accelerating the discovery process.
    
    \item \textbf{Interoperability with Machine Learning Models:} For knowledge graphs to be fully integrated into materials science workflows, research is needed on how to best combine KGs with machine learning (ML) models. This includes developing hybrid systems where KGs are used to inform and guide ML models, enhancing the accuracy and interpretability of predictions made by these models.
    
    \item \textbf{Scalability and Automation:} As the volume of materials data continues to grow, the scalability of knowledge graphs becomes increasingly important. Future research should focus on developing scalable algorithms for building and maintaining large-scale knowledge graphs, as well as automating the process of data extraction and graph construction.
    
    \item \textbf{Graph-Based Reasoning and Inference:} Knowledge graphs enable reasoning and inference by providing a structured framework for exploring relationships and dependencies. Future research should explore advanced graph-based reasoning techniques that allow the extraction of new knowledge from existing materials data, providing insights into material design and performance.
    
    \item \textbf{Standardisation of Data Representation:} The creation of standardised formats for representing materials data in knowledge graphs is essential for ensuring interoperability across different systems and platforms. Research is needed to define common data standards and ontologies for materials science that can be widely adopted.
    
    \item \textbf{Collaboration and Open Data Sharing:} The success of knowledge graphs in materials science relies on collaboration and data sharing across institutions and research communities. Open-access materials databases and collaborative platforms will be essential for building comprehensive and diverse knowledge graphs that span multiple material types and research domains.
    
    \item \textbf{Enhancing Data Curation and Quality Control:} To ensure the reliability of knowledge graphs, future work should focus on improving data curation techniques and implementing robust quality control measures. This includes developing methods for detecting and correcting errors, inconsistencies, and gaps in data to maintain the integrity of the knowledge graph.
    
    \item \textbf{Application to Materials Discovery and Optimisation:} Knowledge graphs can play a crucial role in materials discovery and optimisation by providing a comprehensive framework for exploring material properties and behaviours. Research should focus on how to leverage KGs for designing novel materials with tailored properties and optimising material selection for specific applications.
\end{itemize}

\subsection{Enhanced XAI in Materials Science and Engineering}
One of the key contributions of XAI is its ability to make ML models more reliable and interpretable. Moreover, XAI allows for the identification of model errors by offering more accurate diagnostics. For example, the use of heatmaps to visualise classification errors can help pinpoint issues in the data rather than the model itself. This diagnostic capability is crucial for improving model performance, particularly in complex fields like materials science. Another promising direction for XAI in materials science is example-based explanation, where representative data points are used to clarify the model's reasoning. This approach provides a powerful way to assess the reliability of the model and to better understand how data instances influence predictions, reducing the risk of overestimating model accuracy. Furthermore, the design of intrinsically explainable DNN models, through explainable processing or representation mechanisms, or by integrating domain-specific knowledge, represents a major area of advancement. These models can produce transparent outputs even if their internal processing remains complex, which is particularly relevant for the unique challenges encountered in materials science, such as the lack of ground truth data. Several new research areas are emerging as essential for the future development of XAI in materials science and engineering. These include:

\begin{itemize}
    \item \textbf{Multiscale Explainability:} As materials science involves phenomena across multiple scales, developing methods to explain models that operate on multiple scales is a crucial research frontier. Integrating explanations that bridge these scales could enhance our understanding of how microstructural properties influence macroscopic material behaviour.
    
    \item \textbf{Uncertainty Quantification in XAI:} While XAI aims to make models more interpretable, it is equally important to quantify the uncertainty in both the predictions and the explanations provided by these models. Developing techniques to incorporate uncertainty in XAI will be essential for improving model reliability, especially in fields like materials science where data may be sparse or incomplete.
    
    \item \textbf{Explainable Transfer Learning:} Transfer learning, which allows models trained on one dataset to be applied to another, is increasingly used in materials science. However, ensuring that these transferred models remain explainable is a key challenge. Research into explainable transfer learning methods could enable better generalisation of models across diverse material datasets and provide insights into the domain-specific knowledge embedded in these models.
    
    \item \textbf{Human-AI Collaboration for Hypothesis Generation:} One of the long-term goals of XAI is to foster closer collaboration between human experts and AI models. Future research could focus on developing interactive XAI tools that assist researchers in generating new hypotheses, identifying promising material compositions, and optimising experimental designs based on transparent AI reasoning.
    
    \item \textbf{Explainable Quantum AI Models:} As quantum computing begins to play a role in materials design, the development of explainable quantum AI models is a promising area. These models would allow researchers to interpret quantum simulations and understand the underlying mechanisms.
    
    \item \textbf{Ethical and Fairness Considerations in XAI:} As XAI models become more widely used, ensuring they are not only transparent but also ethical and fair is increasingly important. Future research should explore the ethical implications of AI in materials science, particularly in areas like sustainability and resource allocation, ensuring that models do not inadvertently favour certain materials or methods over others based on biased data.
    
    \item \textbf{Data-Efficient XAI:} Many materials science problems face issues of sparse data, which complicates the training and interpretation of machine learning models. Research into data-efficient XAI methods could help maximise the value of limited data by developing techniques that require fewer samples to generate accurate and interpretable explanations.
\end{itemize}

\subsection{Open Materials Acceleration Platforms (O-MAPs)}
The development of Open Materials Acceleration Platforms (O-MAPs) is a promising direction for accelerating the materials discovery and innovation cycle. By seamlessly integrating high-throughput experimentation, computational methods, and artificial intelligence (AI), O-MAPs can significantly compress the time required for the development of new materials. Several key research and development needs must be addressed to fully realise the potential of O-MAPs in materials science:

\begin{itemize}
    \item \textbf{Integration of High-Throughput Experimentation and Computational Methods:} One of the main challenges in developing O-MAPs is integrating experimental high-throughput techniques with computational methods. Future research should focus on creating frameworks that enable real-time feedback between experimental data and computational predictions, facilitating more efficient exploration of the materials space.
    
    \item \textbf{Automation of Experimental and Computational Workflows:} To achieve the desired speed-up in materials discovery, O-MAPs must automate both experimental and computational workflows. Research is needed to develop automated systems that can handle a variety of tasks, such as sample preparation, data collection, analysis, and simulations, to ensure high-throughput and consistent results.
    
    \item \textbf{AI-Driven Materials Design and Optimisation:} Artificial intelligence (AI), particularly machine learning (ML), can play a crucial role in the acceleration of materials discovery. Future work should focus on developing AI-driven algorithms that can analyse large datasets from high-throughput experiments and computational simulations, identifying correlations and guiding materials design to optimise performance.
    
    \item \textbf{Real-Time Data Integration and Decision-Making:} O-MAPs should enable the integration of real-time data from experiments and simulations to drive decision-making processes. Research should focus on developing systems that allow seamless data flow between experimental setups, computational models, and AI algorithms, ensuring timely and data-informed decisions that accelerate materials development.
    
    \item \textbf{Interoperability and Standardisation:} For O-MAPs to be widely adopted, it is essential to establish standardised protocols and formats for data sharing and integration. Research should be directed towards creating universal data standards, ensuring interoperability between different platforms, databases, and research institutions to facilitate collaborative efforts in materials discovery.
    
    \item \textbf{Scalability and Flexibility of Platforms:} As the demand for materials discovery increases, O-MAPs must be scalable and adaptable to a wide range of material types and experimental conditions. Future research should focus on designing flexible platforms that can scale to handle large volumes of data and accommodate diverse material systems, from small molecules to complex composites.
    
    \item \textbf{Integration of Multi-Scale Modelling and Simulations:} Effective integration of multi-scale modelling and simulations with experimental data is essential for the success of O-MAPs. Research should aim to develop methods for coupling different scales of simulations (e.g., atomic, molecular, meso, and continuum) with experimental data to provide a comprehensive understanding of material behaviours and guide new material discovery.
    
    \item \textbf{Data Analytics and Visualisation Tools:} The vast amounts of data generated by O-MAPs require sophisticated data analytics and visualisation tools. Research should focus on developing tools that can process and analyse large datasets from experiments and simulations, as well as provide intuitive visualisations to help researchers interpret and explore complex data patterns and relationships.
    
    \item \textbf{Collaboration and Open-Source Development:} The development of O-MAPs will benefit from open-source collaboration, enabling researchers from different institutions and industries to contribute to the platform's development. Future research should explore methods for creating open-source O-MAP frameworks that promote collaboration and sharing of resources, tools, and best practices.
    
    \item \textbf{Improving the Accuracy and Reliability of Predictions:} One of the key challenges in O-MAPs is ensuring the accuracy and reliability of predictions made by AI and computational models. Future research should focus on developing methods for validating and verifying predictions through experimental validation, ensuring that AI-driven predictions are reliable and trustworthy.
    
    \item \textbf{Sustainability and Environmental Considerations:} As O-MAPs are developed, it is crucial to consider the environmental impact of materials discovery processes. Research should explore ways to design sustainable materials and reduce the environmental footprint of experimentation, data storage, and computational resources, aligning with the growing emphasis on green chemistry and sustainable manufacturing practices.
    
    \item \textbf{Integration with Industry and Manufacturing:} For O-MAPs to truly impact materials innovation, they must be integrated with industry and manufacturing workflows. Future research should explore how to link O-MAPs with real-world industrial processes, ensuring that newly discovered materials can be rapidly translated from the lab to industrial applications.
\end{itemize}

%%%%%%%%%%%%%%%%%%%%%%%%%%%%%%%%
\subsection{Integration of ML Advanced Techniques}
The integration of machine learning (ML) into materials science has witnessed rapid development, offering unprecedented opportunities for accelerating material discovery, understanding, and optimization. Moving forward, two major research directions are expected to drive the next wave of breakthroughs: (i) hybrid models that synergistically combine physics-based and data-driven approaches, and (ii) active learning frameworks for autonomous experimentation.

\paragraph{Hybrid Models: Bridging Data and Physics}
 Hybrid models aim to embed physical laws and domain knowledge into ML architectures, combining the robustness and interpretability of physics-based models with the flexibility and predictive power of data-driven methods.

\begin{itemize}
    \item \textbf{Physics-Informed Neural Networks (PINNs)}: These models integrate differential equations directly into the training of neural networks~\citep{raissi2019physics}, ensuring that the learned functions satisfy known physical laws. This is particularly effective for simulating materials behavior governed by complex PDEs, such as fracture mechanics or phase transitions~\citep{karniadakis2021physics}.
    
    \item \textbf{Operator Learning and Neural Surrogates}: Operator learning approaches, such as DeepONets~\citep{wang2021learning} and Fourier Neural Operators~\citep{peivaste2025teaching, kapoor2022surrogate}, are being used to approximate solution operators of PDEs, enabling generalization to new boundary conditions or geometries. When combined with high-fidelity simulations, these surrogate models allow rapid exploration of large design spaces in materials design.
    
    \item \textbf{Bayesian and Probabilistic Frameworks}: Bayesian neural networks and Gaussian processes integrated with physical constraints offer a principled way to handle uncertainty~\citep{Wang2020, huang2021learning, gal2016dropout, wang2023uncertainty}. These approaches can prioritize physically meaningful predictions and quantify extrapolation risk.
    
    \item \textbf{Hybrid Multi-Scale Models}: In multiscale modeling, ML can serve as a link between scales, e.g., learning homogenized constitutive laws from microstructural simulations~\citep{bessa2017framework}, or accelerating mesoscale simulations through surrogate-based coarse graining~\citep{Liu2023}.
\end{itemize}

\paragraph{Active Learning and Autonomous Experimentation}

Experimental data acquisition remains a bottleneck in materials discovery. Active learning (AL), coupled with autonomous testing, aims to close the loop between prediction, experimentation, and learning, enabling self-driving laboratories.

\begin{itemize}
\item \textbf{Active Learning Algorithms} AL algorithms iteratively select the most informative experiments to perform, based on uncertainty estimates or expected improvement metrics~\citep{Lookman_ActiveLearning_2019, kusne2020fly}. These methods drastically reduce the number of experiments needed to discover optimal materials.

\item \textbf{Autonomous Materials Design, Synthesis and Discovery by machine learning and robotics:} Recent advances in robotics, high-throughput experimentation, and control systems have enabled the realization of autonomous materials laboratories~\citep{jia2024llmatdesign, szymanski2021toward,boiko2023autonomous,rochester2020autonomous}. These systems integrate ML algorithms that guide synthesis, design and characterization decisions in real-time.

\item \textbf{Reinforcement Learning for Experimental Design} Reinforcement learning (RL) offers a powerful paradigm for sequential decision-making under uncertainty, suitable for multi-step experimental workflows~\citep{Häse2021}. RL agents can optimize experimental strategies by balancing exploration and exploitation.

\item \textbf{Digital Twins:} Digital twins,  can be combined with ML and AI to create dynamic, adaptive experimentation loops~\citep{rathore2021role, gunasegaram2021towards}.
\end{itemize}

%%%%%%%%%%%%%%%%%%%%%%%%%%%%%%%%
\subsection{Standardization}

Standardization in materials modeling, design, and discovery is a cornerstone for advancing the field in an era dominated by data-driven approaches \cite{himanen2019data}. As ML methods increasingly underpin materials innovation, the growing complexity of multivariate, multidimensional, and mixed-media datasets generated from experiments, simulations, and characterization workflows has made unified frameworks essential. Standardization establishes common protocols, metadata standards, and ontological frameworks, enabling seamless data integration, interoperability, and reproducibility across diverse platforms and databases \cite{smith2007obo}. International standards organizations, including NIST and ISO, play crucial roles in developing and maintaining these frameworks, ensuring consistency across academic and industrial applications. This ensures that ML models are trained on high-quality, consistent datasets, accelerating the discovery of novel materials, enhancing property predictions, and optimizing design processes. By bridging gaps between experimental, computational, and data-driven efforts, standardization amplifies ML's impact and sets the stage for scalable, efficient, and collaborative research, driving the future of materials science \cite{agrawal2016perspective}. Understanding the challenges inherent in standardization is crucial, as it not only highlights the existing gaps and inefficiencies but also provides a clear pathway for shaping future directions and developing more effective frameworks that enhance collaboration, reproducibility, and innovation in materials modeling and discovery.

\subsubsection{Challenges in Standardization}
Challenges in standardization within materials modeling, design, and discovery are multifaceted, stemming from the inherent complexity and diversity of data in this field \cite{bayerlein2024semantic}. One of the primary issues is data heterogeneity, as experimental, computational, and ML workflows generate vastly different types of data. Experimental outputs, such as microscopy images, spectrographic analyses, and phase diagrams, differ significantly from computational data like DFT, thermodynamic properties, or molecular simulations \cite{zhao2017ontology}. These datasets often lack uniform formats, consistent metadata structures, and standardized terminologies, making it difficult to achieve seamless integration. For example, even basic properties, such as temperature factors, are represented inconsistently across various databases, leading to inefficiencies and confusion.

Another significant challenge lies in the fragmentation of materials databases. Various resources employ unique schemas, access protocols, and data standards. This lack of alignment results in siloed data ecosystems that hinder cross-referencing and data sharing \cite{curtarolo2013high}. Furthermore, these databases often contain gaps in material properties or metadata, which limits their utility for comprehensive ML applications and slows progress in material discovery. The absence of interoperability and semantic alignment further exacerbates the difficulties in standardization. Currently, no universal framework allows seamless mapping of data relationships across platforms and repositories. Although several tools have begun to address this gap by providing semantic frameworks, their adoption remains limited. Without a shared ontology, researchers face significant obstacles in establishing meaningful connections between material properties, processes, and performance \cite{glauer2024chebifier}. This limits the potential of ML methods, which rely on integrated and semantically coherent datasets for accurate predictions and discovery.

A related issue is the inadequate capture of provenance and metadata, which is critical for ensuring the reproducibility and reliability of results. Provenance data, such as experimental conditions, processing steps, and computational parameters, is often inconsistently recorded or completely missing. This lack of detailed documentation reduces datasets' trustworthiness and undermines their value for training ML models. Without high-quality metadata, models risk being biased or producing inaccurate predictions, further complicating the materials discovery process \cite{gil2016toward, raccuglia2016machine}. The dynamic and evolving nature of materials science presents another challenge. Standardization frameworks often struggle to keep pace with the rapid development of new subfields, methodologies, and technologies. For instance, integrating novel features or emerging experimental techniques into existing frameworks requires significant effort and adaptability. Frameworks that fail to scale or evolve with the needs of the field risk becoming obsolete, which poses a significant barrier to their long-term effectiveness.

Finally, resistance to adopting standardized practices remains a persistent issue. Many researchers and organizations are reluctant to overhaul their established workflows, perceiving the cost and effort of transitioning to standardized systems outweighing the immediate benefits. This resistance slows the progress toward a unified ecosystem and perpetuates the inefficiencies caused by fragmented and inconsistent practices. Without widespread adoption, the potential advantages of standardization in improving reproducibility, interoperability, and efficiency remain unrealized, limiting the impact of machine learning on materials discovery and design.

\subsubsection{Opportunities and Tools for Standardization}

Standardization presents significant opportunities to enhance materials modeling, design, and discovery by addressing the above challenges. By leveraging advanced tools and frameworks, the field can establish a unified foundation that enables more efficient collaboration, integration, and innovation.

\begin{figure}
    \centering
    \includegraphics[width=0.7\linewidth]{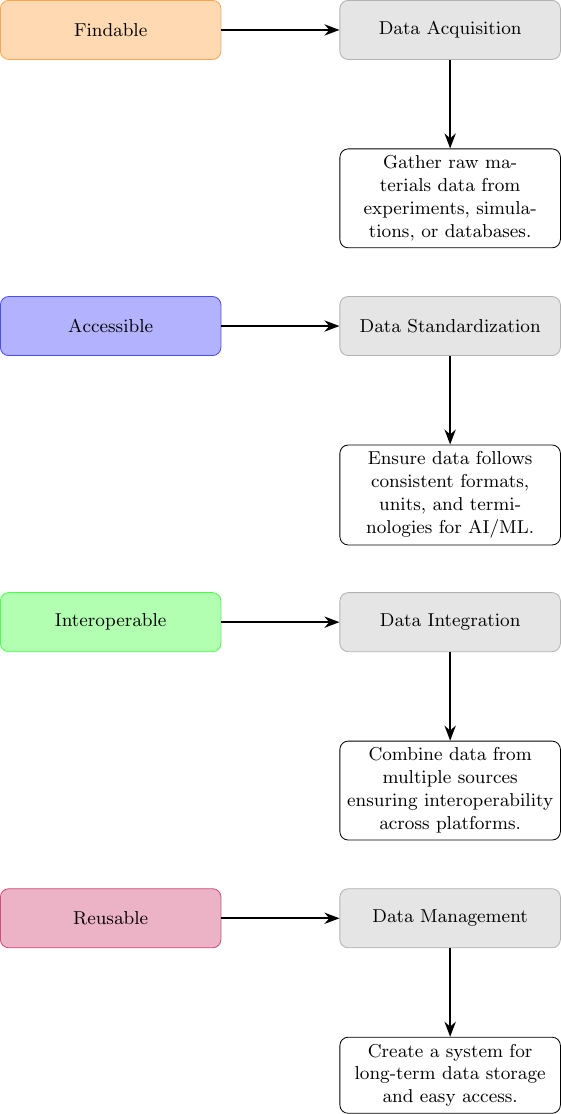}
    \caption{This diagram illustrates the application of the FAIR data principles within the context of materials science. It highlights how each principle maps onto specific steps in a standardised data workflow, from acquisition and standardisation to integration and long-term management. }
    \label{fig:FAIR}
\end{figure}

A primary opportunity lies in leveraging ontologies and semantic knowledge graphs to facilitate data integration and ensure semantic interoperability \cite{valdestilhas2023intersection, zhang2015survey, domingue2011handbook}. Ontologies, such as the PMD Core Ontology (PMDco)\cite{bayerlein2024pmd}, Tensile Test Ontology (TTO) \cite{schilling2024fair}, and Precipitate Geometry Ontology (PGO) \cite{schilling2023pgo} serve as structured frameworks that define domain-specific concepts, relationships, and terminologies. Semantic knowledge graphs, built on these ontologies, enable the representation of complex relationships across diverse datasets, making it possible to link and query information using domain-driven semantics. For instance, the FAIR data infrastructure employs a knowledge graph-based data fusion layer to unify distributed repositories, ensuring that data is findable, accessible, interoperable, and reusable (FAIR) \cite{bayerlein2024semantic}. The integration of FAIR principles is a cornerstone for standardization \cite{schilling2024fair}. FAIR-compliant infrastructures facilitate efficient data storage, semantic annotation, and metadata capture, which are essential for enabling ML and cross-disciplinary research.

Semantic Web technologies such as the Web Ontology Language (OWL) \cite{motik2008owl2}, Resource Description Framework (RDF) \cite{rdf2014}, and SPARQL \cite{rdflib2020sparql} further enhance the usability of standardized datasets. OWL provides a formal mechanism for defining ontologies, while RDF facilitates the creation of machine-readable semantic data. SPARQL enables advanced queries across knowledge graphs, making it possible to discover correlations and insights that are otherwise challenging to uncover. These tools collectively ensure that data is interoperable, reusable, and aligned with FAIR principles. Furthermore, tools such as the Semantic Application Design Language (SADL) \cite{crapo2013toward} and the Semantics Toolkit (SemTK) \cite{cuddihy2018semtk}, further simplify the development and use of knowledge graphs. SADL provides an accessible environment for creating semantic models, enabling domain experts to define ontologies without extensive training in semantic web technologies. SemTK, on the other hand, offers a comprehensive suite of tools for querying, visualizing, and managing knowledge graphs, significantly reducing the complexity of interacting with semantic data. Recently, LLMs have presented an exciting frontier in ontology matching and semantic alignment \cite{ciatto2025large, giglou2024llms4om, Zhao2024UsingLL}. The LLMs4OM \cite{giglou2024llms4om} framework demonstrates how LLMs can augment traditional ontology matching systems by using retrieval-augmented generation (RAG) techniques and advanced prompting methods. This approach enables the efficient alignment of heterogeneous ontologies, further enhancing data integration and knowledge sharing. 

Data-driven frameworks and platforms like MatOnto \cite{Cheung2008TowardsAO} also present significant opportunities. MatOnto integrates heterogeneous relational databases into a unified ontology-based system, enabling semantic queries and expanding the potential for knowledge discovery. Similarly, Material Acceleration Platforms (MAPs) \cite{stier2024materials} incorporate high-throughput experimentation, automated synthesis, and AI-driven data analysis, providing an end-to-end solution for material discovery and design. The FAIR principles are also central to MAPs, ensuring that data generated within these systems is machine-readable and enriched with metadata for improved accessibility and interoperability. These platforms inherently rely on standardized data protocols to integrate diverse workflows effectively, showcasing how standardization directly supports cutting-edge innovation.

The Materials Genome Initiative (MGI) \cite{national2011materials} \url{https://www.mgi.gov/} further exemplifies the role of standardization in advancing materials science. By promoting shared protocols, common data formats, and open-access databases, MGI accelerates the discovery and deployment of new materials. Initiatives like the Open Quantum Materials Database (OQMD), Materials Project, and AFLOW demonstrate how standardized practices can streamline data sharing and enable robust ML applications.

In addition to enabling advanced analytics, standardization fosters collaboration across domains and disciplines. Federated data infrastructures and semantic technologies make it possible to bridge siloed data repositories, creating unified platforms that facilitate interdisciplinary research \cite{gu2024systematic}. This collaborative potential is exemplified by systems that combine experimental data with physics-based equations and ML models, such as compound knowledge graphs. These systems enable researchers to derive new insights by linking disparate datasets and reasoning over the combined knowledge.

Implementing standardization frameworks requires significant investment in training and education. Organizations must develop comprehensive training programs to ensure proper adoption of standardized practices. This includes understanding data quality metrics, validation protocols, and proper documentation practices. The cost implications of implementing standardization frameworks can be substantial, involving technical infrastructure, human resources, and ongoing maintenance. However, the long-term benefits often outweigh these initial investments through improved efficiency, reduced errors, and enhanced collaboration possibilities.

Looking forward, integrating ontologies with AI and ML presents a promising avenue. ML models trained on standardized, high-quality datasets can improve predictive accuracy and uncover complex relationships between material properties and processes. Semantic alignment through ontologies enhances these models by providing context and structured knowledge, thereby advancing data integration and analysis. Tools like AiiDA further complement standardization efforts \cite{huber2020aiida}, which streamline simulation workflows and promote reproducibility. These platforms capture provenance data by automatically tracking and managing the steps involved in computational experiments, ensuring researchers can trace and reproduce results. Future developments in standardization will likely focus on automated validation frameworks, enhanced security protocols, and more sophisticated version control systems to manage evolving datasets and schemas. Integrating blockchain technologies for provenance tracking and smart contracts for automated data validation represents an emerging frontier in materials science standardization.

%%%%%%%%%%%%%%%%%%%%%%%%%%%%%%%%
\subsection{Collaborative Open Platforms} \label{platforms}

%Shared repositories and open-source initiatives.
Modern materials research increasingly relies on collaborative efforts that cross traditional institutional and disciplinary boundaries. Collaborative platforms have become critical for materials modeling and design innovation by encouraging shared knowledge, pooling computational resources, and community-driven development. Two key elements define these platforms: shared repositories, which centralize data and code under open or partially open licenses, and open-source initiatives, which promote transparent, reproducible methodologies and continuous improvement by the larger research community. Both concepts are based on a growing emphasis on metadata standards and FAIR (Findable, Accessible, Interoperable, Reusable) principles, which are essential for achieving efficient data sharing and maximizing the impact of distributed research efforts \cite{tedersoo2021data, scheffler2022fair, groth2020fair}.

Shared and open repositories serve as centralized databases for experimental measurements, computational results, and meta-data, encouraging researchers to deposit, access, and reuse datasets under well-defined frameworks. Examples in the materials domain include:
\begin{itemize}
    \item The Materials Project provides a curated database of computed properties, phase diagrams, and crystal structures, all accessible via an open API \cite{jain2013commentary}. Its centralized format has greatly accelerated materials discovery by reducing duplication of effort and enabling large-scale data analytics.
    \item Open Quantum Materials Database (OQMD), offering thermodynamic and structural data for thousands of inorganic compounds. By furnishing consistent descriptors and thermodynamic information, OQMD empowers researchers to screen candidate materials rapidly \cite{kirklin2015open}.
    \item Materials Cloud, a platform for sharing computational workflows, input files, and derived results. Through integration with the AiiDA framework, Materials Cloud enables full workflow provenance, ensuring traceability and reproducibility \cite{talirz2020materials}
\end{itemize}

Beyond purely materials-focused repositories, general-purpose open data platforms like Figshare or Zenodo host materials-related datasets frequently linked to journal publications. These repositories promote transparency by requiring machine-readable metadata formats, standardized versioning, and the assignment of persistent identifiers (DOIs), confirming the reusability of contributed datasets. 

Open-source initiatives complement shared repositories by making raw data available and the software used to generate and interpret results (simulation packages, ML frameworks, workflow managers, and analysis scripts). This model has been especially effective in materials informatics, where community-driven development encourages rapid iteration and method validation.
Open-source projects contribute to verifying, auditing, and improving published results by disclosing the code and workflows. This improves the reliability of high-throughput studies, machine learning-based property predictions, and complex simulation pipelines. Besides, distributed teams can collaborate in real time, submitting feature enhancements, bug fixes, and specialized modules that expand the capabilities of existing codes. Researchers thus avoid “reinventing the wheel” and focus on domain-specific challenges, such as new descriptors or advanced ML models. Some examples of prominent open-source projects include:
\begin{itemize}
    \item Quantum ESPRESSO \cite{giannozzi2009quantum}, ABINIT \cite{gonze2009abinit}, and LAMMPS \cite{thompson2022lammps}: Widely adopted simulation codes for electronic-structure and atomistic modeling, all developed under community licenses. They often integrate well with data-management frameworks or ML-based force field generators.
    \item AiiDA: A workflow manager and provenance-tracking system that automates large-scale computational campaigns, capturing every step for reproducibility\cite{pizzi2016aiida}.
    \item Matminer: A Python toolkit offering a suite of feature extraction routines and ML utilities, designed to interface seamlessly with materials repositories like The Materials Project \cite{ward2018matminer}.
\end{itemize}

Despite the success of collaborative platforms, some challenges still remain. Data privacy and commercial intellectual property concerns may limit the degree of transparency for specific materials or industrial processes. Furthermore, the diversity of software tools, simulation codes, and experimental measurement protocols makes standardizing data formats across platforms difficult. However, ongoing efforts, such as the OPTIMADE (Open Databases Integration for Materials Design) initiative, aim to harmonize APIs and metadata, allowing different repositories to communicate seamlessly \cite{andersen2021optimade}.
Efforts are also underway to expand documentation and training resources, lowering participation barriers in open-source projects. This includes publicly available tutorials, web-based GUIs, and comprehensive wikis that enable early-career researchers and participants from diverse backgrounds to make meaningful contributions. By addressing these challenges, collaborative platforms can continue to drive faster, more transparent, and efficient materials innovation.

%%%%%%%%%%%%%%%%%%%%%%%%%%%%%%%%
\subsection{Human-AI Collaboration}
% I will work on the strength of each and  points of collaboration and the challenges in the aera of Human-AI

A promising frontier in research lies in the \textbf{systematic integration of human and artificial intelligence} to address increasingly complex scientific and engineering challenges. The synergistic combination of human creativity, adaptability, and critical reasoning with the computational efficiency, precision, and scalability of AI systems offers a unique opportunity to transcend the limitations of either partner alone.

Future investigations should explore \textbf{co-adaptive frameworks} that enable humans and AI to learn from each other iteratively, fostering continuous improvement in both system performance and human understanding. This includes the development of \textbf{interactive AI systems} that can explain their reasoning, adapt to user feedback, and support human decision-making in dynamic environments.

Research should also focus on \textbf{domain-specific implementations} of Human-AI collaboration. In engineering, for instance, future work can deepen the integration of AI in design ideation, real-time simulation, and performance optimization, while preserving the creative and evaluative roles of human experts~\cite{senhora2022machine,zimmermann20183d}. In healthcare, advancing collaborative diagnostic systems, personalized treatment planning, and AI-assisted clinical trials will require robust human-in-the-loop models that maintain interpretability and accountability~\cite{bossen2023batman,wan2024review}.

Moreover, novel \textbf{human-centered AI tools} should be developed to enhance decision-making under uncertainty, especially in critical applications such as autonomous systems, sustainable energy, and materials discovery. Here, AI can propose hypotheses or identify patterns, while humans validate, contextualize, and implement these insights in practice.

\textbf{Domain-specific implementations} of these ideas may include AI-assisted design platforms for microstructure optimization, collaborative digital twins for materials processing, and autonomous experimentation systems guided by expert supervision~\cite{rochester2020autonomous,sparks2020data}. In each case, human experts retain a vital evaluative role—assessing manufacturability, ensuring thermodynamic consistency, and guiding physical interpretation of the results. Ultimately, future research should aim to define \textbf{frameworks, metrics, and best practices} for effective Human-AI teaming, ensuring trust, transparency, and ethical deployment. This collaborative paradigm holds the potential to revolutionize scientific inquiry and technological development across disciplines.

%%%%%%%%%%%%%%%%%%%%%%%%%%%%%%%%
%%%%%%%%%%%%%%%%%%%%%%%%%%%%%%%%
\section{Conclusions}\label{sec-9}

This review has charted the profound and accelerating impact of machine learning across the landscape of materials modeling and design. We have traversed a diverse range of methodologies, data strategies, and applications, demonstrating that ML is now an indispensable component of the modern materials science toolkit. The rapid evolution of techniques, particularly in deep learning, generative models, and specialized architectures like GNNs and Transformers, offers unprecedented capabilities for predicting material properties, discovering novel compounds through inverse design, optimizing complex processes, and tackling grand challenges like sustainability. The imperative for materials scientists is clear: effectively leveraging these tools is becoming essential for driving innovation and maintaining a competitive edge.

Throughout this overview, the critical symbiosis between ML algorithms and materials data has been a recurring theme. Sophisticated methods for data representation—capturing compositional, structural, graph-based, image-based, or even language-inspired features—are fundamental to translating materials knowledge into machine-actionable insights. While significant challenges persist, particularly concerning data scarcity, quality, model interpretability, and computational scalability, these hurdles are increasingly being addressed through dedicated research, community-driven standardization efforts (like FAIR principles and shared platforms), and the development of more robust, physics-informed, and interpretable ML models.

The trajectory points towards an even deeper integration of ML into every facet of materials research and development. The future likely lies in hybrid approaches combining data-driven power with physical constraints, the maturation of autonomous experimental platforms guided by active learning, and enhanced human-AI collaboration where ML augments scientific intuition and creativity. Staying abreast of these developments and effectively integrating ML is crucial for advancing the frontiers of the field. As the pace of innovation continues to quicken, the ability to adeptly apply ML techniques will increasingly differentiate leading research groups and industries. Ultimately, the synergistic fusion of materials domain expertise with sophisticated ML capabilities will define the next era of materials innovation, demanding continuous learning and adaptation from the research community.

\clearpage
\noindent

\noindent
\section*{Acknowledgments}

This review is an outcome of multiple research and innovation projects and collaborative teams. The authors gratefully acknowledge the financial and technical support provided by the following collaborative projects and initiatives.

\begin{itemize}
\item \textbf{SUMO – Sustainable Multi-functional Composites (Grant Number INTER/ GACR/21/16555380):} Sustainable design empowered by materials modelling, semantic interoperability, and multi-criteria optimization. An INTER project funded by the FNR, Luxembourg.\\
\textit{Supported author: Salim Belouettar, Kouider Bendine, Razieh Izadi, Hamidreza Dehghani, Ahmed Makradi, Jakub Lengiewicz and Mohamed El Hachemi}

\item \textbf{HYMAT Porject: PRIDE21/16758661/HYMAT.} A project funded by the FNR under the PRIDE program, focusing on materials for energy.\\
\textit{Supported authors: Iman Peivaste, Salim Belouettar}

\item \textbf{DigiPass CSA – Horizon Europe (Grant No. 101138510):} Coordination and Support Action supporting the implementation and harmonisation of Digital Product Passports (DPP) to enable data-driven sustainability in key industrial sectors.\\
\textit{Supported authors: Salim Belouettar, Halliru Ibrahim, Peter Klein, Martin Hörsch, Heinz Preisig, Natalia Konchakova}

\item \textbf{BatCAT – Battery Cell Assembly Twin (Grant No. 101137725):} A Horizon Europe project focused on creating a digital twin for multiphysics and multiscale simulations integrating both data-driven and physics-based models for battery systems.\\
\textit{Supported authors: Salim Belouettar, Peter Klein, Martin Hörsch}

\item \textbf{COST Action (CA22143)–EuMINe:} European Materials Informatics Network. This COST Action fosters a multidisciplinary community focused on advancing materials informatics and data-driven materials discovery across Europe.\\
\textit{Supported authors: Francesco Mercuri, Iman Peivaste, Salim Belouettar}

\item \textbf{PRIN 2022 PNRR, Project P202278LFC (CUP J53D23015620001):} Funded by the European Union – NextGenerationEU.\\
\textit{Supported author: Nicolas Fantuzzi}

\end{itemize}

\vspace{4mm}

\section*{Author Contributions}

\textbf{Iman Peivaste}\textsuperscript{1,2}: Conceptualization, Methodology, Writing – Review \& Editing. \textbf{Salim Belouettar\textsuperscript{1,*}}: Conceptualization, Methodology, Supervision, Writing, Review \& Editing, Funding Acquisition, Project Administration. \textbf{Francesco Mercuri\textsuperscript{3}}:  Writing, Review \& Editing, Funding Acquisition. \textbf{Nicolas Fantuzzi\textsuperscript{4}}: Conceptualization, Writing, Review \& Editing, Funding Acquisition. \textbf{Mael Belouettar-Mathis\textsuperscript{5}}: Review \& Editing. \textbf{Hamidreza Dehghani\textsuperscript{1}}: Review \& Editing. \textbf{Jakub Lengiewicz\textsuperscript{1}}: Review \& Editing.
 \textbf{Razieh Izadi\textsuperscript{1}}: Review \& Editing.
\textbf{Kouider Bendine\textsuperscript{1}}:  Review \& Editing.
 \textbf{Halliru Ibrahim\textsuperscript{1}}:  Review \& Editing.
 \textbf{Ahmed Makradi\textsuperscript{1}}: Review \& Editing.
\textbf{Martin Hörsch\textsuperscript{6}}: Writing, Review \& Editing, Supervision, Funding Acquisition.
 \textbf{Peter Klein\textsuperscript{7}}: Review \& Editing.
\textbf{Mohamed El Hachemi\textsuperscript{1}}: Review \& Editing.
 \textbf{Heinz A. Preisig\textsuperscript{8}}: Review  \& Editing. \textbf{Yacine Rezgui\textsuperscript{9}}: Review \& Editing.
 \textbf{Natalia Konchakova \textsuperscript{10}}: Review \& Funding Aquisition

\vspace{4mm}

\noindent\textbf{Competing interests}:
The authors declare no competing financial or non-financial interests.

\clearpage
\bibliographystyle{naturemag}. 
\bibliography{main.bib}

@article{kendall2017uncertainties,
  title={What uncertainties do we need in bayesian deep learning for computer vision?},
  author={Kendall, Alex and Gal, Yarin},
  journal={Advances in neural information processing systems},
  volume={30},
  year={2017}
}

@article{wang2021learning,
  title={Learning the solution operator of parametric partial differential equations with physics-informed DeepONets},
  author={Wang, Sifan and Wang, Hanwen and Perdikaris, Paris},
  journal={Science advances},
  volume={7},
  number={40},
  pages={eabi8605},
  year={2021},
  publisher={American Association for the Advancement of Science}
}

@article{lookman2019active,
  title={Active learning in materials science with emphasis on adaptive sampling using uncertainties for targeted design},
  author={Lookman, Turab and Balachandran, Prasanna V. and Xue, Dezhen and Yuan, Ruoqian},
  journal={npj Computational Materials},
  volume={5},
  pages={21},
  year={2019},
  publisher={Nature Publishing Group}
}

@article{rochester2020autonomous,
  title={Autonomous materials synthesis by machine learning and robotics},
  author={Rochester, Christian W. and Yano, Junko and Gregoire, John M.},
  journal={Nature Reviews Materials},
  volume={5},
  pages={1--2},
  year={2020}
}

@article{sparks2020data,
  title={Data-driven autonomous process optimization},
  author={Sparks, Thomas D. and others},
  journal={Accounts of Chemical Research},
  volume={53},
  number={8},
  pages={1895--1905},
  year={2020},
  publisher={ACS Publications}
}

@article{Xie2018,
  author    = {Xie, A. and others},
  title     = {Machine learning for materials science},
  journal   = {Nature Reviews Materials},
  year      = {2018}
}

@article{Wang2020,
  author    = {Wang, P. and others},
  title     = {Bayesian neural networks for modelling composite materials},
  journal   = {Journal of Composite Materials},
  year      = {2020}
}

@article{Nguyen2021,
  author    = {Nguyen, T. and others},
  title     = {Multi-scale modelling and machine learning for composites},
  journal   = {Computational Materials Science},
  year      = {2021}
}

@article{Lee2019,
  author    = {Lee, K. and others},
  title     = {Machine learning models for predicting mechanical properties of composites},
  journal   = {Journal of Materials Science},
  year      = {2019}
}

@article{Zhou2018,
  author    = {Zhou, L. and others},
  title     = {Structure-property relationships in composites using machine learning},
  journal   = {Composites Part B},
  year      = {2018}
}

@article{Yang2020,
  author    = {Yang, B. and others},
  title     = {Convolutional neural networks for microstructure-property predictions of composites},
  journal   = {Materials Design},
  year      = {2020}
}

@article{Wang2021,
  author    = {Wang, Y. and others},
  title     = {Environmental impacts on composite materials predicted using machine learning},
  journal   = {Composites Science and Technology},
  year      = {2021}
}

@article{ogaili2022integration,
  title={Integration of machine learning (ML) and finite element analysis (FEA) for predicting the failure modes of a small horizontal composite blade},
  author={Ogaili, Ahmed Ali Farhan and Hamzah, Mohsin Noori and Jaber, Alaa Abdulhady},
  journal={International Journal of Renewable Energy Research (IJRER)},
  volume={12},
  number={4},
  pages={2168--2179},
  year={2022}
}

@article{zhou2025machine,
  title={A machine learning based multi-scale finite element framework for nonlinear composite materials},
  author={Zhou, Yijing and Semnani, Shabnam J},
  journal={Engineering with Computers},
  pages={1--37},
  year={2025},
  publisher={Springer}
}

@article{yang2021deep,
  title={Deep learning model to predict complex stress and strain fields in hierarchical composites},
  author={Yang, Zhenze and Yu, Chi-Hua and Buehler, Markus J},
  journal={Science Advances},
  volume={7},
  number={15},
  pages={eabd7416},
  year={2021},
  publisher={American Association for the Advancement of Science}
}

@article{sharma2022advances,
  title={Advances in computational intelligence of polymer composite materials: machine learning assisted modeling, analysis and design},
  author={Sharma, Aanchna and Mukhopadhyay, Tanmoy and Rangappa, Sanjay Mavinkere and Siengchin, Suchart and Kushvaha, Vinod},
  journal={Archives of Computational Methods in Engineering},
  volume={29},
  number={5},
  pages={3341--3385},
  year={2022},
  publisher={Springer}
}

@article{liang2025review,
  title={A review on recent applications of machine learning in mechanical properties of composites},
  author={Liang, Yi and Wei, Xinyue and Peng, Yongyue and Wang, Xiaohan and Niu, Xiaoting},
  journal={Polymer Composites},
  volume={46},
  number={3},
  pages={1939--1960},
  year={2025},
  publisher={Wiley Online Library}
}

@article{zhong2022explainable,
  title={Explainable machine learning in materials science},
  author={Zhong, Xiaoting and Gallagher, Brian and Liu, Shusen and Kailkhura, Bhavya and Hiszpanski, Anna and Han, T Yong-Jin},
  journal={npj computational materials},
  volume={8},
  number={1},
  pages={204},
  year={2022},
  publisher={Nature Publishing Group UK London}
}

@article{karniadakis2021physics,
  title={Physics-informed machine learning},
  author={Karniadakis, George Em and Kevrekidis, Ioannis G and Lu, Lu and Perdikaris, Paris and Wang, Sifan and Yang, Liu},
  journal={Nature Reviews Physics},
  volume={3},
  number={6},
  pages={422--440},
  year={2021},
  publisher={Nature Publishing Group}
}

@article{ward2017including,
  title={Including crystal structure attributes in machine learning models of formation energies via Voronoi tessellations},
  author={Ward, Logan and Agrawal, Ankit and Choudhary, Alok and Wolverton, Chris},
  journal={Physical Review B},
  volume={96},
  number={2},
  pages={024104},
  year={2017},
  publisher={APS}
}

@article{tshitoyan2019unsupervised,
  title={Unsupervised word embeddings capture latent knowledge from materials science literature},
  author={Tshitoyan, Vahe and Dagdelen, John and Weston, Leon and Dunn, Alexander and Rong, Ziqin and Kononova, Olga and Persson, Kristin A and Ceder, Gerbrand and Jain, Anubhav},
  journal={Nature},
  volume={571},
  number={7763},
  pages={95--98},
  year={2019},
  publisher={Nature Publishing Group}
}

@article{ghaboussi2006knowledge,
  title={Knowledge-based modeling of material behavior with neural networks},
  author={Ghaboussi, Jamshid and Garrett, James H and Wu, Xin},
  journal={Journal of Engineering Mechanics},
  volume={122},
  number={10},
  pages={1002--1011},
  year={2006},
  publisher={ASCE}
}

@article{xie2018crystal,
  title={Crystal graph convolutional neural networks for an accurate and interpretable prediction of material properties},
  author={Xie, Tian and Grossman, Jeffrey C},
  journal={Physical review letters},
  volume={120},
  number={14},
  pages={145301},
  year={2018},
  publisher={APS}
}

@article{willard2020integrating,
  title={Integrating physics-based modeling with machine learning: A survey},
  author={Willard, Jared and Jia, Xiaowei and Xu, Shaoming and Steinbach, Michael and Kumar, Vipin},
  journal={arXiv preprint arXiv:2003.04919},
  volume={1},
  number={1},
  pages={1--34},
  year={2020}
}

@article{wang2021deep,
  title={Deep learning for surrogate modelling of complex systems},
  author={Wang, Zhen and Wang, Lin and Peng, Rui and Lu, Xuefeng},
  journal={Journal of Computational Physics},
  volume={429},
  pages={109976},
  year={2021}
}

@article{bessa2017framework,
  title={A framework for data-driven analysis of materials under uncertainty: On the fly learning of material laws},
  author={Bessa, Miguel A and Bostanabad, Ramin and Liu, Zhen and Hu, Alan and Apley, Daniel W and Brinson, Catherine and Chen, Wei and Liu, Wing Kam},
  journal={Computational Methods in Applied Mechanics and Engineering},
  volume={320},
  pages={633--667},
  year={2017}
}

@article{ghavamian2020multiscale,
  title={A data-driven framework for the multiscale simulation of history-dependent materials},
  author={Ghavamian, Farid and Simone, Aarnout},
  journal={Computer Methods in Applied Mechanics and Engineering},
  volume={358},
  pages={112619},
  year={2020}
}

@article{raissi2019physics,
  title={Physics-informed neural networks: A deep learning framework for solving forward and inverse problems involving nonlinear partial differential equations},
  author={Raissi, Maziar and Perdikaris, Paris and Karniadakis, George Em},
  journal={Journal of Computational Physics},
  volume={378},
  pages={686--707},
  year={2019}
}

@article{haghighat2021physics,
  title={Physics-informed deep learning for solute transport in heterogeneous porous media},
  author={Haghighat, Ehsan and Raissi, Maziar and Mardani, Mohammad and Juanes, Ruben and Karniadakis, George Em},
  journal={Journal of Hydrology},
  volume={597},
  pages={126190},
  year={2021}
}

@inproceedings{li2022graph,
  title={Mesh-free long-term physics prediction with graph transformers},
  author={Li, Xinyu and Tancik, Matthew and Ng, Ren and Kanazawa, Angjoo},
  booktitle={Advances in Neural Information Processing Systems (NeurIPS)},
  year={2022}
}

@article{huang2020learning,
  title={Learning constitutive relations for viscoelastic solids with physics-informed neural networks},
  author={Huang, Wujie and Darbon, Jérôme and Karniadakis, George Em},
  journal={Computer Methods in Applied Mechanics and Engineering},
  volume={367},
  pages={113002},
  year={2020}
}

@article{ohlson2022assessing,
  title={A framework for strategic discovery of credible neural network surrogate models under uncertainty},
  author={Singh, Pratyush Kumar and Farrell-Maupin, Kathryn A and Faghihi, Danial},
  journal={Computer Methods in Applied Mechanics and Engineering},
  volume={427},
  pages={117061},
  year={2024},
  publisher={Elsevier}
}

@article{huang2021learning,
  title={Model-based Bayesian deep learning architecture for linear inverse problems in computational imaging},
  author={Ekmekci, Canberk and Cetin, Mujdat},
  journal={Electronic Imaging},
  volume={33},
  pages={1--7},
  year={2021},
  publisher={Society for Imaging Science and Technology}
}

@article{willcox2021role,
  title={Machine-learning methods for computational science and engineering},
  author={Frank, Michael and Drikakis, Dimitris and Charissis, Vassilis},
  journal={Computation},
  volume={8},
  number={1},
  pages={15},
  year={2020},
  publisher={MDPI}
}

@inproceedings{blundell2015weight,
  title={Weight uncertainty in neural networks},
  author={Blundell, Charles and Cornebise, Julien and Kavukcuoglu, Koray and Wierstra, Daan},
  booktitle={International Conference on Machine Learning (ICML)},
  pages={1613--1622},
  year={2015}
}

@inproceedings{gal2016dropout,
  title={Dropout as a Bayesian approximation: Representing model uncertainty in deep learning},
  author={Gal, Yarin and Ghahramani, Zoubin},
  booktitle={International Conference on Machine Learning (ICML)},
  pages={1050--1059},
  year={2016}
}

@article{zhang2022quantifying,
  title={Quantifying uncertainty in physics-informed neural networks with noise-robust learning},
  author={Zhang, Zhen and Karniadakis, George Em},
  journal={arXiv preprint arXiv:2202.02303},
  year={2022}
}

@article{yang2021physics,
  title={B-PINNs: Bayesian physics-informed neural networks for forward and inverse PDE problems with noisy data},
  author={Yang, Liu and Meng, Xuhui and Karniadakis, George Em},
  journal={Journal of Computational Physics},
  volume={425},
  pages={109913},
  year={2021}
}

@article{svmcorrosion,
  title={Computer vision and image processing approaches for corrosion detection},
  author={Ali, Ahmad Ali Imran Mohd and Jamaludin, Shahrizan and Imran, Md Mahadi Hasan and Ayob, Ahmad Faisal Mohamad and Ahmad, Sayyid Zainal Abidin Syed and Akhbar, Mohd Faizal Ali and Suhrab, Mohammed Ismail Russtam and Ramli, Mohamad Riduan},
  journal={Journal of Marine Science and Engineering},
  volume={11},
  number={10},
  pages={1954},
  year={2023},
  publisher={MDPI}
}

@article{corrosioncomparison,
  title={An automatic pitting corrosion detection approach for 316L stainless steel},
  author={Jim{\'e}nez--Come, MJ and Turias, IJ and Trujillo, FJ},
  journal={Materials \& Design (1980-2015)},
  volume={56},
  pages={642--648},
  year={2014},
  publisher={Elsevier}
}

@article{fatigueprediction,
  title={Machine learning-based fatigue life prediction of metal materials: Perspectives of physics-informed and data-driven hybrid methods},
  author={Wang, Haijie and Li, Bo and Gong, Jianguo and Xuan, Fu-Zhen},
  journal={Engineering Fracture Mechanics},
  volume={284},
  pages={109242},
  year={2023},
  publisher={Elsevier}
}

@article{tensileprediction,
  title={A machine learning aided interpretable model for rupture strength prediction in Fe-based martensitic and austenitic alloys},
  author={Mamun, Osman and Wenzlick, Madison and Hawk, Jeffrey and Devanathan, Ram},
  journal={Scientific Reports},
  volume={11},
  number={1},
  pages={5466},
  year={2021},
  publisher={Nature Publishing Group UK London}
}

@article{sptprediction,
  title={Investigation on regression model for the force of small punch test using machine learning},
  author={Zhong, Jiru and He, Zhuangzhuang and Guan, Kaishu and Jiang, Tao},
  journal={International Journal of Pressure Vessels and Piping},
  volume={206},
  pages={105031},
  year={2023},
  publisher={Elsevier}
}

@article{pilania2021machine,
  title={Machine learning in materials science: From explainable predictions to autonomous design},
  author={Pilania, Ghanshyam},
  journal={Computational Materials Science},
  volume={193},
  pages={110360},
  year={2021},
  publisher={Elsevier}
}

@article{hodge2004survey,
  title={A survey of outlier detection methodologies},
  author={Hodge, Victoria and Austin, Jim},
  journal={Artificial intelligence review},
  volume={22},
  pages={85--126},
  year={2004},
  publisher={Springer}
}

@article{lawson2021clarity,
  title={CLARITY: comparing heterogeneous data using dissimilarity},
  author={Lawson, Daniel J and Solanki, Vinesh and Yanovich, Igor and Dellert, Johannes and Ruck, Damian and Endicott, Phillip},
  journal={Royal Society Open Science},
  volume={8},
  number={12},
  pages={202182},
  year={2021},
  publisher={The Royal Society}
}

@article{aydilek2012novel,
  title={A novel hybrid approach to estimating missing values in databases using k-nearest neighbors and neural networks},
  author={Aydilek, Ibrahim Berkan and Arslan, Ahmet},
  journal={International Journal of Innovative Computing, Information and Control},
  volume={7},
  number={8},
  pages={4705--4717},
  year={2012}
}

@article{ghoroghi2022advances,
  title={Advances in application of machine learning to life cycle assessment: a literature review},
  author={Ghoroghi, Ali and Rezgui, Yacine and Petri, Ioan and Beach, Thomas},
  journal={The International Journal of Life Cycle Assessment},
  volume={27},
  number={3},
  pages={433--456},
  year={2022},
  publisher={Springer}
}

@article{ghiringhelli2015big,
  author = {Ghiringhelli, Luca M. and Vybiral, Jan and Levchenko, Sergey V. and Draxl, Claudia and Scheffler, Matthias},
  title = {Big Data of Materials Science: Critical Role of the Descriptor},
  journal = {Physical Review Letters},
  volume = {114},
  number = {10},
  pages = {105503},
  year = {2015},
  doi = {10.1103/PhysRevLett.114.105503}
}

@article{schmidt2019recent,
  author = {Schmidt, Jonathan and Marques, Miguel A. L. and Botti, Silvana and Marques, Malgorzata},
  title = {Recent Advances and Applications of Machine Learning in Solid-State Materials Science},
  journal = {npj Computational Materials},
  volume = {5},
  number = {1},
  pages = {1--36},
  year = {2019},
  doi = {10.1038/s41524-019-0221-0}
}

@article{agrawal2016perspective,
  author = {Agrawal, Ankit and Choudhary, Alok},
  title = {Perspective: Materials Informatics and Big Data: Realization of the Fourth Paradigm of Science in Materials Science},
  journal = {APL Materials},
  volume = {4},
  number = {5},
  pages = {053208},
  year = {2016},
  doi = {10.1063/1.4946894}
}

@article{zhang2020machine,
  title={Perspective: Materials informatics and big data: Realization of the “fourth paradigm” of science in materials science},
  author={Agrawal, Ankit and Choudhary, Alok},
  journal={Apl Materials},
  volume={4},
  number={5},
  year={2016},
  publisher={AIP Publishing}
}

@article{mohit2023handling,
  title={Handling aleatoric and epistemic uncertainties in physics-informed neural networks: A survey},
  author={Mohit, Abhinav and Saha, Sudip and Chakraborty, Debasish},
  journal={Archives of Computational Methods in Engineering},
  year={2023},
  note={\url{https://doi.org/10.1007/s11831-023-09831-0}}
}

@article{wang2023uncertainty,
  title={Uncertainty quantification of physics-informed neural networks using adaptive Langevin dynamics-based Bayesian inference},
  author={Wang, Pu and Wang, Shutian and Wu, Qianhong and Zhang, Yuwen and Tian, Yuan and Hu, Yong},
  journal={Engineering Applications of Artificial Intelligence},
  volume={119},
  pages={105781},
  year={2023},
  publisher={Elsevier}
}

@misc{xie2022crystaldiffusionvariationalautoencoder,
	title        = {Crystal Diffusion Variational Autoencoder for Periodic Material Generation},
	author       = {Tian Xie and Xiang Fu and Octavian-Eugen Ganea and Regina Barzilay and Tommi Jaakkola},
	year         = 2022,
	url          = {https://arxiv.org/abs/2110.06197},
	eprint       = {2110.06197},
	archiveprefix = {arXiv},
	primaryclass = {cs.LG}
}

@misc{rombach2022highresolutionimagesynthesislatent,
	title        = {High-Resolution Image Synthesis with Latent Diffusion Models},
	author       = {Robin Rombach and Andreas Blattmann and Dominik Lorenz and Patrick Esser and Björn Ommer},
	year         = 2022,
	url          = {https://arxiv.org/abs/2112.10752},
	eprint       = {2112.10752},
	archiveprefix = {arXiv},
	primaryclass = {cs.CV}
}

@misc{lee2023_2,
	title        = {A Data-Driven Framework for Designing Microstructure of Multifunctional Composites with Deep-Learned Diffusion-Based Generative Models},
	author       = {Kang-Hyun Lee and Hyoung Jun Lim and Gun Jin Yun},
	year         = 2023,
	url          = {https://arxiv.org/abs/2301.09051},
	eprint       = {2301.09051},
	archiveprefix = {arXiv},
	primaryclass = {cond-mat.mtrl-sci}
}

@article{Li_2025,
	title        = {Generative design of crystal structures by point cloud representations and diffusion model},
	author       = {Li, Zhelin and Mrad, Rami and Jiao, Runxian and Huang, Guan and Shan, Jun and Chu, Shibing and Chen, Yuanping},
	year         = 2025,
	month        = jan,
	journal      = {iScience},
	publisher    = {Elsevier BV},
	volume       = 28,
	number       = 1,
	pages        = 111659,
	doi          = {10.1016/j.isci.2024.111659},
	issn         = {2589-0042},
	url          = {http://dx.doi.org/10.1016/j.isci.2024.111659}
}

@misc{lee2023,
	title        = {Microstructure reconstruction using diffusion-based generative models},
	author       = {Kang-Hyun Lee and Gun Jin Yun},
	year         = 2023,
	url          = {https://arxiv.org/abs/2211.10949},
	eprint       = {2211.10949},
	archiveprefix = {arXiv},
	primaryclass = {cond-mat.mtrl-sci}
}

@misc{asperti2020,
	title        = {Balancing reconstruction error and Kullback-Leibler divergence in Variational Autoencoders},
	author       = {Andrea Asperti and Matteo Trentin},
	year         = 2020,
	url          = {https://arxiv.org/abs/2002.07514},
	eprint       = {2002.07514},
	archiveprefix = {arXiv},
	primaryclass = {cs.NE}
}

@misc{cai2017,
	title        = {Multi-Stage Variational Auto-Encoders for Coarse-to-Fine Image Generation},
	author       = {Lei Cai and Hongyang Gao and Shuiwang Ji},
	year         = 2017,
	url          = {https://arxiv.org/abs/1705.07202},
	eprint       = {1705.07202},
	archiveprefix = {arXiv},
	primaryclass = {cs.CV}
}

@misc{zeng2024,
	title        = {Data-efficient and Interpretable Inverse Materials Design using a Disentangled Variational Autoencoder},
	author       = {Cheng Zeng and Zulqarnain Khan and Nathan L. Post},
	year         = 2024,
	url          = {https://arxiv.org/abs/2409.06740},
	eprint       = {2409.06740},
	archiveprefix = {arXiv},
	primaryclass = {cs.LG}
}

@article{ATTARI2023119204,
	title        = {Towards inverse microstructure-centered materials design using generative phase-field modeling and deep variational autoencoders},
	author       = {Vahid Attari and Danial Khatamsaz and Douglas Allaire and Raymundo Arroyave},
	year         = 2023,
	journal      = {Acta Materialia},
	volume       = 259,
	pages        = 119204,
	doi          = {https://doi.org/10.1016/j.actamat.2023.119204},
	issn         = {1359-6454},
	url          = {https://www.sciencedirect.com/science/article/pii/S1359645423005347}
}

@misc{sardeshmukh2024,
	title        = {Material Microstructure Design Using VAE-Regression with Multimodal Prior},
	author       = {Avadhut Sardeshmukh and Sreedhar Reddy and BP Gautham and Pushpak Bhattacharyya},
	year         = 2024,
	url          = {https://arxiv.org/abs/2402.17806},
	eprint       = {2402.17806},
	archiveprefix = {arXiv},
	primaryclass = {cs.LG}
}

@misc{zhang2023,
	title        = {DA-VEGAN: Differentiably Augmenting VAE-GAN for microstructure reconstruction from extremely small data sets},
	author       = {Yichi Zhang and Paul Seibert and Alexandra Otto and Alexander Raßloff and Marreddy Ambati and Markus Kästner},
	year         = 2023,
	url          = {https://arxiv.org/abs/2303.03403},
	eprint       = {2303.03403},
	archiveprefix = {arXiv},
	primaryclass = {cs.LG}
}

@article{Wang_2020,
	title        = {Deep generative modeling for mechanistic-based learning and design of metamaterial systems},
	author       = {Wang, Liwei and Chan, Yu-Chin and Ahmed, Faez and Liu, Zhao and Zhu, Ping and Chen, Wei},
	year         = 2020,
	month        = dec,
	journal      = {Computer Methods in Applied Mechanics and Engineering},
	publisher    = {Elsevier BV},
	volume       = 372,
	pages        = 113377,
	doi          = {10.1016/j.cma.2020.113377},
	issn         = {0045-7825},
	url          = {http://dx.doi.org/10.1016/j.cma.2020.113377}
}

@misc{hashemi2024,
	title        = {Toward Learning Latent-Variable Representations of Microstructures by Optimizing in Spatial Statistics Space},
	author       = {Sayed Sajad Hashemi and Michael Guerzhoy and Noah H. Paulson},
	year         = 2024,
	url          = {https://arxiv.org/abs/2402.11103},
	eprint       = {2402.11103},
	archiveprefix = {arXiv},
	primaryclass = {cs.LG}
}

@article{Taylor2022,
	title        = {Deep Generative Modelling: A Comparative Review of VAEs, GANs, Normalizing Flows, Energy-Based and Autoregressive Models},
	author       = {Bond-Taylor, Sam and Leach, Adam and Long, Yang and Willcocks, Chris G.},
	year         = 2022,
	journal      = {IEEE Transactions on Pattern Analysis and Machine Intelligence},
	volume       = 44,
	number       = 11,
	pages        = {7327--7347},
	doi          = {10.1109/TPAMI.2021.3116668}
}

@article{Ahmad2024,
	title        = {Understanding GANs: fundamentals, variants, training challenges, applications, and open problems},
	author       = {Ahmad, Zeeshan and Jaffri, Zain ul Abidin and Chen, Meng and Bao, Shudi},
	year         = 2024,
	journal      = {Multimedia Tools and Applications},
	doi          = {10.1007/s11042-024-19361-y},
	isbn         = {1573-7721},
	url          = {https://doi.org/10.1007/s11042-024-19361-y},
	da           = {2024/05/14},
	id           = {Ahmad2024},
	ty           = {JOUR}
}

@article{Chun_2020,
	title        = {Deep learning for synthetic microstructure generation in a materials-by-design framework for heterogeneous energetic materials},
	author       = {Chun, Sehyun and Roy, Sidhartha and Nguyen, Yen Thi and Choi, Joseph B. and Udaykumar, H. S. and Baek, Stephen S.},
	year         = 2020,
	month        = aug,
	journal      = {Scientific Reports},
	publisher    = {Springer Science and Business Media LLC},
	volume       = 10,
	number       = 1,
	doi          = {10.1038/s41598-020-70149-0},
	issn         = {2045-2322},
	url          = {http://dx.doi.org/10.1038/s41598-020-70149-0}
}

@article{Jangid2022,
	title        = {3D Grain Shape Generation in Polycrystals Using Generative Adversarial Networks},
	author       = {Jangid, Devendra K. and Brodnik, Neal R. and Khan, Amil and Goebel, Michael G. and Echlin, McLean P. and Pollock, Tresa M. and Daly, Samantha H. and Manjunath, B. S.},
	year         = 2022,
	journal      = {Integrating Materials and Manufacturing Innovation},
	volume       = 11,
	number       = 1,
	pages        = {71--84},
	doi          = {10.1007/s40192-021-00244-1},
	isbn         = {2193-9772},
	url          = {https://doi.org/10.1007/s40192-021-00244-1},
	da           = {2022/03/01},
	id           = {Jangid2022},
	ty           = {JOUR}
}

@article{Hsu2021_,
	title        = {Microstructure Generation via Generative Adversarial Network for Heterogeneous, Topologically Complex 3D Materials},
	author       = {Hsu, Tim and Epting, William K. and Kim, Hokon and Abernathy, Harry W. and Hackett, Gregory A. and Rollett, Anthony D. and Salvador, Paul A. and Holm, Elizabeth A.},
	year         = 2021,
	journal      = {JOM},
	volume       = 73,
	number       = 1,
	pages        = {90--102},
	doi          = {10.1007/s11837-020-04484-y},
	isbn         = {1543-1851},
	url          = {https://doi.org/10.1007/s11837-020-04484-y},
	da           = {2021/01/01},
	id           = {Hsu2021},
	ty           = {JOUR}
}

@misc{ahmad2022chemberta2chemicalfoundationmodels,
	title        = {ChemBERTa-2: Towards Chemical Foundation Models},
	author       = {Walid Ahmad and Elana Simon and Seyone Chithrananda and Gabriel Grand and Bharath Ramsundar},
	year         = 2022,
	url          = {https://arxiv.org/abs/2209.01712},
	eprint       = {2209.01712},
	archiveprefix = {arXiv},
	primaryclass = {cs.LG}
}

@article{Vangala2024,
	title        = {Suitability of large language models for extraction of high-quality chemical reaction dataset from patent literature},
	author       = {Vangala, Sarveswara Rao and Krishnan, Sowmya Ramaswamy and Bung, Navneet and Nandagopal, Dhandapani and Ramasamy, Gomathi and Kumar, Satyam and Sankaran, Sridharan and Srinivasan, Rajgopal and Roy, Arijit},
	year         = 2024,
	journal      = {Journal of Cheminformatics},
	volume       = 16,
	number       = 1,
	pages        = 131,
	doi          = {10.1186/s13321-024-00928-8},
	isbn         = {1758-2946},
	url          = {https://doi.org/10.1186/s13321-024-00928-8},
	da           = {2024/11/26},
	id           = {Vangala2024},
	ty           = {JOUR}
}

@misc{ansari2023agentbasedlearningmaterialsdatasets,
	title        = {Agent-based Learning of Materials Datasets from Scientific Literature},
	author       = {Mehrad Ansari and Seyed Mohamad Moosavi},
	year         = 2023,
	url          = {https://arxiv.org/abs/2312.11690},
	eprint       = {2312.11690},
	archiveprefix = {arXiv},
	primaryclass = {cs.AI}
}

@article{Dong2023,
	title        = {Discovery of 2D Materials using Transformer Network-Based Generative Design},
	author       = {Dong, Rongzhi and Song, Yuqi and Siriwardane, Edirisuriya M. D. and Hu, Jianjun},
	year         = 2023,
	journal      = {Advanced Intelligent Systems},
	volume       = 5,
	number       = 12,
	pages        = 2300141,
	doi          = {https://doi.org/10.1002/aisy.202300141},
	url          = {https://onlinelibrary.wiley.com/doi/abs/10.1002/aisy.202300141},
	eprint       = {https://onlinelibrary.wiley.com/doi/pdf/10.1002/aisy.202300141}
}

@misc{huang2023materialsinformaticstransformerlanguage,
	title        = {Materials Informatics Transformer: A Language Model for Interpretable Materials Properties Prediction},
	author       = {Hongshuo Huang and Rishikesh Magar and Changwen Xu and Amir Barati Farimani},
	year         = 2023,
	url          = {https://arxiv.org/abs/2308.16259},
	eprint       = {2308.16259},
	archiveprefix = {arXiv},
	primaryclass = {cs.LG}
}

@article{KOROLEV2023100803,
	title        = {Accurate, interpretable predictions of materials properties within transformer language models},
	author       = {Vadim Korolev and Pavel Protsenko},
	year         = 2023,
	journal      = {Patterns},
	volume       = 4,
	number       = 10,
	pages        = 100803,
	doi          = {https://doi.org/10.1016/j.patter.2023.100803},
	issn         = {2666-3899},
	url          = {https://www.sciencedirect.com/science/article/pii/S2666389923001587}
}

@misc{chen2024mattergptgenerativetransformermultiproperty,
	title        = {MatterGPT: A Generative Transformer for Multi-Property Inverse Design of Solid-State Materials},
	author       = {Yan Chen and Xueru Wang and Xiaobin Deng and Yilun Liu and Xi Chen and Yunwei Zhang and Lei Wang and Hang Xiao},
	year         = 2024,
	url          = {https://arxiv.org/abs/2408.07608},
	eprint       = {2408.07608},
	archiveprefix = {arXiv},
	primaryclass = {cond-mat.mtrl-sci}
}

@article{Kamal2024,
	title        = {AtomGPT: Atomistic Generative Pretrained Transformer for Forward and Inverse Materials Design},
	author       = {Choudhary, Kamal},
	year         = 2024,
	month        = {07},
	journal      = {The Journal of Physical Chemistry Letters},
	booktitle    = {The Journal of Physical Chemistry Letters},
	publisher    = {American Chemical Society},
	volume       = 15,
	number       = 27,
	pages        = {6909--6917},
	doi          = {10.1021/acs.jpclett.4c01126},
	url          = {https://doi.org/10.1021/acs.jpclett.4c01126},
	da           = {2024/07/11},
	date         = {2024/07/11},
	journal1     = {J. Phys. Chem. Lett.},
	m3           = {doi: 10.1021/acs.jpclett.4c01126},
	ty           = {JOUR},
	year1        = 2024
}

@misc{wei2022crystaltransformerselflearningneural,
	title        = {Crystal Transformer: Self-learning neural language model for Generative and Tinkering Design of Materials},
	author       = {Lai Wei and Qinyang Li and Yuqi Song and Stanislav Stefanov and Edirisuriya M. D. Siriwardane and Fanglin Chen and Jianjun Hu},
	year         = 2022,
	url          = {https://arxiv.org/abs/2204.11953},
	eprint       = {2204.11953},
	archiveprefix = {arXiv},
	primaryclass = {cond-mat.mtrl-sci}
}

@misc{fu2022materialstransformerslanguagemodels,
	title        = {Materials Transformers Language Models for Generative Materials Design: a benchmark study},
	author       = {Nihang Fu and Lai Wei and Yuqi Song and Qinyang Li and Rui Xin and Sadman Sadeed Omee and Rongzhi Dong and Edirisuriya M. Dilanga Siriwardane and Jianjun Hu},
	year         = 2022,
	url          = {https://arxiv.org/abs/2206.13578},
	eprint       = {2206.13578},
	archiveprefix = {arXiv},
	primaryclass = {cond-mat.mtrl-sci}
}

@article{D3DD00137G,
	title        = {Generative adversarial networks and diffusion models in material discovery},
	author       = {Alverson, Michael and Baird, Sterling G. and Murdock, Ryan and Ho, (Enoch) Sin-Hang and Johnson, Jeremy and Sparks, Taylor D.},
	year         = 2024,
	journal      = {Digital Discovery},
	publisher    = {RSC},
	volume       = 3,
	pages        = {62--80},
	doi          = {10.1039/D3DD00137G},
	url          = {http://dx.doi.org/10.1039/D3DD00137G},
	issue        = 1
}

@article{HENKES2022115497,
	title        = {Three-dimensional microstructure generation using generative adversarial neural networks in the context of continuum micromechanics},
	author       = {Alexander Henkes and Henning Wessels},
	year         = 2022,
	journal      = {Computer Methods in Applied Mechanics and Engineering},
	volume       = 400,
	pages        = 115497,
	doi          = {https://doi.org/10.1016/j.cma.2022.115497},
	issn         = {0045-7825},
	url          = {https://www.sciencedirect.com/science/article/pii/S0045782522005126}
}

@article{Lee16092024,
	title        = {Microstructure reconstruction using diffusion-based generative models},
	author       = {Kang-Hyun Lee and Gun Jin Yun},
	year         = 2024,
	journal      = {Mechanics of Advanced Materials and Structures},
	publisher    = {Taylor \& Francis},
	volume       = 31,
	number       = 18,
	pages        = {4443--4461},
	doi          = {10.1080/15376494.2023.2198528},
	url          = {https://doi.org/10.1080/15376494.2023.2198528},
	eprint       = {https://doi.org/10.1080/15376494.2023.2198528}
}

@inproceedings{NEURIPS2020_4c5bcfec,
	title        = {Denoising Diffusion Probabilistic Models},
	author       = {Ho, Jonathan and Jain, Ajay and Abbeel, Pieter},
	year         = 2020,
	booktitle    = {Advances in Neural Information Processing Systems},
	publisher    = {Curran Associates, Inc.},
	volume       = 33,
	pages        = {6840--6851},
	url          = {https://proceedings.neurips.cc/paper_files/paper/2020/file/4c5bcfec8584af0d967f1ab10179ca4b-Paper.pdf},
	editor       = {H. Larochelle and M. Ranzato and R. Hadsell and M.F. Balcan and H. Lin}
}

@inproceedings{Diffusion2015,
	title        = {Deep Unsupervised Learning using Nonequilibrium Thermodynamics},
	author       = {Sohl-Dickstein, Jascha and Weiss, Eric and Maheswaranathan, Niru and Ganguli, Surya},
	year         = 2015,
	month        = {07--09 Jul},
	booktitle    = {Proceedings of the 32nd International Conference on Machine Learning},
	publisher    = {PMLR},
	address      = {Lille, France},
	series       = {Proceedings of Machine Learning Research},
	volume       = 37,
	pages        = {2256--2265},
	url          = {https://proceedings.mlr.press/v37/sohl-dickstein15.html},
	editor       = {Bach, Francis and Blei, David}
}

@misc{kingma2022autoencodingvariationalbayes,
	title        = {Auto-Encoding Variational Bayes},
	author       = {Diederik P Kingma and Max Welling},
	year         = 2022,
	url          = {https://arxiv.org/abs/1312.6114},
	eprint       = {1312.6114},
	archiveprefix = {arXiv},
	primaryclass = {stat.ML}
}

@article{MAL_056,
	title        = {An Introduction to Variational Autoencoders},
	author       = {Diederik P. Kingma and Max Welling},
	year         = 2019,
	journal      = {Foundations and Trends® in Machine Learning},
	volume       = 12,
	number       = 4,
	pages        = {307--392},
	doi          = {10.1561/2200000056},
	issn         = {1935-8237},
	url          = {http://dx.doi.org/10.1561/2200000056}
}

@inproceedings{GANs2014,
	title        = {Generative Adversarial Nets},
	author       = {Goodfellow, Ian and Pouget-Abadie, Jean and Mirza, Mehdi and Xu, Bing and Warde-Farley, David and Ozair, Sherjil and Courville, Aaron and Bengio, Yoshua},
	year         = 2014,
	booktitle    = {Advances in Neural Information Processing Systems},
	publisher    = {Curran Associates, Inc.},
	volume       = 27,
	pages        = {},
	url          = {https://proceedings.neurips.cc/paper_files/paper/2014/file/5ca3e9b122f61f8f06494c97b1afccf3-Paper.pdf},
	editor       = {Z. Ghahramani and M. Welling and C. Cortes and N. Lawrence and K.Q. Weinberger}
}

@article{Liu2023,
	title        = {Data quantity governance for machine learning in materials science},
	author       = {Liu, Yue and Yang, Zhengwei and Zou, Xinxin and Ma, Shuchang and Liu, Dahui and Avdeev, Maxim and Shi, Siqi},
	year         = 2023,
	month        = {05},
	journal      = {National Science Review},
	volume       = 10,
	number       = 7,
	pages        = {nwad125},
	doi          = {10.1093/nsr/nwad125},
	issn         = {2095-5138},
	url          = {https://doi.org/10.1093/nsr/nwad125},
	eprint       = {https://academic.oup.com/nsr/article-pdf/10/7/nwad125/50602961/nwad125.pdf}
}

@misc{kailkhura2019reliableexplainablemachinelearning,
	title        = {Reliable and Explainable Machine Learning Methods for Accelerated Material Discovery},
	author       = {Bhavya Kailkhura and Brian Gallagher and Sookyung Kim and Anna Hiszpanski and T. Yong-Jin Han},
	year         = 2019,
	url          = {https://arxiv.org/abs/1901.02717},
	eprint       = {1901.02717},
	archiveprefix = {arXiv},
	primaryclass = {physics.comp-ph}
}

@misc{chang2022,
	title        = {Towards overcoming data scarcity in materials science: unifying models and datasets with a mixture of experts framework},
	author       = {Rees Chang and Yu-Xiong Wang and Elif Ertekin},
	year         = 2022,
	url          = {https://arxiv.org/abs/2207.13880},
	eprint       = {2207.13880},
	archiveprefix = {arXiv},
	primaryclass = {cond-mat.mtrl-sci}
}

@misc{tian2022,
	title        = {Tackling Data Scarcity with Transfer Learning: A Case Study of Thickness Characterization from Optical Spectra of Perovskite Thin Films},
	author       = {Siyu Isaac Parker Tian and Zekun Ren and Selvaraj Venkataraj and Yuanhang Cheng and Daniil Bash and Felipe Oviedo and J. Senthilnath and Vijila Chellappan and Yee-Fun Lim and Armin G. Aberle and Benjamin P MacLeod and Fraser G. L. Parlane and Curtis P. Berlinguette and Qianxiao Li and Tonio Buonassisi and Zhe Liu},
	year         = 2022,
	url          = {https://arxiv.org/abs/2207.02209},
	eprint       = {2207.02209},
	archiveprefix = {arXiv},
	primaryclass = {cs.LG}
}

@article{jordan2015machine,
	title        = {Machine learning: Trends, perspectives, and prospects},
	author       = {Jordan, Michael I and Mitchell, Tom M},
	year         = 2015,
	journal      = {Science},
	publisher    = {American Association for the Advancement of Science},
	volume       = 349,
	number       = 6245,
	pages        = {255--260}
}

@article{jain2013commentary,
	title        = {Commentary: The Materials Project: A materials genome approach to accelerating materials innovation},
	author       = {Jain, Anubhav and Ong, Shyue Ping and Hautier, Geoffroy and Chen, Wei and Richards, William Davidson and Dacek, Stephen and Cholia, Shreyas and Gunter, Dan and Skinner, David and Ceder, Gerbrand and others},
	year         = 2013,
	journal      = {APL materials},
	publisher    = {AIP Publishing},
	volume       = 1,
	number       = 1
}

@book{Goodfellow-et-al-2016,
	title        = {Deep Learning},
	author       = {Ian Goodfellow and Yoshua Bengio and Aaron Courville},
	year         = 2016,
	publisher    = {MIT Press},
	note         = {\url{http://www.deeplearningbook.org}}
}

@article{Hornik1989,
	title        = {Multilayer feedforward networks are universal approximators},
	author       = {Kurt Hornik and Maxwell Stinchcombe and Halbert White},
	year         = 1989,
	journal      = {Neural Networks},
	volume       = 2,
	number       = 5,
	pages        = {359--366},
	doi          = {https://doi.org/10.1016/0893-6080(89)90020-8},
	issn         = {0893-6080},
	url          = {https://www.sciencedirect.com/science/article/pii/0893608089900208}
}

@article{Cybenko1989,
	title        = {Approximation by superpositions of a sigmoidal function},
	author       = {Cybenko,  G.},
	year         = 1989,
	journal      = {Mathematics of Control,  Signals,  and Systems},
	publisher    = {Springer Science and Business Media LLC},
	volume       = 2,
	number       = 4,
	pages        = {303–314},
	doi          = {10.1007/bf02551274},
	issn         = {1435-568X},
	url          = {http://dx.doi.org/10.1007/BF02551274}
}

@article{krizhevsky2012imagenet,
	title        = {"ImageNet" classification with deep convolutional neural networks},
	author       = {Krizhevsky, Alex and Sutskever, Ilya and Hinton, Geoffrey E},
	year         = 2012,
	journal      = {Advances in neural information processing systems},
	volume       = 25
}

@misc{bengio2009learning,
	title        = {Learning Deep Architectures for {AI}},
	author       = {Bengio, Y},
	year         = 2009,
	publisher    = {Now Publishers Inc}
}

@article{raccuglia2016machine,
	title        = {Machine-learning-assisted materials discovery using failed experiments},
	author       = {Raccuglia, Paul and Elbert, Katherine C and Adler, Philip DF and Falk, Casey and Wenny, Malia B and Mollo, Aurelio and Zeller, Matthias and Friedler, Sorelle A and Schrier, Joshua and Norquist, Alexander J},
	year         = 2016,
	journal      = {Nature},
	publisher    = {Nature Publishing Group UK London},
	volume       = 533,
	number       = 7601,
	pages        = {73--76}
}

@article{klenam2023data,
	title        = {Data science and material informatics in physical metallurgy and material science: An overview of milestones and limitations},
	author       = {Klenam, DEP and Asumadu, TK and Vandadi, M and Rahbar, N and McBagonluri, F and Soboyejo, WO},
	year         = 2023,
	journal      = {Results in Materials},
	publisher    = {Elsevier},
	volume       = 19,
	pages        = 100455
}

@article{national2019decadal,
	title        = {A decadal survey of the social and behavioral sciences: A research agenda for advancing intelligence analysis},
	author       = {National Academies of Sciences and Medicine and Division of Behavioral and Board on Behavioral and Sensory Sciences and Committee on a Decadal Survey of Social and Behavioral Sciences for Applications to National Security},
	year         = 2019,
	publisher    = {National Academies Press}
}

@article{wang2020machine,
	title        = {Machine learning for materials scientists: an introductory guide toward best practices},
	author       = {Wang, Anthony Yu-Tung and Murdock, Ryan J and Kauwe, Steven K and Oliynyk, Anton O and Gurlo, Aleksander and Brgoch, Jakoah and Persson, Kristin A and Sparks, Taylor D},
	year         = 2020,
	journal      = {Chemistry of Materials},
	publisher    = {ACS Publications},
	volume       = 32,
	number       = 12,
	pages        = {4954--4965}
}

@article{LeCun2015,
	title        = {Deep learning},
	author       = {LeCun,  Yann and Bengio,  Yoshua and Hinton,  Geoffrey},
	year         = 2015,
	month        = may,
	journal      = {Nature},
	publisher    = {Springer Science and Business Media LLC},
	volume       = 521,
	number       = 7553,
	pages        = {436–444},
	doi          = {10.1038/nature14539},
	issn         = {1476-4687},
	url          = {http://dx.doi.org/10.1038/nature14539}
}

@misc{bahdanau2016,
	title        = {Neural Machine Translation by Jointly Learning to Align and Translate},
	author       = {Dzmitry Bahdanau and Kyunghyun Cho and Yoshua Bengio},
	year         = 2016,
	url          = {https://arxiv.org/abs/1409.0473},
	eprint       = {1409.0473},
	archiveprefix = {arXiv},
	primaryclass = {cs.CL}
}

@article{HintonEtAlSpeechRecognition_2012,
	title        = {Deep Neural Networks for Acoustic Modeling in Speech Recognition: The Shared Views of Four Research Groups},
	author       = {Hinton, Geoffrey and Deng, Li and Yu, Dong and Dahl, George E. and Mohamed, Abdel-rahman and Jaitly, Navdeep and Senior, Andrew and Vanhoucke, Vincent and Nguyen, Patrick and Sainath, Tara N. and Kingsbury, Brian},
	year         = 2012,
	journal      = {IEEE Signal Processing Magazine},
	volume       = 29,
	number       = 6,
	pages        = {82--97},
	doi          = {10.1109/MSP.2012.2205597}
}

@article{DahlEtAl_DNNsSpeechRecognition_2012,
	title        = {Context-Dependent Pre-Trained Deep Neural Networks for Large-Vocabulary Speech Recognition},
	author       = {Dahl, George E. and Yu, Dong and Deng, Li and Acero, Alex},
	year         = 2012,
	journal      = {IEEE Transactions on Audio, Speech, and Language Processing},
	volume       = 20,
	number       = 1,
	pages        = {30--42},
	doi          = {10.1109/TASL.2011.2134090}
}

@misc{telgarsky2015representationbenefitsdeepfeedforward,
	title        = {Representation Benefits of Deep Feedforward Networks},
	author       = {Matus Telgarsky},
	year         = 2015,
	url          = {https://arxiv.org/abs/1509.08101},
	eprint       = {1509.08101},
	archiveprefix = {arXiv},
	primaryclass = {cs.LG}
}

@inproceedings{Brown_GPT3_NEURIPS_2020,
	title        = {Language Models are Few-Shot Learners},
	author       = {Brown, Tom and Mann, Benjamin and Ryder, Nick and Subbiah, Melanie and Kaplan, Jared D and Dhariwal, Prafulla and Neelakantan, Arvind and Shyam, Pranav and Sastry, Girish and Askell, Amanda and Agarwal, Sandhini and Herbert-Voss, Ariel and Krueger, Gretchen and Henighan, Tom and Child, Rewon and Ramesh, Aditya and Ziegler, Daniel and Wu, Jeffrey and Winter, Clemens and Hesse, Chris and Chen, Mark and Sigler, Eric and Litwin, Mateusz and Gray, Scott and Chess, Benjamin and Clark, Jack and Berner, Christopher and McCandlish, Sam and Radford, Alec and Sutskever, Ilya and Amodei, Dario},
	year         = 2020,
	booktitle    = {Advances in Neural Information Processing Systems},
	publisher    = {Curran Associates, Inc.},
	volume       = 33,
	pages        = {1877--1901},
	url          = {https://proceedings.neurips.cc/paper_files/paper/2020/file/1457c0d6bfcb4967418bfb8ac142f64a-Paper.pdf},
	editor       = {H. Larochelle and M. Ranzato and R. Hadsell and M.F. Balcan and H. Lin}
}

@article{Jumper_AlphaFold_2021,
	title        = {Highly accurate protein structure prediction with AlphaFold},
	author       = {Jumper,  John and Evans,  Richard and Pritzel,  Alexander and Green,  Tim and Figurnov,  Michael and Ronneberger,  Olaf and Tunyasuvunakool,  Kathryn and Bates,  Russ and Žídek,  Augustin and Potapenko,  Anna and Bridgland,  Alex and Meyer,  Clemens and Kohl,  Simon A. A. and Ballard,  Andrew J. and Cowie,  Andrew and Romera-Paredes,  Bernardino and Nikolov,  Stanislav and Jain,  Rishub and Adler,  Jonas and Back,  Trevor and Petersen,  Stig and Reiman,  David and Clancy,  Ellen and Zielinski,  Michal and Steinegger,  Martin and Pacholska,  Michalina and Berghammer,  Tamas and Bodenstein,  Sebastian and Silver,  David and Vinyals,  Oriol and Senior,  Andrew W. and Kavukcuoglu,  Koray and Kohli,  Pushmeet and Hassabis,  Demis},
	year         = 2021,
	month        = jul,
	journal      = {Nature},
	publisher    = {Springer Science and Business Media LLC},
	volume       = 596,
	number       = 7873,
	pages        = {583–589},
	doi          = {10.1038/s41586-021-03819-2},
	issn         = {1476-4687},
	url          = {http://dx.doi.org/10.1038/s41586-021-03819-2}
}

@inproceedings{YosinskiEtAl_TransferLearning_2014,
	title        = {How transferable are features in deep neural networks?},
	author       = {Yosinski, Jason and Clune, Jeff and Bengio, Yoshua and Lipson, Hod},
	year         = 2014,
	booktitle    = {Proceedings of the 27th International Conference on Neural Information Processing Systems - Volume 2},
	location     = {Montreal, Canada},
	publisher    = {MIT Press},
	address      = {Cambridge, MA, USA},
	series       = {NIPS'14},
	pages        = {3320–3328},
	numpages     = 9
}

@article{Bronstein_GometricDL_2017,
	title        = {Geometric Deep Learning: Going beyond Euclidean data},
	author       = {Bronstein, Michael M. and Bruna, Joan and LeCun, Yann and Szlam, Arthur and Vandergheynst, Pierre},
	year         = 2017,
	journal      = {IEEE Signal Processing Magazine},
	volume       = 34,
	number       = 4,
	pages        = {18--42},
	doi          = {10.1109/MSP.2017.2693418}
}

@misc{Kipf_classGCN_2017,
	title        = {Semi-Supervised Classification with Graph Convolutional Networks},
	author       = {Thomas N. Kipf and Max Welling},
	year         = 2017,
	url          = {https://arxiv.org/abs/1609.02907},
	eprint       = {1609.02907},
	archiveprefix = {arXiv},
	primaryclass = {cs.LG}
}

@misc{Velickovic_GraphAttention_2018,
	title        = {Graph Attention Networks},
	author       = {Petar Veličković and Guillem Cucurull and Arantxa Casanova and Adriana Romero and Pietro Liò and Yoshua Bengio},
	year         = 2018,
	url          = {https://arxiv.org/abs/1710.10903},
	eprint       = {1710.10903},
	archiveprefix = {arXiv},
	primaryclass = {stat.ML}
}

@article{Deshpande_MagNET_2024,
	title        = {{MAgNET}: A graph {U-Net} architecture for mesh-based simulations},
	author       = {Saurabh Deshpande and Stéphane P.A. Bordas and Jakub Lengiewicz},
	year         = 2024,
	journal      = {Engineering Applications of Artificial Intelligence},
	volume       = 133,
	pages        = 108055,
	doi          = {https://doi.org/10.1016/j.engappai.2024.108055},
	issn         = {0952-1976},
	url          = {https://www.sciencedirect.com/science/article/pii/S0952197624002136}
}

@article{Deshpande_CNN_UNet_2022,
	title        = {Probabilistic deep learning for real-time large deformation simulations},
	author       = {Saurabh Deshpande and Jakub Lengiewicz and Stéphane P.A. Bordas},
	year         = 2022,
	journal      = {Computer Methods in Applied Mechanics and Engineering},
	volume       = 398,
	pages        = 115307,
	doi          = {https://doi.org/10.1016/j.cma.2022.115307},
	issn         = {0045-7825},
	url          = {https://www.sciencedirect.com/science/article/pii/S004578252200411X}
}

@article{Deshpande_ConvAggrAtt_2023,
	title        = {Convolution,  aggregation and attention based deep neural networks for accelerating simulations in mechanics},
	author       = {Deshpande,  Saurabh and Sosa,  Raúl I. and Bordas,  Stéphane P. A. and Lengiewicz,  Jakub},
	year         = 2023,
	month        = mar,
	journal      = {Frontiers in Materials},
	publisher    = {Frontiers Media SA},
	volume       = 10,
	doi          = {10.3389/fmats.2023.1128954},
	issn         = {2296-8016},
	url          = {http://dx.doi.org/10.3389/fmats.2023.1128954}
}

@misc{redmon_YOLO_2016,
	title        = {You Only Look Once: Unified, Real-Time Object Detection},
	author       = {Joseph Redmon and Santosh Divvala and Ross Girshick and Ali Farhadi},
	year         = 2016,
	url          = {https://arxiv.org/abs/1506.02640},
	eprint       = {1506.02640},
	archiveprefix = {arXiv},
	primaryclass = {cs.CV}
}

@misc{jaegle_PerceiveriIO_2022,
	title        = {Perceiver IO: A General Architecture for Structured Inputs \& Outputs},
	author       = {Andrew Jaegle and Sebastian Borgeaud and Jean-Baptiste Alayrac and Carl Doersch and Catalin Ionescu and David Ding and Skanda Koppula and Daniel Zoran and Andrew Brock and Evan Shelhamer and Olivier Hénaff and Matthew M. Botvinick and Andrew Zisserman and Oriol Vinyals and Joāo Carreira},
	year         = 2022,
	url          = {https://arxiv.org/abs/2107.14795},
	eprint       = {2107.14795},
	archiveprefix = {arXiv},
	primaryclass = {cs.LG}
}

@inproceedings{GlorotBenigo_Initialization_2010,
	title        = {Understanding the difficulty of training deep feedforward neural networks},
	author       = {Glorot, Xavier and Bengio, Yoshua},
	year         = 2010,
	month        = {13--15 May},
	booktitle    = {Proceedings of the Thirteenth International Conference on Artificial Intelligence and Statistics},
	publisher    = {PMLR},
	address      = {Chia Laguna Resort, Sardinia, Italy},
	series       = {Proceedings of Machine Learning Research},
	volume       = 9,
	pages        = {249--256},
	url          = {https://proceedings.mlr.press/v9/glorot10a.html},
	editor       = {Teh, Yee Whye and Titterington, Mike}
}

@misc{He_Initialization_2015,
	title        = {Delving Deep into Rectifiers: Surpassing Human-Level Performance on ImageNet Classification},
	author       = {Kaiming He and Xiangyu Zhang and Shaoqing Ren and Jian Sun},
	year         = 2015,
	url          = {https://arxiv.org/abs/1502.01852},
	eprint       = {1502.01852},
	archiveprefix = {arXiv},
	primaryclass = {cs.CV}
}

@misc{He_ResNets_2015,
	title        = {Deep Residual Learning for Image Recognition},
	author       = {Kaiming He and Xiangyu Zhang and Shaoqing Ren and Jian Sun},
	year         = 2015,
	url          = {https://arxiv.org/abs/1512.03385},
	eprint       = {1512.03385},
	archiveprefix = {arXiv},
	primaryclass = {cs.CV}
}

@misc{Ioffe_BatchNormal_2015,
	title        = {Batch Normalization: Accelerating Deep Network Training by Reducing Internal Covariate Shift},
	author       = {Sergey Ioffe and Christian Szegedy},
	year         = 2015,
	url          = {https://arxiv.org/abs/1502.03167},
	eprint       = {1502.03167},
	archiveprefix = {arXiv},
	primaryclass = {cs.LG}
}

@misc{Kingma_ADAM_2017,
	title        = {Adam: A Method for Stochastic Optimization},
	author       = {Diederik P. Kingma and Jimmy Ba},
	year         = 2017,
	url          = {https://arxiv.org/abs/1412.6980},
	eprint       = {1412.6980},
	archiveprefix = {arXiv},
	primaryclass = {cs.LG}
}

@article{Sherstinsky_RNN_LSTM_2020,
	title        = {Fundamentals of Recurrent Neural Network ({RNN}) and Long Short-Term Memory ({LSTM}) network},
	author       = {Sherstinsky, Alex},
	year         = 2020,
	month        = mar,
	journal      = {Physica D: Nonlinear Phenomena},
	publisher    = {Elsevier BV},
	volume       = 404,
	pages        = 132306,
	doi          = {10.1016/j.physd.2019.132306},
	issn         = {0167-2789},
	url          = {http://dx.doi.org/10.1016/j.physd.2019.132306}
}

@article{Hochreiter_LSTM_1997,
	title        = {Long Short-Term Memory},
	author       = {Hochreiter, Sepp and Schmidhuber, J\"{u}rgen},
	year         = 1997,
	month        = nov,
	journal      = {Neural Comput.},
	publisher    = {MIT Press},
	address      = {Cambridge, MA, USA},
	volume       = 9,
	number       = 8,
	pages        = {1735–1780},
	doi          = {10.1162/neco.1997.9.8.1735},
	issn         = {0899-7667},
	url          = {https://doi.org/10.1162/neco.1997.9.8.1735},
	issue_date   = {November 15, 1997},
	numpages     = 46
}

@inproceedings{Sak_LSTM_2014,
	title        = {Long short-term memory recurrent neural network architectures for large scale acoustic modeling},
	author       = {Sak,  Haşim and Senior,  Andrew and Beaufays,  Fran\c{c}oise},
	year         = 2014,
	month        = sep,
	booktitle    = {Interspeech 2014},
	publisher    = {ISCA},
	series       = {Interspeech 2014},
	doi          = {10.21437/interspeech.2014-80},
	url          = {http://dx.doi.org/10.21437/Interspeech.2014-80},
	collection   = {interspeech 2014}
}

@article{Wu_GNNs_Review_2021,
	title        = {A Comprehensive Survey on Graph Neural Networks},
	author       = {Wu, Zonghan and Pan, Shirui and Chen, Fengwen and Long, Guodong and Zhang, Chengqi and Yu, Philip S.},
	year         = 2021,
	journal      = {IEEE Transactions on Neural Networks and Learning Systems},
	volume       = 32,
	number       = 1,
	pages        = {4--24},
	doi          = {10.1109/TNNLS.2020.2978386}
}

@misc{Xu_MultimodalTransformer_2023,
	title        = {Multimodal Learning with {Transformers}: A Survey},
	author       = {Peng Xu and Xiatian Zhu and David A. Clifton},
	year         = 2023,
	url          = {https://arxiv.org/abs/2206.06488},
	eprint       = {2206.06488},
	archiveprefix = {arXiv},
	primaryclass = {cs.CV}
}

@misc{Dosovitskiy_VisionTransformer_2021,
	title        = {An Image is Worth 16x16 Words: {Transformers} for Image Recognition at Scale},
	author       = {Alexey Dosovitskiy and Lucas Beyer and Alexander Kolesnikov and Dirk Weissenborn and Xiaohua Zhai and Thomas Unterthiner and Mostafa Dehghani and Matthias Minderer and Georg Heigold and Sylvain Gelly and Jakob Uszkoreit and Neil Houlsby},
	year         = 2021,
	url          = {https://arxiv.org/abs/2010.11929},
	eprint       = {2010.11929},
	archiveprefix = {arXiv},
	primaryclass = {cs.CV}
}

@article{Deshpande_GP_2025,
	title        = {Gaussian process regression + deep neural network autoencoder for probabilistic surrogate modeling in nonlinear mechanics of solids},
	author       = {Saurabh Deshpande and Hussein Rappel and Mark Hobbs and Stéphane P.A. Bordas and Jakub Lengiewicz},
	year         = 2025,
	journal      = {Computer Methods in Applied Mechanics and Engineering},
	volume       = 437,
	pages        = 117790,
	doi          = {https://doi.org/10.1016/j.cma.2025.117790},
	issn         = {0045-7825},
	url          = {https://www.sciencedirect.com/science/article/pii/S0045782525000623}
}

@article{Alzubaidi_CNN_Review_2021,
	title        = {Review of deep learning: concepts,  CNN architectures,  challenges,  applications,  future directions},
	author       = {Alzubaidi,  Laith and Zhang,  Jinglan and Humaidi,  Amjad J. and Al-Dujaili,  Ayad and Duan,  Ye and Al-Shamma,  Omran and Santamaría,  J. and Fadhel,  Mohammed A. and Al-Amidie,  Muthana and Farhan,  Laith},
	year         = 2021,
	month        = mar,
	journal      = {Journal of Big Data},
	publisher    = {Springer Science and Business Media LLC},
	volume       = 8,
	number       = 1,
	doi          = {10.1186/s40537-021-00444-8},
	issn         = {2196-1115},
	url          = {http://dx.doi.org/10.1186/s40537-021-00444-8}
}

@inbook{Frazier_BayesianOptimiz_2016,
	title        = {Bayesian Optimization for Materials Design},
	author       = {Frazier, Peter I. and Wang, Jialei},
	year         = 2016,
	booktitle    = {Information Science for Materials Discovery and Design},
	publisher    = {Springer International Publishing},
	address      = {Cham},
	pages        = {45--75},
	doi          = {10.1007/978-3-319-23871-5_3},
	isbn         = {978-3-319-23871-5},
	url          = {https://doi.org/10.1007/978-3-319-23871-5_3},
	editor       = {Lookman, Turab and Alexander, Francis J. and Rajan, Krishna}
}

@article{Lookman_ActiveLearning_2019,
	title        = {Active learning in materials science with emphasis on adaptive sampling using uncertainties for targeted design},
	author       = {Lookman,  Turab and Balachandran,  Prasanna V. and Xue,  Dezhen and Yuan,  Ruihao},
	year         = 2019,
	month        = feb,
	journal      = {npj Computational Materials},
	publisher    = {Springer Science and Business Media LLC},
	volume       = 5,
	number       = 1,
	doi          = {10.1038/s41524-019-0153-8},
	issn         = {2057-3960},
	url          = {http://dx.doi.org/10.1038/s41524-019-0153-8}
}

@article{wan2024review,
	title        = {Review of Human--Robot Collaboration in Robotic Surgery},
	author       = {Wan, Qianyi and Shi, Yudong and Xiao, Xilin and Li, Xiaojian and Mo, Hangjie},
	year         = 2024,
	journal      = {Advanced Intelligent Systems},
	publisher    = {Wiley Online Library},
	pages        = 2400319
}

@article{senhora2022machine,
	title        = {Machine learning for topology optimization: Physics-based learning through an independent training strategy},
	author       = {Senhora, Fernando V and Chi, Heng and Zhang, Yuyu and Mirabella, Lucia and Tang, Tsz Ling Elaine and Paulino, Glaucio H},
	year         = 2022,
	journal      = {Computer Methods in Applied Mechanics and Engineering},
	publisher    = {Elsevier},
	volume       = 398,
	pages        = 115116
}

@article{zimmermann20183d,
	title        = {A 3D, performance-driven generative design framework: automating the link from a 3D spatial grammar interpreter to structural finite element analysis and stochastic optimization},
	author       = {Zimmermann, Luca and Chen, Tian and Shea, Kristina},
	year         = 2018,
	journal      = {AI EDAM},
	publisher    = {Cambridge University Press},
	volume       = 32,
	number       = 2,
	pages        = {189--199}
}

@article{bossen2023batman,
	title        = {Batman and robin in healthcare knowledge work: Human-AI collaboration by clinical documentation integrity specialists},
	author       = {Bossen, Claus and Pine, Kathleen H},
	year         = 2023,
	journal      = {ACM Transactions on Computer-Human Interaction},
	publisher    = {ACM New York, NY},
	volume       = 30,
	number       = 2,
	pages        = {1--29}
}

@book{advanced_materials_for_societal_implementation_2023,
	title        = {Advanced Materials for Societal Implementation},
	year         = {2023},
	doi          = {10.3390/books978-3-0365-6131-8}
}

@article{bojan_podgornik_2023,
	title        = {Advanced materials and research for the green future},
	author       = {Bojan Podgornik},
	year         = {2023},
	journal      = {Materiali in Tehnologije},
	volume       = {57},
	number       = {1},
	doi          = {10.17222/mit.2022.717}
}

@book{colin_tong_2019,
	title        = {Introduction to Materials for Advanced Energy Systems},
	author       = {Colin Tong},
	year         = {2019}
}

@article{raabe2023accelerating,
	title        = {Accelerating the design of compositionally complex materials via physics-informed artificial intelligence},
	author       = {Raabe, Dierk and Mianroodi, Jaber Rezaei and Neugebauer, J{\"o}rg},
	year         = 2023,
	journal      = {Nature Computational Science},
	publisher    = {Nature Publishing Group US New York},
	volume       = 3,
	number       = 3,
	pages        = {198--209}
}

@article{bauer2024roadmap,
	title        = {Roadmap on Data-Centric Materials Science},
	author       = {Bauer, Stefan and Benner, Peter and Bereau, Tristan and Blum, Volker and Boley, Mario and Carbogno, Christian and Catlow, Richard and Dehm, Gerhard and Eibl, Sebastian and Ernstorfer, Ralph and others},
	year         = 2024,
	journal      = {Modelling and Simulation in Materials Science and Engineering}
}

@article{himanen2019data,
	title        = {Data-driven materials science: status, challenges, and perspectives},
	author       = {Himanen, Lauri and Geurts, Amber and Foster, Adam Stuart and Rinke, Patrick},
	year         = 2019,
	journal      = {Advanced Science},
	publisher    = {Wiley Online Library},
	volume       = 6,
	number       = 21,
	pages        = 1900808
}

@article{rodrigues2021big,
	title        = {Big data and machine learning for materials science},
	author       = {Rodrigues, Jose F and Florea, Larisa and De Oliveira, Maria CF and Diamond, Dermot and Oliveira, Osvaldo N},
	year         = 2021,
	journal      = {Discover Materials},
	publisher    = {Springer},
	volume       = 1,
	pages        = {1--27}
}

@article{carla_p__gomes__2019,
	title        = {Artificial intelligence for materials discovery},
	author       = {Carla P. Gomes and Bart Selman and John M. Gregoire},
	year         = {2019},
	journal      = {Mrs Bulletin},
	publisher    = {Springer International Publishing},
	volume       = {44},
	number       = {7},
	pages        = {538--544},
	doi          = {10.1557/MRS.2019.158}
}

@incollection{michael_mckerns_2018,
	title        = {Is Automated Materials Design and Discovery Possible},
	author       = {Michael McKerns},
	year         = {2018},
	publisher    = {Springer, Cham},
	pages        = {15--58},
	doi          = {10.1007/978-3-319-99465-9_2}
}

@article{peivaste2024rapid,
	title        = {Rapid and accurate predictions of perfect and defective material properties in atomistic simulation using the power of 3D CNN-based trained artificial neural networks},
	author       = {Peivaste, Iman and Ramezani, Saba and Alahyarizadeh, Ghasem and Ghaderi, Reza and Makradi, Ahmed and Belouettar, Salim},
	year         = 2024,
	journal      = {Scientific Reports},
	publisher    = {Nature Publishing Group UK London},
	volume       = 14,
	number       = 1,
	pages        = 36
}

@article{machine_ramprasad_2017,
	title        = {Machine learning in materials informatics: recent applications and prospects},
	author       = {Ramprasad, Rampi and Batra, Rohit and Pilania, Ghanshyam and Mannodi-Kanakkithodi, Arun and Kim, Chiho},
	year         = 2017,
	journal      = {npj Computational Materials},
	doi          = {10.1038/S41524-017-0056-5},
	litmapsid    = 55980137
}

@article{materials_zhu_2022,
	title        = {Materials Data toward Machine Learning: Advances and Challenges.},
	author       = {Zhu, Linggang and Zhou, Jian and Sun, Zhimei},
	year         = 2022,
	journal      = {Journal of Physical Chemistry Letters},
	doi          = {10.1021/ACS.JPCLETT.2C00576},
	litmapsid    = 277680763
}

@article{materials_liu_2017,
	title        = {Materials discovery and design using machine learning},
	author       = {Liu, Yue and Zhao, Tianlu and Ju, Wangwei and Shi, Siqi},
	year         = 2017,
	journal      = {Journal of Materiomics},
	doi          = {10.1016/J.JMAT.2017.08.002},
	litmapsid    = 118415027
}

@article{machine_gubernatis_2018,
	title        = {Machine learning in materials design and discovery: Examples from the present and suggestions for the future},
	author       = {Gubernatis, J. and Lookman, T.},
	year         = 2018,
	journal      = {PHYSICAL REVIEW MATERIALS},
	doi          = {10.1103/PHYSREVMATERIALS.2.120301},
	litmapsid    = 65307825
}

@article{recent_choudhary_2022,
	title        = {Recent advances and applications of deep learning methods in materials science},
	author       = {Choudhary,  Kamal and DeCost,  Brian and Chen,  Chi and Jain,  Anubhav and Tavazza,  Francesca and Cohn,  Ryan and Park,  Cheol Woo and Choudhary,  Alok and Agrawal,  Ankit and Billinge,  Simon J. L. and Holm,  Elizabeth and Ong,  Shyue Ping and Wolverton,  Chris},
	year         = 2022,
	month        = apr,
	journal      = {npj Computational Materials},
	publisher    = {Springer Science and Business Media LLC},
	volume       = 8,
	number       = 1,
	doi          = {10.1038/s41524-022-00734-6},
	issn         = {2057-3960},
	url          = {http://dx.doi.org/10.1038/s41524-022-00734-6}
}

@article{big_zhou_2019,
	title        = {Big Data Creates New Opportunities for Materials Research: A Review on Methods and Applications of Machine Learning for Materials Design},
	author       = {Zhou, Teng and Song, Zhen and Sundmacher, Kai},
	year         = 2019,
	journal      = {Engineering},
	doi          = {10.1016/J.ENG.2019.02.011},
	litmapsid    = 199510643
}

@article{opportunities_morgan_2020,
	title        = {Opportunities and Challenges for Machine Learning in Materials Science},
	author       = {Morgan, D. and Jacobs, R.},
	year         = 2020,
	journal      = {Annual review of materials research (Print)},
	doi          = {10.1146/ANNUREV-MATSCI-070218-010015},
	litmapsid    = 275978061
}

@article{curtarolo2013high,
	title        = {The high-throughput highway to computational materials design},
	author       = {Curtarolo, Stefano and Hart, Gus LW and Nardelli, Marco Buongiorno and Mingo, Natalio and Sanvito, Stefano and Levy, Ohad},
	year         = 2013,
	journal      = {Nature materials},
	publisher    = {Nature Publishing Group UK London},
	volume       = 12,
	number       = 3,
	pages        = {191--201}
}

@article{stein2019progress,
	title        = {Progress and prospects for accelerating materials science with automated and autonomous workflows},
	author       = {Stein, Helge S and Gregoire, John M},
	year         = 2019,
	journal      = {Chemical science},
	publisher    = {Royal Society of Chemistry},
	volume       = 10,
	number       = 42,
	pages        = {9640--9649}
}

@article{himanen2020dscribe,
	title        = {DScribe: Library of descriptors for machine learning in materials science},
	author       = {Himanen, Lauri and J{\"a}ger, Marc OJ and Morooka, Eiaki V and Canova, Filippo Federici and Ranawat, Yashasvi S and Gao, David Z and Rinke, Patrick and Foster, Adam S},
	year         = 2020,
	journal      = {Computer Physics Communications},
	publisher    = {Elsevier},
	volume       = 247,
	pages        = 106949
}

@article{ong2013python,
	title        = {Python Materials Genomics (pymatgen): A robust, open-source python library for materials analysis},
	author       = {Ong, Shyue Ping and Richards, William Davidson and Jain, Anubhav and Hautier, Geoffroy and Kocher, Michael and Cholia, Shreyas and Gunter, Dan and Chevrier, Vincent L and Persson, Kristin A and Ceder, Gerbrand},
	year         = 2013,
	journal      = {Computational Materials Science},
	publisher    = {Elsevier},
	volume       = 68,
	pages        = {314--319}
}

@article{machine_butler_2018,
	title        = {Machine learning for molecular and materials science},
	author       = {Butler, K. and Davies, D. and Cartwright, H. and Isayev, O. and Walsh, A.},
	year         = 2018,
	journal      = {Nature},
	doi          = {10.1038/S41586-018-0337-2},
	litmapsid    = 276379428
}

@article{machine_gu_2019,
	title        = {Machine learning for renewable energy materials},
	author       = {Gu, Geun Ho and Noh, Juhwan and Kim, Inkyung and Jung, Yousung},
	year         = 2019,
	journal      = {Journal of Materials Chemistry A},
	doi          = {10.1039/C9TA02356A},
	litmapsid    = 224129254
}

@article{exploiting_anand_2022,
	title        = {Exploiting Machine Learning in Multiscale Modelling of Materials},
	author       = {Anand, G. and Ghosh, Swarnava and Zhang, Liwei and Anupam, Angesh and Freeman, Colin L. and Ortner, Christoph and Eisenbach, Markus and Kermode, James R. and Anand, G. and Ghosh, Swarnava and Zhang, Liwei and Anupam, Angesh and Freeman, Colin L. and Ortner, Christoph and Eisenbach, Markus and Kermode, James R.},
	year         = 2022,
	journal      = {Journal of The Institution of Engineers (India): Series D},
	doi          = {10.1007/S40033-022-00424-Z},
	litmapsid    = 250801774
}

@article{evolution_gupta_2023,
	title        = {Evolution of artificial intelligence for application in contemporary materials science},
	author       = {Gupta, Vishu and Liao, W. and Choudhary, Alok Ratan and Agrawal, Ankit},
	year         = 2023,
	journal      = {MRS Communications},
	doi          = {10.1557/S43579-023-00433-3},
	litmapsid    = 264803186
}

@article{generalpurpose_ward_2016,
	title        = {A general-purpose machine learning framework for predicting properties of inorganic materials},
	author       = {Ward, Logan and Agrawal, Ankit and Choudhary, Alok and Wolverton, Chris},
	year         = 2016,
	journal      = {npj computational materials},
	doi          = {10.1038/NPJCOMPUMATS.2016.28},
	litmapsid    = 35399942
}

@article{dft_schleder_2019,
	title        = {From DFT to machine learning: recent approaches to materials science–a review},
	author       = {Schleder, G. R. and Padilha, A. C. and Acosta, C. M. and Costa, M. and Fazzio, A.},
	year         = 2019,
	journal      = {Journal of Physics: Materials},
	doi          = {10.1088/2515-7639/AB084B},
	litmapsid    = 140426065
}

@article{szymanski2023autonomous,
	title        = {An autonomous laboratory for the accelerated synthesis of novel materials},
	author       = {Szymanski, Nathan J and Rendy, Bernardus and Fei, Yuxing and Kumar, Rishi E and He, Tanjin and Milsted, David and McDermott, Matthew J and Gallant, Max and Cubuk, Ekin Dogus and Merchant, Amil and others},
	year         = 2023,
	journal      = {Nature},
	publisher    = {Nature Publishing Group UK London},
	volume       = 624,
	number       = 7990,
	pages        = {86--91}
}

@article{akhshik2022prediction,
	title        = {Prediction of greenhouse gas emissions reductions via machine learning algorithms: Toward an artificial intelligence-based life cycle assessment for automotive lightweighting},
	author       = {Akhshik, Masoud and Bilton, Amy and Tjong, Jimi and Singh, Chandra Veer and Faruk, Omar and Sain, Mohini},
	year         = 2022,
	journal      = {Sustainable Materials and Technologies},
	publisher    = {Elsevier},
	volume       = 31,
	pages        = {e00370}
}

@article{lin2023advancing,
	title        = {Advancing the industrial circular economy: the integrative role of machine learning in resource optimization},
	author       = {Lin, KY and Wei, SH},
	year         = 2023,
	journal      = {Journal of green economy and low-carbon development},
	volume       = 2,
	number       = 3,
	pages        = {122--136}
}

@article{dean2023interpretable,
	title        = {Interpretable machine learning for materials design},
	author       = {Dean, James and Scheffler, Matthias and Purcell, Thomas AR and Barabash, Sergey V and Bhowmik, Rahul and Bazhirov, Timur},
	year         = 2023,
	journal      = {Journal of Materials Research},
	publisher    = {Springer},
	volume       = 38,
	number       = 20,
	pages        = {4477--4496}
}

@article{johnson2020invited,
	title        = {Invited review: Machine learning for materials developments in metals additive manufacturing},
	author       = {Johnson, NS and Vulimiri, PS and To, AC and Zhang, X and Brice, CA and Kappes, BB and Stebner, AP},
	year         = 2020,
	journal      = {Additive Manufacturing},
	publisher    = {Elsevier},
	volume       = 36,
	pages        = 101641
}

@article{peivaste2023data,
	title        = {Data-driven analysis and prediction of stable phases for high-entropy alloy design},
	author       = {Peivaste, Iman and Jossou, Ericmoore and Tiamiyu, Ahmed A},
	year         = 2023,
	journal      = {Scientific Reports},
	publisher    = {Nature Publishing Group UK London},
	volume       = 13,
	number       = 1,
	pages        = 22556
}

@article{yifan_wang_phase_2023,
	title        = {Phase {Prediction} and {Visualized} {Design} {Process} of {High} {Entropy} {Alloys} via {Machine} {Learned} {Methodology}},
	author       = {{Yifan Wang} and {Jianxin Hou} and {Junhua You} and {Keqiang Qiu} and {S. Zhang} and {Jianqiang Wang}},
	year         = 2023,
	month        = jan,
	journal      = {Metals},
	doi          = {10.3390/MET13020283},
	url          = {https://www.mdpi.com/2075-4701/13/2/283#B21-metals-13-00283}
}

@article{yong_liu_focused_2021,
	title        = {A focused review on machine learning aided high-throughput methods in high entropy alloy},
	author       = {{Yong Liu} and {Jingchuan Zhu}},
	year         = 2021,
	month        = oct,
	journal      = {Journal of Alloys and Compounds},
	doi          = {10.1016/J.JALLCOM.2021.160295}
}

@article{wei_zhong_insights_2022,
	title        = {Insights on phase formation from thermodynamic calculations and machine learning of 2436 experimentally measured high entropy alloys},
	author       = {{Wei Zhong} and {Ji-Cheng Zhao}},
	year         = 2022,
	month        = apr,
	journal      = {Journal of Alloys and Compounds},
	doi          = {10.1016/J.JALLCOM.2022.165173}
}

@article{ujjawal_kumar_jaiswal_machine_2021,
	title        = {Machine learning approach to predict new multiphase high entropy alloys},
	author       = {{Ujjawal Kumar Jaiswal} and {Rahul M R}},
	year         = 2021,
	month        = may,
	journal      = {Scripta Materialia},
	doi          = {10.1016/J.SCRIPTAMAT.2021.113804}
}

@article{wanlin_guo_machine-learning_2019,
	title        = {Machine-learning model for predicting phase formations of high-entropy alloys},
	author       = {{Wanlin Guo}},
	year         = 2019,
	month        = sep,
	journal      = {Physical Review Materials},
	doi          = {10.1103/PHYSREVMATERIALS.3.095005}
}

@article{zhu2022phase,
	title        = {Phase formation prediction of high-entropy alloys: a deep learning study},
	author       = {Zhu, Wenhan and Huo, Wenyi and Wang, Shiqi and Wang, Xu and Ren, Kai and Tan, Shuyong and Fang, Feng and Xie, Zonghan and Jiang, Jianqing},
	year         = 2022,
	journal      = {journal of materials research and technology},
	publisher    = {Elsevier},
	volume       = 18,
	pages        = {800--809}
}

@article{seok_yeong_byeon_deep_2021,
	title        = {Deep learning-based phase prediction of high-entropy alloys: {Optimization}, generation, and explanation},
	author       = {{Seok Yeong Byeon} and {Hyoung Seop Kim} and {Hyungyu Jin} and {Seung Chul Lee}},
	year         = 2021,
	month        = jan,
	journal      = {Materials \& Design},
	doi          = {10.1016/J.MATDES.2020.109260}
}

@article{mitra2022machine,
	title        = {Machine learning based approach for phase prediction in high entropy borides},
	author       = {Mitra, Rahul and Bajpai, Anurag and Biswas, Krishanu},
	year         = 2022,
	journal      = {Ceramics International},
	publisher    = {Elsevier},
	volume       = 48,
	number       = 12,
	pages        = {16695--16706}
}

@article{hu2024interpretable,
	title        = {Interpretable machine learning predictions for efficient perovskite solar cell development},
	author       = {Hu, Jinghao and Chen, Zhengxin and Chen, Yuzhi and Liu, Hongyu and Li, Wenhao and Wang, Yanan and Peng, Lin and Liu, Xiaolin and Lin, Jia and Chen, Xianfeng and others},
	year         = 2024,
	journal      = {Solar Energy Materials and Solar Cells},
	publisher    = {Elsevier},
	volume       = 271,
	pages        = 112826
}

@article{fang2022machine,
	title        = {Machine learning accelerates the materials discovery},
	author       = {Fang, Jiheng and Xie, Ming and He, Xingqun and Zhang, Jiming and Hu, Jieqiong and Chen, Yongtai and Yang, Youcai and Jin, Qinglin},
	year         = 2022,
	journal      = {Materials Today Communications},
	publisher    = {Elsevier},
	volume       = 33,
	pages        = 104900
}

@article{dhaliwal2022machine,
	title        = {Machine learned interatomic potentials using random features},
	author       = {Dhaliwal, Gurjot and Nair, Prasanth B and Singh, Chandra Veer},
	year         = 2022,
	journal      = {npj Computational Materials},
	publisher    = {Nature Publishing Group UK London},
	volume       = 8,
	number       = 1,
	pages        = 7
}

@article{sparks2016data,
	title        = {Data mining our way to the next generation of thermoelectrics},
	author       = {Sparks, Taylor D and Gaultois, Michael W and Oliynyk, Anton and Brgoch, Jakoah and Meredig, Bryce},
	year         = 2016,
	journal      = {Scripta Materialia},
	publisher    = {Elsevier},
	volume       = 111,
	pages        = {10--15}
}

@article{kim2017materials,
	title        = {Materials synthesis insights from scientific literature via text extraction and machine learning},
	author       = {Kim, Edward and Huang, Kevin and Saunders, Adam and McCallum, Andrew and Ceder, Gerbrand and Olivetti, Elsa},
	year         = 2017,
	journal      = {Chemistry of Materials},
	publisher    = {ACS Publications},
	volume       = 29,
	number       = 21,
	pages        = {9436--9444}
}

@book{Rasmussen2005GP,
	title        = {Gaussian Processes for Machine Learning},
	author       = {Rasmussen,  Carl Edward and Williams,  Christopher K. I.},
	year         = 2005,
	month        = nov,
	publisher    = {The MIT Press},
	doi          = {10.7551/mitpress/3206.001.0001},
	isbn         = 9780262256834,
	url          = {http://dx.doi.org/10.7551/mitpress/3206.001.0001}
}

@misc{businesswire2019,
	title        = {Survey: 96\% of Enterprises Encounter Training Data Quality and Labeling Challenges in Machine Learning Projects},
	author       = {{BusinessWire}},
	year         = 2019,
	url          = {https://www.businesswire.com},
	note         = {Accessed: 2025-01-02}
}

@misc{ibm2020,
	title        = {{AI} and the Future of Data: Addressing Key Data Quality Challenges for Successful Machine Learning},
	author       = {{IBM}},
	year         = 2020,
	url          = {https://www.ibm.com/downloads/documents/us-en/10a99803abafd842},
	note         = {Accessed: 2025-01-02}
}

@inproceedings{chen2016xgboost,
	title        = {Xgboost: A scalable tree boosting system},
	author       = {Chen, Tianqi and Guestrin, Carlos},
	year         = 2016,
	booktitle    = {Proceedings of the 22nd acm sigkdd international conference on knowledge discovery and data mining},
	pages        = {785--794}
}

@article{liang2022advances,
	title        = {Advances, challenges and opportunities in creating data for trustworthy AI},
	author       = {Liang, Weixin and Tadesse, Girmaw Abebe and Ho, Daniel and Fei-Fei, Li and Zaharia, Matei and Zhang, Ce and Zou, James},
	year         = 2022,
	journal      = {Nature Machine Intelligence},
	publisher    = {Nature Publishing Group UK London},
	volume       = 4,
	number       = 8,
	pages        = {669--677}
}

@book{breiman2017classification,
	title        = {Classification and regression trees},
	author       = {Breiman, Leo},
	year         = 2017,
	publisher    = {Routledge}
}

@article{breiman2001random,
	title        = {Random forests},
	author       = {Breiman, Leo},
	year         = 2001,
	journal      = {Machine learning},
	publisher    = {Springer},
	volume       = 45,
	pages        = {5--32}
}

@article{xu2023small,
	title        = {Small data machine learning in materials science},
	author       = {Xu, Pengcheng and Ji, Xiaobo and Li, Minjie and Lu, Wencong},
	year         = 2023,
	journal      = {npj Computational Materials},
	publisher    = {Nature Publishing Group UK London},
	volume       = 9,
	number       = 1,
	pages        = 42
}

@article{pilania2013accelerating,
	title        = {Accelerating materials property predictions using machine learning},
	author       = {Pilania, Ghanshyam and Wang, Chenchen and Jiang, Xun and Rajasekaran, Sanguthevar and Ramprasad, Ramamurthy},
	year         = 2013,
	journal      = {Scientific reports},
	publisher    = {Nature Publishing Group UK London},
	volume       = 3,
	number       = 1,
	pages        = 2810
}

@article{jha2018elemnet,
	title        = {Elemnet: Deep learning the chemistry of materials from only elemental composition},
	author       = {Jha, Dipendra and Ward, Logan and Paul, Arindam and Liao, Wei-keng and Choudhary, Alok and Wolverton, Chris and Agrawal, Ankit},
	year         = 2018,
	journal      = {Scientific reports},
	publisher    = {Nature Publishing Group UK London},
	volume       = 8,
	number       = 1,
	pages        = 17593
}

@article{guo2021artificial,
	title        = {Artificial intelligence and machine learning in design of mechanical materials},
	author       = {Guo, Kai and Yang, Zhenze and Yu, Chi-Hua and Buehler, Markus J},
	year         = 2021,
	journal      = {Materials Horizons},
	publisher    = {Royal Society of Chemistry},
	volume       = 8,
	number       = 4,
	pages        = {1153--1172}
}

@article{tabor2018accelerating,
	title        = {Accelerating the discovery of materials for clean energy in the era of smart automation},
	author       = {Tabor, Daniel P and Roch, Lo{\"\i}c M and Saikin, Semion K and Kreisbeck, Christoph and Sheberla, Dennis and Montoya, Joseph H and Dwaraknath, Shyam and Aykol, Muratahan and Ortiz, Carlos and Tribukait, Hermann and others},
	year         = 2018,
	journal      = {Nature reviews materials},
	publisher    = {Nature Publishing Group UK London},
	volume       = 3,
	number       = 5,
	pages        = {5--20}
}

@article{shi2006unsupervised,
	title        = {Unsupervised learning with random forest predictors},
	author       = {Shi, Tao and Horvath, Steve},
	year         = 2006,
	journal      = {Journal of Computational and Graphical Statistics},
	publisher    = {Taylor \& Francis},
	volume       = 15,
	number       = 1,
	pages        = {118--138}
}

@article{wang2022machine,
	title        = {Machine learning guided dopant selection for metal oxide-based photoelectrochemical water splitting: the case study of Fe2O3 and CuO},
	author       = {Wang, Zhiliang and Gu, Yuang and Zheng, Lingxia and Hou, Jingwei and Zheng, Huajun and Sun, Shijing and Wang, Lianzhou},
	year         = 2022,
	journal      = {Advanced Materials},
	publisher    = {Wiley Online Library},
	volume       = 34,
	number       = 10,
	pages        = 2106776
}

@article{cortes1995support,
	title        = {Support-Vector Networks},
	author       = {Cortes, Corinna and Vapnik, Vladimer},
	year         = 1995,
	journal      = {Machine Learning},
	publisher    = {Kluwer Academic Publishers},
	volume       = 20,
	number       = 1,
	pages        = {273--297}
}

@article{liu2020gaussian,
	title        = {When Gaussian process meets big data: A review of scalable GPs},
	author       = {Liu, Haitao and Ong, Yew-Soon and Shen, Xiaobo and Cai, Jianfei},
	year         = 2020,
	journal      = {IEEE transactions on neural networks and learning systems},
	publisher    = {IEEE},
	volume       = 31,
	number       = 11,
	pages        = {4405--4423}
}

@article{bostanabad2019globally,
	title        = {Globally approximate gaussian processes for big data with application to data-driven metamaterials design},
	author       = {Bostanabad, Ramin and Chan, Yu-Chin and Wang, Liwei and Zhu, Ping and Chen, Wei},
	year         = 2019,
	journal      = {Journal of Mechanical Design},
	publisher    = {American Society of Mechanical Engineers},
	volume       = 141,
	number       = 11,
	pages        = 111402
}

@article{kanagawa2018gaussian,
	title        = {Gaussian processes and kernel methods: A review on connections and equivalences},
	author       = {Kanagawa, Motonobu and Hennig, Philipp and Sejdinovic, Dino and Sriperumbudur, Bharath K},
	year         = 2018,
	journal      = {arXiv preprint arXiv:1807.02582}
}

@article{kingsford2008decision,
	title        = {What are decision trees?},
	author       = {Kingsford, Carl and Salzberg, Steven L},
	year         = 2008,
	journal      = {Nature biotechnology},
	publisher    = {Nature Publishing Group US New York},
	volume       = 26,
	number       = 9,
	pages        = {1011--1013}
}

@article{li2019prediction,
	title        = {Prediction of material removal rate for chemical mechanical planarization using decision tree-based ensemble learning},
	author       = {Li, Zhixiong and Wu, Dazhong and Yu, Tianyu},
	year         = 2019,
	journal      = {Journal of Manufacturing Science and Engineering},
	publisher    = {American Society of Mechanical Engineers},
	volume       = 141,
	number       = 3,
	pages        = {031003}
}

@article{rokach2005decision,
	title        = {Decision trees},
	author       = {Rokach, Lior and Maimon, Oded},
	year         = 2005,
	journal      = {Data mining and knowledge discovery handbook},
	publisher    = {Springer},
	pages        = {165--192}
}

@article{huo2019semi,
	title        = {Semi-supervised machine-learning classification of materials synthesis procedures},
	author       = {Huo, Haoyan and Rong, Ziqin and Kononova, Olga and Sun, Wenhao and Botari, Tiago and He, Tanjin and Tshitoyan, Vahe and Ceder, Gerbrand},
	year         = 2019,
	journal      = {Npj Computational Materials},
	publisher    = {Nature Publishing Group UK London},
	volume       = 5,
	number       = 1,
	pages        = 62
}

@article{ward2016general,
	title        = {A general-purpose machine learning framework for predicting properties of inorganic materials},
	author       = {Ward, Logan and Agrawal, Ankit and Choudhary, Alok and Wolverton, Christopher},
	year         = 2016,
	journal      = {npj Computational Materials},
	publisher    = {Nature Publishing Group},
	volume       = 2,
	number       = 1,
	pages        = {1--7}
}

@article{hearst1998support,
	title        = {Support vector machines},
	author       = {Hearst, Marti A. and Dumais, Susan T and Osuna, Edgar and Platt, John and Scholkopf, Bernhard},
	year         = 1998,
	journal      = {IEEE Intelligent Systems and their applications},
	publisher    = {IEEE},
	volume       = 13,
	number       = 4,
	pages        = {18--28}
}

@article{lu2013using,
	title        = {Using support vector machine for materials design},
	author       = {Lu, Wen-Cong and Ji, Xiao-Bo and Li, Min-Jie and Liu, Liang and Yue, Bao-Hua and Zhang, Liang-Miao},
	year         = 2013,
	journal      = {Advances in Manufacturing},
	publisher    = {Springer},
	volume       = 1,
	pages        = {151--159}
}

@article{scholkopf2001estimating,
	title        = {Estimating the support of a high-dimensional distribution},
	author       = {Sch{\"o}lkopf, Bernhard and Platt, John C and Shawe-Taylor, John and Smola, Alex J and Williamson, Robert C},
	year         = 2001,
	journal      = {Neural computation},
	publisher    = {MIT Press One Rogers Street, Cambridge, MA 02142-1209, USA journals-info~…},
	volume       = 13,
	number       = 7,
	pages        = {1443--1471}
}

@article{ccevik2015support,
	title        = {Support vector machines in structural engineering: a review},
	author       = {{\c{C}}evik, Abdulkadir and Kurto{\u{g}}lu, Ahmet Emin and Bilgehan, Mahmut and G{\"u}l{\c{s}}an, Mehmet Eren and Albegmprli, Hasan M},
	year         = 2015,
	journal      = {Journal of Civil Engineering and Management},
	publisher    = {Taylor \& Francis},
	volume       = 21,
	number       = 3,
	pages        = {261--281}
}

@article{khakurel2021machine,
	title        = {Machine learning assisted prediction of the Young’s modulus of compositionally complex alloys},
	author       = {Khakurel, Hrishabh and Taufique, MFN and Roy, Ankit and Balasubramanian, Ganesh and Ouyang, Gaoyuan and Cui, Jun and Johnson, Duane D and Devanathan, Ram},
	year         = 2021,
	journal      = {Scientific reports},
	publisher    = {Nature Publishing Group UK London},
	volume       = 11,
	number       = 1,
	pages        = 17149
}

@article{li2019computational,
	title        = {Computational screening of new perovskite materials using transfer learning and deep learning},
	author       = {Li, Xiang and Dan, Yabo and Dong, Rongzhi and Cao, Zhuo and Niu, Chengcheng and Song, Yuqi and Li, Shaobo and Hu, Jianjun},
	year         = 2019,
	journal      = {Applied Sciences},
	publisher    = {MDPI},
	volume       = 9,
	number       = 24,
	pages        = 5510
}

@article{meredig2018can,
	title        = {Can machine learning identify the next high-temperature superconductor? Examining extrapolation performance for materials discovery},
	author       = {Meredig, Bryce and Antono, Erin and Church, Carena and Hutchinson, Maxwell and Ling, Julia and Paradiso, Sean and Blaiszik, Ben and Foster, Ian and Gibbons, Brenna and Hattrick-Simpers, Jason and others},
	year         = 2018,
	journal      = {Molecular Systems Design \& Engineering},
	publisher    = {Royal Society of Chemistry},
	volume       = 3,
	number       = 5,
	pages        = {819--825}
}

@article{sun2020machine,
	title        = {Machine-learning-accelerated screening of hydrogen evolution catalysts in MBenes materials},
	author       = {Sun, Xiang and Zheng, Jingnan and Gao, Yijing and Qiu, Chenglong and Yan, Yilong and Yao, Zihao and Deng, Shengwei and Wang, Jianguo},
	year         = 2020,
	journal      = {Applied Surface Science},
	publisher    = {Elsevier},
	volume       = 526,
	pages        = 146522
}

@article{sun2020covalency,
	title        = {Covalency competition dominates the water oxidation structure--activity relationship on spinel oxides},
	author       = {Sun, Yuanmiao and Liao, Hanbin and Wang, Jiarui and Chen, Bo and Sun, Shengnan and Ong, Samuel Jun Hoong and Xi, Shibo and Diao, Caozheng and Du, Yonghua and Wang, Jia-Ou and others},
	year         = 2020,
	journal      = {Nature catalysis},
	publisher    = {Nature Publishing Group UK London},
	volume       = 3,
	number       = 7,
	pages        = {554--563}
}

@article{choudhary2018machine,
	title        = {Machine learning with force-field-inspired descriptors for materials: Fast screening and mapping energy landscape},
	author       = {Choudhary, Kamal and DeCost, Brian and Tavazza, Francesca},
	year         = 2018,
	journal      = {Physical review materials},
	publisher    = {APS},
	volume       = 2,
	number       = 8,
	pages        = {083801}
}

@article{oliynyk2016high,
	title        = {High-throughput machine-learning-driven synthesis of full-Heusler compounds},
	author       = {Oliynyk, Anton O and Antono, Erin and Sparks, Taylor D and Ghadbeigi, Leila and Gaultois, Michael W and Meredig, Bryce and Mar, Arthur},
	year         = 2016,
	journal      = {Chemistry of Materials},
	publisher    = {ACS Publications},
	volume       = 28,
	number       = 20,
	pages        = {7324--7331}
}

@article{ge2020predicted,
	title        = {Predicted optimal bifunctional electrocatalysts for the hydrogen evolution reaction and the oxygen evolution reaction using chalcogenide heterostructures based on machine learning analysis of in silico quantum mechanics based high throughput screening},
	author       = {Ge, Lei and Yuan, Hao and Min, Yuxiang and Li, Li and Chen, Shiqian and Xu, Lai and Goddard III, William A},
	year         = 2020,
	journal      = {The Journal of Physical Chemistry Letters},
	publisher    = {ACS Publications},
	volume       = 11,
	number       = 3,
	pages        = {869--876}
}

@article{seeger2004gaussian,
	title        = {Gaussian processes for machine learning},
	author       = {Seeger, Matthias},
	year         = 2004,
	journal      = {International journal of neural systems},
	publisher    = {World Scientific},
	volume       = 14,
	number       = {02},
	pages        = {69--106}
}

@incollection{rasmussen2003gaussian,
	title        = {Gaussian processes in machine learning},
	author       = {Rasmussen, Carl Edward},
	year         = 2003,
	booktitle    = {Summer school on machine learning},
	publisher    = {Springer},
	pages        = {63--71}
}

@article{Ghaboussi1991,
	title        = {Knowledge‐Based Modeling of Material Behavior with Neural Networks},
	author       = {Ghaboussi,  J. and Garrett,  J. H. and Wu,  X.},
	year         = 1991,
	month        = jan,
	journal      = {Journal of Engineering Mechanics},
	publisher    = {American Society of Civil Engineers (ASCE)},
	volume       = 117,
	number       = 1,
	pages        = {132–153},
	doi          = {10.1061/(asce)0733-9399(1991)117:1(132)},
	issn         = {1943-7889},
	url          = {http://dx.doi.org/10.1061/(ASCE)0733-9399(1991)117:1(132)}
}

@article{Lefik2003,
	title        = {Artificial neural network as an incremental non-linear constitutive model for a finite element code},
	author       = {Lefik,  M. and Schrefler,  B.A.},
	year         = 2003,
	month        = jul,
	journal      = {Computer Methods in Applied Mechanics and Engineering},
	publisher    = {Elsevier BV},
	volume       = 192,
	number       = {28–30},
	pages        = {3265–3283},
	doi          = {10.1016/s0045-7825(03)00350-5},
	issn         = {0045-7825},
	url          = {http://dx.doi.org/10.1016/S0045-7825(03)00350-5}
}

@article{Lefik2009,
	title        = {Artificial Neural Networks in numerical modelling of composites},
	author       = {Lefik,  M. and Boso,  D.P. and Schrefler,  B.A.},
	year         = 2009,
	month        = may,
	journal      = {Computer Methods in Applied Mechanics and Engineering},
	publisher    = {Elsevier BV},
	volume       = 198,
	number       = {21–26},
	pages        = {1785–1804},
	doi          = {10.1016/j.cma.2008.12.036},
	issn         = {0045-7825},
	url          = {http://dx.doi.org/10.1016/j.cma.2008.12.036}
}

@article{Unger2009,
	title        = {Neural networks as material models within a multiscale approach},
	author       = {Unger,  J\"{o}rg F. and K\"{o}nke,  Carsten},
	year         = 2009,
	month        = oct,
	journal      = {Computers \&amp; Structures},
	publisher    = {Elsevier BV},
	volume       = 87,
	number       = {19–20},
	pages        = {1177–1186},
	doi          = {10.1016/j.compstruc.2008.12.003},
	issn         = {0045-7949},
	url          = {http://dx.doi.org/10.1016/j.compstruc.2008.12.003}
}

@article{li2021combining,
	title        = {Combining machine learning and high-throughput experimentation to discover photocatalytically active organic molecules},
	author       = {Li, Xiaobo and Maffettone, Phillip M and Che, Yu and Liu, Tao and Chen, Linjiang and Cooper, Andrew I},
	year         = 2021,
	journal      = {Chemical Science},
	publisher    = {Royal Society of Chemistry},
	volume       = 12,
	number       = 32,
	pages        = {10742--10754}
}

@article{bai2019accelerated,
	title        = {Accelerated discovery of organic polymer photocatalysts for hydrogen evolution from water through the integration of experiment and theory},
	author       = {Bai, Yang and Wilbraham, Liam and Slater, Benjamin J and Zwijnenburg, Martijn A and Sprick, Reiner Sebastian and Cooper, Andrew I},
	year         = 2019,
	journal      = {Journal of the American Chemical Society},
	publisher    = {ACS Publications},
	volume       = 141,
	number       = 22,
	pages        = {9063--9071}
}

@article{gao2021machine,
	title        = {Machine learning-optimized synthesis of doped TiO2 with improved photocatalytic performance: A multi-step workflow supported by designed wet-lab experiments},
	author       = {Gao, Bowen and Sun, Mingxuan and Ding, Zhipeng and Liu, Wenzhu},
	year         = 2021,
	journal      = {Journal of Alloys and Compounds},
	publisher    = {Elsevier},
	volume       = 881,
	pages        = 160534
}

@article{fathinia2016development,
	title        = {Development of kinetic models for photocatalytic ozonation of phenazopyridine on TiO2 nanoparticles thin film in a mixed semi-batch photoreactor},
	author       = {Fathinia, Mehrangiz and Khataee, Alireza and Aber, Soheil and Naseri, Abdolhossein},
	year         = 2016,
	journal      = {Applied Catalysis B: Environmental},
	publisher    = {Elsevier},
	volume       = 184,
	pages        = {270--284}
}

@article{karim2020coupling,
	title        = {Coupling high-throughput experiments and regression algorithms to optimize PGM-free ORR electrocatalyst synthesis},
	author       = {Karim, Mohammad Rezaul and Ferrandon, Magali and Medina, Samantha and Sture, Elliot and Kariuki, Nancy and Myers, Deborah J and Holby, Edward F and Zelenay, Piotr and Ahmed, Towfiq},
	year         = 2020,
	journal      = {ACS Applied Energy Materials},
	publisher    = {ACS Publications},
	volume       = 3,
	number       = 9,
	pages        = {9083--9088}
}

@article{schutt2018schnet,
	title        = {Schnet--a deep learning architecture for molecules and materials},
	author       = {Sch{\"u}tt, Kristof T and Sauceda, Huziel E and Kindermans, P-J and Tkatchenko, Alexandre and M{\"u}ller, K-R},
	year         = 2018,
	journal      = {The Journal of Chemical Physics},
	publisher    = {AIP Publishing},
	volume       = 148,
	number       = 24
}

@article{gasteiger2020directional,
	title        = {Directional message passing for molecular graphs},
	author       = {Gasteiger, Johannes and Gro{\ss}, Janek and G{\"u}nnemann, Stephan},
	year         = 2020,
	journal      = {arXiv preprint arXiv:2003.03123}
}

@article{gasteiger2021gemnet,
	title        = {Gemnet: Universal directional graph neural networks for molecules},
	author       = {Gasteiger, Johannes and Becker, Florian and G{\"u}nnemann, Stephan},
	year         = 2021,
	journal      = {Advances in Neural Information Processing Systems},
	volume       = 34,
	pages        = {6790--6802}
}

@article{choudhary2021atomistic,
	title        = {Atomistic line graph neural network for improved materials property predictions},
	author       = {Choudhary, Kamal and DeCost, Brian},
	year         = 2021,
	journal      = {npj Computational Materials},
	publisher    = {Nature Publishing Group UK London},
	volume       = 7,
	number       = 1,
	pages        = 185
}

@article{chen2022universal,
	title        = {A universal graph deep learning interatomic potential for the periodic table},
	author       = {Chen, Chi and Ong, Shyue Ping},
	year         = 2022,
	journal      = {Nature Computational Science},
	publisher    = {Nature Publishing Group US New York},
	volume       = 2,
	number       = 11,
	pages        = {718--728}
}

@article{hsu2022efficient,
	title        = {Efficient and interpretable graph network representation for angle-dependent properties applied to optical spectroscopy},
	author       = {Hsu, Tim and Pham, Tuan Anh and Keilbart, Nathan and Weitzner, Stephen and Chapman, James and Xiao, Penghao and Qiu, S Roger and Chen, Xiao and Wood, Brandon C},
	year         = 2022,
	journal      = {npj Computational Materials},
	publisher    = {Nature Publishing Group UK London},
	volume       = 8,
	number       = 1,
	pages        = 151
}

@article{batatia2022design,
	title        = {The design space of e (3)-equivariant atom-centered interatomic potentials},
	author       = {Batatia, Ilyes and Batzner, Simon and Kov{\'a}cs, D{\'a}vid P{\'e}ter and Musaelian, Albert and Simm, Gregor NC and Drautz, Ralf and Ortner, Christoph and Kozinsky, Boris and Cs{\'a}nyi, G{\'a}bor},
	year         = 2022,
	journal      = {arXiv preprint arXiv:2205.06643}
}

@article{geiger2022e3nn,
	title        = {e3nn: Euclidean neural networks},
	author       = {Geiger, Mario and Smidt, Tess},
	year         = 2022,
	journal      = {arXiv preprint arXiv:2207.09453}
}

@article{batzner20223,
	title        = {E (3)-equivariant graph neural networks for data-efficient and accurate interatomic potentials},
	author       = {Batzner, Simon and Musaelian, Albert and Sun, Lixin and Geiger, Mario and Mailoa, Jonathan P and Kornbluth, Mordechai and Molinari, Nicola and Smidt, Tess E and Kozinsky, Boris},
	year         = 2022,
	journal      = {Nature communications},
	publisher    = {Nature Publishing Group UK London},
	volume       = 13,
	number       = 1,
	pages        = 2453
}

@article{liao2022equiformer,
	title        = {Equiformer: Equivariant graph attention transformer for 3d atomistic graphs},
	author       = {Liao, Yi-Lun and Smidt, Tess},
	year         = 2022,
	journal      = {arXiv preprint arXiv:2206.11990}
}

@article{dwivedi2021graph,
	title        = {Graph neural networks with learnable structural and positional representations},
	author       = {Dwivedi, Vijay Prakash and Luu, Anh Tuan and Laurent, Thomas and Bengio, Yoshua and Bresson, Xavier},
	year         = 2021,
	journal      = {arXiv preprint arXiv:2110.07875}
}

@inproceedings{chen2022structure,
	title        = {Structure-aware transformer for graph representation learning},
	author       = {Chen, Dexiong and O’Bray, Leslie and Borgwardt, Karsten},
	year         = 2022,
	booktitle    = {International Conference on Machine Learning},
	pages        = {3469--3489},
	organization = {PMLR}
}

@article{louis2020graph,
	title        = {Graph convolutional neural networks with global attention for improved materials property prediction},
	author       = {Louis, Steph-Yves and Zhao, Yong and Nasiri, Alireza and Wang, Xiran and Song, Yuqi and Liu, Fei and Hu, Jianjun},
	year         = 2020,
	journal      = {Physical Chemistry Chemical Physics},
	publisher    = {Royal Society of Chemistry},
	volume       = 22,
	number       = 32,
	pages        = {18141--18148}
}

@article{yan2022periodic,
	title        = {Periodic graph transformers for crystal material property prediction},
	author       = {Yan, Keqiang and Liu, Yi and Lin, Yuchao and Ji, Shuiwang},
	year         = 2022,
	journal      = {Advances in Neural Information Processing Systems},
	volume       = 35,
	pages        = {15066--15080}
}

@article{chen2019graph,
	title        = {Graph networks as a universal machine learning framework for molecules and crystals},
	author       = {Chen, Chi and Ye, Weike and Zuo, Yunxing and Zheng, Chen and Ong, Shyue Ping},
	year         = 2019,
	journal      = {Chemistry of Materials},
	publisher    = {ACS Publications},
	volume       = 31,
	number       = 9,
	pages        = {3564--3572}
}

@article{wu2021representing,
	title        = {Representing long-range context for graph neural networks with global attention},
	author       = {Wu, Zhanghao and Jain, Paras and Wright, Matthew and Mirhoseini, Azalia and Gonzalez, Joseph E and Stoica, Ion},
	year         = 2021,
	journal      = {Advances in Neural Information Processing Systems},
	volume       = 34,
	pages        = {13266--13279}
}

@article{gong2023examining,
	title        = {Examining graph neural networks for crystal structures: limitations and opportunities for capturing periodicity},
	author       = {Gong, Sheng and Yan, Keqiang and Xie, Tian and Shao-Horn, Yang and Gomez-Bombarelli, Rafael and Ji, Shuiwang and Grossman, Jeffrey C},
	year         = 2023,
	journal      = {Science Advances},
	publisher    = {American Association for the Advancement of Science},
	volume       = 9,
	number       = 45,
	pages        = {eadi3245}
}

@article{Bartok2012OnRC,
	title        = {On representing chemical environments},
	author       = {Albert P. Bart'ok and Risi Kondor and G{\'a}bor Cs{\'a}nyi},
	year         = 2012,
	journal      = {Physical Review B},
	volume       = 87,
	pages        = 184115,
	url          = {https://api.semanticscholar.org/CorpusID:118375156}
}

@article{de2016comparing,
	title        = {Comparing molecules and solids across structural and alchemical space},
	author       = {De, Sandip and Bart{\'o}k, Albert P and Cs{\'a}nyi, G{\'a}bor and Ceriotti, Michele},
	year         = 2016,
	journal      = {Physical Chemistry Chemical Physics},
	publisher    = {Royal Society of Chemistry},
	volume       = 18,
	number       = 20,
	pages        = {13754--13769}
}

@article{behler2007generalized,
	title        = {Generalized neural-network representation of high-dimensional potential-energy surfaces},
	author       = {Behler, J{\"o}rg and Parrinello, Michele},
	year         = 2007,
	journal      = {Physical review letters},
	publisher    = {APS},
	volume       = 98,
	number       = 14,
	pages        = 146401
}

@article{behler2011atom,
	title        = {Atom-centered symmetry functions for constructing high-dimensional neural network potentials},
	author       = {Behler, J{\"o}rg},
	year         = 2011,
	journal      = {The Journal of chemical physics},
	publisher    = {AIP Publishing},
	volume       = 134,
	number       = 7
}

@article{behler2014representing,
	title        = {Representing potential energy surfaces by high-dimensional neural network potentials},
	author       = {Behler, J{\"o}rg},
	year         = 2014,
	journal      = {Journal of Physics: Condensed Matter},
	publisher    = {IOP Publishing},
	volume       = 26,
	number       = 18,
	pages        = 183001
}

@article{Unke2021,
	title        = {Machine Learning Force Fields},
	author       = {Unke,  Oliver T. and Chmiela,  Stefan and Sauceda,  Huziel E. and Gastegger,  Michael and Poltavsky,  Igor and Sch\"{u}tt,  Kristof T. and Tkatchenko,  Alexandre and M\"{u}ller,  Klaus-Robert},
	year         = 2021,
	month        = mar,
	journal      = {Chemical Reviews},
	publisher    = {American Chemical Society (ACS)},
	volume       = 121,
	number       = 16,
	pages        = {10142–10186},
	doi          = {10.1021/acs.chemrev.0c01111},
	issn         = {1520-6890},
	url          = {http://dx.doi.org/10.1021/acs.chemrev.0c01111}
}

@article{rupp2012fast,
	title        = {Fast and accurate modeling of molecular atomization energies with machine learning},
	author       = {Rupp, Matthias and Tkatchenko, Alexandre and M{\"u}ller, Klaus-Robert and Von Lilienfeld, O Anatole},
	year         = 2012,
	journal      = {Physical review letters},
	publisher    = {APS},
	volume       = 108,
	number       = 5,
	pages        = {058301}
}

@article{montavon2013machine,
	title        = {Machine learning of molecular electronic properties in chemical compound space},
	author       = {Montavon, Gr{\'e}goire and Rupp, Matthias and Gobre, Vivekanand and Vazquez-Mayagoitia, Alvaro and Hansen, Katja and Tkatchenko, Alexandre and M{\"u}ller, Klaus-Robert and Von Lilienfeld, O Anatole},
	year         = 2013,
	journal      = {New Journal of Physics},
	publisher    = {IOP Publishing},
	volume       = 15,
	number       = 9,
	pages        = {095003}
}

@article{bartok2017machine,
	title        = {Machine learning unifies the modeling of materials and molecules},
	author       = {Bart{\'o}k, Albert P and De, Sandip and Poelking, Carl and Bernstein, Noam and Kermode, James R and Cs{\'a}nyi, G{\'a}bor and Ceriotti, Michele},
	year         = 2017,
	journal      = {Science advances},
	publisher    = {American Association for the Advancement of Science},
	volume       = 3,
	number       = 12,
	pages        = {e1701816}
}

@article{simine2020predicting,
	title        = {Predicting optical spectra for optoelectronic polymers using coarse-grained models and recurrent neural networks},
	author       = {Simine, Lena and Allen, Thomas C and Rossky, Peter J},
	year         = 2020,
	journal      = {Proceedings of the National Academy of Sciences},
	publisher    = {National Acad Sciences},
	volume       = 117,
	number       = 25,
	pages        = {13945--13948}
}

@article{webb2020targeted,
	title        = {Targeted sequence design within the coarse-grained polymer genome},
	author       = {Webb, Michael A and Jackson, Nicholas E and Gil, Phwey S and de Pablo, Juan J},
	year         = 2020,
	journal      = {Science advances},
	publisher    = {American Association for the Advancement of Science},
	volume       = 6,
	number       = 43,
	pages        = {eabc6216}
}

@article{patel2022featurization,
	title        = {Featurization strategies for polymer sequence or composition design by machine learning},
	author       = {Patel, Roshan A and Borca, Carlos H and Webb, Michael A},
	year         = 2022,
	journal      = {Molecular Systems Design \& Engineering},
	publisher    = {Royal Society of Chemistry},
	volume       = 7,
	number       = 6,
	pages        = {661--676}
}

@article{bhattacharya2022predicting,
	title        = {Predicting aggregate morphology of sequence-defined macromolecules with recurrent neural networks},
	author       = {Bhattacharya, Debjyoti and Kleeblatt, Devon C and Statt, Antonia and Reinhart, Wesley F},
	year         = 2022,
	journal      = {Soft Matter},
	publisher    = {Royal Society of Chemistry},
	volume       = 18,
	number       = 27,
	pages        = {5037--5051}
}

@article{goswami2021deep,
	title        = {Deep learning based approach for prediction of glass transition temperature in polymers},
	author       = {Goswami, Subhasish and Ghosh, Rajdeep and Neog, Arohan and Das, Bitopan},
	year         = 2021,
	journal      = {Materials Today: Proceedings},
	publisher    = {Elsevier},
	volume       = 46,
	pages        = {5838--5843}
}

@article{mikolov2013efficient,
	title        = {Efficient estimation of word representations in vector space},
	author       = {Mikolov, Tomas},
	year         = 2013,
	journal      = {arXiv preprint arXiv:1301.3781},
	volume       = 3781
}

@article{vaswani2017attention,
	title        = {Attention is all you need},
	author       = {Vaswani, A},
	year         = 2017,
	journal      = {Advances in Neural Information Processing Systems}
}

@article{cheng2009combining,
	title        = {Combining instance-based learning and logistic regression for multilabel classification},
	author       = {Cheng, Weiwei and H{\"u}llermeier, Eyke},
	year         = 2009,
	journal      = {Machine Learning},
	publisher    = {Springer},
	volume       = 76,
	pages        = {211--225}
}

@article{Jaeger2018Mol2vecUM,
	title        = {Mol2vec: Unsupervised Machine Learning Approach with Chemical Intuition},
	author       = {Sabrina Jaeger and Simone Fulle and Samo Turk},
	year         = 2018,
	journal      = {Journal of chemical information and modeling},
	volume       = {58 1},
	pages        = {27--35},
	url          = {https://api.semanticscholar.org/CorpusID:34512664}
}

@article{wang2019symbolic,
	title        = {Symbolic regression in materials science},
	author       = {Wang, Yiqun and Wagner, Nicholas and Rondinelli, James M},
	year         = 2019,
	journal      = {MRS Communications},
	publisher    = {Cambridge University Press},
	volume       = 9,
	number       = 3,
	pages        = {793--805}
}

@article{sathish2020analysis,
	title        = {Analysis and modelling of dissimilar materials welding based on K-nearest neighbour predictor},
	author       = {Sathish, T and Rangarajan, S and Muthuram, A and Kumar, R Praveen},
	year         = 2020,
	journal      = {Materials Today: Proceedings},
	publisher    = {Elsevier},
	volume       = 21,
	pages        = {108--112}
}

@article{li2018discrimination,
	title        = {Discrimination of soft tissues using laser-induced breakdown spectroscopy in combination with k nearest neighbors (kNN) and support vector machine (SVM) classifiers},
	author       = {Li, Xiaohui and Yang, Sibo and Fan, Rongwei and Yu, Xin and Chen, Deying},
	year         = 2018,
	journal      = {Optics \& laser technology},
	publisher    = {Elsevier},
	volume       = 102,
	pages        = {233--239}
}

@article{chaabene2020machine,
	title        = {Machine learning prediction of mechanical properties of concrete: Critical review},
	author       = {Chaabene, Wassim Ben and Flah, Majdi and Nehdi, Moncef L},
	year         = 2020,
	journal      = {Construction and Building Materials},
	publisher    = {Elsevier},
	volume       = 260,
	pages        = 119889
}

@article{liu2020machine,
	title        = {A machine learning approach to fracture mechanics problems},
	author       = {Liu, Xing and Athanasiou, Christos E and Padture, Nitin P and Sheldon, Brian W and Gao, Huajian},
	year         = 2020,
	journal      = {Acta Materialia},
	publisher    = {Elsevier},
	volume       = 190,
	pages        = {105--112}
}

@article{pentos2022evaluation,
	title        = {Evaluation of multiple linear regression and machine learning approaches to predict soil compaction and shear stress based on electrical parameters},
	author       = {Pento{\'s}, Katarzyna and Mbah, Jasper Tembeck and Pieczarka, Krzysztof and Niedba{\l}a, Gniewko and Wojciechowski, Tomasz},
	year         = 2022,
	journal      = {Applied Sciences},
	publisher    = {MDPI},
	volume       = 12,
	number       = 17,
	pages        = 8791
}

@article{Honda2019SMILESTP,
	title        = {SMILES Transformer: Pre-trained Molecular Fingerprint for Low Data Drug Discovery},
	author       = {Shion Honda and Shoi Shi and Hiroki R. Ueda},
	year         = 2019,
	journal      = {ArXiv},
	volume       = {abs/1911.04738},
	url          = {https://api.semanticscholar.org/CorpusID:207863425}
}

@article{ying2021transformers,
	title        = {Do transformers really perform badly for graph representation?},
	author       = {Ying, Chengxuan and Cai, Tianle and Luo, Shengjie and Zheng, Shuxin and Ke, Guolin and He, Di and Shen, Yanming and Liu, Tie-Yan},
	year         = 2021,
	journal      = {Advances in neural information processing systems},
	volume       = 34,
	pages        = {28877--28888}
}

@article{irwin2022chemformer,
	title        = {Chemformer: a pre-trained transformer for computational chemistry},
	author       = {Irwin, Ross and Dimitriadis, Spyridon and He, Jiazhen and Bjerrum, Esben Jannik},
	year         = 2022,
	journal      = {Machine Learning: Science and Technology},
	publisher    = {IOP Publishing},
	volume       = 3,
	number       = 1,
	pages        = {015022}
}

@article{magar2022crystal,
	title        = {Crystal twins: self-supervised learning for crystalline material property prediction},
	author       = {Magar, Rishikesh and Wang, Yuyang and Barati Farimani, Amir},
	year         = 2022,
	journal      = {npj Computational Materials},
	publisher    = {Nature Publishing Group UK London},
	volume       = 8,
	number       = 1,
	pages        = 231
}

@article{cao2023moformer,
	title        = {Moformer: self-supervised transformer model for metal--organic framework property prediction},
	author       = {Cao, Zhonglin and Magar, Rishikesh and Wang, Yuyang and Barati Farimani, Amir},
	year         = 2023,
	journal      = {Journal of the American Chemical Society},
	publisher    = {ACS Publications},
	volume       = 145,
	number       = 5,
	pages        = {2958--2967}
}

@article{li2022estimating,
	title        = {Estimating the thermal conductivity of soils using six machine learning algorithms},
	author       = {Li, Kai-Qi and Liu, Yong and Kang, Qing},
	year         = 2022,
	journal      = {International Communications in Heat and Mass Transfer},
	publisher    = {Elsevier},
	volume       = 136,
	pages        = 106139
}

@article{spellings2018machine,
	title        = {Machine learning for crystal identification and discovery},
	author       = {Spellings, Matthew and Glotzer, Sharon C},
	year         = 2018,
	journal      = {AIChE Journal},
	publisher    = {Wiley Online Library},
	volume       = 64,
	number       = 6,
	pages        = {2198--2206}
}

@article{cover1967nearest,
	title        = {Nearest neighbor pattern classification},
	author       = {Cover, Thomas and Hart, Peter},
	year         = 1967,
	journal      = {IEEE transactions on information theory},
	publisher    = {IEEE},
	volume       = 13,
	number       = 1,
	pages        = {21--27}
}

@inproceedings{sharma2020prediction,
	title        = {Prediction of the fracture toughness of silicafilled epoxy composites using K-nearest neighbor (KNN) method},
	author       = {Sharma, Aanchna and Madhushri, Priyanka and Kushvaha, Vinod and Kumar, Anand},
	year         = 2020,
	booktitle    = {2020 international conference on computational performance evaluation (ComPE)},
	pages        = {194--198},
	organization = {IEEE}
}

@article{arunadevi2023comparison,
	title        = {Comparison of k-nearest neighbor \& artificial neural network prediction in the mechanical properties of aluminum alloys},
	author       = {Arunadevi, M and Rani, Monisha and Sibinraj, R and Chandru, MK and Prasad, C Durga},
	year         = 2023,
	journal      = {Materials Today: Proceedings},
	publisher    = {Elsevier}
}

@article{kramer2013k,
	title        = {K-nearest neighbors},
	author       = {Kramer, Oliver and Kramer, Oliver},
	year         = 2013,
	journal      = {Dimensionality reduction with unsupervised nearest neighbors},
	publisher    = {Springer},
	pages        = {13--23}
}

@inproceedings{sambasivan2021everyone,
	title        = {“Everyone wants to do the model work, not the data work”: Data Cascades in High-Stakes AI},
	author       = {Sambasivan, Nithya and Kapania, Shivani and Highfill, Hannah and Akrong, Diana and Paritosh, Praveen and Aroyo, Lora M},
	year         = 2021,
	booktitle    = {proceedings of the 2021 CHI Conference on Human Factors in Computing Systems},
	pages        = {1--15}
}

@misc{aflow,
	title        = {Automatic Flow for Materials Discovery},
	author       = {{AFLOW}},
	year         = 2025,
	url          = {http://aflowlib.org/},
	note         = {[Accessed: January 2025]}
}

@misc{citrine,
	title        = {Citrine Informatics},
	author       = {{Citrine Informatics}},
	year         = 2025,
	url          = {https://citrination.com/},
	note         = {[Accessed: January 2025]}
}

@misc{nomad,
	title        = {NOMAD: FAIR Data Sharing for Materials Science},
	author       = {{NOMAD Repository}},
	year         = 2025,
	url          = {https://nomad-lab.eu/},
	note         = {[Accessed: January 2025]}
}

@article{chithrananda2020chemberta,
	title        = {ChemBERTa: large-scale self-supervised pretraining for molecular property prediction},
	author       = {Chithrananda, Seyone and Grand, Gabriel and Ramsundar, Bharath},
	year         = 2020,
	journal      = {arXiv preprint arXiv:2010.09885}
}

@article{butler2018machine,
	title        = {Machine learning for molecular and materials science},
	author       = {Butler, Keith T and Davies, Daniel W and Cartwright, Hugh and Isayev, Olexandr and Walsh, Aron},
	year         = 2018,
	journal      = {Nature},
	publisher    = {Nature Publishing Group UK London},
	volume       = 559,
	number       = 7715,
	pages        = {547--555}
}

@article{kim2014material,
	title        = {Material pixel-based process planning for layered manufacturing of heterogeneous objects},
	author       = {Kim, Hong Seok and Shin, Ki-Hoon},
	year         = 2014,
	journal      = {International journal of precision engineering and manufacturing},
	publisher    = {Springer},
	volume       = 15,
	pages        = {2421--2427}
}

@article{webel2018new,
	title        = {A new analysis approach based on Haralick texture features for the characterization of microstructure on the example of low-alloy steels},
	author       = {Webel, Johannes and Gola, Jessica and Britz, Dominik and M{\"u}cklich, Frank},
	year         = 2018,
	journal      = {materials Characterization},
	publisher    = {Elsevier},
	volume       = 144,
	pages        = {584--596}
}

@article{velichko2008unambiguous,
	title        = {Unambiguous classification of complex microstructures by their three-dimensional parameters applied to graphite in cast iron},
	author       = {Velichko, A and Holzapfel, C and Siefers, A and Schladitz, K and M{\"u}cklich, F},
	year         = 2008,
	journal      = {Acta Materialia},
	publisher    = {Elsevier},
	volume       = 56,
	number       = 9,
	pages        = {1981--1990}
}

@article{decost2017exploring,
	title        = {Exploring the microstructure manifold: image texture representations applied to ultrahigh carbon steel microstructures},
	author       = {DeCost, Brian L and Francis, Toby and Holm, Elizabeth A},
	year         = 2017,
	journal      = {Acta Materialia},
	publisher    = {Elsevier},
	volume       = 133,
	pages        = {30--40}
}

@article{chen2022application,
	title        = {Application of edge detection based on hexagonal image structure to delamination detection of carbon fiber reinforced polymer material},
	author       = {Chen, YongQiang and Luo, Kai and Chen, Liang and Weng, Haobo and Liang, Wei},
	year         = 2022,
	journal      = {Smart Materials and Structures},
	publisher    = {IOP Publishing},
	volume       = 31,
	number       = 4,
	pages        = {045006}
}

@inproceedings{vincent2009descriptive,
	title        = {A descriptive algorithm for sobel image edge detection},
	author       = {Vincent, O Rebecca and Folorunso, Olusegun and others},
	year         = 2009,
	booktitle    = {Proceedings of informing science \& IT education conference (InSITE)},
	volume       = 40,
	pages        = {97--107}
}

@inproceedings{rong2014improved,
	title        = {An improved CANNY edge detection algorithm},
	author       = {Rong, Weibin and Li, Zhanjing and Zhang, Wei and Sun, Lining},
	year         = 2014,
	booktitle    = {2014 IEEE international conference on mechatronics and automation},
	pages        = {577--582},
	organization = {IEEE}
}

@inproceedings{rani2020edge,
	title        = {Edge detection in scanning electron microscope (SEM) images using various algorithms},
	author       = {Rani, G Elizabeth and Murugeswari, R and Rajini, N},
	year         = 2020,
	booktitle    = {2020 4th International Conference on Intelligent Computing and Control Systems (ICICCS)},
	pages        = {401--405},
	organization = {IEEE}
}

@inproceedings{venkatesh2022analysis,
	title        = {Analysis of Textural Variations in Cerebellum in Brain to Identify Alzheimers by using Haralicks in Comparison with Gray Level Co-occurrence Matrix (GLRLM)},
	author       = {Venkatesh, U and Balachander, Bhuvaneswari},
	year         = 2022,
	booktitle    = {ICIPTM Conference Proceedings},
	volume       = 2,
	pages        = {549--556},
	organization = {IEEE}
}

@article{jivani2022skeletal,
	title        = {Skeletal-based microstructure representation and featurization through descriptors},
	author       = {Jivani, Devyani and Wodo, Olga},
	year         = 2022,
	journal      = {Computational Materials Science},
	publisher    = {Elsevier},
	volume       = 214,
	pages        = 111668
}
\end{document}